\documentclass[11pt,a4paper,twoside,openright]{book}
\usepackage{bbding}
\usepackage[english]{babel}
\usepackage[pdftex]{graphicx}
\usepackage{amssymb,latexsym,amsmath,amsfonts,mathtools}
\usepackage{lscape}
\usepackage{booktabs}
\usepackage{empheq}
\usepackage{stmaryrd}
\usepackage{longtable}
\usepackage{bold-extra} 
\usepackage{anyfontsize} 
\usepackage{bbm}
\usepackage{indentfirst}
\usepackage{fancyhdr}
\usepackage{cite}
\usepackage[title,titletoc]{appendix}
\usepackage{float}
\usepackage{subfigure}
\usepackage{array}
\usepackage{multirow}
\usepackage{dcolumn}
\usepackage{bm}
\usepackage{color}
\usepackage{titlesec} 
\titleformat{\chapter}[display]
  {\bfseries\LARGE\scshape}
  {\filright\MakeUppercase{\chaptertitlename} \thechapter}
  {1ex}
  {\titlerule\vspace{1ex}\filleft}
  [\vspace{1ex}\titlerule]
\titlespacing*{\chapter}{1pt}{3.3cm}{1.5cm}
\titlespacing*{\section}{1pt}{1cm}{.5cm}
\usepackage{hyperref}
\usepackage{changepage}
\usepackage{geometry}
\usepackage{datetime}
\usepackage[T1]{fontenc}
\hypersetup{colorlinks}
\hypersetup{
    citecolor=blue,
    linkcolor=blue,
    urlcolor=blue,
    linktoc={page}
}
\usepackage[capitalize]{cleveref}
\definecolor{myblue}{rgb}{.85, .85, 1}
\newcommand*\mybluebox[1]{\colorbox{myblue}{\hspace{.5em}#1\hspace{.5em}}}
\newenvironment{emphequation}{\empheq[box=\mybluebox]{equation}}{\endempheq}
\newenvironment{emphalign}{\empheq[box=\mybluebox]{align}}{\endempheq}
\newcommand{\unit}[1]{\ {\rm #1}}
\newcommand{\onlinecite}[1]{Ref.~\citen{#1}}

\newcommand{\ket}[1]{\left|{#1}\right>}
\newcommand{\bra}[1]{\left<{#1}\right|}
\newcommand{\sket}[1]{|{#1}\rangle}
\newcommand{\sbra}[1]{\langle{#1}|}

\newcommand{\nodag}{{\phantom{\dagger}}}
\newcommand{\nod}{{\phantom{\dagger}}}
\newcommand{\rket}[1]{\!\!\parallel\!\!{#1}\rangle}
\newcommand{\rbra}[1]{\langle{#1}\!\!\parallel\!\!}
\newcommand{\ClebschGordan}[6]{\langle#1#2;#3#4|#5#6\rangle}
\newcommand{\ThreeJSymbol}[6]{{\scriptstyle\left(\begin{matrix}#1 & #3 & #5 \\#2 & #4 & #6 \\\end{matrix}\right)}}

\graphicspath{{./figures/}}
\hypersetup{
pdftitle ={Theoretical approach to direct resonant inelastic x-ray scattering on magnets and superconductors -- Pasquale Marra}, 
pdfauthor={Pasquale Marra},
pdfsubject={Direct resonant inelastic x-ray scattering on magnets and superconductors},
pdfkeywords={RIXS, cuprates, pnictides, x-ray spectroscopy, resonant inelastic x-ray spectroscopy, superconductivity, orbital physics, magnetism},
pdfcreationdate={\pdfdate}
}
\begin{document}


\newgeometry{centering,bottom=2cm,top=2cm}
\begin{titlepage}
\begin{adjustwidth*}{-1.5cm}{-1.5cm}
\begin{center}
\begin{tabular*}{\textwidth+3cm}{l @{\extracolsep{\fill}} r}
\includegraphics[height=20mm]{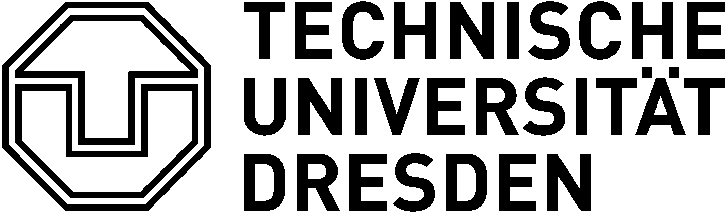}&
\includegraphics[height=20mm]{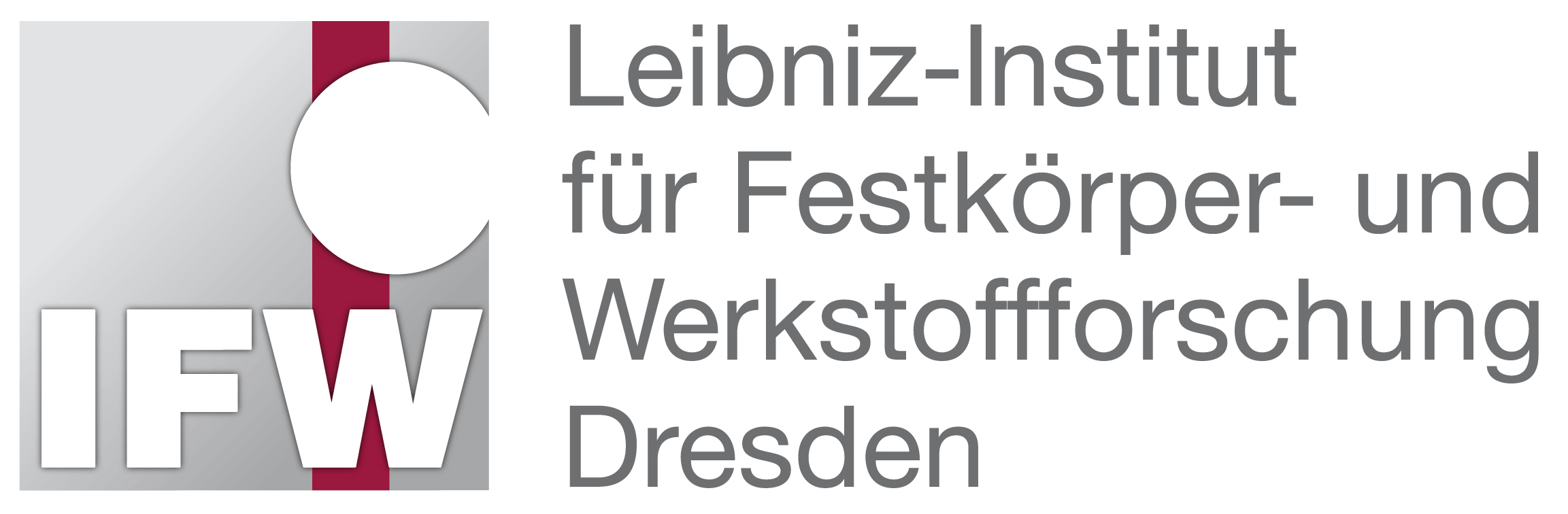}\\[.5cm]
\end{tabular*}
\rule[0.5ex]{\linewidth}{2pt}\vspace*{-\baselineskip}\vspace{3.2pt}
\rule[0.5ex]{\linewidth}{1pt}\\[\baselineskip]
{\bfseries\scshape\fontsize{25.5}{29}\selectfont
Theoretical approach to Direct\\
Resonant Inelastic X-Ray Scattering \\
on Magnets and Superconductors\\
}
\quad\\[3mm]
\rule[0.5ex]{\linewidth}{1pt}\vspace*{-\baselineskip}\vspace{3.2pt}
\rule[0.5ex]{\linewidth}{2pt}\\
\vspace{13mm}
{\bfseries\scshape\fontsize{22.5}{29}\selectfont
Pasquale Marra
}\\
\vspace{30mm}
{
\LARGE
Dissertation\\
zur Erlangung des akademischen Grades\\
Doctor rerum naturalium (Dr. rer. nat.)\\
}
\vspace{40mm}
{\LARGE
Fachrichtung Physik\\
Fakult\"at f\"ur Mathematik und Naturwissenschaften\\
Technische Universit\"at Dresden\\
}
\vspace{10mm}
{\bfseries\scshape\LARGE\selectfont
Februar 2015\\
}
\end{center}
\end{adjustwidth*}
\end{titlepage}

\newpage
{
\thispagestyle{empty}
\vspace*{\fill}
\Large
\quad\\
Erstgutachter: {\bfseries Prof. Dr. Jeroen van den Brink}\\
Zweitgutachter: {\bfseries Prof. Dr. Berndt B\"uchner}\\
\\
Eingereicht am {\bfseries 26. Februar 2015}\\
Disputation am {\bfseries 26. Oktober 2015}\\
\quad\\
This revised version: {\bfseries May 10, 2016}\\
}

{\clearpage\thispagestyle{empty}
\vspace*{\fill}
\centering
\em\Large
La Nature est un temple o\`u de vivants piliers\\
Laissent parfois sortir de confuses paroles\\
\vspace*{\fill}
}
\newpage\thispagestyle{empty}\mbox{}\newpage

\restoregeometry
\newgeometry{bottom=3.5cm}
\pagenumbering{roman}

\frontmatter

\tableofcontents\markboth{}{}
\listoffigures\markboth{}{}
\listoftables\markboth{}{}

\chapter*{List of symbols}
\begin{longtable}{cl}

$\hbar$ & Planck constant\\
$k_B$ & Boltzmann constant\\
&\\[-3mm]

$\delta (E)$ & Dirac delta\\
$\delta_{ij}$ & Kronecker delta\\
$\epsilon_{\lambda\mu\nu}$ & Levi-Civita symbol \\
$\ClebschGordan{l_1}{m_1}{l_2}{l_2}{l_3}{m_3}$ & Clebsch-Gordan coefficients\\[.5mm]
$\ThreeJSymbol{l_1}{m_1}{l_2}{m_2}{l_3}{m_3}$ & Wigner $3j$ symbol\\
$\lambda,\mu,\nu$ & Cartesian components of a vector $x,y,z$\\
$q=0,\pm1$ & spherical components of a vector\\
$\sigma$, $\tau$ & electron spin\\
$\sigma^x$, $\sigma^y$, $\sigma^z$ & Pauli matrices\\
&\\[-3mm]

$Z$ & atomic number\\
$n$, $l$, $m$ & principal, azimuthal, and magnetic quantum number\\
$\Delta l$, $\Delta m$ & variation of the azimuthal and magnetic quantum numbers\\
$s$, $s_z$ & eigenvalues of the spin operator\\
$j$, $j_z$ & eigenvalues of the total angular momentum operator\\
$\hat{\bf L}$ & orbital angular momentum operator\\
$\hat{\bf S}$ & spin operator\\
$\hat{\bf J}$ & total angular momentum operator\\
$Y_l^m(\theta,\phi)$ & spherical harmonics\\
$Y_{lm}(\theta,\phi)$ & tesseral spherical harmonics\\
$R_{nl}(r)$ & radial wave functions\\
$e_g$, $t_{2g}$ & orthorombic (tetrahedral) crystal field levels\\
&\\[-3mm]

$T_c$ & critical temperature\\
$\cal Z$ & partition function\\
$t$ & hopping parameter\\
$U$ & electron-electron (Hubbard) repulsion\\
$\Delta_{pd}$ & energy required to hop an electron from the ligand\\& to the metal ion\\
$J$, $J_{ij}$ & superexchange constant\\
$\hat{\cal H}_{so}$ & spin-orbit coupling Hamiltonian\\
$\lambda_{so}$ & spin-orbit coupling parameter\\
$\Delta E_{so}$ & spin-orpit coupling splitting\\
$N$ & number of lattice sites\\
${\bf R}_i$ & position of the ion site $i$\\
$\hat{\bf r}$, $\hat{\bf p}$ & position and momentum operators\\
$\hat{\bf r}_i$, $\hat{\bf p}_i$ & position and momentum operators of an electron\\& close to the ion site $i$\\
&\\[-3mm]

${\bf k}$ / ${\bf k}'$& momentum of the incident/scattered photon\\
${\bf q}$ & transferred momentum $\bf k-k'$\\
$\hbar\omega_{\bf k}$ / $\hbar\omega_{\bf k'}$ & energy of the incident/scattered photon\\
$\hbar\omega$ & transferred energy (energy loss) $\hbar\omega_{\bf k}-\hbar\omega_{\bf k'}$\\
${\bf e}$ / ${\bf e}'$& polarization of the incident/scattered photon\\
$I=\dfrac{d^2 \sigma}{d\omega d\Omega}$ & double differential cross section\\[2mm]
$\ket{g}$ & ground state\\
$\ket{i}$ & initial state (if not the ground state)\\
$\ket{n}$ & intermediate state of RIXS process\\
$\ket{f}$ & final state (excited state)\\
$E_g$ & ground state energy\\
$E_i$ & initial state energy (if not the ground state)\\
$E_n$ & intermediate state energy\\
$E_f$ & final state energy\\
$K$, $L$, $M$ & absorption edges\\
$\Gamma$ & core hole broadening\\
$\Delta t$ & core hole lifetime\\
$\Delta E$ & energy fluctuation\\
$\hat{\bf A}$ & electromagnetic vector potential\\
$\hat{\cal D}$, $\hat{\cal D}'$ & optical transition operator\\
$\hat{D}_i$, $\hat{D}$ & dipole operator \\
$\hat{\cal O}$ & RIXS operator\\
$\hat{O}_i$ & local RIXS operator\\
$\hat{\cal H}_0$ & unperturbed Hamiltonian\\
$\hat{G}$ & intermediate state propagator\\
$\hat{G}_0$ & intermediate state propagator (unperturbed)\\
$\hat{\cal H}_c$ & core hole Hamiltonian $\hat{\cal H}_c$\\
$U_c$ & core hole potential\\
$\alpha(\omega_{\bf k})$, $\beta(\omega_{\bf k})$ & resonance functions\\

\end{longtable}

\chapter*{Common abbreviations}
\begin{longtable}{cl}

BCS  &  Bardeen, Cooper, and Schrieffer theory  \\
DSF  & dynamical structure factor \\

&\\

AF &  antiferromagnetic, antiferromagnet \\
FM &  ferromagnetic, ferromagnet \\

A-AF & ferromagnetic planes with antiferromagnetic coupling along \\
& the perpendicular direction \\

&\\

AO & alternating orbital \\
FO & ferroorbital  \\
C-AO & alternating orbitals within planes with same orbitals stacked \\
& along the perpendicular direction\\
G-AO & isotropic three dimensional alternating orbitals\\
OL & orbital liquid  \\
&\\

RIXS  & resonant inelastic x-ray scattering  \\
XAS  & x-ray absorption spectroscopy   \\

STM  & scanning tunnelling microscopy \\

ARPES  & angle-resolved photoemission spectroscopy   \\
&\\

\end{longtable}

\chapter{Abstract}\markboth{Abstract}{}
The capability to probe the dispersion of elementary spin, charge, orbital, and lattice excitations has positioned resonant inelastic x-ray scattering (RIXS) at the forefront of photon science. 
In this work, we will investigate how RIXS can contribute to a deeper understanding of the orbital properties and of the pairing mechanism in unconventional high-temperature superconductors. 
In particular, we will show how direct RIXS spectra of magnetic excitations can reveal long-range orbital correlations in transition metal compounds, by discriminating different kind of orbital order in magnetic and antiferromagnetic systems.
Moreover, we will show how RIXS spectra of quasiparticle excitations in superconductors can measure the superconducting gap magnitude, and reveal the presence of nodal points and phase differences of the superconducting order parameter on the Fermi surface. 
This can reveal the properties of the underlying pairing mechanism in unconventional superconductors, in particular cuprates and iron pnictides, discriminating between different superconducting order parameter symmetries, such as $s$, $d$ (singlet pairing) and $p$~wave (triplet pairing). 

\mainmatter
\renewcommand{\sectionmark}[1]{\markright{\thesection.\ #1}}%
\renewcommand{\chaptermark}[1]{\markboth{\thechapter.\ #1}{}}%

\chapter{Introduction}
{
Superconductivity, a state of matter characterized by zero electrical resistance and by the expulsion of the magnetic field below a certain critical temperature $T_c$, was discovered by Onnes~\cite{Onnes1911} in 1911.
After many decades, a satisfactory theoretical understanding of the phenomenon was established in the 50s, with the work of Ginzburg and Landau~\cite{Ginzburg1950}, and the one of Bardeen, Cooper, and Schrieffer~\cite{Bardeen1957a,Bardeen1957b}. 
In particular in the latter work, the authors presented the first microscopic theory of superconductivity, known as BCS, which was able to fully explain the phenomenon in \emph{conventional} superconductors. 
The BCS theory describes the superconducting state as a condensation of Cooper pairs, that is, bound states of two electrons, driven by a weak attractive interaction. 
In particular, the electron condensate is described via a mean field order parameter $\Delta$, which is proportional to the density of the Cooper pairs. 
In conventional superconductors (e.g., mercury, niobium, lead), the attractive interaction between electrons is due to the electron-phonon coupling, and the critical temperatures are in general low ($<10 \unit{K}$). 
However, Bednorz and M\"uller~\cite{Bednorz1986} in 1986 found that a certain cuprate compound was superconducting below an unexpectedly high critical temperature of $30\unit{K}$. 
After this breakthrough, many other cuprate compounds were discovered to be superconducting, with critical temperatures as high as 138 K, well above the boiling point of liquid nitrogen. 
The exceptional high $T_c$ of these superconductors cannot be explained in terms of the electron-phonon coupling. 
Cuprate superconductors were the first class of unconventional and high $T_c$ superconductors, and were considered for many decades as exceptional~\cite{Tsuei2000,Lee2006}.
Nevertheless, a new class of unconventional superconductors was recognized recently, with the discovery of superconductivity in an iron based pnictide at the critical temperature of 26~K by Kamihara~\cite{Kamihara2008}. 
The pnictide superconductors are characterized by a moderately high critical temperature (up to 56 K), which is believed to require a mechanism beyond the electron-phonon coupling~\cite{Paglione2010,Stewart2011}. 
As the microscopical mechanism of the electron pairing in unconventional superconductors is yet to be fully understood, superconductivity remains at the frontier of condensed matter physics, even after hundred years since its discovery. 

Despite a somewhat similar crystal structure, with superconducting planes stacked and interleaved by isolating planes, iron pnictide and cuprate superconductors are very different systems. 
In fact, in the undoped nonsuperconducting phase, cuprates are charge-transfer insulators with antiferromagnetic order, with strong Coulomb interaction and localized electron states. 
In the normal state, pnictides are instead antiferromagnetic metals. 
Therefore, whereas cuprates are a typical example of strongly correlated systems, in pnictides correlations appear to play a less important role. 
Nevertheless, these two classes of unconventional superconductors share some similarities, in particular: 
\begin{itemize}
\item
The presence of the spin and of the orbital degrees of freedom of the $3d$ electron states in the valence band. 
Since the seminal work of Kugel and Khomskii~\cite{Kugel1982} in the 1980s, it has been known that orbital degrees of freedom can play a crucial role in strongly correlated transition metal compounds. 
In fact, even if the orbital degree of freedom do not seem to play a role in superconducting cuprates, many other cuprate compounds show a non trivial interplay between orbital and magnetic order~\cite{Imada1998}. 
On the other hand, pnictide superconductors are characterized by a variation of the orbital content along the Fermi surface, which have been proposed to be of direct relevance to superconductivity in these compounds~\cite{Millis1995,Moreo2009,Shimojima2011}. 
\item
The presence of an unconventional superconducting order parameter characterized by a strong momentum dependence along the Fermi surface. 
Since the electron-phonon coupling is widely believed to be inadequate to explain the high critical temperature of unconventional superconductors, it is natural to expect that electronic correlations play a role in the pairing~\cite{Leggett2006}. 
In this case, the superconducting order parameter, which describes the momentum-dependent coupling strength between bounded electrons, shows a variation along the Fermi surface~\cite{Kirtley1995,Tsuei1997,Damascelli2003} which reflects the presence of strong correlations being present at rather short distances. 
Moreover, while the order parameter of conventional superconductors has the same phase throughout the momentum space ($s$-wave symmetry), that of unconventional superconductors exhibits a sign reversal between momenta on the Fermi surface connected by the characteristic wave vector ${\bf Q}_{\rm AF}$ of spin fluctuations~\cite{Kuroki2001}. 
As a consequence of that, in cuprates the superconducting order parameter vanishes at nodal points on the Fermi surface ($d$-wave). 
A sign reversal can also occur without the presence of nodal points, as is the case of iron pnictides~\cite{Mazin2008,Kuroki2008} ($s_\pm$ wave), where the Fermi surface consists of disconnected hole and electron pockets~\cite{Hirschfeld2011,Chubukov2012,Hosono2015}.
\end{itemize}

In the light of this, the main focus of this thesis will be on how to probe the orbital degree of freedom and the superconducting order parameter in cuprates and iron pnictides by resonant inelastic x-ray scattering (RIXS). 
RIXS is a photon-in photon-out core-hole spectroscopy which allow one to measure the energy and the dispersion of low energy elementary excitations in condensed matter systems. 
More precisely, the RIXS cross section, measured as a function of the energy loss and of the transferred momentum of the scattered radiation, is a direct probe of the excitation spectra and, as a consequence, of the symmetry properties of the electronic system under study. 
The aim of this work is to investigate how RIXS can contribute to a deeper understanding of the orbital properties and of the pairing mechanism in unconventional high $T_c$ superconductors. 
In order to do this, we will formulate some theoretical predictions on the general properties and on the symmetry of RIXS spectra of magnetic excitations in orbital systems, and of quasiparticle excitations in the superconducting phase. 
Although the main focus will be on cuprate and iron pnictide superconductors, nonetheless the theoretical description of the symmetry properties of RIXS spectra will be grounded on general models which are relevant to other unconventional superconductors and strongly correlated systems. 

To introduce these concepts in more detail, in the next Sections of this Introduction we will give a short overview on the orbital degrees of freedom (\cref{sec:Intro-Orbital,sec:Intro-Order}) and on the relationship between the pairing mechanism and the superconducting order parameter (\cref{sec:Intro-Superconducting}) in unconventional superconductors. 
Finally, we will highlight the importance of RIXS in the study of elementary excitations in solid state systems (\cref{sec:Intro-RIXS}), and we will set the framework of this work in the last Section of this Introduction (\cref{sec:Intro-Overview}).

\section{The orbital degree of freedom}
\label{sec:Intro-Orbital}

The orbital degree of freedom is of paramount importance to understand the properties of a number of transition metal compounds and superconductors.
In a crystal, single ion states of localized electrons are determined by the so-called crystal field, which is induced by the electrostatic interaction with the surrounding ions. 
In particular, transition metal ions tend to form coordination complexes, which consist of a periodic arrangement of metal ions surrounded by a set of binding anions, generally known as ligands. 
The transition metal $3d$ electrons strongly interact with the electron states of the ligand ions, and the energy levels of the valence electrons strongly depend on the number of ligands, known as coordination number, and on the geometry of the coordination complex. 
Depending on the characteristic symmetry of the coordination complex and on the crystal field strength, the degeneracy of the electron levels within the same shell (same principal and azimuthal quantum numbers) is partially or totally removed. 
The different energy levels which correspond to different orbital angular momentum states are known as orbitals.
Different orbital states have different symmetry properties and their energy are determined by the crystal field properties. 
Therefore, localized electrons in a crystal can be described in terms of the spin, of the principal and azimuthal quantum number, and of the crystal field energy level, i.e., the orbital, they occupy. 

\begin{figure}[t!]
\centering
\includegraphics[width=1\textwidth]{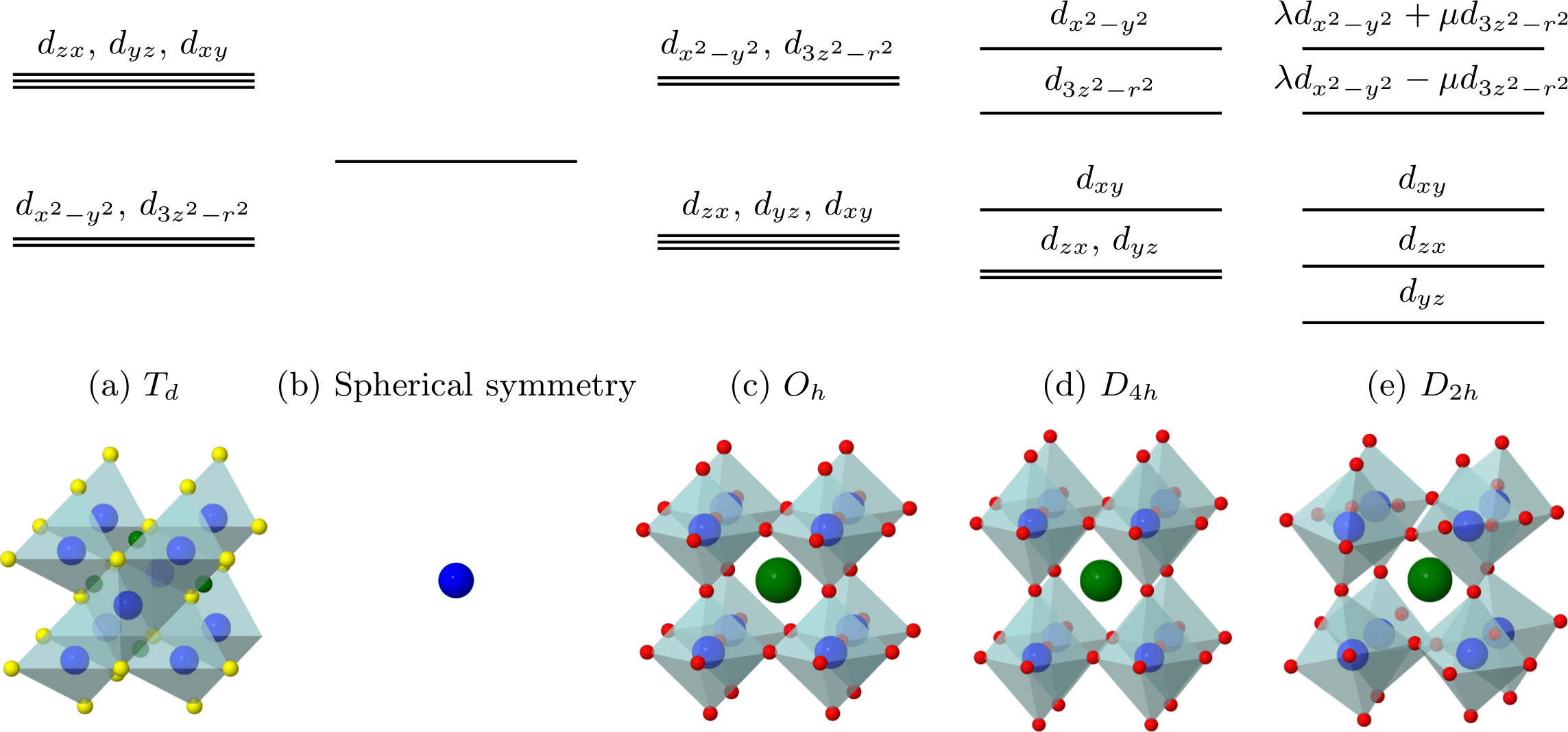}
\caption[Energy levels of $3d$ electrons of transition metal ions in different crystal environments]{
Energy levels of $3d$ electrons of transition metal ions in different crystal environments. 
From left to right: 
(a) crystal structure of a tetrahedral transition metal complex (tetrahedral symmetry $T_d$), i.e., the iron based superconductor LiFeAs, with iron ions (blue) surrounded by six ligand arsenic ions (yellow) at the vertex of a tetrahedron, 
(b) isolated transition metal ion (full spherical symmetry),
(c) cubic perovskite structure of a octahedral transition metal complex (octahedral symmetry $O_h$), i.e., a cuprate compound with copper ions (blue) surrounded by eight oxygen ions (red) at the vertex of a perfect octahedron, 
(d) deformed perovskite with elongated octahedra (square bipyramidal symmetry $D_{4h}$), 
and (e) deformed perovskite with tilted and deformed octahedra (bipyramidal symmetry $D_{2h}$).
}
\label{fig:PerovskitePnictide}
\end{figure}

In cuprates, copper ions form a coordination complex with eight surrounding ligand oxygen ions at the vertices of an octahedron, generally embedded in a perovskite structure. 
In cubic perovskites, the ligand oxygens form a perfect octahedron, and hence the coordination complex has a full octahedral symmetry ($O_h$), while in tetragonal and orthorhombic perovskite the octahedron is deformed and the symmetry is lowered to square bipyramidal ($D_{4h}$) and bipyramidal ($D_{2h}$) respectively. 
The different perovskite structures are shown in \cref{fig:PerovskitePnictide}, along with the symmetry group of the coordination complex and with the crystal field levels of valence $3d$ electrons. 
As one can see, the lower the symmetry, the lower is the degeneracy of the energy levels. 
In fact, a full octahedral symmetry splits the energy levels of $d$ electrons into two degenerate levels, the $t_{2g}$ orbitals, which include any combination of the orbitals $d_{xy}$, $d_{yz}$, and $d_{zx}$, and the $e_g$ orbitals, which include the orbitals $d_{x^2-y^2}$ and $d_{3z^2-r^2}$. 
In this case, $t_{2g}$ orbitals correspond to the lowest energy electron states. 
This follows from some simple physical considerations, i.e., by considering the shape of $t_{2g}$ and $e_g$ orbitals and of the charge distribution of the surrounding ions. 
In fact, the lobes of $e_g$ orbitals are closer to the surrounding ions than the ones of the $t_{2g}$ orbitals, and as a consequence the electrostatic repulsion and the energy of such orbitals are higher in the $e_g$ case. 
If the symmetry is lowered, i.e., if the octahedron formed by the ligand oxygen ions is deformed into a square bypiramid or to a rectangular bypiramid, the $e_g$ and $t_{2g}$ orbital states splits in energy and the degeneracy is further removed. 

In iron pnictides instead, iron ions form a coordination complex with six surrounding pnictogen ions (e.g, arsenic) at the vertex of a tetrahedron. 
In the iron based superconductor LiFeAs, for example, the iron ion is surrounded by six arsenic ions at the vertex of a perfect tetrahedron, and the coordination complex have a full tetrahedral symmetry ($T_d$). 
In \cref{fig:PerovskitePnictide} is shown the crystal structure LiFeAs, along with the crystal field levels of the valence $3d$ electrons. 
In this case, the single ion electron energies are splitted into two degenerate levels, the lowest energy $e$ orbitals ($d_{x^2-y^2}$ and $d_{3z^2-r^2}$) and the $t_2$ orbitals ($d_{xy}$, $d_{yz}$, and $d_{zx}$). 

\section{Orbital order in correlated systems}
\label{sec:Intro-Order}

In many correlated systems, the crystal field do not remove completely the degeneracy of the lowest energy states, or the energy difference between different orbital states is very small. 
As a consequence, a superexchange mechanism between adjacent sites can emerge and induce an orbital ordered ground state, which is a periodic repetition of orbital states in the crystal lattice, and is the analogue of spin ordered states, such as ferromagnets or antiferromagnets. 
These systems can show in principle a new form of elementary excitation, called orbitons, which corresponds to the collective modes of the local orbital degrees of freedom. 
The superexchange mechanism between nearest neighbor transition metal ions is usually described in terms of the Kugel-Khomskii spin-orbital model~\cite{Kugel1982}, i.e., by an effective spin-orbital Hamiltonian, which represent the second order perturbation expansion  of the related multi-band Hubbard model with respect to the Hubbard on-site Coulomb repulsion. 
An idealized high-symmetry version of the Kugel-Khomskii Hamiltonian for transition metal $3d$ electrons reads~\cite{vandenBrink1998}
\begin{emphequation}\label{eq:KugelKhomskiiHamiltonian}
	\hat{\cal H}=\sum_{\langle i,j \rangle} 
	J_1\ \hat{\bf S}_i \cdot \hat{\bf S}_j
	+J_2\ \hat{\bf T}_i \cdot \hat{\bf T}_j
	+4J_3\ \hat{\bf S}_i \cdot \hat{\bf S}_j\ \hat{\bf T}_i \cdot \hat{\bf T}_j
	,
\end{emphequation}
where $\hat{\bf S}_i$ and $\hat{\bf T}_i$ are the spin and pseudospin operators acting on the nearest neighbor sites $\langle i,j\rangle$, with the latter defined as
\begin{align}\label{eq:Pseudospin}
	\hat{\bf T}_i^+&=\sum\limits_{\sigma} a^\dag_{i\sigma} b^\nodag_{i\sigma}, \quad
	\hat{\bf T}_i^-=\sum\limits_\sigma b^\dag_{i\sigma} a^\nodag_{i\sigma}, \nonumber\\
	\hat{\bf T}_i^z&=\frac12\sum\limits_\sigma 
	\left(a^\dag_{i\sigma} a^\nodag_{i\sigma}-b^\dag_{i\sigma} b^\nodag_{i\sigma}\right)
	,
\end{align}
where the operators $a^\dag_{i\sigma}$ and $b^\dag_{i\sigma}$ ($a^\nodag_{i\sigma}$ and $b^\nodag_{i\sigma}$) create (annihilate) an electron state respectively in the orbital $a$ and $b$ at the lattice site $i$ with spin $\sigma$. 

The effective Hamiltonian in \cref{eq:KugelKhomskiiHamiltonian} can be viewed as a generalization of the Heisenberg Hamiltonian where the exchange parameters are no longer constants but are expressed in terms of pseudospin operators which represent the orbital character of valence electrons. 
This gives rise to a rather complex, and interesting, interplay between orbital and spin degrees of freedom. 
In fact, the spin-orbital model in \cref{eq:KugelKhomskiiHamiltonian} allows both spin and orbital ordering and, even more interestingly, can exhibit, besides the ordinary spin waves, also new types of elementary excitations.
These include orbital waves (orbitons) and, in principle, spin-orbital excitations in which the spin and orbital degrees of freedom are coupled. 
In this model, the orbital and spin dynamics are a direct consequence of the superexchange mechanism and of the orbital degeneracy. 
However, in the presence of large Jahn-Teller interactions, the orbital interactions prevail, and therefore one can assume that the system realizes an orbital ordered state which is largely independent from spin fluctuations. 
In this case, the orbital degree of freedom can be integrated out, e.g., using the mean field approximation $\hat{\bf T}_i \cdot \hat{\bf T}_j\approx\langle \hat{\bf T}_i \cdot \hat{\bf T}_j \rangle$. 
Hence, the ground state and the low energy excitations of the system can be described by a mean field spin-only Hamiltonian
\begin{emphequation}\label{eq:GeneralHeisenbergHamiltonian}
	\hat{\cal H}=\sum_{\langle i,j \rangle} 
	J_{ij}\ \hat{\bf S}_i \cdot \hat{\bf S}_j+K_{ij},
\end{emphequation}
where the effective spin exchange interaction constants are $J_{ij}=J_1+4 J_3\langle \hat{\bf T}_i \cdot \hat{\bf T}_j \rangle$ and $K_{ij}=J_2\langle \hat{\bf T}_i \cdot \hat{\bf T}_j \rangle$. 
Hence, the orbital interactions will determine not only the nature of the orbital order but also the nature of the spin ground state and excitations. 
Some examples of orbital ordered states include the ferroorbital order (same orbital on each site) and the alternating orbital order (alternating orbitals every other site). 

\begin{figure}
	\centering
	\subfigure[$C$-AO/$A$-AF]{\includegraphics[scale=.5]{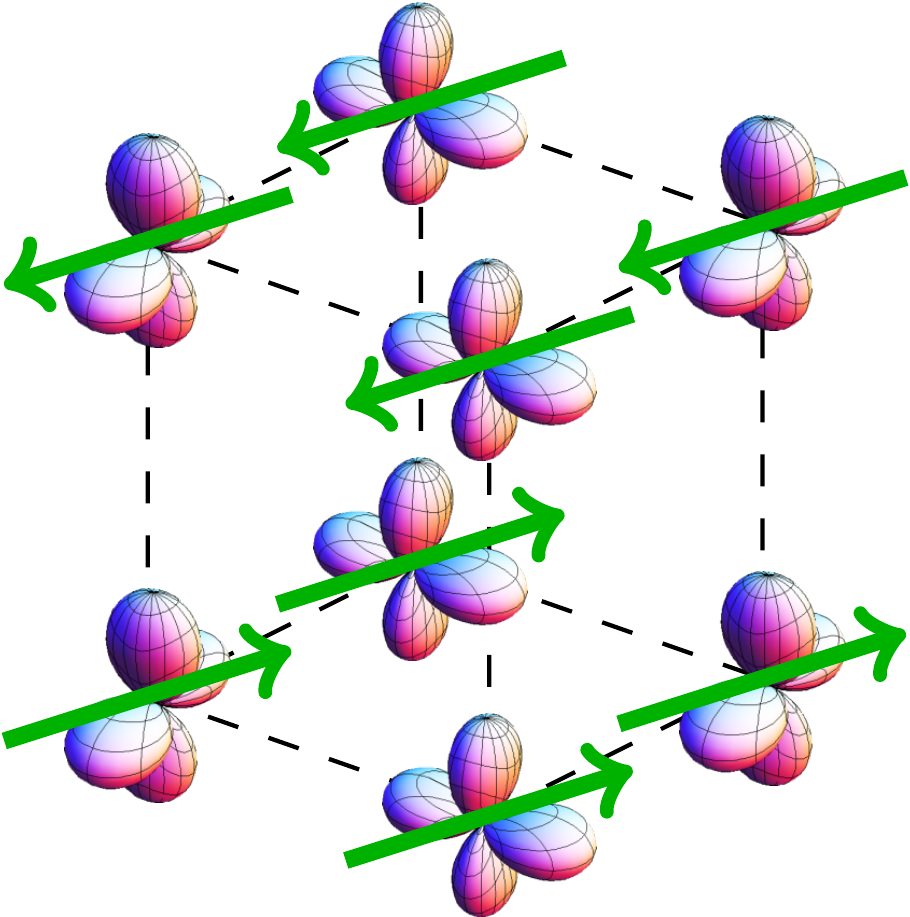}}\qquad\qquad
	\subfigure[$G$-AO/$A$-AF]{\includegraphics[scale=.5]{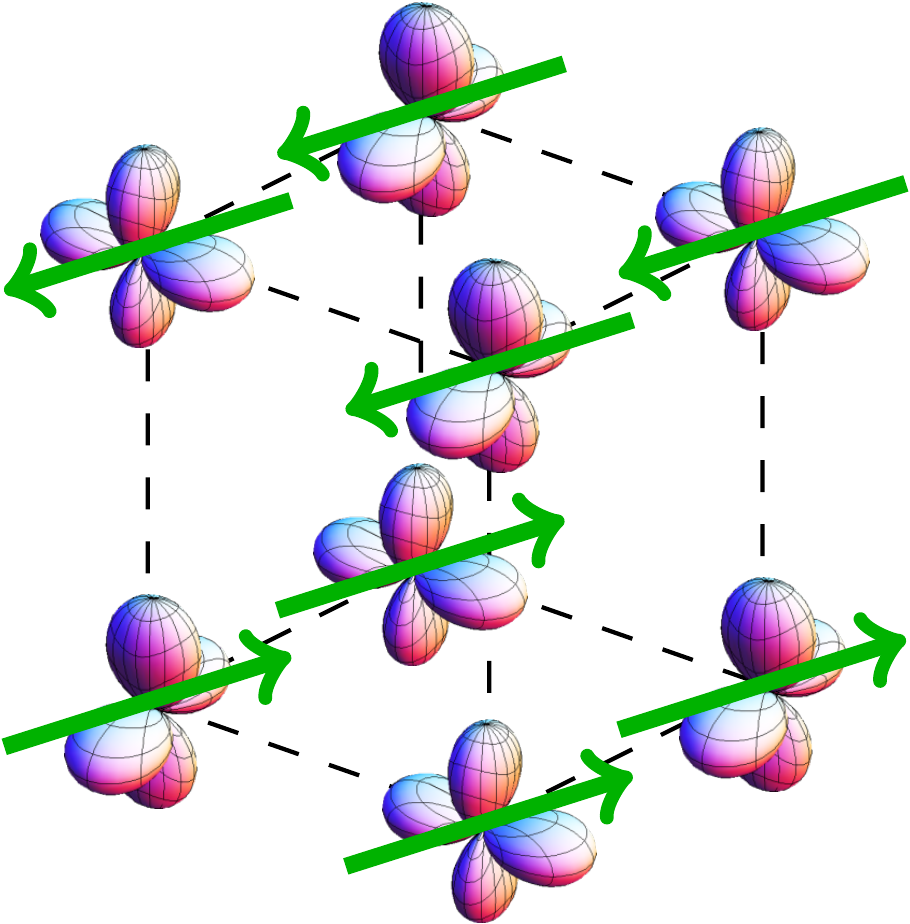}}
	\caption[Antiferromagnetic state with orbital order]{
The magnetic $A$-AF state with orbital $C$-AO and $G$-AO orders, as realized respectively in the (d)-type and (a)-type KCuF$_3$ polytypes below the N\'eel temperature $T_N\approx38\unit{K}$. 
}
	\label{fig:KCuF}
\end{figure}

A paradigmatic example of the coexistence of magnetic and orbital order is the cuprate compound KCuF$_3$~\cite{Towler1995,Feiner1997,Oles2005}. 
Below the N\'eel temperature $T_N\approx38 \unit{K}$, the system exhibits a long-range magnetic order, the so called $A$-AF state, with ferromagnetic planes $xy$ and antiferromagnetic coupling along the $z$ direction~\cite{Satija1980,Lake2000}. 
Moreover, the two different polytype structures realized below the critical temperature $T_S\approx 800\unit{K}$ show two different kind of long-range orbital order, respectively the $C$-AO order in the (d)-type polytype, i.e., alternating orbitals within the $xy$ planes with the same orbitals stacked along the $z$ direction, and the $G$-AO order in the (a)-type polytype, i.e., isotropically alternating orbitals in the three dimensions~\cite{Satija1980}. 
The magnetic and orbital orders realized in the cuprate compound KCuF$_3$ below the critical temperature $T_N\approx38 \unit{K}$ are shown in \cref{fig:KCuF}. 

More generally, the comprehension of orbital physics is of fundamental importance, either in the strongly correlated cuprates, which show in many cases a tendency towards orbital ordering, or in iron pnictides, where the type of orbital ordering or its lack is heavily debated. 
Besides, orbital ordering and orbital correlations have been proposed to be of direct relevance to spectacular phenomena such as colossal magnetoresistance in manganites and, in general, the orbital degree of freedom is believed to play a key role in other strongly correlated systems like, for instance, titanium and vanadium  oxides, where different theoretical scenarios have been proposed --- a rather exotic orbital liquid phase~\cite{Khaliullin2000,Khaliullin2001} or a classical alternating orbital-ordered state~\cite{Pavarini2004,Raychaudhury2007}. 
Yet, the precise nature of correlated orbital states and their existence is intensely debated, which to large extent is due to the fact that orbital correlations turn out to be very difficult to access experimentally. 
For instance, neutron scattering is almost not sensitive to the orbital symmetries of the ground state, in particular in orbital systems where the angular momentum is quenched by the crystal field~\cite{Kim2011}. 
Traditional x-ray diffraction instead is dominated by the scattering from the atomic core electrons, whereas resonant x-ray diffraction~\cite{Elfimov1999,Benedetti2001}, in particular in the soft x-ray regime, suffers from a very limited scattering phase space, making Bragg scattering only possible for special orbital superstructures that have large spatial periodicities~\cite{Wilkins2003}. 
Therefore, a direct experimental access to orbital order or orbital excitations would be of great help in unravelling the puzzling properties of many orbital systems.

\section{Unconventional superconductivity}
\label{sec:Intro-Superconducting}

In conventional superconductors, the superconducting pairing is mediated by the electron-phonon coupling, which is largely independent from the electron momenta. 
As a consequence, the superconducting order parameter is uniform in momentum space and the energy gap is isotropic. 
This case is referred to as $s$-wave superconductivity.
In unconventional superconductors, the pairing mechanism goes beyond the electron-phonon coupling, and the electron-electron attractive potential is in general dependent on the electron momenta. 
For this reason, the superconducting order parameter $\Delta_{\bf k}$ is in general not uniform in momentum space, and the energy gap is not isotropic, with the possible presence of nodal points on the Fermi surface, i.e., points where the order parameter is zero. 
This is the case, for example, in cuprate superconductors, where the strong electron correlations seem to be responsible of the pairing mechanism. 
In these systems, the Fermi surface shows a magnetic instability corresponding to the nesting vector ${\bf Q}_{\rm AF}=(\pi,\pi)$, i.e., the ordering vector of the antiferromagnetic order in the normal phase. 
In the superconducting phase, the nesting vector connects momenta with an opposite phase of the superconducting order parameter, which has nodes on the Fermi surface. 
The characteristic order parameter dependence in the momentum space, referred to as $d$-wave, is shown in \cref{fig:OP-d}. 

In pnictides, the Fermi surface shows two distinct branches, respectively with an electron-like and a hole-like dispersion, connected by a nesting vector which correspond to a magnetic instability towards spin density wave fluctuations. 
Also in this case, the superconducting order parameter show a sign reversal between points of the Fermi surface connected by the antiferromagnetic ordering vector ${\bf Q}_{\rm AF}$. 
However, the nesting vector connects momenta on different branches of the Fermi surface, and there are no nodal points. 
This case is referred to as $s_\pm$~wave superconductivity, and the typical dependence of the order parameter in the momentum space is shown in \cref{fig:OP-spm}. 
However, in pnictides the nesting of the Fermi surface seems to be not a mandatory condition for the realization of the superconducting state. 
For instance, in the LiFeAs superconductor, angle-resolved photoemission spectroscopy (ARPES) measurements have shown~\cite{Borisenko2010} how the nesting of the electron and hole pockets is marginal, due to the different relative sizes of the two branches of the Fermi surface. 
In this case, the superconducting pairing mechanism is not clear, and even the momentum dependence of the order parameter is debated. 
In fact, while there is a general agreement about the realization of a spin-singlet $s_\pm$~wave superconductivity~\cite{Mazin2008} in other iron based superconductors, where nesting dominates the low energy properties, the nature of the superconducting state in LiFeAs is debated. 
Different scenarios have been proposed in place of the $s_\pm$~wave pairing, e.g., an $s_{++}$~wave superconductivity, driven by the critical $d$-orbital fluctuations induced by moderate electron-phonon interactions~\cite{Kontani2010}, or even a $p$-wave spin-triplet pairing driven by ferromagnetic fluctuations~\cite{Brydon2011}. 
The experimental momentum dependence of the superconducting gap measured by ARPES~\cite{Borisenko2010,Borisenko2012} is consistent, in principle, with spin-singlet superconductivity, both with or without sign reversal ($s_\pm$ or $s_{++}$~wave), as well as with spin-triplet pairing. 
In particular, in \cref{fig:OP-spm,fig:OP-spp} are shown two different scenarios which seems to be consinstent with the measured superconducting gap, namely the $s_\pm$ wave, shown in \cref{fig:OP-spm}, with a sign reversal of the order parameter between the electron and hole pockets, and the $s_{++}$ wave, shown in \cref{fig:OP-spp} which is not uniform along the Fermi surface but nevertheless preserves the sign of the order parameter between the two different branches of the Fermi surface. 
On the other hand, while the singlet pairing is supported by some neutron scattering experiments~\cite{Taylor2011}, the unusual shape of the Fermi surface and the momentum dependency of the superconducting gap measured by ARPES~\cite{Borisenko2012} is in conflict with the $s_\pm$~wave symmetry. 
Moreover, scanning tunnelling microscopy (STM) of the quasiparticle interference~\cite{Hanke2012} are consistent either with a $p$~wave spin-triplet state or with a singlet pairing mechanism with a more complex order parameter ($s+\imath d$~wave). 
Whereas ARPES has been proven to be powerful in measuring the momentum dependence of the superconducting gap on the Fermi surface, it should be noted here that ARPES, since not sensitive to the order parameter phase, can neither distinguish between these two scenarios, nor between singlet and triplet pairing, i.e., between even ($\Delta_{\bf k}=\Delta_{-\bf k}$) and odd ($\Delta_{\bf k}=-\Delta_{-\bf k}$) symmetry of the order parameter. 
This lack of experimental and theoretical agreement on the pairing mechanism in the LiFeAs superconductor shows the importance of probing the order parameter momentum dependence, and in particular to determine the presence of a sign reversal of the order parameter phase. 

\begin{figure}
	\centering
	\subfigure[$d$-wave]{\label{fig:OP-d}\includegraphics[scale=.5]{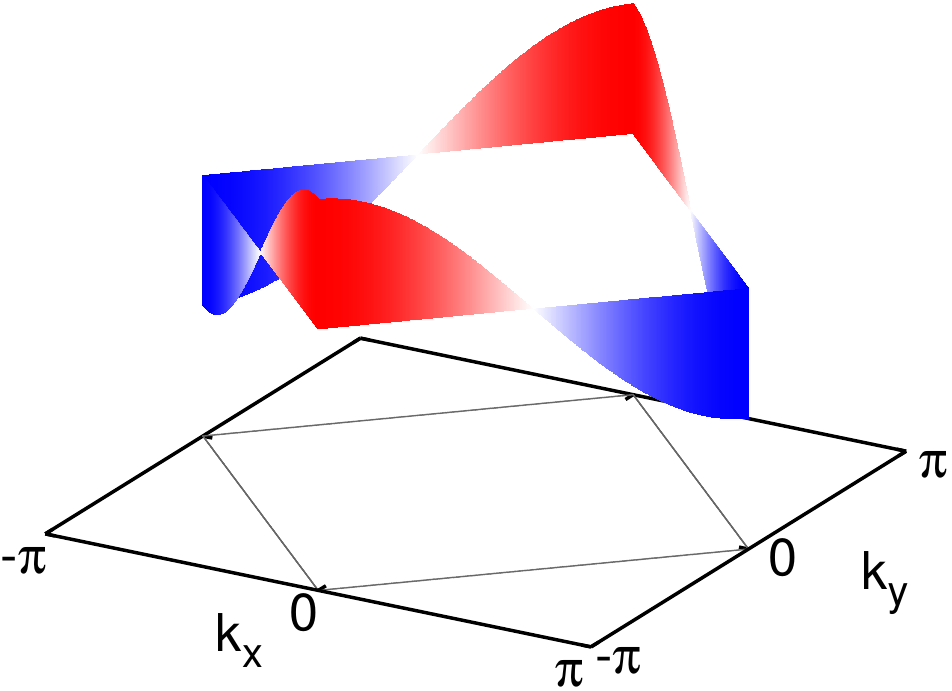}}%
	\subfigure[$s_{\pm}$~wave]{\label{fig:OP-spm}\includegraphics[scale=.5]{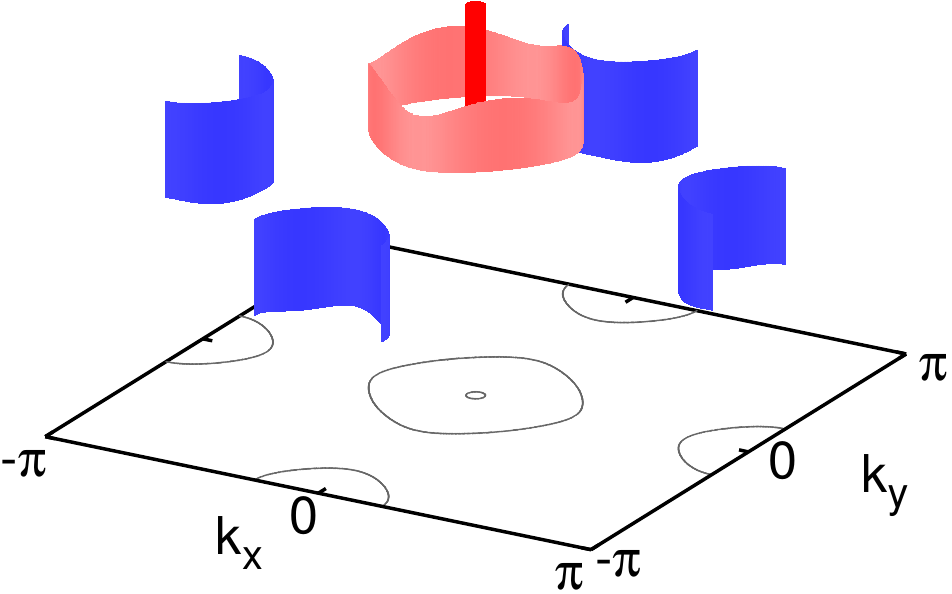}}%
	\subfigure[$s_{++}$~wave]{\label{fig:OP-spp}\includegraphics[scale=.5]{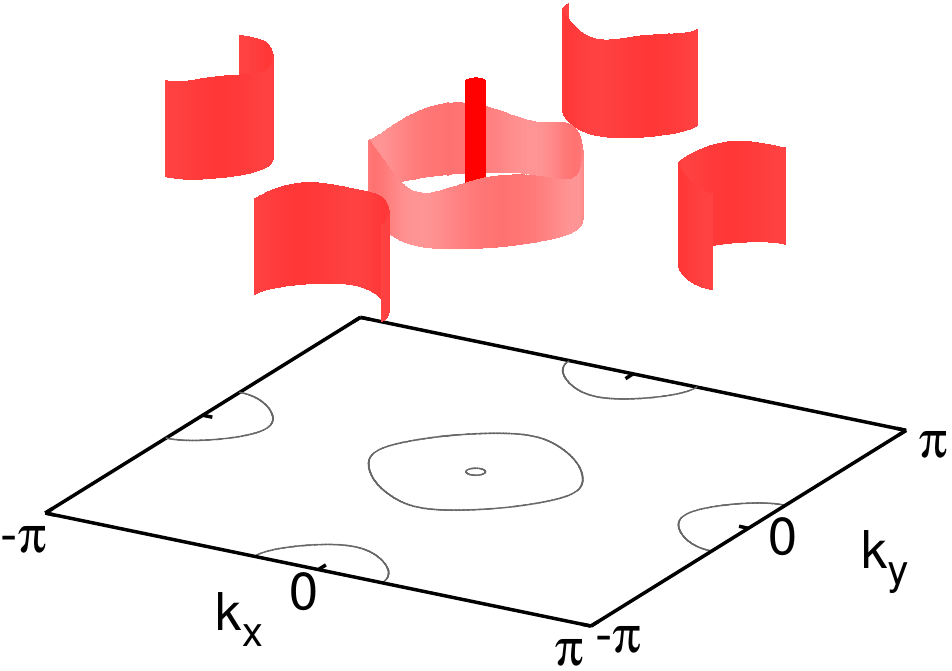}}%
	\caption[Possible symmetries of the superconducting order parameter in high $T_c$ superconductors]{
(a) $d$-wave superconducting order parameter in high $T_c$ cuprate superconductors, 
(b) $s_\pm$~wave order parameter in pnictides, and 
(c) $s_{++}$~wave order parameter, a possible scenario for the LiFeAs superconductor. 
ARPES measurements of the superconducting gap in LiFeAs (not shown) are consistent with both $s_\pm$ and $s_{++}$ superconductivity~\cite{Borisenko2012}. 
}
\end{figure}

Experimentally, the superconducting order parameter can be characterized by probing the quasiparticle spectra of the superconducting phase, i.e., the spectra of the elementary excitations of the electron condensate. 
In fact, the quasiparticle spectra give a direct access to the order parameter momentum dependence and, in particular, can reveal the presence of gapless excitations which are the fingerprints of the existence of nodes on the Fermi surface. 
Moreover, in order to fully characterize the symmetry of the order parameter and the nature of the pairing mechanism, it is desirable to be able to probe not only the magnitude, but also the phase of the order parameter along the Fermi surface. 
Josephson junctions experiments~\cite{Tsuei2000}, composite superconducting loops~\cite{Chen2010}, and STM~\cite{Hoffman2002,McElroy2003,Hanaguri2007,Kohsaka2008,Hanaguri2009} are examples of phase-sensitive experimental techniques which are able to distinguish between sign-preserving and sign-reversing excitations on the Fermi surface. 
It should be noted however that the correct interpretation of the quasiparticle spectra relies on a detailed knowledge of the low energy properties of the system, such as the band structure, correlations, or the type of the scattering impurities, the lack of which can prevent a direct characterization of the superconducting state.

\section{Resonant inelastic x-ray scattering}
\label{sec:Intro-RIXS}

In this thesis we will explore from a theoretical standpoint an alternative method to probe orbital correlations and quasiparticle excitations in unconventional superconductors, i.e., by means of RIXS spectroscopy. 
In fact, the capability to probe the dispersion of elementary spin~\cite{Hill2008,Braicovich2009,Braicovich2010,LeTacon2011,Kim2012,Dean2012}, charge~\cite{Ghiringhelli2004,Matsubara2005}, orbital~\cite{Ulrich2009,Schlappa2012}, and lattice~\cite{Yavas2010} excitations has recently positioned RIXS at the forefront of photon science. 
Compared to the available spectroscopic methods, such as scanning-tunnelling spectroscopy, photoemission spectroscopy, optical spectroscopy, or inelastic neutron scattering, RIXS uniquely combines the advantages of bulk sensitivity, momentum and energy resolution, while at the same time requiring only small sample volumes. 

RIXS appear to be a good candidate to detect orbital correlations, since the RIXS scattering process directly couples the orbital degree of freedom of $d$ electrons with the photon polarization. 
Indeed, $dd$ excitations, i.e., transitions between different $d$ orbitals, have been observed by RIXS in many systems, and in particular in transition metal oxides~\cite{Ghiringhelli2005,Ghiringhelli2007}. 
However, although the RIXS scattering process is in principle sensitive to orbital correlations, a direct and clear observation of orbital order or of orbital modes (orbitons) is still missing. 
On the other hand, direct RIXS has been successfully employed to probe magnetic excitations (magnons) in cuprates~\cite{Hill2008,Braicovich2010}. 
At a fundamental level, the sensitivity of direct RIXS to magnetic excitations is due to the strong spin-orbit coupling deep in the core hole $2p$ shell. 
This suggests the possibility to probe the symmetry of the orbital long-range order indirectly. 
More precisely, one can investigate whether RIXS spectral intensities of \emph{magnetic} excitations are sensitive to the underlying \emph{orbital} order or, equivalently, whether \emph{orbital} correlations leave a discernible fingerprint in the RIXS spectra of \emph{magnetic} excitations. 

On the other hand, the energy resolution of RIXS has reached $\approx30\unit{meV}$ in the hard x-ray regime~\cite{Kim2012}, and is projected to reach 11 meV at the Cu $L_{2,3}$ edges at the NSLS-II (Brookhaven National Laboratory) presently under construction~\cite{NSLS-II_design_report}. 
This brings the RIXS energy resolution well into the regime of the energy gap of cuprate superconductors, which stretches out to 119 meV for mercury based high $T_c$ systems~\cite{Yu2009}. 
Consequently the fundamental question arises of how the superconducting pairing leaves its fingerprints in RIXS spectra.
In particular, one can investigate whether the quasiparticle spectra of the superconducting state, i.e., the spectra of particle-hole excitations of the BCS condensate, contribute to the low-energy range of RIXS spectra. 
In this case, one could ask oneself whether and how RIXS is sensitive to the superconducting gap and, moreover, whether RIXS could discriminate between nodal and nodeless superconductivity and between different pairing symmetry in unconventional superconductors.

\section{Overview}
\label{sec:Intro-Overview}

In \cref{ch:RIXS} we will give s short overview of the theory of RIXS and, in particular, on direct RIXS at transition metal edges in correlated $3d$ electron systems. 
In particular in \cref{sec:RIXS-DSF} we will show how direct RIXS is able to probe the charge and spin dynamical structure factor (DSF) in $3d$ electron systems.
Consequently, in \cref{ch:Orbitals} we will apply the theoretical methods developed in the previous Chapter to study the direct RIXS spectra of magnetic excitations in strongly correlated systems with orbital degree of freedom, and in particular in two and three dimensional cuprates. 
The main result there will be the sensitivity of the \emph{magnetic} RIXS spectra to the \emph{orbital} ground state underneath, which allow one to discriminate between different orbital ground states, e.g., alternating orbital order against ferroorbital order or orbital liquid state. 
The method proposed is of direct relevance to two dimensional systems, e.g., K$_2$CuF$_4$ or Cs$_2$AgF$_4$ with ferromagnetic layers and predicted, but not verified, alternating orbital order~\cite{Hidaka1983, Wu2007, McLain2006}, as well as to three dimensional transition metal oxides with orbital degree of freedom, such as LaMnO$_3$, KCuF$_3$, LaTiO$_3$ or LaVO$_3$~\cite{Oles2005}. 

In \cref{ch:BCS-Cuprates} we will concentrate instead to the RIXS spectra of quasiparticle excitations in superconductors.
We will show how RIXS 
is able to directly access the quasiparticle excitations of the superconducting condensate, to probe superconducting order parameter, and to disclose the properties of the superconducting pairing. 
In particular, the main finding will be that RIXS spectra are sensitive not only to the magnitude but also to the phase of the superconducting order parameter. 
This will allow one to distinguish between different superconducting pairing symmetries (e.g., $s$-wave or $d$-wave). 
In \cref{ch:BCS-Pnictides} we will extend the theory of the previous Chapter to look at the symmetry properties of the superconducting order parameter in pnictides. 
Through the theoretical evaluation of the RIXS cross section of the quasiparticle excitations, we will show that a measure of RIXS spectra at the energy scale of the superconducting gap 
can shed some light on the pairing mechanism of pnictides, being able to distinguish between singlet and triplet pairing superconductivity. 
Compared with other experimental techniques, such as STM, RIXS has the advantage of simplicity, since the interpretations of RIXS spectra does not rely on the detail of the low energy properties of the material. 
Finally, in \cref{ch:Conclusions}, we will summarize and conclude this work.

}
\chapter[Theory of direct resonant inelastic x-ray scattering]{Theory of direct resonant inelastic x-ray scattering\footnotemark{}}
\footnotetext{Part of this chapter has been published in Refs.~\citen{Marra2012} and \citen{Marra2012b}}
{
\label{ch:RIXS}

Resonant inelastic x-ray scattering (RIXS)~\cite{Blume1985,Kotani2001,Schulke2007,deGroot2008,Ament2011} is a photon-in photon-out x-ray spectroscopy, in which the incident photon resonantly excites a core hole state, that consequently decays by the emission of a scattered photon. 
The scattering as a whole is inelastic, which means that the energy of the scattered photon is generally lower than the energy of the incident one.
The resulting energy loss is transferred to the system, which is left into an excited state. 
Besides, since the incident and scattered x-ray momenta do not in general coincide, this excited state has a definite momentum, which is comparable with the crystal momentum of conduction electrons. 
Therefore, RIXS can probe the energy, the dispersion and, more generally, the symmetry properties of low energy elementary excitations in condensed matter systems. 

The study of the ground state and of the excitation spectra of a physical system is the very aim of RIXS, as well as of any spectroscopy. 
This~Chapter will draw a quick overview of RIXS spectroscopy, in particular on its~capability to probe elementary excitations and to investigate the physical properties of condensed matter systems, and on the theoretical framework which describes the RIXS scattering process in the light of which RIXS spectra can be analyzed and understood. 
In particular, \cref{sec:RIXS} will be concerned with the general description of the RIXS scattering process, the general theoretical formulation of the RIXS cross section via the dipole approximation, and the difference between direct and indirect RIXS\@. e
Afterwards, \cref{sec:direct-RIXS} will give some details on the fast collision approximation in the case of strong spin-orbit coupling in the core shell, which allow one to disclose the relation between direct RIXS spectra and DSF of magnetic and charge excitations (\cref{sec:RIXS-DSF}). 
Based on these considerations, in \cref{sec:RIXS-Diagram} we will present a diagrammatic way to evaluate direct RIXS scattering amplitudes in $3d$ systems, which will be employed, as an application, to calculate RIXS intensities of $dd$ excitations in cuprates.

\section{Resonant inelastic x-ray scattering (RIXS)}
\label{sec:RIXS}

\subsection{The RIXS process}
\begin{figure}[t]
	\centering
	\includegraphics[width=.6\textwidth]{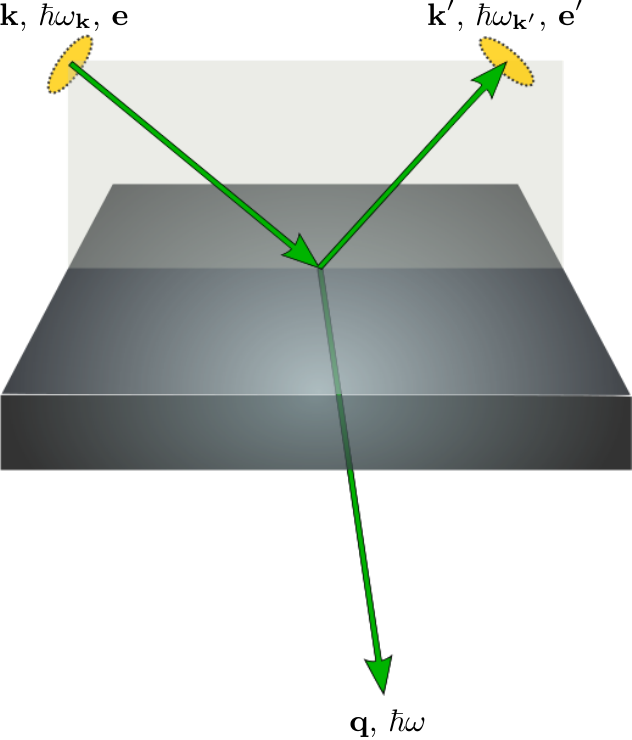}
	\caption[A simple cartoon of a typical RIXS experiment]{
A simple cartoon of a typical RIXS experiment. 
The incident x-ray radiation with energy $\hbar\omega_{\bf k}$, momentum $\bf k$, and polarization $\bf e$ is shone on the sample, and the scattered radiation is collected under a fixed direction, and resolved in energy. 
The RIXS cross section is therefore measured as a function of the scattered radiation energy $\hbar\omega_{\bf k'}$, momentum $\bf k'$, and, in principle, polarization $\bf e'$. 
In the process, a finite energy $\hbar\omega=\hbar\omega_{\bf k}-\hbar\omega_{\bf k'}$ and momentum $\bf q=k-k'$, and angular momentum $\bf J$ 
may be transferred into the system. 
}
	\label{fig:exp}
\end{figure}

\begin{figure}[t]
	\centering
	\includegraphics[width=1\textwidth]{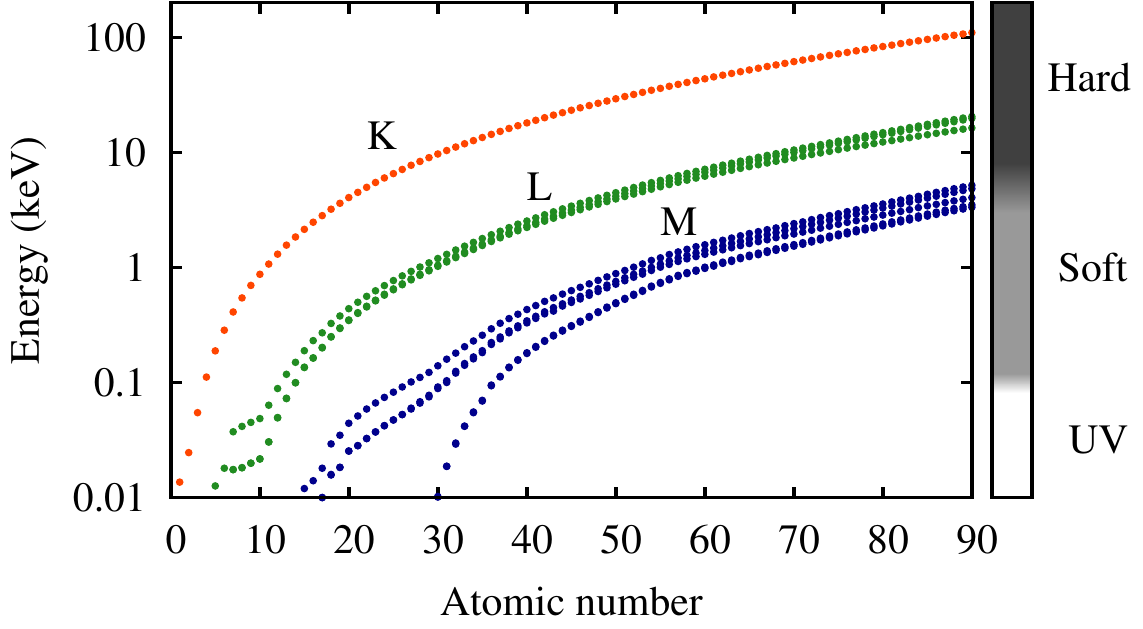}
	\caption[Absorption edges as a function of the atomic number]{
Absorption edges as a function of the atomic number~\cite{Bearden1967,Robinson1974}. 
The $K$, $L$, and $M$ edges correspond to the creation of a core hole in a core shell with principal quantum number $n=1,2,3$. 
The energy spans several orders of magnitudes (the energy scale is logarithmic), from the extreme ultraviolet ($\lesssim 100 \unit{eV}$) to soft x-ray ($0.1-5 \unit{keV}$) up to the hard x-ray region ($\gtrsim 5\unit{keV}$). 
}
	\label{fig:edges}
\end{figure}

In a typical photon scattering experiment, an incident photon beam is shone on the sample and the scattered radiation is collected. 
The scattering process is described in \cref{fig:exp}. 
In RIXS spectroscopy, the incident radiation is an x-ray photon beam either in the soft or in the hard x-ray range. 
The incident x-ray energy is chosen to resonate with one of the absorption edges of the system, i.e., the energy required to scatter off a core electron to an excited state above the Fermi level. 
This energy is largely determined by the atomic shell where the core hole state is created, and therefore depends, for a given atom, on the principal and azimuthal quantum number of the core electron state which is scattered by the incident x-ray photon. 
The x-ray absorption edges are shown in \cref{fig:edges} as a function of the atomic number. 
Therefore, as well as other core hole spectroscopies, RIXS can probe different absorption edges in the same physical system. 
Specifically, the incident photon energy can be tuned in principle to any of the absorption edges of any of the different atoms of the system. 
By means of that, RIXS can probe different degrees of freedom and therefore, in some cases, different properties or even different kind of elementary excitations of the system. 
However, unless other core hole spectroscopies, the scattered radiation is resolved in energy, and therefore one can disentangle the inelastic scattering processes from the elastic response of the system. 
The different scattering channels may or may not be elastic, i.e., the final state may or may not coincide to the initial state of the system. 
In the case of inelastic scattering, energy conservation implies that a finite energy $\hbar\omega$ is transferred into the system, i.e., the system is left into an excited state. 
For this reason, RIXS allows one to investigate the excitation spectra of the system, and in particular to probe several kind of low energy elementary excitations. 
Moreover, RIXS is momentum resolved, i.e., the cross section is measured as a function of both the incident and the scattered radiation momenta. 
In fact the incident and scattered radiation momenta do not in general coincide, and therefore a finite momentum $\bf q=k-k'$ is transferred into the system, as shown in \cref{fig:exp}. 
For typical x-ray energies employed in RIXS, the photon momentum is comparable with the crystal momentum of solid state electrons, and therefore the transferred momentum $\bf q$ can be mapped into the Brillouin zone of the system. 
For this reason, RIXS can probe not only the energy spectra of elementary excitations, but also the momentum dispersion of these excitations, which is a unique feature among other photon scattering spectroscopies. 
As a consequence, the experimental spectra, i.e., the RIXS cross section measured as a function of the energy loss $\hbar\omega=\hbar\omega_{\bf k}-\hbar\omega_{\bf k'}$ and of the transferred momentum $\bf q=k-k'$, are directly related to the spectra of elementary excitations and to the physical properties of the system. 

On the other hand, RIXS intensities can be measured as a function of the incident and scattered photon polarizations. 
A fully polarized x-ray beam contains photons which have the same spin direction (circular polarization) or, more generally, which are in a coherent superposition of the two spin eigenstates (linear or elliptical polarization). 
The conservation of angular momentum implies that any finite change in polarization between the incident and the scattered photon must correspond to a finite angular momentum transferred into the system. 
In particular, the spin-orbit coupling in the core hole intermediate state can induce a magnetic or an orbital excitation, i.e., the photon angular momentum can be transferred to the spin or to the orbital angular momentum of the scattered electron. 
For this reason, the polarization dependence of the RIXS cross section may allow one to disentangle the symmetry properties of the elementary excitations and of the ground state of the system. 
Some common choices for polarization of the incident radiation are linear polarizations, e.g., perpendicular or parallel to the scattering plane, or circular polarizations. 
Normally, the polarization of the scattered radiation is not resolved and the energy spectra contain contributions from all the polarizations which are perpendicular to the scattered radiation momentum. 
Nevertheless, in some state of the art experiments~\cite{Braicovich2007,Ishii2011} has been proven that linear polarization analysis of the scattered radiation is possible.

\subsection{Absorption edges}

In RIXS experiments, the incident x-ray energy is chosen to resonate with an absorption edge of the system, that is, the energy required to scatter off a core electron to an excited state above the Fermi level. 
The energy of each absorption edge coincides with the transition energy between the core and the valence shell, and it is determined by the atomic shell where the core hole state is created, and in particular depends on the principal and azimuthal quantum number of the core electron state which is scattered by the incident x-ray photon (see \cref{fig:edges}). 
The $K$, $L$, and $M$ edges correspond respectively to the excitations of a core hole in an atomic shell with principal quantum number $n=1,2,3$. 
For $n>1$, the edge structure is divided into sublevels which correspond to different azimuthal quantum numbers $l=1,2,3$ ($s,p,d$ shells). 
Furthermore, within a given atomic shell with azimuthal quantum number $l>0$, the spin orbit coupling removes the spin degeneracy of the core hole state and, therefore, two distinct sublevels appear, which are characterized by a total angular momentum $j=l\pm1/2$. 
A progressive subscript label the different absorption edges for atomic shells with principal quantum number $n>0$. 
The resonant edges in transition metal oxides, for example, are typically chosen at the ligand (oxygen) $K$ edge or at the transition metal $K$, $L$, or at the $M$ absorption edges. 

As one can see from \cref{fig:edges}, the atomic absorption edges span several orders of magnitude in the electromagnetic spectrum, from the less energetic edges of the lightest elements down in the ultraviolet region ($\lesssim 100 \unit{eV}$), to the soft x-ray ($0.1-5 \unit{keV}$), and up to the hard x-ray region ($\gtrsim 5\unit{keV}$) of the $K$ edges and some of the $L$ edges of the heavier elements. 
For example, the oxygen $K$ edge is in the soft x-ray region, as well as the $L$ edges of transition metals of the 4th period ($3d$ valence shell), while the transition metal $K$ edges are in the hard region of the x-ray spectrum. 
Different experimental setups and instrumentations have been developed specifically for soft and hard x-ray RIXS spectroscopy. 
In fact, while soft x-ray are easily absorbed by air, hard x-rays can traverse relatively thick objects without any substantial attenuation. 
Moreover, the different wavelengths require different kind of optics, i.e., Bragg crystals like silicon or germanium monocrystals for hard x-ray, and artificial periodic structures like diffraction gratings for soft x-ray spectrometers. 

\subsection{The Kramers-Heisenberg equation}

Microscopically, the RIXS process can be described as an inelastic scattering of x-ray photons with matter. 
The incident x-ray photon resonantly excites a core level state into an unoccupied state in the valence band. 
This intermediate state is highly energetic, with a core hole and an excited electron state respectively far below and above the Fermi level, and is therefore unstable, with a lifetime of the orders of femtoseconds. 
The recombination of the core hole and the excited electron, with the concurrent emission of a scattered photon, allows the system to relax into the initial state (elastic) or into a low energy excited state (inelastic scattering). 
In particular, the inelastic scattering processes allow one to probe the elementary excitations of the system. 

The interaction between electromagnetic radiation and matter is described by the theory of quantum electrodynamics. 
In the low energy limit, this interaction can be treated perturbatively, and the scattering of x-ray photons with an electronic system can be described by the Fermi's Golden Rule, which gives the transition amplitudes of the allowed low energy scattering processes. 
In the non-relativistic limit, the second order terms of the perturbation expansion dominates in the case of resonant scattering, while the first order terms contains the remnant low energy non-resonant processes. 
Neglecting the magnetic coupling between the photon field and the magnetic moment of the electron, and taking only the second order resonant terms of the perturbation expansion, one obtains the Kramers-Heisenberg equation~\cite{Kramers1925,Sakurai2006,Blume1985,Schulke2007}, which gives the cross section of x-ray photons with matter at any resonant edge of the system. 
The double-differential RIXS cross section is therefore given by the Kramers-Heisenberg equation, which at zero temperature reads
\begin{emphequation}\label{eq:KramersHeisenberg1st}
	\frac{d^2 \sigma}{d\omega d\Omega}\propto
	\sum\limits_{f}
	\left|
	\sum\limits_n 
	\frac{\bra{f} \hat{\cal D}^{\prime\dag} \ket{n}\bra{n} \hat{\cal D} \ket{g}}
	{E_g+\hbar\omega_{\bf k}-E_n+\imath\Gamma}
	\right|^2
	\delta (E_g+\hbar\omega-E_f)
	,
\end{emphequation}
where $\ket{g}$ is the ground state with energy $E_g$, $\ket{f}$ is any of the final states of the system with energy $E_f$, $\ket{n}$ is any of the intermediate core hole states with energy $E_n$, and $\hbar\omega=\hbar\omega_{\bf k}-\hbar\omega_{\bf k'}$ is the transferred photon energy. 
The core hole broadening $\Gamma$ is assumed to be independent of the intermediate state energy, and ranges from 1 eV of transition metal edges to 8 eV of actinides $L$ edges, which correspond to a core hole lifetime $\Delta t=\hbar/2\Gamma$ from 2 fs down to 0.2 fs. 
The optical transition operator $\hat{\cal D}=\hat{\bf p}\cdot\hat{\bf A}$ describes the dominant term of the electron-photon interaction in the low energy perturbative expansion of the quantum electrodynamics. 
Expanding the vector potential $\hat{\bf A}$ in terms of plane waves around any lattice site $i$, the optical transition operator $\hat{\cal D}$ ($\hat{\cal D}'$) can be written as a function of the momentum ${\bf k}$ (${\bf k}'$) and polarization ${\bf e}$ (${\bf e}'$) of the incident (scattered) photon, as~\cite{Ament2011}
\begin{equation}\label{eq:DipoleOperator}
	\hat{\cal D}= \frac{1}{\sqrt{N}}
	\sum\limits_{i=1}^N e^{\imath{\bf k}\cdot (\hat{\bf r}_i+\bf{R}_i)} {\bf e}\cdot\hat{\bf p}_i
	,
\end{equation}
where $\hat{\bf r}_i$ and $\hat{\bf p}_i$ are the position and momentum operators of the electron respect to the lattice site ${\bf R}_i$.

If one introduces the intermediate state propagator $\hat{G}$, which describes the propagation of the core hole in the intermediate state, defined by the Green's function
\begin{equation}\label{eq:Propagator}
	\hat{G}=
	\sum\limits_n \frac{\ket{n}\bra{n}}{E_g+\hbar\omega_{\bf k}-E_n+\imath\Gamma}=
	\frac{1}{E_g+\hbar\omega_{\bf k}-\hat{\cal H}+\imath\Gamma}
	,
\end{equation}
where $\hat{\cal H}$ is the Hamiltonian of the system describing the core hole eigenstates $\ket{n}$, one can rewrite the RIXS cross section as
\begin{emphequation}\label{eq:KramersHeisenberg}
	\frac{d^2 \sigma}{d\omega d\Omega}\propto
	\sum\limits_{f}
	|\bra{f} \hat{\cal O} \ket{g}|^2
	\delta (E_g+\hbar\omega-E_f)
	,
\end{emphequation}
where the RIXS scattering operator is defined by $\hat{\cal O}=\hat{\cal D}^{\prime\dag} \hat{G} \hat{\cal D}$.
\begin{figure}[t]
	\centering
	\includegraphics[width=0.8\textwidth]{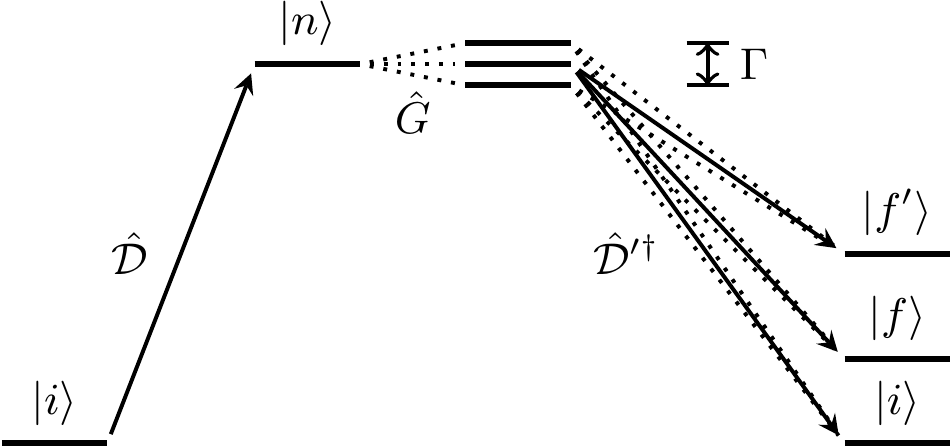}
	\caption[Schematic representation of the RIXS scattering process]{
Schematic representation of the RIXS scattering process. 
The transition operator $\hat{\cal D}$ excites the system into a core hole state, which therefore propagates via the Green's function $\hat{G}$, and is eventually annihilated by the operator $\hat{\cal D}^{\prime\dag}$. 
In the intermediate state, virtual transitions have a finite probability in an energy range comparable with the core hole broadening $\Gamma$. 
The final state may or may not coincide with the initial state of the system. 
}
	\label{fig:kramers}
\end{figure}
The action of the RIXS scattering operator $\hat{\cal O}$, and the scattering process described by \cref{eq:KramersHeisenberg} are illustrated in \cref{fig:kramers}. 
The optical transition operator $\hat{\cal D}$ represents the photoelectric scattering of the incident photon with the electronic system. 
The action of this operator is to promote a core electron into a valence state, i.e., to create a core hole and an additional valence electron into the system. 
The Green's function $\hat{G}$ describes the propagation of this excited state, which is in general not an eigenstate of the system. 
At this stage, there is a finite probability that the system will undergo virtual transitions into nearby intermediate states, as long as the energy fluctuation is small compared with the core hole broadening ($\Delta E<\Gamma$). 
Finally, the optical transition operator $\hat{\cal D}^{\prime\dag}$ recombines the core hole with an electron in the valence band. 
If the scattering is inelastic, the final state does not coincide with the initial state of the system. 

Alternatively, using the representation of the Dirac delta function as $\delta(E)=\Im\lim_{\eta \to 0^+}(E+\imath\eta)^{-1}$ one finds that the RIXS cross section can be written as
\begin{align}\label{eq:KramersHeisenbergAlt}
	\frac{d^2 \sigma}{d\omega d\Omega}&\propto
	\Im\lim_{\eta \to 0^+}
	\sum\limits_{f}
	\frac{\bra{g} \hat{\cal O}^\dag\ket{f}\bra{f}\hat{\cal O} \ket{g}}{E_g+\hbar\omega-E_f+\imath\eta}
	\nonumber\\&=
	\Im\lim_{\eta \to 0^+}
	\bra{g} \hat{\cal O}^\dag \frac{1}{E_g+\hbar\omega-\hat{\cal H}+\imath\eta} \hat{\cal O} \ket{g}
	,
\end{align}
where $\hat{\cal H}$ is again the system Hamiltonian and $\eta$ is the broadening of the elementary excitations of the system.

Equation \eqref{eq:KramersHeisenberg} is still valid at finite temperature, provided that the excitation spectrum of the system is gapped, and that the temperature is low enough for thermal fluctuations to be negligible, i.e., in the case where $|E_f-E_g|\gg k_B T$ for any excited state $\ket{f}$ of the system. 
If this is not the case, \cref{eq:KramersHeisenberg} has to be generalized considering the statistical combination of all possible initial states of the system, as
\begin{equation}
	\frac{d^2 \sigma}{d\omega d\Omega}\propto
	\sum\limits_{if}
	\frac1{\cal Z}e^{-E_i/{k_B T}}
	|\bra{f} \hat{\cal O} \ket{i}|^2
	\delta (E_i-E_f+\hbar\omega)
	,
\end{equation}
where $\cal Z$ is the partition function and $\ket{i}$ is any of the states accessible to the system at temperature $T$. 

\subsection{X-ray absorption spectroscopy}

Closely related to RIXS is the x-ray absorption spectroscopy (XAS)~\cite{deGroot2008}. 
In a XAS experiment, an incident x-ray beam is shone on the sample and the absorption cross section is measured. 
If the incident photon energy is close to an absorption edge, i.e., the energy required to scatter off an electron from a core hole shell, the XAS spectra will provide information about the core hole states and on the ground state of the system. 
Close to an absorption edge, the absorption cross section of the XAS process can be derived from the Fermi's Golden Rule at the first order of the electron-photon interaction, and reads
\begin{equation}\label{XAS}
	\frac{d \sigma}{d\omega}\propto
	\sum\limits_{n}
	|\bra{g} \hat{\cal D} \ket{n}|^2
	\delta (E_i-E_n+\hbar\omega)
	.
\end{equation}
where $|g\rangle$ is the ground state with energy $E_g$, and $|n\rangle$ is any of the core hole states of the system with energy $E_n$, and $\hbar\omega$ is the incident photon energy, and $\hat{\cal D}$ is the dipole operator defined in \cref{eq:DipoleOperator}. 
Loosely speaking, the XAS absorption process can be described as the first half of a RIXS scattering process, since the final states $\ket{n}$ in the XAS absorption cross section in \cref{XAS} coincide with the intermediate states in the RIXS cross section in \cref{eq:Propagator}. 
The main similarity between XAS and RIXS is indeed the presence of a core hole state, and the fact that the absorption edges probed by XAS spectra correspond in RIXS to the resonant edges at which the incident photon energy is tuned. 
The measure of the XAS spectra is therefore the necessary premise of any RIXS experiment, since the choice of the resonant energy is of paramount importance in RIXS spectroscopy. 
On the other hand, the understanding of the core hole state and of XAS edges is the necessary premise of any theoretical description of RIXS spectra.

\subsection{Dipole approximation and local RIXS operator}\label{sec:Dipole}

The RIXS scattering operator depends on the scattering geometry, i.e., on the incident and scattered photon polarizations and momenta, and, at a more fundamental level, on the ground state and on the excited states of the system. 
Since the aim of the RIXS spectroscopy is to probe the excitations spectra and the ground state of a physical system, it would be desirable to disentangle the geometrical dependence of the RIXS cross section from the fundamental response of the physical system. 
In order to do so, one can introduce the dipole approximation, which drastically simplifies the form of the optical transition operator, and allows one to factorize out easily the geometrical dependence of the RIXS cross section. 
In general, on can expand the optical transition operator around the single ion site as
\begin{equation}
	\hat{\cal D}=
	\frac{1}{\sqrt{N}}
	\sum\limits_{i=1}^N e^{\imath{\bf k}\cdot {\bf R}_i} {\bf e}\cdot
	\hat{\bf D}_i
	,
	\text{\qquad\qquad with\ \ }
	\hat{\bf D}_i=
	e^{\imath{\bf k}\cdot \hat{\bf r}_i}
	\hat{\bf p}_i
	,
\end{equation}
and therefore, one can further expand the local optical transition operator $\hat{D}_i$ using the multipole expansion
\begin{equation}\label{eq:Multipole}
	\hat{\bf D}_i=\sum\limits_{n=0}^\infty\frac{\left(\imath{\bf k}\cdot \hat{\bf r}_i\right)^{n}}{n!} \hat{\bf p}_i
	.
\end{equation}
The dipole approximation corresponds to the zero order approximation of the multipole expansion, i.e., assuming $e^{\imath{\bf k}\cdot\hat{\bf r}_i}\approx 1$, and thereby the optical transition operator reduces to the momentum operator at the single ion site
\begin{equation}\label{eq:DipoleApproximation}
	\hat{\bf D}_i=e^{\imath{\bf k}\cdot \hat{\bf r}_i}\hat{\bf p}_i\approx \hat{\bf p}_i
	.
\end{equation}
Therefore, in the dipole approximation, the matrix element of the optical transition operator are nonzero only for transitions between states with azimuthal quantum number $l$ and $l\pm1$, and in particular one has 
\begin{equation}
	\bra{n'l'm'}\hat{\bf D}_i\ket{nlm}\ne0 \Leftrightarrow \Delta l=\pm1 \text{ and } \Delta m=0,\pm1
	,
\end{equation}
where $\Delta l= l'-l$ and $\Delta m= m'-m$, as follows from the selection rules of spherical vector operators (see also \cref{eq:SelectionRulesSphericalVectors,eq:DipoleOperatorSpherical}). 
Moreover, the component $\hat{D}_z$ allows nonzero matrix elements between eigenstates with the same magnetic quantum number $\Delta m=0$, while the components $\hat{D}_x$ and $\hat{D}_y$ allows nonzero matrix elements for $\Delta m=\pm1$. 
The next term of the multipole expansion, i.e., the quadrupole operator $\left(\imath{\bf k}\cdot \hat{\bf r}_i\right) \hat{\bf p}_i$ allows instead transitions between states with angular momentum $l$ and $l\pm2$. 
Since the dipole term of the multipole expansion is, in general, the dominant term of the RIXS cross section, hereafter the dipole approximation will be used throughout, and the local optical transition operator will be referred as the dipole operator. 
\begin{figure}[t]
	\centering
	\includegraphics[width=0.8\textwidth]{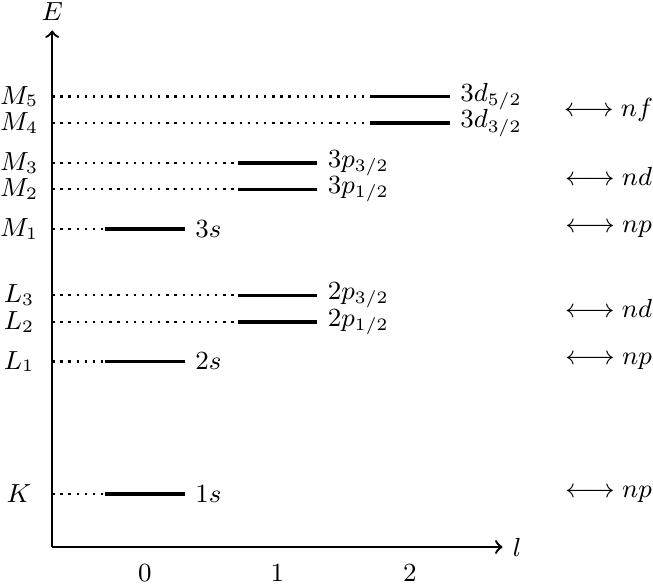}
	\caption[Dipole-allowed scattering processes]{
Schematic representations of the dipole-allowed scattering processes. 
The absorption edges $K$, $L$, and $M$ correspond a core hole in the core shell with principal quantum number $n=1,2,3$.
For $n>1$, the edge structure is divided into sublevels ($L_1$ and $L_{2,3}$ for $n=2$, and $M_1$, $M_{2,3}$ and $M_{4,5}$ for $n=3$) which correspond to different azimuthal quantum numbers $l=1,2,3$ ($s,p,d$ shells). 
Moreover, for $l>1$ the spin-orbit coupling splits further the edge structure in sublevels corresponding to the total angular momentum $j=l\pm1/2$. 
The dipole operators allows transitions between the core hole levels and valence electron states with $\Delta l=\pm 1$, e.g., $ns\leftrightarrow n'\!p$, $np\leftrightarrow n'\!d$, and $nd\leftrightarrow n'\!f$. 
}
	\label{fig:selection}
\end{figure}

Using the Wigner-Eckart theorem (see \cref{sec:WignerEckartTheorem}), one can obtain an explicit expression of the dipole operator $\hat{\bf D}_i=\hat{\bf p}_i$ in the basis of orbital angular momentum eigenstates, as in \cref{eq:DipoleOperatorCartesian}. 
From \cref{eq:DipoleOperatorSpherical,eq:DipoleOperatorCartesian} follows that the momentum and position operators, although not proportional (they satisfy the canonical commutation relation $[\hat{r}_\lambda,\hat{p}_\lambda]=\imath \hbar$) have the same matrix elements between angular momentum eigenstates, up to constants and complex conjugacy. 
For this reason, it is common to substitute the momentum operator with the position operator~\cite{Ament2011} in the multipole expansion of the optical transition operator in \cref{eq:Multipole}, and thus having $\hat{\bf D}_i\approx \hat{\bf r}_i$ in the dipole approximation limit. 
Indeed, matrix elements of the momentum and of the position operator are equivalent with respect to the evaluation of RIXS scattering amplitudes. 

The dipole approximation allows one to remove the trivial dependence from the incident and scattered photon momenta. 
In fact, assuming that the core hole is created and annihilated at the same lattice site~\cite{Haverkort2010}, the RIXS operator can be rewritten as
\begin{align}\label{eq:RIXSOperator}
	\hat{\cal O}
	&=
	\frac1N
	\sum\limits_{i,j=1}^N e^{\imath({\bf k}\cdot {\bf R}_i-{\bf k'}\cdot {\bf R}_j)}
	\left({\bf e}^{\prime}\cdot\hat{\bf D}_j\right)^\dag \hat{G} \left({\bf e}\cdot\hat{\bf D}_i\right)
	\nonumber\\&\approx
	\frac1N
	\sum\limits_{i=1}^N e^{\imath{\bf q}\cdot {\bf R}_i}
	\left({\bf e}^{\prime*}\cdot\hat{\bf D}_i^\dag\right) \hat{G} \left({\bf e}\cdot\hat{\bf D}_i\right)
	,
\end{align}
where the local dipole operators do not depend on the photon momenta, and where $\bf q=k-k'$ is the transferred momentum. 
Therefore, the RIXS cross section can be rewritten in terms of the sum of local transition amplitudes as
\begin{emphequation}
\label{eq:KramersHeisenbergLocal}
	\frac{d^2 \sigma}{d\omega d\Omega}\propto
	\sum\limits_{f}
	\left|
	\frac1N
	\sum\limits_{i=1}^N e^{\imath{\bf q}\cdot {\bf R}_i}
	\bra{f} 
	\hat{O}_i
	\ket{g}
	\right|^2
	\delta (E_g-E_f+\hbar\omega)
	,
\end{emphequation}
where the local RIXS operator $\hat{O}_i$ is defined by
\begin{equation}\label{eq:RIXSOperatorLocal}
	\hat{O}_i
	=
	\left({\bf e}^{\prime*} \cdot \hat{\bf D}_i^{\dag} \right)
	\hat{G}
	\left({\bf e} \cdot \hat{\bf D}_i\right)
	.
\end{equation}
Moreover, expanding the polarizations $\bf e$ and the dipole operator $\hat{\bf D}_i$ in their components, one can factorize out the polarization dependence as
\begin{equation}\label{eq:RIXSOperatorLocal2nd}
	\hat{O}_i
	=
	\sum\limits_{\lambda\mu} e_\mu^{\prime*}
	\left(\hat{D}_\mu^{\dag} \hat{G} \hat{D}_\lambda\right)
	e_\lambda
	=
	{\bf e}^{\prime\dag} \cdot
	\llbracket\hat{D}_\mu^{\dag} \hat{G} \hat{D}_\lambda\rrbracket
	\cdot{\bf e}
	.
\end{equation}
where the lattice site index $i$ is dropped out, for the sake of simplicity. 
In the last term, ${\bf e}$ and the conjugate transposed ${\bf e}^{\prime\dag}$ are considered respectively as column and row vectors, while $\llbracket\hat{D}_\mu^{\dag} \hat{G} \hat{D}_\lambda\rrbracket$ is a $3\times3$ matrix with elements $\hat{D}_\mu^{\dag} \hat{G} \hat{D}_\lambda$. 
The RIXS cross section therefore does not depend separately on the incident and scattered photon momenta, but only on the transferred momentum, which corresponds to the intrinsic momentum of the elementary excitations of the system. 

\subsection{Direct and indirect RIXS}

Inelastic scattering occurs when the energy of the final state does not coincide with the initial energy of the system. 
In the RIXS process, the transition to the final excited state can be either \emph{direct} or \emph{indirect}~\cite{Brink2005,Brink2006}. 
The difference between direct and indirect scattering is illustrated in \cref{fig:direct_indirect}. 
In a direct scattering, the incident photon excites a core electron into an unoccupied state of a partially filled valence band slightly above the Fermi level. 
Afterwards, an electron from an occupied valence state slightly below the Fermi level decays and annihilates the core hole, and a photon is emitted. 
In this case, the scattering is inelastic and the system is left into an excited state, \emph{directly} induced by the electron-photon interaction. 
In an indirect scattering instead, the incident photon excites the core electron into an unoccupied state far above the Fermi level. 
After that, this highly energetic electron state decays to fill the core hole. 
If no other interactions are present, the final state of the process coincide with the initial state of the system, and therefore inelastic scattering precesses cannot occur. 
However, in the highly excited and unstable intermediate state, the core hole and the excited electron are strongly interacting with each other and with the valence electron of the system. 
In particular, the core hole Coulomb interaction is usually the strongest and dominates the intermediate state dynamics. 
This strong interaction, usually described as a core hole potential $U_c$, can produce an excitation in the valence band of the system. 
In this way, the system is left into an excited state, which is not induced by the mere electron-photon interaction, but it is \emph{indirectly} produced by the core hole potential. 

\begin{figure}[t]
	\centering
	\subfigure[Direct RIXS]{\includegraphics[scale=.93]{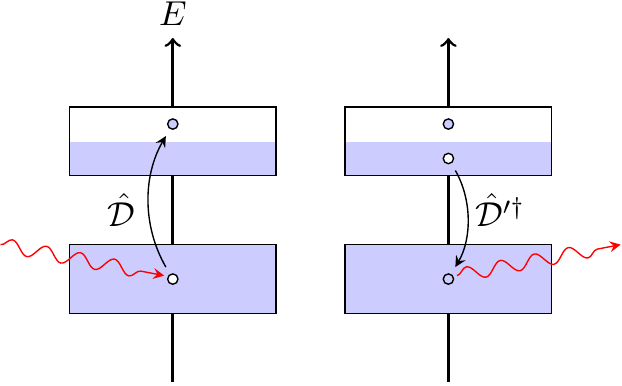}}
	\,
	\subfigure[Indirect RIXS]{\includegraphics[scale=.93]{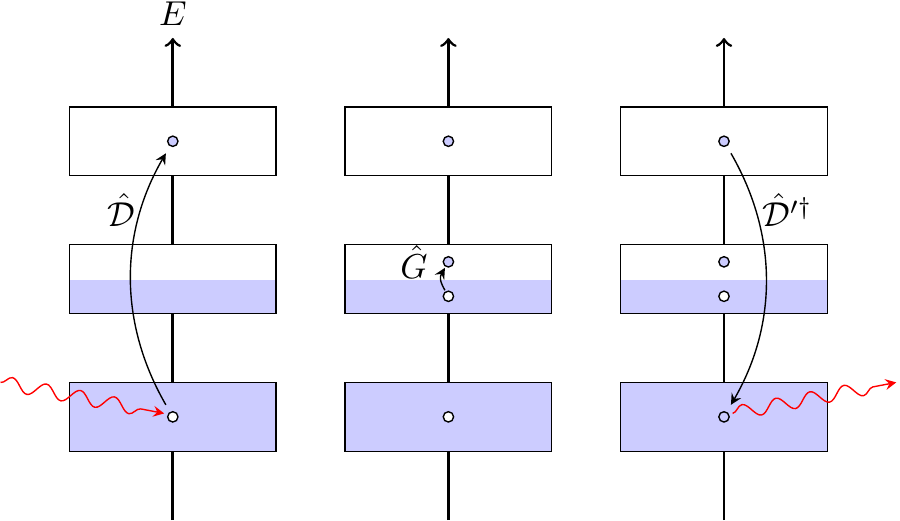}}
	\caption[Direct and indirect RIXS scattering]{
(a) 
In a direct scattering process, the incident photon excites a core electron into an unoccupied state of a partially filled valence band. 
Afterwards, an electron from another occupied valence state decays and annihilates the core hole, and a photon is emitted. 
The scattering is inelastic and the system is left into an excited state.
(b) 
In an indirect scattering process, the incident photon excites the core electron into an unoccupied state far above the Fermi level. 
After that, this highly energetic electron state decays to fill the core hole. 
If no other interactions are present, only elastic scattering is possible. 
However, in intermediate state, the strong interaction between the core hole and valence electrons can induce excitations in the final state of the system. 
}
	\label{fig:direct_indirect}
\end{figure}
In light of the above, it is natural to decompose the Hamiltonian of the system, which contains in principle all the relevant interactions, as $\hat{\cal H}=\hat{\cal H}_0+\hat{\cal H}_c$, i.e., in an unperturbed Hamiltonian $\hat{\cal H}_0$, containing the low energy interactions governing the ground state and the excited states of the system, and the core hole Hamiltonian $\hat{\cal H}_c$ containing the interaction between the core hole, the excited electron and the valence electrons of the system in the intermediate state of the scattering process. 
Introducing the unperturbed propagator 
\begin{equation}\label{eq:DirectPropagator}
	\hat{G}_0=\frac1{E_g+\hbar\omega_{\bf k}-\hat{\cal H}_0+\imath\Gamma}
	,
\end{equation}
which satisfies the identity $\hat{G}^{-1}=\hat{G}_0^{-1}-{\cal H}_c$, the intermediate state propagator can be decomposed as $\hat{G}=\hat{G}_0+\hat{G}_0\hat{\cal H}_c\hat{G}$.
As a consequence, the RIXS scattering operator can be written as
\begin{emphequation}\label{eq:RIXSoperatorDirectIndirect}
	\hat{\cal O}=
	\hat{\cal D}^{\prime\dag} \hat{G}_0 \hat{\cal D}
	+
	\hat{\cal D}^{\prime\dag} \hat{G}_0 \hat{\cal H}_c\hat{G} \hat{\cal D}
	,
\end{emphequation}
where the two terms describe respectively the direct and the indirect scattering processes. 
If the direct transition between the core hole and the valence state is dipole allowed, the direct term does not vanish and is by far the leading order contribution to the RIXS cross section. 
Although the indirect scattering term still contributes to the total RIXS scattering, these higher order processes are generally weaker and can be neglected to the leading order. 
Therefore, in this case one talks about direct RIXS scattering, and the RIXS scattering operator reduces to
\begin{equation}\label{eq:RIXSoperatorDirect}
	\hat{\cal O}\approx
	\hat{\cal D}^{\prime\dag} \hat{G}_0 \hat{\cal D}
	\text{\qquad(Direct)}
	.
\end{equation}
Conversely, if the transition between the core hole and the valence state is forbidden by the dipole selection rules, the direct term contains higher order multipole transitions, which are typically weak, so that the main contributions to the inelastic scattering are only given by indirect processes. 
This is the case of the indirect RIXS scattering, where the RIXS scattering operator becomes
\begin{equation}\label{eq:RIXSoperatorIndirect}
	\hat{\cal O}\approx
	\hat{\cal D}^{\prime\dag} \hat{G}_0\hat{\cal H}_c\hat{G} \hat{\cal D}
	\text{\qquad(Indirect)}
	.
\end{equation}
Direct and indirect RIXS scattering are therefore qualitatively different, since the main contributions to the RIXS cross section are provided by two distinct scattering mechanisms. 
In both cases, the RIXS cross section depends on the combined action of optical transitions and intermediate state propagator. 
However, in the direct RIXS process, the elementary excitations in the final state are directly produced by dipole transitions, whereas in indirect RIXS, they are indirectly induced by the core hole propagator. 
Therefore, in direct RIXS, the cross section depends crucially on the optical transition amplitudes and selection rules, whereas in indirect RIXS the main role is played by the core hole propagation. 
Without the core hole potential, in fact, the indirect RIXS cross section simply vanishes. 
Nevertheless, the intermediate state dynamic does also play a role in the direct RIXS scattering, being responsible, for example, of the non-vanishing scattering amplitudes of magnetic excitations at the transition metal edges. 

Direct RIXS scattering occurs at the oxygen $K$ edge, where the dipole allowed transitions $1s\leftrightarrow2p$ directly excite the $2p$ valence electrons, and at the transition metal $L_{2,3}$ and $M_{2,3}$ edges, where excitations in the $3d$ valence shell are directly created by the dipole transitions $2p\leftrightarrow3d$ and $3p\leftrightarrow3d$ respectively. 
Indirect RIXS scattering occurs instead at the transition metal $K$, $L_1$, and $M_1$ edges, where the dipole allowed transitions are respectively $1s\leftrightarrow4p$, $2s\leftrightarrow4p$, and $3s\leftrightarrow4p$, while direct quadrupole transitions $ns\leftrightarrow3d$ are largely negligible.
In these cases, an highly energetic $4p$ electron state is excited in the intermediate state altogether with a core hole in the $1s$, $2s$, or $3s$ shell, which exerts a strong core hole potential that indirectly excites the $3d$ valence states of the system. 

\section{Direct RIXS cross section}
\label{sec:direct-RIXS}

\subsection{The spin-orbit coupling}

The sensitivity of direct RIXS scattering at transition metal edges to the orbital and to the magnetic degrees of freedom relies at a fundamental level on the strong spin-orbit coupling deep in the $2p$ core hole shell. 
In fact, since the direct interaction between the photon field and the electron spin and orbital angular momentum is very weak (photons carry no magnetic moment), the dominant photoelectric transitions (dipole transitions) do not allow a direct transfer of angular momentum to the electron system. 
Nevertheless, at the transition metal ions $L_{2,3}$ edges, neither the spin or the orbital angular momentum are conserved in the intermediate state of the RIXS process, due to the strong spin-orbit coupling in the core hole $2p$ shell. 
This allows a spin flip and a consequent angular momentum transition of the core hole, which results in non-vanishing transition amplitudes of magnetic and orbital transitions~\cite{Braicovich2010}. 

Within a given atomic shell, the angular momentum coupling between the spin and the orbital angular momentum of the core hole is described by the spin-orbit Hamiltonian in the form
\begin{emphequation}\label{eq:SOcoupling}
	\hat{\cal H}_{so}=-\lambda_{so} \hat{\bf L}\cdot\hat{\bf S}=-\frac{\lambda_{so}}{2}(\hat{\bf J}^2-\hat{\bf L}^2-\hat{\bf S}^2)
\end{emphequation}
where $\hat{\bf S}$ and $\hat{\bf L}$ and $\hat{\bf J}=\hat{\bf L}+\hat{\bf S}$ are the spin, the orbital, and the total angular momentum of the core hole, $\lambda_{so}$ is an effective parameter which describes the strength of the interaction and depends on the atom and on the atomic shell considered, and where the minus sign derives from the fact that the core hole can be described as an effective particle with negative charge. 
From the commutation relations between the angular momentum operators, it follows that the eigenstates of the spin-orbit Hamiltonian are simultaneous eigenstates of the operators $\hat{\bf J}^2$, $\hat{\bf L}^2$, $\hat{\bf S}^2$, and of the component $\hat{J}_z$ of the total angular momentum. 
Therefore, the eigenstates $\ket{n,j,j_z,l,s}$ of the spin-orbit Hamiltonian are described by the quantum numbers $j$, $j_z$, $l$, and $s=1/2$, where one has $|l-1/2|\le j\le|l+1/2|$ and $|j_z|\le j$ from the usual rules of addition of angular momenta, and where $n$ represents all the other quantum numbers which define the state of the system.  
The eigenstates of the spin-orbit Hamiltonian are not, however, eigenstates of the components $\hat{L}_z$ and $\hat{S}_z$, i.e., the magnetic quantum number $m$ and the spin projection $s_z$ are not good quantum numbers anymore. 
As a consequence, neither the spin nor the angular momentum, but only the total angular momentum is conserved in the scattering process. 

\begin{figure}[t]
	\centering
	\includegraphics[width=1\textwidth]{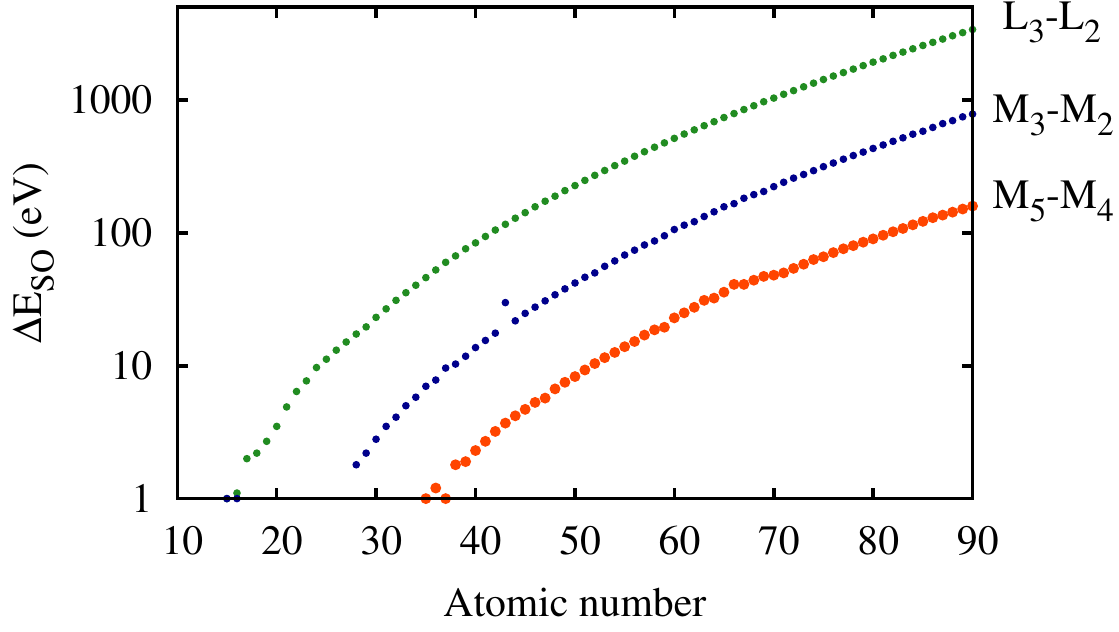}
	\caption[Spin-orbit splitting as a function of the atomic number]{
Spin-orbit splitting as a function of the atomic number $Z<90$, in the $2p$, $3p$, and $3d$ shells, corresponding to the energy differences between absorption edges~\cite{Bearden1967,Robinson1974}, respectively $L_3$ and $L_2$, $M_3$ and $M_2$, and $M_5$ and $M_4$. 
The energy spans several orders of magnitudes (the energy scale is logarithmic) from $2.2\unit{eV}$ and $19.6\unit{eV}$ of the copper $3p$ and $2p$ shells respectively, up to $1000\unit{eV}$ of the heavy atoms $2p$ shells. 
}
	\label{fig:spin-orbit}
\end{figure}

In the case of $l=0$, i.e., if the core hole is in any of the $ns$ atomic shells, the spin-orbit interaction vanishes, and the core hole states reduce, if other magnetic interactions are negligible, to the two degenerate eigenstates with $s_z=\pm1/2$. 
However, in the case $l>0$, e.g., if the core hole is in any of the $np$ or $nd$ atomic shells, the spin-orbit interaction does not vanish, and the eigenstates $\ket{n,j,j_z,l,s}$ have energy
\begin{equation}\label{eq:SpinOrbitLevels}
E_j=
	\begin{cases}
		\phantom{-}\dfrac{\lambda_{so}}2\hbar^2\ (l+1) &\text{for } j=l-\frac12,\\
	&\\
		-\dfrac{\lambda_{so}}2\hbar^2\ l   &\text{for } j=l+\frac12,\\
	\end{cases}
\end{equation}
which does not depend on the direction $\hat{J}_z$, but only on the modulus square of the total angular momentum $\hat{\bf J}^2$. 
The energy level splitting due to the spin-orbit coupling within a given atomic shell is therefore
\begin{equation}
	\Delta E_{so}=\frac{\lambda_{so}}2\hbar^2\ (2l+1)
\end{equation}
The spin-orbit coupling is therefore responsible of the energy splitting of the absorption edges corresponding to the same atomic core shell, as one can see in \cref{fig:selection}. 
Moreover, in \cref{fig:spin-orbit} are shown the values of the spin-orbit splitting $\Delta E_{so}$ for atoms with $Z<100$, in the $2p$, $3p$, and $3d$ shells, which correspond respectively to the energy differences between the $L_3$ and $L_2$, the $M_3$ and $M_2$, and the $M_5$ and $M_4$ absorption edges. 

\subsection{Fast collision approximation}

If one assumes that the energy scale of elementary excitations under study is negligible respect to the core hole broadening, and that the spin-orbit coupling is nonzero, the unperturbed intermediate state propagator in \cref{eq:DirectPropagator} describing the direct RIXS scattering becomes
\begin{equation}\label{eq:G0Direct}
	\hat{G}_0=\frac1{E_g+\hbar\omega_{\bf k}-(E_0+\hat{\cal H}_{so})+\imath\Gamma}
\end{equation}
where $\hat{\cal H}_{so}$ the spin-orbit coupling as defined in \cref{eq:SOcoupling}, and $E_0=E_{nl}$ the dominant energy term of the core hole state, which depends only on the principal and azimuthal quantum numbers of the core hole shell.  
If direct scattering is allowed by the dipole selection rules, indirect scattering processes are negligible and therefore one can neglect the interaction term $\hat{\cal H}_c$ between the valence electrons and the core hole. 

At this point, it is useful to introduce some further approximations in the expression of the intermediate state propagator. 
In order to do so, one should at first recognize what is the dominant interaction of the problem and, consequently, approximate the other relevant interactions as a small perturbation or, at the first order, neglect them. 
The relevant energy scales in the intermediate core hole state are the spin-orbit coupling, the core hole broadening, and the energy of the elementary excitations under study. 
Typically, the core hole broadening is of the order of $\Gamma\approx1\unit{eV}$, which is larger than the energy scale of most of the elementary excitations which are accessible with RIXS\@. 
For instance, phonons, with typical energies up to $\approx 100\unit{eV}$, and magnetic excitations, up to $\approx400\unit{eV}$, have an energy scale which is approximately one order of magnitude smaller than the core hole broadening. 
On the other hand, $dd$ excitations and charge transfer excitations have an energy scale that is comparable, to some extent, with the core hole broadening. 
Moreover, while in the case of $K$, $L_1$, and $M_1$, the spin-orbit coupling vanishes, at any other edge the spin-orbit coupling easily overcomes the core hole broadening, being one order of magnitude larger, e.g., in the case of $L_{2,3}$ transition metal edges, whereas in some other cases, e.g., at the transition metal $M_{2,3}$ edges, the spin-orbit splitting and the core hole broadening are comparable. 

In what follows, the energy scale of the elementary excitations under study will be considered small compared with the core hole broadening. 
In this case, the intermediate state propagator will reduce to a resonant factor which can be factorized out. 
This approximation scheme is known as fast collision approximation~\cite{Luo1993,vanVeenendaal2006}. 
In particular, the rest of this Section will be concerned with the case of weak spin-orbit coupling, whereas the next Section will generalize the approximation scheme to the case in which the spin-orbit coupling overcomes, or is comparable to, the core hole broadening. 

If the spin-orbit coupling is weak, or vanishing (e.g., at the $K$, $L_1$, and at the $M_1$ edges), and if the energy scale of the elementary excitations under study is small, the core hole broadening is the dominant energy scale, and the core hole state degeneracy can be considered negligible. 
Therefore, at the resonance one has $|E_g+\hbar\omega_{\bf k}-E_n|\ll\Gamma$ for any of the core hole level $E_n$, and the intermediate state propagator in \cref{eq:G0Direct} can be approximated as~\cite{Luo1993,vanVeenendaal2006}
\begin{emphequation}\label{eq:FCA}
	\hat{G}_0
	\approx\sum\limits_n\frac{\ket{n}\bra{n}}{\imath\Gamma}=\frac1{\imath\Gamma}
	,
\end{emphequation}
The intermediate state propagator is thus substituted by a resonance factor, and as a consequence, all the possible intermediate states $\ket{n}$ contribute to the scattering amplitudes, which are therefore determined only by the dipole transitions. 
For this reason, for example, the fast collision approximation yields a vanishing cross section in the case of indirect scattering processes. 
Although the fast collision approximation can be a useful and simple description of direct RIXS scattering, it may be in some cases a very crude approximation, as in the just mentioned case of indirect scattering processes. 
Whereas the fast collision approximation is a zero order approximation of the intermediate state propagator in terms of the core hole broadening, it is natural to extend this expansion in order to include corrections to higher orders. 
To this end, the ultrafast core-hole lifetime expansion~\cite{Brink2005,Brink2006} has been developed, which provides a low-energy expansion of the intermediate state propagator in terms of the core hole broadening. 
If one defines a resonance factor as $\Delta(\omega_{\bf k})=\hbar\omega_{\bf k}-\hbar\omega_{res}+\imath\Gamma$, where $\hbar\omega_{res}\approx E_n$ is the resonant edge of the RIXS process, \cref{eq:G0Direct} can be written as 
\begin{equation}\label{UCL1st}
	\hat{G}_0=
	\frac1{\Delta(\omega_{\bf k})}\left[{1-\dfrac{\hat{\cal H}_0-E_g-\hbar\omega_{res}}{\Delta(\omega_{\bf k})}}\right]^{-1}
	.
\end{equation}
If one assumes that the core hole broadening is the relevant energy scale in the intermediate state, one has $|\hat{\cal H}_0-E_g-\hbar\omega_{res}|\ll{|\Delta(\omega_{\bf k})|}$, and hence one can expand \cref{UCL1st} as a Taylor series and obtain the ultrafast core-hole lifetime expansion of the intermediate state propagator which reads~\cite{Ament2007}
\begin{emphequation}\label{UCL}
	\hat{G}_0=
	\frac1{\Delta(\omega_{\bf k})}\sum\limits_{i=0}^\infty
	\left[{\dfrac{\hat{\cal H}_0-E_g-\hbar\omega_{res}}{\Delta(\omega_{\bf k})}}\right]^i
	,
\end{emphequation}
where at the zero order one regains the fast collision approximation in \cref{eq:FCA}. 

\subsection{Strong spin-orbit coupling in the core hole state}

If the spin-orbit coupling in the core hole intermediate state is the dominant energy scale (e.g., at the $L_{2,3}$ and $M_{2,3}$ transition metal edges), or at least comparable to the core hole broadening, and if the energy scale of the elementary excitations under study is negligible, it is reasonable to expand at first the intermediate state propagator in terms of the spin-orbit coupling. 
After that, one can apply the fast collision approximation separately to the two core hole states in \cref{eq:SpinOrbitLevels} with total angular momentum $j=l\pm1/2$, and eventually factorize out the corresponding resonant factors, neglecting any further substructure of the core hole energy levels. 
In fact, neglecting the spin-orbit coupling leads to vanishing scattering amplitudes of spin-flip excitations, and may fail to reproduce the correct scattering amplitudes of orbital angular momentum transitions, since the dipole transitions alone do not allow spin-flip and orbital momentum transitions in the core hole state. 
Therefore, if the core hole broadening and the spin-orbit coupling energy scales dominate, and the low energy elementary excitation energy is negligible, the intermediate state propagator in \cref{eq:G0Direct} can be decomposed in terms of the total angular momentum eigenstates of the spin-orbit Hamiltonian \ref{eq:SOcoupling} as
\begin{equation}\label{eq:SOexpansion1st}
	\hat{G}_0=
	\sum\limits_{j=l\pm\frac12}
	\frac{\Pi_j}{\hbar(\omega_{\bf k}-\omega_j)+\imath\Gamma}
	,
\end{equation}
where $\hbar\omega_j=E_g+E_0+E_j$ is the resonant edge, $\Pi_j=\sum_{|j_z|<j}\ket{n,j,j_z}\bra{n,j,j_z}$ the projector operator corresponding to the core hole states with total angular momentum $j=l\pm1/2$, and $l>0$ is the azimuthal quantum number of the core hole state. 
The analytical expression of the projector $\Pi_j$ can be obtained directly from its definition, summing over all the eigenstates with total angular momentum $j$ and $-j<j_z<j$. 
However, the projector operator is completely defined by its action on the total angular momentum eigenstates ${\Pi_j}\ket{n,j',j_z}=\delta_{jj'}\ket{n,j',j_z}$, which immediately leads to\footnote{
Hereafter, the factor $\hbar$ in the definition of angular momentum and spin operators will be dropped out, for the sake of simplicity. 
}
\begin{align}\label{eq:ProjectorJ}
	\Pi_{j=l\pm\frac12}=&\pm\frac1{2l+1}\left[{\hat{\bf J}^2-\left(l\mp\frac12\right)\left(l\mp\frac12+1\right)}\right]
	\nonumber\\
	&=\frac{2}{2l+1}\left[2(j-l)\hat{\bf L}\cdot\hat{\bf S}+\frac12\left({j+\frac12}\right)\right]
	.
\end{align}

\begin{table}[t]
	\centering
\begin{tabular*}{\textwidth}{c|c| l @{\extracolsep{\fill}} l}
		spin-orbit strength &resonant edge&\multicolumn{2}{c}{
		$\hat{G}_0=\alpha(\omega_{\bf k})\hat{\bf L}\cdot\hat{\bf S}+\beta(\omega_{\bf k})$}\\[1mm]
		\hline&&&\\[-3mm]
	$\Delta E_{so}\gg \Gamma$ & $\omega_{j=l\pm1/2}$ & 
	$\alpha(\omega_{\bf k})=\pm\dfrac1{\imath\Gamma}$ & $\beta(\omega_{\bf k})=\dfrac{j+1/2}{2\imath\Gamma}$ \\[3mm]
	$\Delta E_{so}\ll \Gamma$ & $\omega_{l-1/2}\simeq\omega_{l+1/2}$ & 
	$\alpha(\omega_{\bf k})=0$ & $\beta(\omega_{\bf k})=\dfrac{2l+1}{2\imath\Gamma}$ \\
	\end{tabular*}
\caption[Resonant functions in the limits of strong and weak spin-orbit coupling]{
Resonant functions $\alpha(\omega_{\bf k})$ and $\beta(\omega_{\bf k})$ as defined in \cref{eq:ResonantFunctions} which appear in the intermediate state propagator in \cref{eq:SOexpansion} in the limits of strong ($\Delta E_{so}\gg \Gamma$) and weak $\Delta E_{so}\ll \Gamma$ spin-orbit coupling. 
}
	\label{tab:FCA-SO}
\end{table}

Hence, summing over the two total angular momentum states, the intermediate state propagator in \cref{eq:SOexpansion1st} can be written as
\begin{emphequation}\label{eq:SOexpansion}
	\hat{G}_0=
	\alpha(\omega_{\bf k})\hat{\bf L}\cdot\hat{\bf S}
	+
	\beta(\omega_{\bf k})
	,
\end{emphequation}
which represents the intermediate state propagator expansion in terms of the spin-orbit coupling, and where the resonance functions $\alpha(\omega_{\bf k})$ and $\beta(\omega_{\bf k})$ are defined up to constant prefactors as
\begin{equation}\label{eq:ResonantFunctions}
	\begin{aligned}
	\alpha(\omega_{\bf k})&=
	\dfrac1{\Delta_{l+\frac12}(\omega_{\bf k})}-
	\dfrac1{\Delta_{l-\frac12}(\omega_{\bf k})}
	,\\
	\beta(\omega_{\bf k})&=
	\dfrac{(l+1)/2}{\Delta_{l+\frac12}(\omega_{\bf k})}+
	\dfrac{l/2}{\Delta_{l-\frac12}(\omega_{\bf k})}
	,
	\end{aligned}
\end{equation}
where $\Delta_j(\omega_{\bf k})={\hbar(\omega_{\bf k}-\omega_j)+\imath\Gamma}$ is the resonance factor corresponding to the resonant edge $\omega_j$. 
As stated before, this expansion is valid only in the case where the energy scale of the elementary excitations under study are negligible respect to the spin-orbit coupling of the core hole state and to the core hole broadening. 

In the case where the spin-orbit coupling overcomes the core hole broadening, i.e., if $\Delta E_{so}\gg \Gamma$, and the incident photon energy is tuned to one of the two absorption edges $\omega_{\bf k}=\omega_j$ with total angular momentum $j=l\pm\frac12$, the corresponding term in \cref{eq:SOexpansion} dominates while the other become negligible since $|\Delta E_{so}+\imath\Gamma|\gg|\Gamma|$, and therefore one has
\begin{equation}\label{eq:SOexpansion_H_SO>Gamma}
	\hat{G}_0\approx
	\begin{cases}
	\dfrac1{\imath\Gamma}\left({{-\hat{\bf L}\cdot\hat{\bf S}+\dfrac{l}2}}\right)
	&\text{ for } \omega_{\bf k}=\omega_{l-\frac12},\\
	&\\
	\dfrac1{\imath\Gamma}\left({{\hat{\bf L}\cdot\hat{\bf S}+\dfrac{l+1}2}}\right)
	&\text{ for } \omega_{\bf k}=\omega_{l+\frac12},\\
	\end{cases}
\end{equation}
which corresponds to the resonant functions $\alpha(\omega_{\bf k})=\pm1/{\imath\Gamma}$ and $\beta(\omega_{\bf k})=\frac{(j+1/2)}/{2\imath\Gamma}$ in \cref{eq:SOexpansion}. 
On the other hand, if the spin-orbit coupling is negligible compared with the core hole broadening, i.e., if $ \Gamma\gg\Delta E_{so}$, both terms in \cref{eq:SOexpansion} contribute and since in this case one has $|\hbar(\omega_{\bf k}-\omega_j)+\imath\Gamma|\approx|\Gamma|$, the intermediate propagator simply reduces to the form $\hat{G}_0\approx1/{\imath\Gamma}$ as in \cref{eq:FCA}, which corresponds to $\alpha(\omega_{\bf k})=0$ and $\beta(\omega_{\bf k})=\frac{(2l+1)}/{2\imath\Gamma}$ in \cref{eq:SOexpansion}. 
Eventually, if the core hole broadening and the spin-orbit coupling are comparable, both terms in \cref{eq:SOexpansion} have to be considered, and no further approximation can be employed. 
In \cref{fig:FCA-SO} the values of the resonant functions $\alpha(\omega_{\bf k})$ and $\beta(\omega_{\bf k})$ are summarized in the case of strong ($\Delta E_{so}\gg \Gamma$) and weak $\Delta E_{so}\ll \Gamma$ spin-orbit coupling. 

\begin{figure}[t]
	\centering
	\subfigure[$\Gamma\ll\Delta E_{so}$]{\includegraphics[width=.32\textwidth]{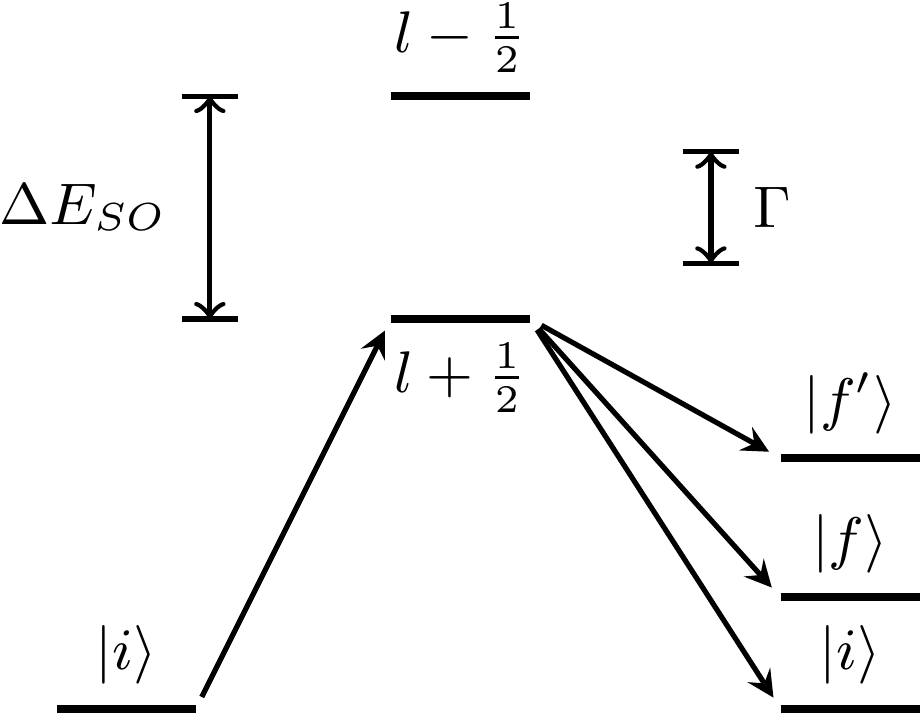}\label{fig:FCA-SO-SO>G}}
	\,
	\subfigure[$\Gamma\sim\Delta E_{so}$]{\includegraphics[width=.32\textwidth]{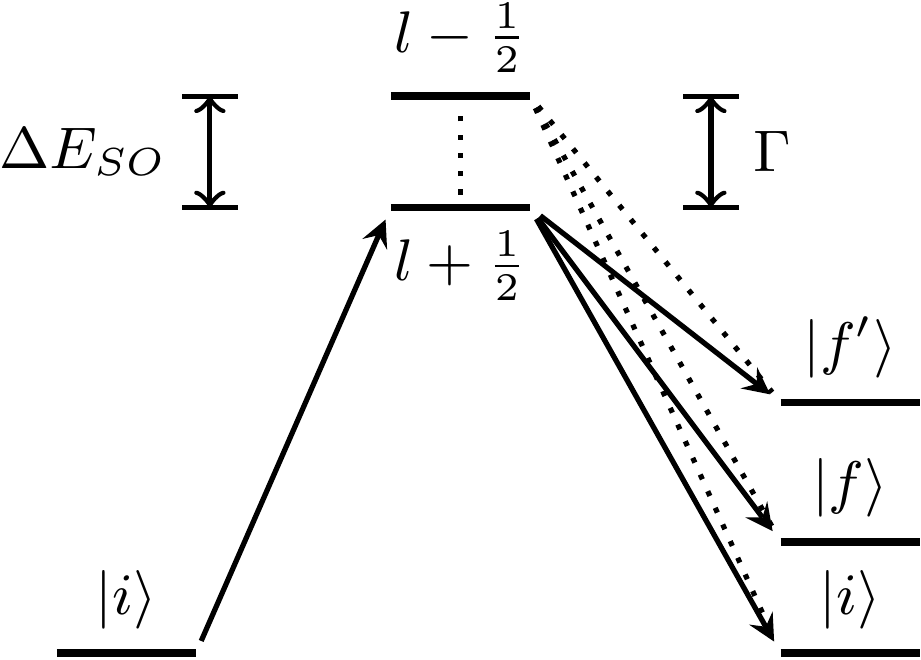}\label{fig:FCA-SO-SO=G}}
	\,
	\subfigure[$\Gamma\gg\Delta E_{so}$]{\includegraphics[width=.32\textwidth]{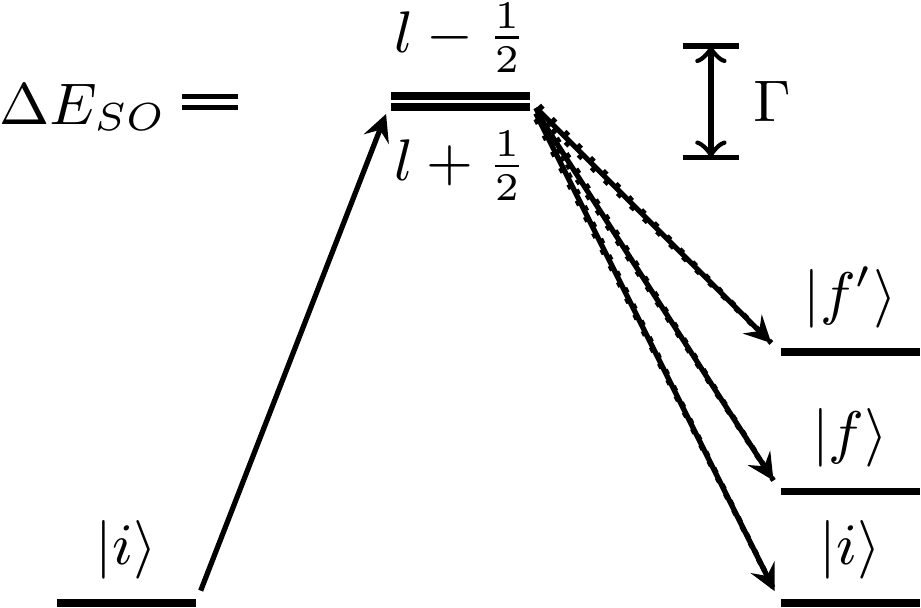}\label{fig:FCA-SO-G>SO}}
	\caption[RIXS scattering intermediate state]{
In the short living intermediate state, virtual transitions with $\Delta E\approx\Gamma$ have a finite probability. 
(a) In the case $\Delta E_{so}\gg \Gamma$, the core hole state has a definite energy, since the transition energy $\Delta E_{so}$ between states with different total angular momentum $j$ is much larger than the core hole broadening $\Gamma$. 
(b) If $\Delta E_{so}\approx\Gamma$ instead, transitions between states with different total angular momentum are allowed, and both states contribute to the scattering amplitude. 
(c) If $\Delta E_{so}\ll \Gamma$, the two core hole states contribute with the same weight. 
}
	\label{fig:FCA-SO}
\end{figure}

The different physical scenarios which arise in the cases above can be understood in terms of uncertainty relations and quantum interference of the core hole states. 
In fact, the intermediate state is a short living non-stationary state, and for this reason it cannot have a definite energy. 
According to the time-energy uncertainty relation (which is the analogous of the Heisenberg principle for position and momentum), the uncertainty in the core hole state energy $\Delta E$ is related to the core hole lifetime $\Delta t$ and, as a consequence, to the core hole broadening $\Gamma$ by
\begin{equation}
	\Delta E \Delta t \approx \frac\hbar2 \quad\Rightarrow\quad \Delta E \approx \Gamma
	.
\end{equation}
In the case $\Delta E_{so}\gg \Gamma$, the core hole state has a definite energy and total angular momentum, since the uncertainty $\Delta E$ of the core hole energy is much smaller than the transition energy $\Delta E_{so}$ between states with different total angular momentum $j$, as shown in \cref{fig:FCA-SO-SO>G}. 
In this case, the intermediate state propagator in \cref{eq:SOexpansion_H_SO>Gamma} do not allow transitions between the two core hole states with different $j=l\pm\frac12$, whereas spin-flip and orbital angular momentum transitions are allowed, since neither spin nor orbital angular momentum are good quantum numbers of the core hole states. 
Conversely, if $\Delta E_{so}\approx\Gamma$, the uncertainty principle allows transitions between core hole states with different energy and total angular momentum, and these different states will contribute to the scattering amplitude and interfere with each other, as shown in \cref{fig:FCA-SO-SO=G}. 
In particular, if $\Delta E_{so}\ll \Gamma$, the two core hole states contribute with the same weight, as in \cref{fig:FCA-SO-G>SO}. 
In this case, contributions which correspond to a spin-flip or to an orbital angular momentum transition will interfere destructively, resulting in a vanishing resonant function $\alpha(\omega_{\bf k})=0$ in \cref{eq:ResonantFunctions}, and therefore the intermediate state propagator reduces to a simple resonance factor as in \cref{eq:FCA}.

\subsection{The dynamical structure factor and RIXS}
\label{sec:RIXS-DSF}

\label{sec:DSF_W}

In the case of direct RIXS, the local operator in \cref{eq:RIXSOperatorLocal} for a given incident and scattered photon polarizations can be expanded in the basis of the local electron spin $\tau$ as
\begin{equation}\label{eq:DSF_W_local}
	\hat{O}_i=
	\sum_{\psi\psi'\tau\tau'}
	W^{\tau\tau'}_{{\bf e}{\bf e}'}(\psi,\psi') c^\dag_{i\psi'\tau'}c^\nod_{i\psi\tau}
	,
\end{equation}
where the operators $c^\dag_{i\psi\tau}$ and $c^\nodag_{i\psi\tau}$ create and annihilate an electron state at the lattice site $i$ with spin $\tau$, and where $\psi$ denotes any other degree of freedom, e.g., the orbital occupancy. 
The local transition amplitudes can be calculated by using $W^{\tau\tau'}_{{\bf e}{\bf e}'}(\psi,\psi')=\bra{\psi'\tau'}\hat{O}_i\ket{\psi\tau}$. 
By a Fourier transform of the local electron operators $c^\nod_{i\psi\tau}$ and $c^\dag_{i\psi\tau}$ and using the definition of the RIXS operator $\hat{\cal O}$ in \cref{eq:RIXSOperator}, one obtains
\begin{equation}\label{eq:DSF_W}
	\hat{\cal O}=
	\frac1N\ 
	\sum_{\psi\psi'\tau\tau'}
	W^{\tau\tau'}_{{\bf e}{\bf e}'}(\psi,\psi') c^\dag_{{\bf k+q}\psi'\tau'}c^\nod_{{\bf k}\psi\tau}
	.
\end{equation}
The expansion of the RIXS operators in \cref{eq:DSF_W_local,eq:DSF_W} relate the RIXS scattering process to the local transition amplitudes $W^{\tau\tau'}_{{\bf e}{\bf e}'}(\psi,\psi')$, which depends only on the photon polarizations and on the local degrees of freedom of the system. 
In particular \cref{eq:DSF_W} relates the \emph{bulk} response of the scattering process to the \emph{local} response. 
Moreover, using the RIXS operator expansion of \cref{eq:DSF_W}, the RIXS cross section in \cref{eq:KramersHeisenberg} can be rewritten in terms of correlation functions of the valence electrons, which characterize the excitation spectra of the system. 

Let us specialize however, for the sake of simplicity, to the case of a system with local spin $s=1/2$, and consider only local spin transitions within the same orbital state $\phi$. 
In this case, as one can see from \cref{eq:SOexpansion}, the spin-orbit coupling in the intermediate state of a direct RIXS process can flip the spin of the core electron, and as a consequence, the final and the initial spin of the valence electron state can differ by $\Delta\tau=\tau'-\tau=0,\pm1/2$. 
Therefore \cref{eq:DSF_W_local} can be simplified by introducing the density and the spin operators respectively as $\hat{\rho}_{i\psi}=\sum_\tau c^\dag_{i\psi\tau}c^\nod_{i\psi\tau}$ and $\hat{S}^\lambda_{i\psi}=\sum_{\tau\tau'}\sigma^\lambda_{\tau\tau'} c^\dag_{i\psi\tau}c^\nod_{i\psi\tau'}$ with $\sigma^\lambda=\sigma^x, \sigma^y, \sigma^z$ the Pauli matrices, in the form
\begin{emphequation}\label{eq:DSF_W_local_12}
	\hat{O}_i=
	W^{0}_{{\bf e}{\bf e}'}			\hat{\rho}_{i\psi}+
	{\bf W}_{{\bf e}{\bf e}'}	\cdot	\hat{\bf S}_{i\psi}
	,
\end{emphequation}
where the new transition amplitudes $W^{0}_{{\bf e}{\bf e}'}$ and ${\bf W}_{{\bf e}{\bf e}'}$ are defined by
\begin{equation}\label{eq:W_general}
	W^\lambda_{{\bf e}{\bf e}'}=\frac12\ \sum_{\tau\tau'} \sigma^\lambda_{\tau\tau'}W^{\tau\tau'}_{{\bf e}{\bf e}'}(\psi,\psi),
\end{equation}
where $\lambda=0,x,y,z$ and $\sigma^0$ is the identity matrix.
Again, by a Fourier transform of the local electron operators and using the definition of the RIXS operator $\hat{O}$ in \cref{eq:RIXSOperator}, or one obtains in this case
\begin{align}\label{eq:DSF_W_12}
	\hat{\cal O}=\frac1N
	&\left(
	W^{0}_{{\bf e}{\bf e}'}			\hat{\rho}_{{\bf q}\psi}+
	{\bf W}_{{\bf e}{\bf e}'}	\cdot	\hat{\bf S}_{{\bf q}\psi}
	\right)
	,
\end{align}
where $\hat{\rho}_{{\bf q}\psi}=\sum_{{\bf k}\tau} c^\dag_{{\bf k+q}\psi\tau}c^\nod_{{\bf k}\psi\tau}$ and $\hat{S}^\lambda_{{\bf q}\psi}=\sum_{{\bf k}\tau\tau'}\sigma^\lambda_{\tau\tau'} c^\dag_{{\bf k+q}\psi\tau}c^\nod_{{\bf q}\psi\tau'}$. 
Therefore, the RIXS cross section in \cref{eq:KramersHeisenberg}, using the expansion of the RIXS operator in \cref{eq:DSF_W_12}, can be rewritten in a compact form as 
\begin{emphalign}\label{eq:KramersHeisenberg_DSF}
	\frac{d^2 \sigma}{d\omega d\Omega}\propto
	\sum_{\lambda\mu}
	W^{\lambda}_{{\bf e}{\bf e}'}
	W^{\mu*}_{{\bf e}{\bf e}'}
	\chi^{\lambda\mu}
	,
\end{emphalign}
where the correlation functions $\chi^{\lambda\mu}$ are defined as
\begin{align}\label{eq:DSF_correlation}
	\chi^{\lambda\mu}=&\sum_f
	\bra{g}\hat{\rho}^{\mu}_{{\bf q}\psi}\ket{f}\bra{f}\hat{\rho}^{\lambda}_{{\bf q}\psi}\ket{g}\delta (E_g+\hbar\omega-E_f)
	\nonumber\\
	=&\lim_{\eta\rightarrow0}
	\bra{g}\hat{\rho}^{\mu}_{{\bf q}\psi}
	\left(\sum_f\frac{\ket{f}\bra{f}}{E_g+\hbar\omega-E_f+\imath\eta}\right)
	\hat{\rho}^{\lambda}_{{\bf q}\psi}\ket{g}
	,
\end{align}
with $\hat{\rho}^{0}_{{\bf q}\psi}\equiv\hat{\rho}_{{\bf q}\psi}$ and $\hat{\rho}^\lambda_{{\bf q}\psi}\equiv\hat{S}^\lambda_{{\bf q}\psi}$ for $\lambda=x,y,z$. 
In particular, for $\lambda=\mu$ the correlation functions in \cref{eq:DSF_correlation} coincide with the charge and spin DSF of the valence electrons, respectively
\begin{align}\label{eq:DSF_DSF}
	\chi^{0}\equiv\chi^{00}=&\sum_f
	|\bra{f}\hat{\rho}_{{\bf q}\psi}\ket{g}|^2\delta (E_g+\hbar\omega-E_f)
	,\nonumber\\
	\chi^{\lambda}\equiv\chi^{\lambda\lambda}=&\sum_f
	|\bra{f}\hat{S}^{\lambda}_{{\bf q}\psi}\ket{g}|^2\delta (E_g+\hbar\omega-E_f)
	\qquad \lambda=x,y,z
	,
\end{align}
Therefore, direct RIXS is a probe of the correlation functions and of the charge and spin DSF of the system under study. 

\subsection{Graphical representations of the RIXS operator}
\label{sec:RIXS-Diagram}

\label{sec:pd_Transitions}
\begin{figure}[t!]
\centering
\includegraphics[scale=1.7]{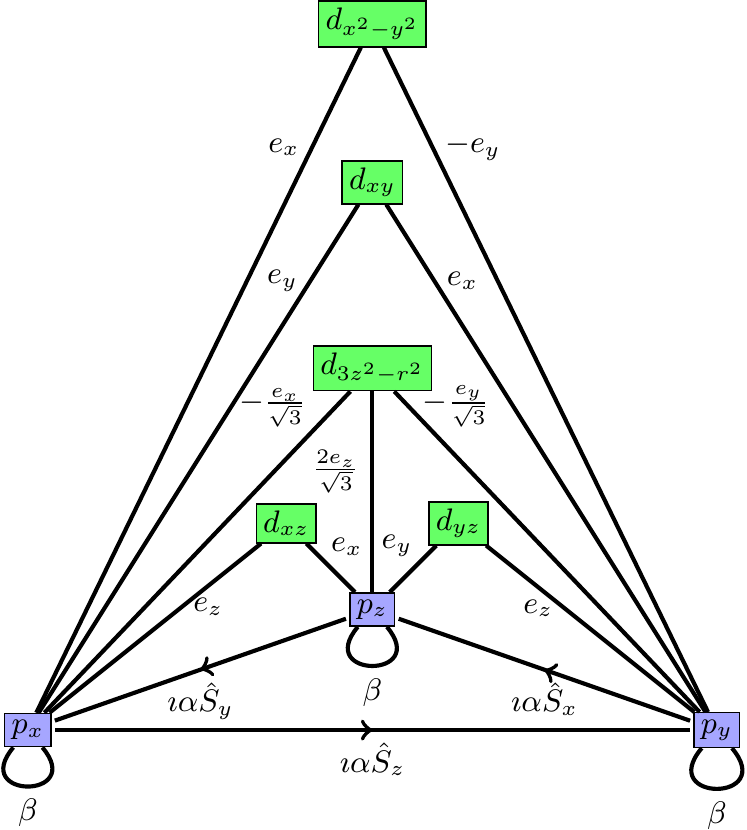}
 \caption[Schematic representation of the direct RIXS operator]{
Schematic representation of the direct RIXS operator $\hat{O}_i$ on a single site at $L_{2,3}$ or $M_{2,3}$ edges.
To calculate the matrix elements of the operator between the initial and final $d$ electron states, one needs to sum over all possible paths connecting them via a three step process, and multiplying at each step by the factor indicated, which is respectively proportional to: (i) a component of the incident photon polarization $\bf e$, (ii) one of the resonant functions $\beta=\beta(\omega_{\bf k})$, or $\alpha=\alpha(\omega_{\bf k})$ times a component of the spin operator $\hat{\bf S}$, with a plus or minus sign respectively for steps along or opposite to the direction of the arrows (iii) the complex conjugate of a component of the scattered polarization $\bf e'$. 
The three steps correspond respectively to the dipole transition $p\rightarrow d$ (annihilation of a core $p$ electron and creation of a $d$ electron), the propagation of the core hole $p\rightarrow p$ (creation and annihilation of $p$ electrons), and to the dipole transition $d\rightarrow p$ (annihilation of a $d$ electron and creation of a $p$ electron).
}
\label{fig:RIXSoperatorDirect_l1}
\end{figure}

In this Section, we will specialize the expression for the local RIXS scattering operator in \cref{eq:RIXSOperatorLocal} in the case of direct scattering, at transition metal $L_{2,3}$ and $M_{2,3}$ absorption edges. 
To do so, we will employ the dipole approximation in \cref{eq:DipoleApproximation}, neglect indirect processes, and assume a core hole propagator in the form of \cref{eq:G0Direct}, i.e., considering the core hole broadening and the spin-orbit coupling as the dominant energy scales. 
Moreover, since electron states in condensed matter have in general a lower symmetry than spherical, due to, e.g., the crystal field generated by the surrounding charge distribution, we will express the relevant single particle operators in terms of tesseral harmonics states, defined in \cref{sec:TesseralHarmonics}, which correspond to the usual set of electronic orbitals employed to describe $p$, $d$, and $f$ electrons in condensed matter systems. 
This will lead to a simple diagrammatic representation of the RIXS scattering operator. 

\label{sec:pd_Transitons}

In the case of direct RIXS at the $L_{2,3}$ and $M_{2,3}$ metal edges, the core hole is in a $np$ shell ($l=1$), and the direct scattering occurs via the dipole allowed $np\leftrightarrow n'\!d$ transitions.
The relevant energy scales are the spin-orbit coupling and the core hole broadening and, therefore, using \cref{eq:SOexpansion} the local RIXS scattering operator in \cref{eq:RIXSOperatorLocal} becomes
\begin{emphequation}\label{eq:RIXSoperatorDirect_l12}
	\hat{O}_i=
	\sum\limits_{\lambda\mu} e_\mu^{\prime*} 
	\ \hat{D}^\dag_\mu
	\left[
	\alpha(\omega_{\bf k})
	\hat{\bf L}\cdot\hat{\bf S}
	+
	\beta(\omega_{\bf k})
	\right]
	\hat{D}_\lambda
	e_\lambda
	,
\end{emphequation}
where the resonance functions $\alpha(\omega_{\bf k})$ and $\beta(\omega_{\bf k})$ are defined as in \cref{eq:ResonantFunctions}. 

Using the explicit expression of the orbital angular momentum operator in the corresponding core hole shell and of the dipole operator for the dipole allowed $\Delta l=\pm1$ transitions in \cref{tab:MomentumDipole} of \cref{sec:WignerEckartAppendix}, one obtains the general expression of the local RIXS operator in \cref{eq:RIXSoperatorDirect_l12} for direct RIXS at the $L_{2,3}$ and $M_{2,3}$ metal edges. 
The action of the local RIXS operator for direct scattering $np\leftrightarrow n'\!d$ transitions ($L_{2,3}$ and $M_{2,3}$ edges) is shown in the diagram in \cref{fig:RIXSoperatorDirect_l1}. 

\subsection[$dd$ excitations in cuprates]{$dd$ excitations in cuprates}

To illustrate how to calculate the direct RIXS cross section via \cref{eq:KramersHeisenbergLocal}, in particular using the local transition amplitudes in \cref{eq:DSF_W_local} and the diagrammatic representation of the local RIXS operator in \cref{fig:RIXSoperatorDirect_l1}, we will evaluate in this Section the RIXS spectra at the $L_{2,3}$ copper edges of $dd$ excitation in cuprates.
The energy of the electron levels of a single ion in a crystal are determined by the crystal field, which is induced by the charge distribution of the surrounding ions. 
As a consequence, the degeneracy of the electron levels within the same shell (same principal and azimuthal quantum numbers) is partially or totally removed, depending on the characteristic symmetry of the crystal and on the crystal field strength. 
Therefore, valence electrons occupy different orbital states, with different energies and symmetry properties. 
Crystal field excitations concern local transitions between non-degenerate electron levels of the same ion within the valence shell. 
In particular, $dd$ excitations are crystal field transitions between different $d$ orbitals, i.e., non-degenerate $3d$ electron levels of the same ion, with energies which are of the order of electronvolts in transition metal oxydes. 
Due to the strong spin-orbit coupling in the core hole state, the direct RIXS scattering couples directly with the spin and orbital degrees of freedom of the local ions. 
In fact, $dd$ excitations have been observed in many systems, and in particular in the paradigmatic charge transfer insulator NiO~\cite{Ishii2001} and in cuprates~\cite{Ghiringhelli2004,Ghiringhelli2007}. 
Since crystal field excitations are well localized excited states, they have little or no dispersion and can be well understood in a single ion approximation, i.e., by considering the scattering amplitude as a sum of contributions which do not depend on the ion site. 
In this case, for small transferred momenta, and for an energy loss equal to the excitation energy of the particular orbital state considered, the RIXS intensity in \cref{eq:KramersHeisenbergLocal} reduces to 
\begin{emphequation}
\label{eq:RIXS_dd}
	I({\bf q\simeq0},\hbar\omega=E_{\psi'})\propto
	\sum_{\tau\tau'}	\left|W^{\tau\tau'}_{{\bf e}{\bf e}'}(\psi,\psi')\right|^2
	,
\end{emphequation}
where $\psi$ and $\psi'$ are respectively the orbital ground state of the local ion and the excited state with energy $E_{\psi'}$, and where one assumes that the different spin states are nearly degenerate compared to the crystal field splitting. 
Therefore, the RIXS cross section of $dd$ excitation is completely determined by the local transition amplitudes $W^{\tau\tau'}_{{\bf e}{\bf e}'}(\psi,\psi')$. 

\begin{figure}[ht!]
	\includegraphics[width=1\textwidth]{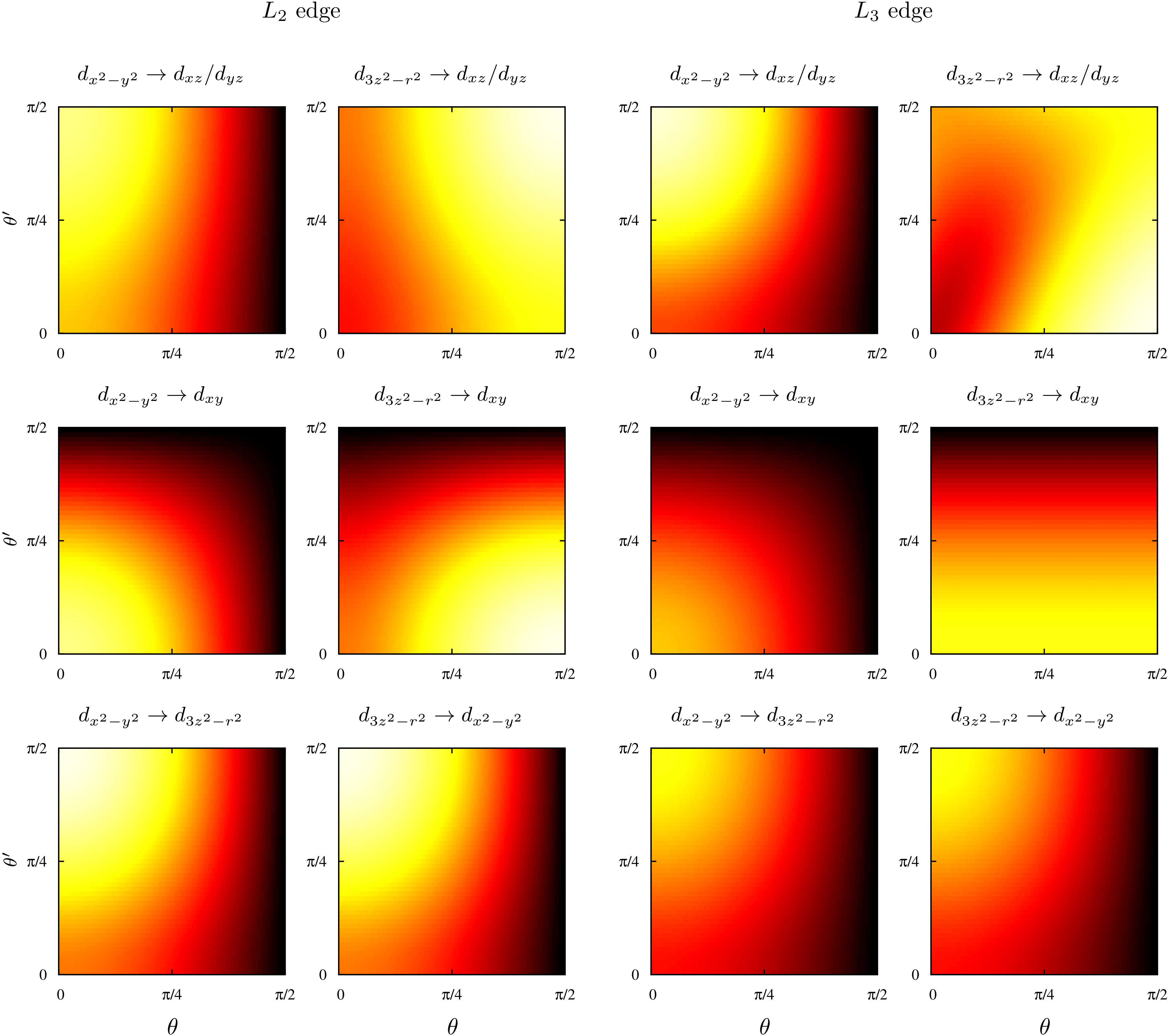}
\caption[RIXS spectra of $dd$ excitations in cuprates at the copper $L_{2,3}$ edges]{
RIXS spectra of $dd$ excitations in cuprates at the copper $L_{2,3}$ edges, calculated via \cref{eq:RIXS_dd}, corresponding to orbital transitions $\psi\rightarrow\psi'$ for two orbital ground states $\psi=d_{x^2-y^2}, d_{3z^2-r^2}$ and for different excited states $\psi'$, as a function of the incident and scattered photon polarization directions. 
Intensities are summed over the spin degrees of freedom. 
The incident and scattered polarization angles $\theta$ and $\theta'$ are considered between the polarization direction and the $z$ lattice axis, such that ${\bf e}=(\cos{\theta}/\sqrt{2},\cos{\theta}/\sqrt{2},\sin{\theta})$ and ${\bf e}'=(\cos{\theta'}/\sqrt{2},\cos{\theta'}/\sqrt{2},\sin{\theta'})$. 
}
\label{fig:dd_CuO}
\end{figure}

In cuprates, the copper atoms are doubly ionized and form an orthogonal complex with the six surrounding ligand oxygen ions, embedded in a perovskite crystal lattice. 
As a consequence, the electronic configuration $3d^9$ has one single hole in the valence shell, generally in one of the $e_g$ orbitals for cubic and tetragonal perovskites ($O_h$ or $D_{4h}$ symmetries in \cref{fig:PerovskitePnictide}), or in a linear combinations of the two orbitals $d_{x^2-y^2}$ and $d_{3z^2-r^2}$ in orthorhombic perovskites ($D_{2h}$ symmetry in \cref{fig:PerovskitePnictide}). 
Therefore, it is much simpler to describe the RIXS scattering process in these systems in the hole picture. 
Within the hole picture, the incident photon excites the hole in the $3d$ shell to the core $2p$ shell, which consequently decays into a scattered photon and a hole back to the $3d$ shell, possibly in a different orbital state in the case of inelastic scattering. 
In the light of this, it is straightforward to calculate via \cref{fig:RIXSoperatorDirect_l1} the local transition amplitudes $W^{\tau\tau'}_{{\bf e}{\bf e}'}(\psi,\psi')$ between any possible initial and excited states $\ket{\tau\psi}$ and $\ket{\tau'\psi'}$, which are given in \cref{tab:W_CuO}. 

The RIXS cross section of $dd$ excitations in cuprates can be calculated via \cref{eq:RIXS_dd} using the local transition amplitudes in \cref{tab:W_CuO}, considering an orbital ground state with a hole in the $d_{s^2-y^2}$ or in the $d_{3z^2-r^2}$ orbital, and with the final state in any of the remnant orbital states. 
In \cref{fig:dd_CuO} are shown the direct RIXS spectra of $dd$ excitations in cuprates at the copper $L_{2,3}$ edges, for the two orbital ground states considered, as a function of the incident and scattered photon polarizations. 
The polarization directions are chosen to have one component along the $z$ axes another along the bisector line of the $x$ and $y$ axes. 
As one can see, the polarization dependence of the RIXS spectra of $dd$ excitations is different for the two different orbital ground state considered, in particular for $dd$ excitations corresponding to a transition to any of the $t_{2g}$ orbitals. 
This allow one in principle to discriminate between different orbital ground states by looking at the excitation spectra of the system. 

\begin{landscape}
\begin{table}[t]
	\centering
	\scriptsize
\newcommand{\ID}{
$
\begin{array}{c|c}
\tau=\tau'&\psi=\psi'\\[3mm]
\hline
d_{x^2-y^2}&	\beta \left(e_x e_x^{\prime*}+e_y e_y^{\prime*}\right)					\\
&		+\frac{\imath\alpha}{2} \sigma_z \left(e_y e_x^{\prime*}-e_x e_y^{\prime*}\right)	\\[3mm]
d_{xz}&		\beta \left(e_x e_x^{\prime*}+e_z e_z^{\prime*}\right) 					\\[3mm]
d_{3z^2-r^2}&	\frac{\beta}{3} \left(e_x e_x^{\prime*}+e_y e_y^{\prime*}+4 e_z e_z^{\prime*}\right)	\\
&		+\frac{\imath\alpha}{6} \sigma_z \left(e_x e_y^{\prime*}-e_y e_x^{\prime*}\right)	\\[3mm]
d_{yz}&		\beta \left(e_y e_y^{\prime*}+e_z e_z^{\prime*}\right)					\\[3mm]
d_{xy}&		\beta \left(e_x e_x^{\prime*}+e_y e_y^{\prime*}\right)					\\
&		+\frac{\imath\alpha}{2} \sigma_z \left(e_y e_x^{\prime*}-e_x e_y^{\prime*}\right)	\\
\multicolumn{2}{c}{\quad}\\
\tau\ne\tau'&\psi=\psi'\\[3mm]
\hline
d_{x^2-y^2}&	0\\[3mm]
d_{xz}&		\frac{\alpha}{2} \sigma_z \left(e_z e_x^{\prime*}-e_x e_z^{\prime*}\right)	\\[3mm]
d_{3z^2-r^2}&	\frac{\alpha}{3} \sigma_z \left(e_z e_x^{\prime*}-e_x e_z^{\prime*}\right)	\\
&		+\frac{\imath\alpha}{3} \left(e_z e_y^{\prime*}-e_y e_z^{\prime*}\right)	\\[3mm]
d_{yz}&		\frac{\imath\alpha}{2} \left(e_z e_y^{\prime*}-e_y e_z^{\prime*}\right)		\\[3mm]
d_{xy}&		0\\
\multicolumn{2}{c}{\quad}\\[-1.1mm]
\end{array}
$
}
\newcommand{\dd}{
$
\begin{array}{c|cccc}
\tau=\tau'&d_{xz}&d_{3z^2-r^2}&d_{yz}&d_{xy}\\[3mm]
\hline
d_{x^2-y^2}
&
 \beta e_x e_z^{\prime*}+\frac{\imath\alpha}{2} \sigma_z e_y e_z^{\prime*} &
 \frac{\beta}{\sqrt{3}} \left(e_y e_y^{\prime*}-e_x e_x^{\prime*}\right) &
 -\beta e_y e_z^{\prime*}+\frac{\imath\alpha}{2} \sigma_z e_x e_z^{\prime*} &
 \beta \left(e_x e_y^{\prime*}-e_y e_x^{\prime*}\right)\\
&
&
-\frac{\imath\alpha}{2\sqrt{3}} \sigma_z \left(e_x e_y^{\prime*}+e_y e_x^{\prime*}\right) &
&
+\frac{\imath\alpha}{2} \sigma_z \left(e_x e_x^{\prime*}+e_y e_y^{\prime*}\right) \\[3mm]
d_{xz}
&
&
 \frac{\beta}{\sqrt{3}} \left(2 e_x e_z^{\prime*}-e_z e_x^{\prime*}\right)&
 \beta e_x e_y^{\prime*}+\frac{\imath\alpha}{2} \sigma_z e_z e_z^{\prime*} &
 \beta e_z e_y^{\prime*}+\frac{\imath\alpha}{2} \sigma_z e_z e_x^{\prime*} \\
&
&
-\frac{\imath\alpha}{2\sqrt{3}} \sigma_z e_z e_y^{\prime*} &
&
\\[3mm]
d_{3z^2-r^2}
&
&
&
 \frac{\beta}{\sqrt{3}} \left(2 e_z e_y^{\prime*}-e_y e_z^{\prime*}\right)&
 -\frac{\beta}{\sqrt{3}} \left(e_x e_y^{\prime*}+e_y e_x^{\prime*}\right)\\
&
&
&
-\frac{\imath\alpha}{2\sqrt{3}} \sigma_z e_x e_z^{\prime*} &
+\frac{\imath\alpha}{2\sqrt{3}} \sigma_z \left(e_y e_y^{\prime*}-e_x e_x^{\prime*}\right) \\[3mm]
d_{yz}
&
&
&
&
 \beta e_z e_x^{\prime*}-\frac{\imath\alpha}{2} \sigma_z e_z e_y^{\prime*} 
\\
\multicolumn{5}{c}{\quad}\\[3.8mm]
\tau\ne\tau'&d_{xz}&d_{3z^2-r^2}&d_{yz}&d_{xy}\\[3mm]
\hline
d_{x^2-y^2}
&
 \frac{\alpha}{2} (\sigma_z e_x-\imath e_y) e_x^{\prime*} &
 \frac{\alpha}{\sqrt{3}} (\sigma_z e_x-\imath e_y) e_z^{\prime*} &
 \frac{\alpha}{2} (\sigma_z e_x-\imath e_y) e_y^{\prime*} &
 0 \\
\\[3mm]
d_{xz}
&
&
 \frac{\alpha}{2 \sqrt{3}} \sigma_z \left(e_x e_x^{\prime*}+2 e_z e_z^{\prime*}\right) &
 \frac{\alpha}{2} \left(\sigma_z e_z e_y^{\prime*}-\imath e_x e_z^{\prime*}\right) &
 -\frac{\alpha}{2} e_x \left(\sigma_z e_y^{\prime*}+\imath e_x^{\prime*}\right) \\
&
&
+ \frac{\imath\alpha}{2 \sqrt{3}} e_x e_y^{\prime*} &
&
\\[3mm]
d_{3z^2-r^2}
&
&
&
 -\frac{\alpha}{2 \sqrt{3}} \sigma_z e_x e_y^{\prime*} &
 -\frac{\alpha}{\sqrt{3}} e_z \left(\sigma_z e_y^{\prime*}+\imath e_x^{\prime*}\right) \\
&
&
&
 -\frac{\imath\alpha}{2 \sqrt{3}} \left(2 e_z e_z^{\prime*}+e_y e_y^{\prime*}\right) &
\\[3mm]
d_{yz}
&
&
&
&
 -\frac{\alpha}{2} e_y \left(\sigma_z e_y^{\prime*}+\imath e_x^{\prime*}\right) 
\\[-.2mm]
\end{array}
$
}
\begin{tabular}{lr}
\ID&\dd
\end{tabular}
	\caption[Transition amplitudes at the copper $L_{2,3}$ edges]{
Transition amplitudes $W^{\tau\tau'}_{\bf e e'}(\psi\psi')=\bra{\psi'\tau'}\hat{O_i}\ket{\psi\tau}$ at the Cu$^{+2}$ $L_{2,3}$ edges corresponding to the matrix element of the local RIXS operator $\hat{O}_i$ between states with the same orbital ($\psi=\psi'$) or between different initial and the final orbital states $\psi\ne\psi'$ (rows and columns on the table), and with the same spin ($\tau=\tau'$) or different spin ($\tau\ne\tau'$), as a function of the components of the incident and scattered photon polarizations $e_\lambda$, $e_\lambda'$. 
The resonant functions $\alpha=\alpha(\omega_{\bf k})$ and $\beta=\beta(\omega_{\bf k})$ are defined as in \cref{eq:ResonantFunctions}, while $\sigma_z=\pm1$ respectively for up and down spin of the initial state. 
Please note that one has $W^{\tau\tau'}_{\bf e e'}(\psi\psi')=W^{\tau'\tau}_{\bf e' e}(\psi'\psi)^*$. 
}
	\label{tab:W_CuO}
\end{table}
\end{landscape}

}
\chapter[Unraveling orbital correlations via magnetic RIXS]{Unraveling orbital correlations via magnetic RIXS\footnotemark{}}
\footnotetext{Part of this chapter has been published in \onlinecite{Marra2012}}
{
\label{ch:Orbitals}

Although orbital degrees of freedom are a factor of fundamental importance in strongly correlated transition metal compounds, orbital correlations and dynamics remain very difficult to access, in particular by neutron scattering. 
However, RIXS has been proven successful in measuring spin excitations in various cuprates~\cite{Braicovich2009,Braicovich2010,Bisogni2009,Schlappa2009,Guarise2010,LeTacon2011}, nickelates~\cite{Ghiringhelli2009}, and even iron based superconductors~\cite{Hancock2010}. 
In this Chapter, we will show via a direct calculation of scattering amplitudes in a general setting how the polarization-dependent intensity of \emph{magnetic} RIXS directly provides an insight into the \emph{orbital} correlations in the ground state of correlated materials. 
In particular, we will present a general overview of the interplay between orbital and magnetic degrees of freedom in correlated systems (\cref{sec:Magnetic-orbital}), we will evaluate the RIXS cross section of magnetic excitations in the presence of orbital order (\cref{sec:Magnetic-RIXS}), and consequently we will specialize to the case of two and three dimensional cuprates (\cref{sec:Magnetic-RIXS-orbital}), to show how magnetic RIXS can discriminate the orbital ground state in these materials. 
In contrast to neutron scattering, the intensity of the magnetic excitations in RIXS depends very sensitively on the symmetry of the orbitals that spins occupy and on photon polarizations. 
In particular, we will verify that RIXS discriminates between different orbital states, e.g., the alternating orbital order against the ferroorbital order or the orbital liquid state. 
This method is applicable to any orbital-active material that has distinct dispersive spectral features in its spin structure factor $\chi^z({\bf q},\omega)$, for instance, due to the presence of magnetic modes arising from long-range magnetic ordering.

\section{Magnetic systems with orbital degrees of freedom}
\label{sec:Magnetic-orbital}
\subsection{Orbital order}
In correlated transition metal systems, the interaction between spin and orbital degrees of freedom, arising from the superexchange mechanism between nearest neighbor transition metal ions, is usually described in terms of the Kugel-Khomskii spin-orbital model~\cite{Kugel1982}.
Assuming that the transition metal electrons occupy a set of degenerate $3d$ orbitals, the superexchange interaction can be described by the effective spin-orbital Hamiltonian \ref{eq:KugelKhomskiiHamiltonian} in \cref{sec:Intro-Order}, which represents the second order perturbation expansion of the related multi-band Hubbard model with respect to the Hubbard on-site Coulomb repulsion. 
In the presence of large Jahn-Teller interactions, orbital interactions prevail over spin fluctuations, and in the mean field limit one can assume a ground state characterized by a long-range orbital order with an effective low energy Hamiltonian as in \cref{eq:GeneralHeisenbergHamiltonian}, which can be viewed as a generalization of the Heisenberg Hamiltonian where the exchange parameters depends on the orbital ground state. 
Some examples of orbital ordered states include the ferroorbital order (same orbital on each site) and the alternating orbital order (alternating orbitals every other site), respectively with $\langle \hat{\bf T}_i \cdot \hat{\bf T}_j \rangle=\pm1/4$, and defined by
\begin{equation}\label{eq:FOAO}
	\hat{T}^z_i \ket{\rm FO}=\frac12\ket{\rm FO}, \qquad 
	\hat{T}^z_i \ket{\rm AO}=\pm\frac12\ket{\rm AO},
\end{equation}
where the sign takes opposite values every other lattice site in the alternating orbital ordered state. 
In these cases the exchange parameters do not depend on the lattice site and become $J_{ij}=J=J_1\pm J_3$ and $K_{ij}=K=\pm{J_2}/4$, and therefore, neglecting the constant term, \cref{eq:GeneralHeisenbergHamiltonian} becomes
\begin{emphequation}\label{eq:HeisenbergHamiltonian}
	\hat{\cal H}= J\sum_{\langle i,j \rangle} 
	\hat{\bf S}_i \cdot \hat{\bf S}_j,
\end{emphequation}
which is nothing else than an effective Heisenberg Hamiltonian with nearest neighbor interaction, where the exchange parameter $J$ and, as a consequence, the ferromagnetic or antiferromagnetic nature of the spin ground state, is determined by the superexchange interactions between nearest neighbor transition metal ions. 
Examples of magnetic and orbital orders which are considered in this Chapter are shown in \cref{fig:FMAFxFOAO,fig:CAOGAO}. 
Nevertheless, one can consider a more general class of orbital orders defined by
\begin{equation}\label{eq:OQ}
	\hat{T}^z_i \ket{O(\bf\bar{Q})}=\frac12 e^{\imath{\bf\bar{Q}}\cdot{\bf R}_i}\ket{O(\bf \bar{Q})}
\end{equation}
where $\bf\bar{Q}$ is the ordering vector of the orbital ordered ground state\footnote{
In order to guarantee the hermiticity of the pseudospin operators $\hat{T}^z_i$, the ordering vector $\bf\bar{Q}$ is assumed to have components which are integer multiples of $\pi$. }
, which coincides with ${\bf\bar{Q}}=0$ and ${\bf\bar{Q}}=(\pi,\pi,\pi)$ in the case, e.g., of three dimensional ferroorbital and alternating orbital orders. 
However, \cref{eq:OQ} can describe more complex types of orbital order, such as, for instance, a state with ${\bf\bar{Q}}=(\pi,\pi,0)$, having alternating orbitals on $xy$ planes and same orbitals on the the $z$ direction. 

Besides, one can consider also the case of an orbital liquid (OL) ground state, where the orbital occupancies fluctuate in a similar way as spins in a spin liquid state, and with vanishing expectation value of the pseudospin operator
\begin{equation}\label{eq:OL}
	\langle \hat{T}^z_i \rangle_{\rm OL}=0.
\end{equation}

In typical alternating orbital and ferroorbital states with large crystal field Jahn-Teller interactions, it is reasonable~\cite{Wohlfeld2011} to assume a mean field description of the orbital degrees of freedom, and describe the system with an effective Heisenberg Hamiltonian as in \cref{eq:GeneralHeisenbergHamiltonian} or \cref{eq:HeisenbergHamiltonian}. 
In the orbital liquid state however, the mean field treatment of the orbital degrees of freedom may be questionable, but the experimental results suggest that this approach is valid as well~\cite{Oles2005}. 
However, if this is not the case, the nature of the elementary excitations in such systems is not clear~\cite{Wohlfeld2011b}. 

\subsection{Magnon dispersion in a ferromagnet}

The elementary excitations of the Heisenberg model of \cref{eq:GeneralHeisenbergHamiltonian} for both the ferromagnetic ($J_{ij}<0$) and antiferromagnetic ($J_{ij}>0$) cases are quantized spin excitations which are known as magnons, or spin waves. 
This magnon excitations correspond to Goldstone modes which arise as a consequence of the spontaneous symmetry breaking of the rotational symmetry of the ferromagnetic and antiferromagnetic ground states. 
In the rest of this Section, we will derive the dispersion of the magnetic excitations in ferromagnets, antiferromagnets, and, in general, in magnetic systems with mixed ferro and antiferromagnetic character. 

To obtain the excitation spectra of a magnetic system one can at first rewrite \cref{eq:HeisenbergHamiltonian} in terms of spin ladder operators, as
\begin{equation}\label{eq:HeisenbergHamiltonianLadder}
	\hat{\cal H}= \sum_{\langle i,j \rangle} J_{ij}
	\left[
	\frac12\left(\hat{S}^+_i \hat{S}^-_j +\hat{S}^-_i \hat{S}^+_j\right)
	+\hat{S}^z_i \hat{S}^z_j
	\right] 
	.
\end{equation}
The spin ladder operators commute at different sites, resembling the commutation algebra of boson operators, whereas at the same site one has $[\hat{S}^+_i \hat{S}^-_i]=2\hat{S}^z_i$. 
For this reason, spin ladder operators can be approximated as bosons, if there is no interaction nor superpositions of magnetic excitations, i.e., if their density is small~\cite{Khomskii2010}. 

In the ferromagnetic case $J_{ij}<0$, the ground states of the Heisenberg Hamiltonian are defined as the states where all the spins are parallel to a fixed direction. 
These ground states are degenerate and, although the Heisenberg Hamiltonian is invariant respect to spin rotations, the realization of a single ground state breaks the rotational symmetry. 
Therefore, one can assume that the system is in the ground state $\ket{\rm FM}$ with all the spin parallel to a quantization axis\footnote{The spin quantization axis do not have, in general, to coincide to the quantization axis of the orbital states, nor to any of the ion lattice axes.}
$z$, and with $\hat{S}^z_i\ket{\rm FM}=s\ket{\rm FM}$, where $s$ is the spin quantum number. 
The ordered ferromagnetic state is therefore described by an ordering vector ${\bf Q}=0$ and by an order parameter which is defined as the total spin of the system, i.e., as the magnetization ${\bf M}_0=\sum_i {\bf S}_i$. 
To calculate the magnon spectra, one can introduce the boson creation and annihilation operators $a^\dag_i$ and $a_i$ via the so called Holstein-Primakoff transformation
\begin{align}\label{eq:HolsteinPrimakoff}
	\hat{S}^+_i&=({2s-\alpha^\dag_i \alpha^\nodag_i})^{\frac12}\alpha^\nodag_i,\nonumber\\
	\hat{S}^-_i&=\alpha^\dag_i ({2s-\alpha^\dag_i \alpha^\nodag_i})^{\frac12},\\
	\hat{S}^z_i&=s-\alpha^\dag_i \alpha^\nodag_i,\nonumber
\end{align}
where one assumes that the boson operators satisfy the canonical commutation rule $[\alpha^\nodag_i,\alpha^\dag_j]=\delta_{ij}$. 
This transformation is canonical, in the sense that the canonical commutation rules of the spin operators directly follows from the canonical commutation rules of the boson operators. 
However, the Fock space described by the new operators contains states with an unlimited number of boson at any lattice site, in particular states with a number of bosons $n_i>2s$ which correspond to states with a component of the spin on the quantization axis $s_z=s-n_i<-s$. 
These states are not physical and, for this reason, one should restrict the Fock space only to states with a number of bosons $0\le n_i\le 2s$ at any lattice site. 

Considering only the linear and the quadratic terms of \cref{eq:HolsteinPrimakoff}, i.e., taking $\hat{S}^+_i\approx \alpha^\nodag_i$ and $\hat{S}^-_i\approx \alpha^\dag_i$, and neglecting higher order terms, the spin Hamiltonian \ref{eq:HeisenbergHamiltonianLadder} can be rewritten in terms of boson operators as
\begin{equation}
	\hat{\cal H}= \sum_{\langle i,j \rangle} J_{ij}
	\left[
	s\left(
	 \alpha^\dag_i \alpha^\nod_j + \alpha^\nod_i \alpha^\dag_j 
	-\alpha^\dag_i \alpha^\nod_i - \alpha^\dag_j \alpha^\nod_j
	\right)+s^2\right]
	,
\end{equation}
up to higher order terms. 
In this approximation, one considers only the harmonic terms of the Hamiltonian, and therefore neglect any interaction between boson, which arise from the neglected higher order terms. 
If one assumes that the exchange constants $J_{ij}$ are translational invariant, i.e., depend only on the lattice displacement ${\bf r}_{ij}={\bf R}_j-{\bf R}_i$ and not separately on the lattice sites, a Fourier transform of the boson operators diagonalizes the Hamiltonian as
\begin{emphequation}\label{eq:FerromagneticHamiltonian}
	\hat{\cal H}= \sum_{\bf k} \omega_{\bf k} \alpha^\dag_{\bf k} \alpha^\nod_{\bf k}
	= \sum_{\bf k} s|J_{0}|\left(1-J_{\bf k}/J_{0}\right) \alpha^\dag_{\bf k} \alpha^\nod_{\bf k}
	,
\end{emphequation}
up to a constant term, and where $J_{\bf k}=\sum_j J_{ij} e^{\imath {\bf k}\cdot{\bf r}_{ij}}$ and $J_0=\sum_j J_{ij}$. 
Therefore, the excited states of a ferromagnet have the form 
\begin{equation}\label{eq:FerroMagnon}
	\alpha^\dag_{\bf k}\ket{\rm FM}
	\approx\frac1{\sqrt N}\sum_i e^{\imath {\bf k}\cdot {\bf R}_i} \hat{S}^-_i\ket{\rm FM}
	,
\end{equation}
with dispersion $\omega_{\bf k}=s(J_{\bf k}-J_0)$. 
The magnon dispersion in a ferromagnet is quadratic for small momenta ${\bf k}\rightarrow 0$. 
This is a consequence of the fact that the order parameter operator of a ferromagnet, i.e., the magnetization $\hat{\bf M}_0=\sum_i \hat{\bf S}_i$, does commute with the Heisenberg Hamiltonian. 
The ground state $\ket{\rm FM}$ coincides with the boson vacuum, since \cref{eq:FerroMagnon} implies that $\alpha_{\bf k}\ket{\rm FM}=0$. 

In particular, in the case of isotropic exchange constants $J_{ij}=J<0$ which are nonzero only for nearest neighbor sites, the magnon dispersion becomes
\begin{equation}\label{eq:MagnonDispersion}
	\omega_{\bf k}=zs|J|(1-\gamma_{\bf k})
	,
\end{equation}
where $\gamma_{\bf k}=J_{\bf k}/J_{0}=\sum_{j=1}^z e^{\imath {\bf k}\cdot{\bf r}_{ij}}/z$ with the sum over the $z$ nearest neighbors lattice sites $j$. 
\Cref{fig:Magnon} shows the magnon dispersion of a ferromagnet with $s=1/2$ on a two dimensional square lattice, where $\gamma_{\bf k}=\frac12 (\cos{k_x}+\cos{k_y})$ and $z=4$. 

\subsection{Magnon dispersion in an antiferromagnet}

In the antiferromagnetic case instead, e.g., in the case where the nearest neighbor exchange constants $J_{ij}>0$ dominate, the ground state and the magnetic excitations of the Heisenberg Hamiltonian can be assumed, as a first approximation, as perturbations to the N\'eel state, defined as an ordered state where the spin are parallel and antiparallel every other site to a fixed direction $z$, with $\hat{S}^z_i\ket{\text{N\'eel}}=\pm s\ket{\text{N\'eel}}$. 
This ordered state is described by an ordering vector which is, e.g., ${\bf Q}=(\pi,\pi,\pi)$ in a three dimensional lattice, and by an order parameter defined as the total staggered magnetization $\hat{\bf M}_{\bf Q}=\sum_i e^{\imath {\bf Q}\cdot{\bf R}_i}\hat{\bf S}_i$. 
However, it should be noted that the N\'eel state is neither the ground state nor an eigenstate of the Heisenberg Hamiltonian in \cref{eq:HeisenbergHamiltonian}, as become clear by considering the action of the Hamiltonian on this state. 

In the light of this, it is convenient to divide the spin lattice into two sublattices $A$ and $B$, corresponding to the spin up and down of the N\'eel state, and rotate the spin operators in the second one by $2\pi$ around the $x$ axis, and therefore having $\hat{S}^{\pm}_i\rightarrow\hat{S}^{\mp}_i$ and $\hat{S}^{z}_i\rightarrow -\hat{S}^{z}_i$ for spin operators on the sublattice $B$. 
More precisely, one assume the transformation
\begin{align}\label{eq:SpinRotation}
	\hat{S}^\pm_i&\rightarrow\frac12\left[\left(1+e^{\imath{\bf Q}\cdot{\bf R}_i}\right)\hat{S}^\pm_i+\left(1-e^{\imath{\bf Q}\cdot{\bf R}_i}\right)\hat{S}^\mp_i\right]
	,\nonumber\\
	\hat{S}^z_i&\rightarrow e^{\imath{\bf Q}\cdot{\bf R}_i}\hat{S}^z_i,
\end{align}
where ${\bf Q}$ is the ordering vector of the antiferromagnetic N\'eel state. 
The Hamiltonian \ref{eq:HeisenbergHamiltonianLadder} therefore becomes 
\begin{equation}\label{eq:HeisenbergHamiltonianLadderAntiFerro}
	\hat{\cal H}= \sum_{\langle i,j \rangle} J_{ij}
	\left[
	\frac12\left(\hat{S}^+_i \hat{S}^+_j +\hat{S}^-_i \hat{S}^-_j\right)
	-\hat{S}^z_i \hat{S}^z_j
	\right]
	.
\end{equation}
At this point, one can introduce the boson operators $\alpha^\dag_i$ and $\alpha^\nod_i$ via the Holstein-Primakoff transformation as in \cref{eq:HolsteinPrimakoff}, and therefore obtain the bosonic Hamiltonian
\begin{equation}
	\hat{\cal H}= \sum_{\langle i,j \rangle} J_{ij}
	\left[
	s\left(
	 \alpha^\nod_i \alpha^\nod_j + \alpha^\dag_i \alpha^\dag_j 
	+\alpha^\dag_i \alpha^\nod_i + \alpha^\dag_j \alpha^\nod_j
	\right)
	-s^2
	\right]
	,
\end{equation}
again neglecting higher order terms. 
In terms of Fourier transformed boson operators the Hamiltonian becomes
\begin{equation}\label{eq:AntiFerromagneticHamiltonianNonDiagonal}
	\hat{\cal H}= \frac12 s \sum_{\bf k} 
	\left[
	J_{\bf k} 
	\left(\alpha^\nod_{\bf-k} \alpha^\nod_{\bf k} + \alpha^\dag_{\bf k} \alpha^\dag_{\bf-k}\right) 
	+ J_0
	\left(\alpha^\dag_{\bf k} \alpha^\nod_{\bf k} + \alpha^\nod_{\bf-k} \alpha^\dag_{\bf-k}\right) 
	\right]
	,
\end{equation}
up to a constant term, and where again $J_{\bf k}=\sum_j J_{ij} e^{\imath {\bf k}\cdot{\bf r}_{ij}}$ and $J_0=\sum_j J_{ij}$. 
The transformed Hamiltonian is not diagonal yet, but only mixes states with momenta $\bf k$ and $\bf -k$. 
To diagonalize the Hamiltonian, one can introduce a new set of boson operators $\beta^\dag_{\bf k}$ and $\beta^\nod_{\bf k}$ defined by the Bogoliubov transformation
\begin{align}\label{eq:BogoliubovAntiFerro}
	\alpha^\nod_{\bf k}&=u_{\bf k}\beta^\nod_{\bf k}-v_{\bf k}\beta^\dag_{\bf -k}
	,
\end{align}
assuming the Bogoliubov factors $u_{\bf k}=u_{-{\bf k}}$ and $v_{\bf k}=v_{-{\bf k}}$ as real parameters, and with the condition $|u_{\bf k}|^2-|v_{\bf k}|^2=1$, which ensures that the new operators satisfy the the canonical commutation relation $[\beta^\nod_{\bf k},\beta^\dag_{\bf k}]=1$. 
\begin{figure}[t]
	\centering
	\includegraphics[width=.8\textwidth]{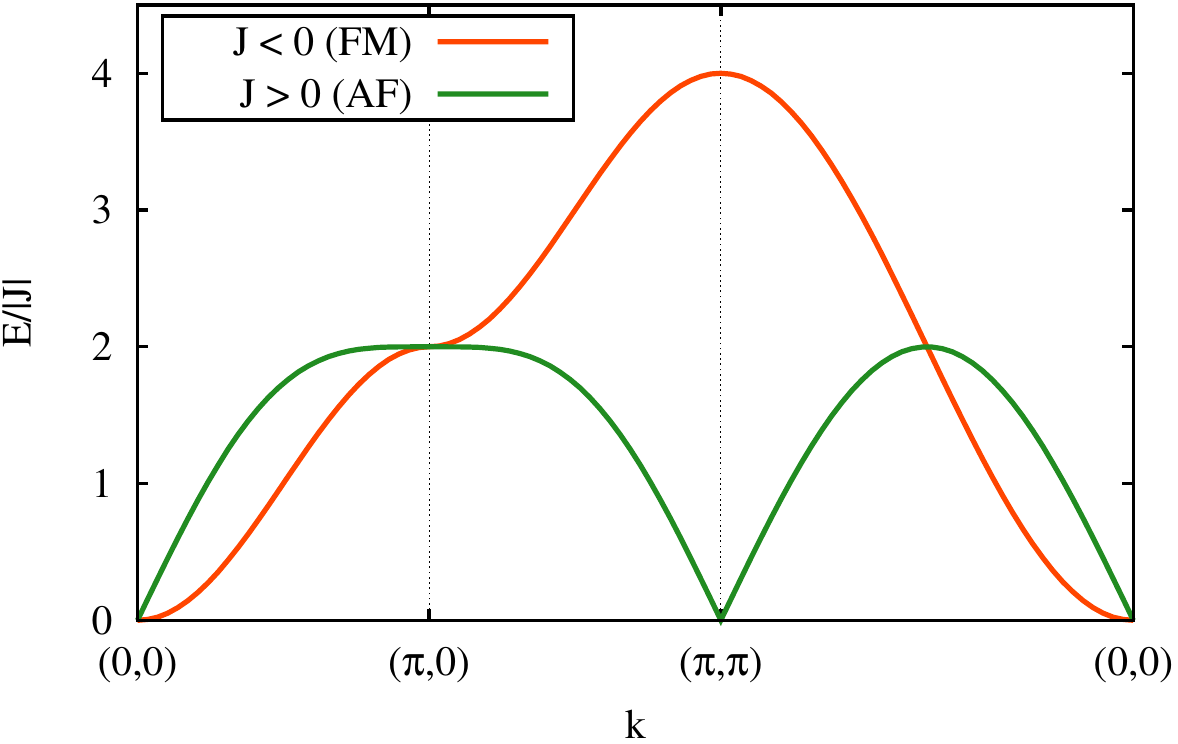}
	\caption[Magnon dispersion in a ferromagnet and an antiferromagnet]{
Magnon dispersion for a two dimensional magnet with $s=1/2$ on a square lattice along a high symmetry path in the Brillouin zone, respectively for ferromagnetic $J<0$ and antiferromagnetic $J>0$ coupling. 
}
	\label{fig:Magnon}
\end{figure}
The Hamiltonian in \cref{eq:AntiFerromagneticHamiltonianNonDiagonal} is diagonalized by the Bogoliubov transformation in \cref{eq:BogoliubovAntiFerro} in the case where the identity $(u_{\bf k}^2+v_{\bf k}^2)J_{\bf k}=2u_{\bf k}v_{\bf k}J_0$ is satisfied. 
Since the canonical condition $u_{\bf k}^2-v_{\bf k}^2=1$ is analogous to the hyperbolic functions identity $\cosh^2{\theta}-\sinh^2{\theta}$=1, one can parametrize the Bogoliubov factors as $u_{\bf k}=\cosh\theta_{\bf k}$ and $v_{\bf k}=\sinh\theta_{\bf k}$ with $\theta_{\bf k}=\theta_{\bf -k}$ a real parameter. 
Therefore, the Hamiltonian \ref{eq:AntiFerromagneticHamiltonianNonDiagonal} is diagonalized by the Bogoliubov transformation in \cref{eq:BogoliubovAntiFerro} in the case where $\tanh 2\theta_{\bf k}=J_{\bf k}/J_0$, which leads to
\begin{equation}\label{eq:ukvk}
	u_{\bf k}=\sqrt{\frac1{2\sqrt{1-\gamma_{\bf k}^2}}+\frac12}
	,\qquad
	v_{\bf k}=\frac{\gamma_{\bf k}}{|\gamma_{\bf k}|}\sqrt{\frac1{2\sqrt{1-\gamma_{\bf k}^2}}-\frac12}
	,
\end{equation}
with $\gamma_{\bf k}=J_{\bf k}/J_{0}$.
The Hamiltonian is therefore diagonal in terms of the Bogoliubov bosons, in the form
\begin{emphequation}\label{eq:AntiFerromagneticHamiltonian}
	\hat{\cal H}= \sum_{\bf k} \Omega_{\bf k} \left(\beta^\dag_{\bf k} \beta^\nod_{\bf k}+\frac12\right)
	= \sum_{\bf k} s J_0 
	\sqrt{1-\gamma_{\bf k}^2} \left(\beta^\dag_{\bf k} \beta^\nod_{\bf k}+\frac12\right)
	.
\end{emphequation}
The ground state of the antiferromagnet is defined by the boson vacuum $\beta_{\bf k}\ket{\rm AF}=0$, which does not coincide with the N\'eel state, while the excited states are defined as
\begin{align}\label{eq:AntiMagnon}
	\beta^\dag_{\bf k}\ket{\rm AF}=
	\frac1{\sqrt N}\sum_i \frac12 
	\Big[
	e^{\imath{\bf k}\cdot{\bf R}_i}
&	\left(u_{\bf k}+v_{\bf k}\right)\left( \hat{S}^-_i + \hat{S}^+_i \right)+
	\nonumber\\
	e^{\imath\left(\bf{k+Q}\right)\cdot{\bf R}_i}
&	\left(u_{\bf k}-v_{\bf k}\right)\left( \hat{S}^-_i - \hat{S}^+_i \right)
	\Big]
	\ket{\rm AF}
	,
\end{align}
with dispersion $\Omega_{\bf k}=s J_0 \sqrt{1-\gamma_{\bf k}^2}$, which is linear for small momenta ${\bf k}\rightarrow 0$. 
The linearity of the magnon dispersion around the high symmetry point ${\bf k}=0$ is a consequence of the fact that, in contrast to the ferromagnetic case, the order parameter of an antiferromagnet, i.e., the staggered magnetization $\hat{\bf M}_{\bf Q}$, does not commute in this case with the Heisenberg Hamiltonian. 

In particular, in the case of isotropic exchange constants $J_{ij}=J>0$ which are nonzero only for nearest neighbor sites, the magnon dispersion in \cref{eq:AntiFerromagneticHamiltonian} becomes
\begin{equation}\label{eq:AntiMagnonDispersion}
	\omega_{\bf k}=zs J\sqrt{1-\gamma_{\bf k}^2}
	,
\end{equation}
where $\gamma_{\bf k}=\sum_{j=1}^z e^{\imath {\bf k}\cdot{\bf t}_{ij}}/z$ with the sum over the $z$ nearest neighbors lattice sites $j$. 
\Cref{fig:Magnon} shows the magnon dispersion for an antiferromagnet with $s=1/2$ on a two dimensional square lattice, where $\gamma_{\bf k}=\frac12 (\cos{k_x}+\cos{k_y})$ and $z=4$. 

\subsection{Anisotropic magnetic coupling}

In this Section, we will assume that the exchange constants depend only on the distance ${\bf r}_{ij}={\bf R}_{j}-{\bf R}_{i}$ between the two lattice points involved, but have a different sign on the plane $xy$ and along the $z$ direction. 
In this case, the ground state of the system can be assumed to be a N\'eel-like ordered state $\ket{\text{N\'eel}({\bf Q})}$, characterized by an ordering vector $\bf Q$, in such a way that
\begin{equation}\label{eq:GeneralMagneticOrder}
	\hat{S}_i^z\ket{\text{N\'eel}({\bf Q})}=e^{\imath {\bf Q}\cdot{\bf R}_i}s\ket{\text{N\'eel}({\bf Q})}
	,
\end{equation}
where the hermicity of the spin operator mandates that the exponential factor takes only real values. 
For this reason, the local spin component along the local quantization axis has to be either $\pm s$ and, as a consequence, the components of the ordering vector $\bf Q$ have to be integer multiples of $\pi$, e.g., ${\bf Q}=(0,\pi,0)$ or ${\bf Q}=(\pi,\pi,0)$. 
In this case, the spin operator defined by the transformation in \cref{eq:SpinRotation} are physical, and can be used to describe the excitation of the system respect to the ordered state defined in \cref{eq:GeneralMagneticOrder}. 
As in the case of the N\'eel state of an antiferromagnet however, this ordered state is not, in general, either a ground state or an eigenstate of the Heisenberg Hamiltonian in \cref{eq:HeisenbergHamiltonian}. 
Nevertheless, this ordered state describes a wide range of systems where up and down spins alternate on the lattice, e.g., with an antiferromagnetic alignment on some directions, and ferromagnetic in others. 
However, it does not describes systems where the magnetic unit cell is larger than the one of a pure antiferromagnet, such as in the case of, e.g., helimagnetic order, or more exotic incommensurate orders. 

Therefore, rotating the spin operator as in \cref{eq:SpinRotation}, using the fact that $e^{2{\bf Q}\cdot{\bf R_i}}=1$ for the ordering vectors considered, and applying the Holstein-Primakoff transformation as in \cref{eq:HolsteinPrimakoff}, the Hamiltonian \ref{eq:HeisenbergHamiltonian} can be rewritten in terms of the boson operators as
\begin{align}\label{eq:HeisenbergHamiltonianBosonGeneral}
	\hat{\cal H}= \frac14s\sum_{\langle i,j \rangle}
	J_{ij}\Big[
	&
	\left(1+e^{\imath{\bf Q}\cdot{\bf r}_{ji}}\right)
	\left(\alpha^\nod_i \alpha^\dag_j +\alpha^\dag_i \alpha^\nod_j\right)
	+
	\left(1-e^{\imath{\bf Q}\cdot{\bf r}_{ji}}\right)
	\left(\alpha^\nod_i \alpha^\nod_j +\alpha^\dag_i \alpha^\dag_j\right)
	+\nonumber\\
	-&
	2 e^{\imath{\bf Q}\cdot{\bf r}_{ji}}
	\left(\alpha^\dag_i \alpha^\nod_i+\alpha^\dag_j \alpha^\nod_j\right)
	\Big]
	,
\end{align}
neglecting constant and higher order terms. 
In terms of the Fourier transformed boson operators, the Hamiltonian becomes
\begin{align}\label{eq:HeisenbergHamiltonianFourierGeneral}
	\hat{\cal H}= \frac{1}{4}s\sum_{\bf k}
	\Big[
	&
	\left(J_{\bf k}-J_{\bf k+Q}\right)
	\left(\alpha^\nod_{-{\bf k}} \alpha^\nod_{\bf k} +\alpha^\dag_{\bf k} \alpha^\dag_{-{\bf k}}\right)
	\nonumber\\
	+&
	\left(J_{\bf k}+J_{\bf k+Q}-2J_{\bf Q}\right)
	\left(\alpha^\dag_{\bf k} \alpha^\nod_{\bf k} +\alpha^\nod_{-{\bf k}} \alpha^\dag_{-{\bf k}}\right)
	\Big]
	,
\end{align}
which closely resembles \cref{eq:AntiFerromagneticHamiltonianNonDiagonal} if one defines $\widetilde{J}_{\bf k}=\frac12\left(J_{\bf k}-J_{\bf k+Q}\right)$ and $\widetilde{J}_0=\frac12\left(J_{\bf k}+J_{\bf k+Q}-2J_{\bf Q}\right)$. 
Therefore the Hamiltonian \ref{eq:HeisenbergHamiltonianFourierGeneral} is diagonalized by the Bogoliubov transformation in \cref{eq:BogoliubovAntiFerro} with the Bogoliubov factors defined as in \cref{eq:ukvk} with the substitution $\gamma_{\bf k}\rightarrow\widetilde{\gamma}_{\bf k}=\widetilde{J}_{\bf k}/\widetilde{J}_{0}$, in the form
\begin{emphequation}\label{eq:GeneralMagnonDispersion}
	\hat{\cal H}= \sum_{\bf k} \varepsilon_{\bf k} \left(\beta^\dag_{\bf k} \beta^\nodag_{\bf k}+\frac12\right)
	= \sum_{\bf k} s \widetilde{J}_0 
	\sqrt{1-\widetilde{\gamma}_{\bf k}^2} \left(\beta^\dag_{\bf k} \beta^\nodag_{\bf k}+\frac12\right)
	.
\end{emphequation}
The ground state of the system is again defined as the boson vacuum $\beta_{\bf k}\ket{S({\bf Q})}=0$, while the excited states $\beta^\dag_{\bf k}\ket{S({\bf Q})}$ have dispersion $\varepsilon_{\bf k}=s \widetilde{J}_0 \sqrt{1-\widetilde{\gamma}_{\bf k}^2}$. 
In particular, in the case of a pure ferromagnetic ($J_{ij}<0$) or antiferromagnetic ($J_{ij}>0$) system, the magnon dispersion in \cref{eq:GeneralMagnonDispersion} reduces respectively to \cref{eq:MagnonDispersion,eq:AntiMagnonDispersion}. 

As an example, one can consider a three dimensional cubic lattice with $s=1/2$, with different combinations of ferromagnetic and antiferromagnetic exchange constants in the $xy$ plane and on the perpendicular direction $z$. 
For example, in the case of a system with ferromagnetic exchange constant $J_{xy}<0$ in the $xy$ plane, and antiferromagnetic $J_z>0$ along the $z$ direction, whose ground state can be approximated as an ordered state with ordering vector ${\bf Q}=(0,0,\pi)$, the magnon dispersion in \cref{eq:GeneralMagnonDispersion} becomes
\begin{equation}\label{eq:MagnonFFA}
	\varepsilon_{\bf k}=\sqrt{\left(J_z-2J_{xy}+J_{xy}\cos{k_x}+J_{xy}\cos{k_y}\right)^2-J_z^2\cos^2{k_z}}
	.
\end{equation}
Alternatively, in the case of a system with an antiferromagnetic exchange constant $J_{xy}>0$ in the $xy$ plane, and ferromagnetic $J_z<0$ along the $z$ direction, with ordering vector ${\bf Q}=(\pi,\pi,0)$, the magnon dispersion in \cref{eq:GeneralMagnonDispersion} becomes
\begin{equation}\label{eq:MagnonAAF}
	\varepsilon_{\bf k}=\sqrt{\left(2J_{xy}-J_z+J_{z}\cos{k_z}\right)^2-\left(J_{xy}\cos{k_x}+J_{xy}\cos{k_y}\right)^2}
	.
\end{equation}
In all these cases, the magnon dispersion is linear near the symmetry point ${\bf k}=0$, as well as the pure antiferromagnetic case, since also in this case the order parameter, i.e., the staggered magnetization $\hat{\bf M}_{\bf Q}$ do not commute with the Hamiltonian. 

\section{Magnetic RIXS cross section}
\label{sec:Magnetic-RIXS}
\subsection{Magnetic RIXS in orbital systems}

RIXS is particularly apt to probe the properties of strongly correlated electrons, for instance in transition metal oxides~\cite{Ament2011}. 
With an incident x-ray beam of energy $\hbar\omega_{\bf k}$ and momentum ${\bf k}$ an electron is resonantly excited from a core level into the valence shell. 
At the transition metal $L_{2,3}$ edges this involves a $2p\leftrightarrow3d$ dipole-allowed transition. 
In this intermediate state, the spin of the $2p$ core hole is not conserved, as the very large spin-orbit interactions strongly couple the spin and orbital momentum of the core hole. 
A spin flip in the core allows the subsequent recombination of the core hole with a $3d$ electron that has a spin opposite to the electron state that was originally excited into the $3d$ shell. 
The energy $\hbar\omega_{{\bf k}'}$ and momentum ${\bf k}'$ of the scattered x-ray photon resulting from this recombination are then related to a spin excitation with energy $\hbar\omega=\hbar\omega_{{\bf k}'}-\hbar\omega_{\bf k}$ and momentum ${\bf q}={\bf k}-{\bf k}'$. 

In order to calculate the direct RIXS cross section of magnetic excitations at the transition metal $L_{2,3}$ edges, we will use the \cref{eq:KramersHeisenberg} and the expansion of the RIXS operator $\hat{\cal O}={\frac1{\sqrt{N}}}\sum_i e^{\imath {\bf q}\cdot{\bf R}_i}\hat{O}_i$ in terms of the local RIXS operator $\hat{O}_i$ as defined in \cref{sec:Dipole}. 
Moreover, since the focus here is on the direct scattering processes at the transition metal $L_{2,3}$ edges, we will employ the fast collision approximation in the case of strong spin-orbit coupling as in \cref{eq:SOexpansion_H_SO>Gamma} and the expansion of the local RIXS operator in the case of $p\leftrightarrow d$ transitions as in \cref{sec:pd_Transitions}. 

Magnetic excitations in a ferromagnetic or in an antiferromagnetic systems are excited states which are a linear combinations of local spin flip states, as one can see from \cref{eq:FerroMagnon,eq:AntiMagnon}. 
Moreover, in the limit where the local degrees of freedom can integrated out, one restrict oneself only to purely magnetic excitations, i.e., in which the local orbital momentum is conserved.
Therefore, the local transition amplitudes, i.e., the matrix elements of the local RIXS operator $\hat{O}_i$, which are relevant for the calculation of the RIXS cross section of purely magnetic excitations are those between states with the same orbital occupancy and opposite spin, that is
\begin{align}\label{eq:LocalTransitionAmplitudes}
	W^+_{{\bf e}{\bf e}'}(d)=\bra{d \uparrow} \hat{O}_i \ket{d \downarrow}
	,\quad
	W^-_{{\bf e}{\bf e}'}(d)=\bra{d \downarrow} \hat{O}_i \ket{d \uparrow}
	,
\end{align}
with $W^+_{{\bf e}{\bf e}'}(d)=W^{-}_{{\bf e}{\bf e}'}(d)^*$, and where $\ket{d \sigma}$ denotes the local electron state with orbital $d$ and spin $\sigma$. 
Hence, the local RIXS operator in \cref{eq:RIXSOperatorLocal} can be rewritten in terms of this local transition amplitudes as
\begin{emphequation}\label{eq:DSF_W_local_FMAF}
	\hat{O}_i=
	\sum\limits_{\sigma=\pm}\sum\limits_{d}
	\hat{n}_i(d)W^\sigma_{{\bf e}{\bf e}'}(d)
	\hat{S}^\sigma_i
	,
\end{emphequation}
where the number operator $\hat{n}_i(d)=\sum_{\sigma}d^\dag_{i\sigma} d^\nod_{i\sigma}$ counts the occupancy of the orbital state $d$ for valence electrons in the $3d$ shell on the lattice site $i$. 
Note that \cref{eq:DSF_W_local_FMAF} follows directly from \cref{eq:DSF_W_local_12} in \cref{sec:DSF_W} when one considers only spin flip transitions within the same orbital state. 
If one consider only two active orbitals $d=a,b$, and introduces the pseudospin operator as in \cref{eq:Pseudospin}, the RIXS operator $\hat{\cal O}$ becomes
\begin{equation}\label{eq:RIXSOperatorPseudospin}
	\hat{\cal O}
	={\frac1{\sqrt N}}\sum\limits_i e^{\imath {\bf q}\cdot{\bf R}_i}
	\sum\limits_{\sigma=\pm}\left[
	\left(\frac12 + T^z_i\right) W^\sigma_{{\bf e}{\bf e}'}(a)+
	\left(\frac12 - T^z_i\right) W^\sigma_{{\bf e}{\bf e}'}(b)
	\right]\hat{S}^\sigma_i
	,
\end{equation}
where one has $T^z_i\ket{d\sigma}=\pm\frac12\ket{d\sigma}$ respectively for $d=a,b$. 
The local transition amplitudes $W^\pm_{{\bf e}{\bf e}'}(d)$ in \cref{eq:LocalTransitionAmplitudes} strongly depend on the photon polarizations and on the orbital occupancy at the lattice site $i$. 
This is because the local RIXS operator is defined in terms of the dipole operator as in \cref{eq:RIXSOperatorLocal}, which inherently depends on the orbital angular momentum of the scattered valence electrons~\cite{Haverkort2010,Haverkort2010b,deGroot1990}. 

\subsection{Orbital dependence of the RIXS operator}

As stated above the orbital dependence of local transition amplitudes $W^\pm_{{\bf e}{\bf e}'}(d)$ is generic to any orbital system. 
Nevertheless, to be explicit, we will show how this dependence arises in the simple case of direct RIXS scattering at the $L_{2,3}$ transition metal edges. 
In this case in fact, the direct scattering processes dominate since the transitions $2p\leftrightarrow3d$ between the valence $3d$ electrons and the core hole state in the $2p$ shell are dipole allowed.
Therefore, using the fast collision approximation in the case of strong spin-orbit coupling as in \cref{eq:SOexpansion} and the definition of the local RIXS operator in \cref{eq:RIXSOperatorLocal}, the local transition amplitudes in \cref{eq:LocalTransitionAmplitudes} become
\begin{equation}\label{eq:LocalTransitionAmplitudesExpanded}
	W^+_{{\bf e}{\bf e}'}(d)=W^-_{{\bf e}{\bf e}'}(d)^*=
	\bra{d \uparrow} 
	\left({\bf e}^{\prime*} \cdot \hat{\bf D}_i^{\dag} \right)
	\hat{\bf L}\cdot\hat{\bf S}
	\left({\bf e} \cdot \hat{\bf D}_i\right)
	\ket{d \downarrow}
	,
\end{equation}
where $\hat{\bf D}_i\approx \hat{\bf p}_i$ in the dipole approximation, $\hat{\bf L}$ and $\hat{\bf S}$ are the orbital angular momentum and the spin of the core hole, and where the constant term in the propagator as well as constant prefactors in \cref{eq:SOexpansion} are neglected. 
The compact expression of the core hole propagator and of the local RIXS operator can be further expanded as in \cref{eq:RIXSoperatorDirect_l12}, and leads to the schematic representation in \cref{fig:RIXSoperatorDirect_l1}. 
While the intermediate state propagator $\hat{G}_0\propto\hat{\bf L}\cdot\hat{\bf S}$ brings the spin dependence due to the spin-orbit coupling in the $2p$ core hole states, the dipole operator $\hat{\bf D}_i$ acts in a different way depending on the orbital occupancy on the lattice site $i$.
Therefore the local transition amplitudes $W^\pm_{{\bf e}{\bf e}'}(d)$ strongly depends on the orbital symmetry of the ground state. 

Having analyzed the inherent dependence of the local transition amplitudes $W^\pm_{{\bf e}{\bf e}'}(d)$ on the orbital occupancy of the \emph{single} lattice site, one can now investigate how the RIXS operator $\hat{\cal O}$ in \cref{eq:RIXSOperatorPseudospin} acts on the orbital ground state of the \emph{bulk}. 
In general, since the pseudospin operator $T^z_{\bf i}\ket{\rm FO}=\frac12\ket{\rm FO}$ at any lattice site in the ferroorbital state while $T^z_{\bf i}\ket{\rm AO}=\pm\frac12\ket{\rm AO}$ every other site in the alternating orbital state, the RIXS operator $\hat{\cal O}$ acts differently on different orbital ground states. 
Hereafter, we will consider different orbital ground states in magnetic systems on a two dimensional lattice: a ferroorbital (FO) order with the same orbital occupied on each site, an alternating orbital (AO) order with $a$ ($b$) orbitals occupied on sublattice $A$ ($B$), both defined in \cref{eq:FOAO}, a more general example of orbital ordered state $O({\bf\bar{Q}})$ characterized by an orbital ordering vector $\bf\bar{Q}$, defined in \cref{eq:OQ}, and finally an orbital liquid (OL) ground state with the occupancies of $a$ and $b$ orbitals fluctuating similarly to the up and down spins in the spin liquid state, as in \cref{eq:OL}. 
Eventually, the RIXS cross section of magnetic excitations can be calculated via \cref{eq:KramersHeisenberg} in terms of the transition amplitudes $\bra{g}\hat{\cal O}\ket{f}$, i.e., the matrix elements of the RIXS operator between the system ground state, and the magnetic excited final states $\alpha^\dag_{\bf q}\ket{\rm FM}$ and $\beta^\dag_{\bf q}\ket{\rm AF}$ defined in \cref{eq:FerroMagnon,eq:AntiMagnon} respectively for ferromagnetic and antiferromagnetic systems. 
In this way, we will show how the orbital ground state affects the RIXS spectra of magnetic excitations. 

\subsection{Ferromagnetic case}\label{sec:FM}

In the case of a ferromagnetic system, the only spin operators which are relevant for the calculation of the RIXS transition amplitudes are the operators $\hat{S}^-_i$, since one has $\hat{S}^+_i\ket{\rm FM}=0$. 
Moreover, using the definition of ferroorbital order in \cref{eq:FOAO}, and the definition of magnetic excitations in a ferromagnet in terms of Holstein-Primakoff bosons as in \cref{eq:FerroMagnon}, the action of the RIXS operator in \cref{eq:RIXSOperatorPseudospin} on a ferromagnet with ferroorbital order is
\begin{equation}\label{eq:RIXS-FMFO}
	\hat{\cal O}\ket{\rm FM{\otimes} FO}=
	W^-_{{\bf e}{\bf e}'}(a)
	\alpha^\dag_{\bf q}
	\ket{\rm FM{\otimes} FO}
	.
\end{equation}

On the other hand, considering the alternating orbital order as in \cref{eq:FOAO}, the action of the RIXS operator in \cref{eq:RIXSOperatorPseudospin} on a system with ferromagnetic and alternating orbital order becomes
\begin{align}\label{eq:RIXS-FMAO}
	\hat{\cal O}\ket{\rm FM{\otimes} AO}
	=\frac12\Big\{&
	\left[	W^-_{{\bf e}{\bf e}'}(a) + W^-_{{\bf e}{\bf e}'}(b)	\right]
	\alpha^\dag_{\bf q}
	+
	\nonumber\\
	&\left[	W^-_{{\bf e}{\bf e}'}(a) - W^-_{{\bf e}{\bf e}'}(b)	\right]
	\alpha^\dag_{\bf q+Q}
	\Big\}
	\ket{\rm FM{\otimes} AO}
	,
\end{align}
where $\bf Q=\bar{Q}$ is the ordering vector of the alternating orbital state as well as the antiferromagnetic ordered state. 
Since the alternating orbital order breaks the physical equivalence of the lattice sites, with the two sublattices corresponding to different occupied orbitals $a$ and $b$, the magnetic and the orbital Brillouin zones do not longer coincide, and therefore a new branch of the magnon dispersion appears, corresponding to excitations in the form $\alpha^\dag_{\bf q+Q}$ with energy $\omega_{\bf q+Q}$, due to the backfolding in the orbital Brillouin zone. 
The new branch of the magnon dispersion corresponds to an optical mode, since $\omega_{\bf q+Q}=4|J|$ for ${\bf q}=0$, and gains a finite scattering intensity $\propto\vert W^-_{{\bf e}{\bf e}'}(a)-W^-_{{\bf e}{\bf e}'}(b)\vert^2$ as long as the spin flip amplitudes for orbitals $a$ and $b$ differ. 

In the case of an orbital liquid state instead, the two active orbitals fluctuate and the expectation value of the pseudospin operator vanishes and therefore, approximating the action of the operator with its mean value, \cref{eq:RIXSOperatorPseudospin} becomes
\begin{equation}\label{eq:RIXS-FMOL}
	\hat{\cal O}\ket{\rm FM{\otimes} OL}
	=\frac12\left[	W^-_{{\bf e}{\bf e}'}(a) + W^-_{{\bf e}{\bf e}'}(b)	\right]
	\alpha^\dag_{\bf q}\ket{\rm FM{\otimes} OL}
	.
\end{equation}
In this case there is no optical mode, since the orbital liquid state does not break the translation symmetry of the lattice. 

Finally, using \cref{eq:KramersHeisenberg}, the RIXS cross section can be directly calculated in terms of the transition amplitudes $\bra{g}\hat{\cal O}\ket{f}$. 
In \cref{tab:FMAFxFOAOOL_cs} is shown the explicit form of the RIXS cross section of magnetic excitations in a ferromagnet in the case of ferroorbital, alternating orbital order, and orbital liquid state. 

\subsection{Antiferromagnetic case}\label{sec:AF}

\begin{table}
\centering
\begin{tabular*}{\textwidth}{l| @{\extracolsep{\fill}} l}
	$\rm FM{\otimes}FO$&
	$\vert W^-_{{\bf e}{\bf e}'}(a)\vert ^2	\delta \left( \omega - \omega_{\bf q} \right)$
	\\[2mm]
	$\rm FM{\otimes}AO$&
	$\vert W^-_{{\bf e}{\bf e}'}(a) + W^-_{{\bf e}{\bf e}'}(b)\vert ^2	\delta \left( \omega - \omega_{\bf q} \right)+
	\vert W^-_{{\bf e}{\bf e}'}(a) - W^-_{{\bf e}{\bf e}'}(b)\vert ^2	\delta \left( \omega - \omega_{{\bf q}+{\bf Q}} \right)$
	\\[2mm]
	$\rm FM{\otimes}OL$&
	$\vert W^-_{{\bf e}{\bf e}'}(a) + W^-_{{\bf e}{\bf e}'}(b)\vert ^2	\delta \left( \omega - \omega_{\bf q}\right)$
	\\[5mm]
	$\rm AF{\otimes}FO$&
	$\vert  W^-_{{\bf e}{\bf e}'} (a) \vert ^2  \left( u_{\bf q} - v_{\bf q} \right)^2
	\delta \left( \omega - \Omega_{\bf q} \right)$
	\\[2mm]
	$\rm AF{\otimes}AO$&
	$\vert	 \left[ W^-_{{\bf e}{\bf e}'}(a) + W^+_{{\bf e}{\bf e}'}(b) \right] u_{\bf q} 
		-\left[ W^+_{{\bf e}{\bf e}'}(a) + W^-_{{\bf e}{\bf e}'}(b) \right] v_{\bf q} \vert^2
		 \delta \left( \omega - \Omega_{\bf q} \right)+$
	\\[1mm]&
	$\vert	 \left[ W^-_{{\bf e}{\bf e}'}(a) - W^+_{{\bf e}{\bf e}'}(b) \right] u_{{\bf q}+{\bf Q}}
		-\left[ W^+_{{\bf e}{\bf e}'}(a) - W^-_{{\bf e}{\bf e}'}(b) \right] v_{{\bf q}+{\bf Q}} \vert ^2 
		 \delta \left(\omega-\Omega_{{\bf q}+{\bf Q}} \right)$
	\\[2mm]
	$\rm AF{\otimes}OL$&
	$\vert  W^-_{{\bf e}{\bf e}'} (a) + W^-_{{\bf e}{\bf e}'} (b) \vert ^2  \left( u_{\bf q} - v_{\bf q} \right)^2
	\delta \left( \omega - \Omega_{\bf q} \right)$
\end{tabular*}
\caption[Magnetic RIXS cross sections for alternating orbital, ferroorbital, and orbital liquid states in ferromagnetic and antiferromagnetic systems]{
Magnetic RIXS cross sections for alternating orbital (AO), ferroorbital (FO), and orbital liquid (OL) states in ferromagnetic (FM) and antiferromagnetic (AF) systems. 
Constant factors are omitted. 
}
\label{tab:FMAFxFOAOOL_cs}
\end{table}

In the case of an antiferromagnetic system, using the definition of ferroorbital order in \cref{eq:FOAO}, and consequently applying the rotation of the spin operators in \cref{eq:SpinRotation}, the Holstein-Primakoff and the Bogoliuov transformation defined in \cref{eq:HolsteinPrimakoff,eq:BogoliubovAntiFerro}, the action of the RIXS operator in \cref{eq:RIXSOperatorPseudospin} on an antiferromagnet with ferroorbital order becomes
\begin{equation}\label{eq:RIXS-AFFO}
	\hat{\cal O}\ket{\rm AF{\otimes} FO}=
	W^-_{{\bf e}{\bf e}'}(a) \left(u_{\bf q} - v_{\bf q} \right) \beta^\dag_{\bf q}
	\ket{\rm AF{\otimes} FO}
	,
\end{equation}
with $u_{\bf q}$ and $v_{\bf q}$ defined as in \cref{eq:ukvk}. 
In this case, the RIXS operator leads to vanishing intensities at ${\bf q}\rightarrow0$, in agreement with~\onlinecite{Ament2009}. 

On the other hand, considering the alternating orbital order as in \cref{eq:FOAO}, and again applying the spin rotation, the Holstein-Primakoff and the Bogoliuov transformation, the action of the RIXS operator in \cref{eq:RIXSOperatorPseudospin} on an antiferromagnet with alternating orbital order becomes
\begin{align}\label{eq:RIXS-AFAO}
	\hat{\cal O}\ket{\rm AF{\otimes} AO}
	=\frac12\Big\{&
	\left[	W^-_{{\bf e}{\bf e}'}(a) + W^+_{{\bf e}{\bf e}'}(b)	\right]	u_{\bf q} \beta^\dag_{\bf q}
	\nonumber\\-&
	\left[	W^+_{{\bf e}{\bf e}'}(a) + W^-_{{\bf e}{\bf e}'}(b)	\right]	v_{\bf q} \beta^\dag_{\bf q}
	\nonumber\\+&
	\left[	W^-_{{\bf e}{\bf e}'}(a) - W^+_{{\bf e}{\bf e}'}(b)	\right]	u_{\bf q+Q} \beta^\dag_{\bf q+Q}
	\nonumber\\-&
	\left[	W^+_{{\bf e}{\bf e}'}(a) - W^-_{{\bf e}{\bf e}'}(b)	\right]	v_{\bf q+Q} \beta^\dag_{\bf q+Q}
	\Big\}
	\ket{\rm AF{\otimes} AO}
	.
\end{align}
In contrast with the ferroorbital case, in this case the RIXS operator leads in general to a nonvanishing intensity at ${\bf q}\rightarrow0$ as a result of the alternating orbital ordering. 
Moreover, in the usual case of an antiferromagnet described by an Heisenberg Hamiltonian with only nearest neighbor interaction on a cubic or on a square lattice, one has $\Omega_{\bf q+Q}=\Omega_{\bf q}$, and therefore the magnetic excitations $\beta^\dag_{\bf q}$ and $\beta^\dag_{\bf q+Q}$ are degenerate. 
As a consequence, no additional branch appears in the magnon dispersion, in contrast to the ferromagnetic case (although any corrections to the Heisenberg model for which $\Omega_{\bf q+Q}\neq\Omega_{\bf q}$ will give rise to a new branch in the dispersion).

Eventually, in the case of an orbital liquid state, approximating again the action of the pseudospin operator with its mean value, \cref{eq:RIXSOperatorPseudospin} becomes
\begin{equation}\label{eq:RIXS-AFOL}
	\hat{\cal O}\ket{\rm AF{\otimes} OL}=
	\frac12\left[	W^-_{{\bf e}{\bf e}'}(a) + W^-_{{\bf e}{\bf e}'}(b)	\right] 
	\left(u_{\bf q} \beta^\dag_{\bf q} - v_{\bf q} \beta^\dag_{\bf -q} \right)
	\ket{\rm AF{\otimes} OL}
	.
\end{equation}
As in the ferroorbital case, RIXS intensities vanish at ${\bf q}\rightarrow0$. 

Again, using \cref{eq:KramersHeisenberg}, the RIXS cross section can be directly calculated in terms of the transition amplitudes $\bra{g}\hat{\cal O}\ket{f}$. 
In \cref{tab:FMAFxFOAOOL_cs} is shown the explicit form of the RIXS cross section of magnetic excitations in a antiferromagnet in the case of ferroorbital, alternating orbital order, and orbital liquid state. 

\subsection{General case}

\begin{table}
\newcommand{\pp}{\phantom{+}}
\centering
\begin{tabular*}{\textwidth}{l| @{\extracolsep{\fill}} l}
	$S({\bf Q}){\otimes}O({\bf\bar{Q}})$&
	$\pp\Re{\left[ W^-_{{\bf e}{\bf e}'}(a) + W^-_{{\bf e}{\bf e}'}(b) \right]}^2 \left(u_{\bf q} - v_{\bf q} \right)^2 
	\delta \left( \omega - \varepsilon_{\bf q} \right)+$
	\\[1mm]&
	$\pp\Re{\left[ W^-_{{\bf e}{\bf e}'}(a) - W^-_{{\bf e}{\bf e}'}(b) \right]}^2 \left(u_{\bf q+\bar{Q}} - v_{\bf q+\bar{Q}} \right)^2
	\delta \left( \omega - \varepsilon_{\bf q+\bar{Q}} \right)+$
	\\[1mm]&
	$  -\Im{\left[ W^-_{{\bf e}{\bf e}'}(a) + W^-_{{\bf e}{\bf e}'}(b) \right]}^2 \left(u_{\bf q+Q} + v_{\bf q+Q} \right)^2
	\delta \left( \omega - \varepsilon_{\bf q+Q} \right)+$
	\\[1mm]&
	$  -\Im{\left[ W^-_{{\bf e}{\bf e}'}(a) - W^-_{{\bf e}{\bf e}'}(b) \right]}^2 \left(u_{\bf q+Q+\bar{Q}} + v_{\bf q+Q+\bar{Q}} \right)^2
	\delta \left( \omega - \varepsilon_{\bf q+Q+\bar{Q}} \right)$
	\\[2mm]
	$S({\bf Q}){\otimes}\text{OL}$&
	$\pp\Re{\left[ W^-_{{\bf e}{\bf e}'}(a) + W^-_{{\bf e}{\bf e}'}(b) \right]}^2 \left(u_{\bf q} - v_{\bf q} \right)^2 
	\delta \left( \omega - \varepsilon_{\bf q} \right)+$
	\\[1mm]&
	$  -\Im{\left[ W^-_{{\bf e}{\bf e}'}(a) + W^-_{{\bf e}{\bf e}'}(b) \right]}^2 \left(u_{\bf q+Q} + v_{\bf q+Q} \right)^2
	\delta \left( \omega - \varepsilon_{\bf q+Q} \right)$
\end{tabular*}
\caption[Magnetic RIXS cross sections for a system with magnetic and orbital order]{
Magnetic RIXS cross sections for a system with magnetic order (ordering vector $\bf Q$), respectively with orbital order (ordering vector $\bf\bar{Q}$) and orbital liquid state (OL). 
Constant factors are omitted. 
}
\label{tab:SQOQ_cs}
\end{table}

Finally, one can consider the case of a generic spin system with orbital degrees of freedom described by the Heisenberg Hamiltonian~\ref{eq:HeisenbergHamiltonian}, assuming an orbital ordered state $\ket{O({\bf\bar{Q}})}$ described by \cref{eq:OQ} with orbital ordering vector $\bf\bar{Q}$ and the spin ground state of \cref{eq:GeneralMagnonDispersion} with a spin ordering vector $\bf Q$. 
In this case, using the definition of orbital order and consequently applying the rotation of the spin operators in \cref{eq:SpinRotation}, the Holstein-Primakoff and the Bogoliuov transformation defined in \cref{eq:HolsteinPrimakoff,eq:BogoliubovAntiFerro}, the action of the RIXS operator in \cref{eq:RIXSOperatorPseudospin} on a generic spin-orbital system becomes
\begin{align}\label{eq:RIXS-SQOQ}
	\hat{\cal O}
	\ket{S({\bf Q}){\otimes}O({\bf\bar{Q}})}=
	\frac12	\Big\{
	&
	\Re{\left[ W^-_{{\bf e}{\bf e}'}(a) + W^-_{{\bf e}{\bf e}'}(b) \right]}\!\!
	\left(u_{\bf q} - v_{\bf q} \right)\!
	\beta^\dag_{\bf q}
	\nonumber\\
	+&
	\Re{\left[ W^-_{{\bf e}{\bf e}'}(a) - W^-_{{\bf e}{\bf e}'}(b) \right]}\!\!
	\left(u_{\bf q+\bar{Q}} - v_{\bf q+\bar{Q}} \right)\!
	\beta^\dag_{\bf q+\bar{Q}}
	\nonumber\\
	+\imath&
	\Im{\left[ W^-_{{\bf e}{\bf e}'}(a) + W^-_{{\bf e}{\bf e}'}(b) \right]}\!\!
	\left(u_{\bf q+Q} + v_{\bf q+Q} \right)\!
	\beta^\dag_{\bf q+Q}
	\nonumber\\
	+\imath&
	\Im{\left[ W^-_{{\bf e}{\bf e}'}(a) - W^-_{{\bf e}{\bf e}'}(b) \right]}\!\!
	\left(u_{\bf q+Q+\bar{Q}} + v_{\bf q+Q+\bar{Q}} \right)\!
	\beta^\dag_{\bf q+Q+\bar{Q}}
	\Big\}
	\nonumber\\
	\times&\ket{S({\bf Q}){\otimes}O({\bf\bar{Q}})}
	,
\end{align}
where $u_{\bf q}$ and $v_{\bf q}$ are defined as in \cref{eq:ukvk} with $\gamma_{\bf q}\rightarrow\widetilde{\gamma}_{\bf q}=\widetilde{J}_{\bf q}/\widetilde{J}_{0}$, and using the fact that $W^+(d)=W^-(d)^*$. 

In the case of an orbital liquid state instead, the mean value of the pseudospin operator vanishes and therefore in the mean field approximation the terms in \cref{eq:RIXS-SQOQ} with $\bf\bar{Q}$ do not contribute, and the RIXS operator becomes
\begin{align}\label{eq:RIXS-SQOL}
	\hat{\cal O}
	\ket{S({\bf Q}){\otimes}{\rm OL}}=
	\frac12	\Big\{
	&
	\Re{\left[ W^-_{{\bf e}{\bf e}'}(a) + W^-_{{\bf e}{\bf e}'}(b) \right]} \left(u_{\bf q} - v_{\bf q} \right) \beta^\dag_{\bf q}
	\nonumber\\
	+\imath&
	\Im{\left[ W^-_{{\bf e}{\bf e}'}(a) + W^-_{{\bf e}{\bf e}'}(b) \right]} \left(u_{\bf q+Q} + v_{\bf q+Q} \right) \beta^\dag_{\bf q+Q}
	\Big\}
	\nonumber\\
	\times&
\ket{S({\bf Q}){\otimes}{\rm OL}}
	,
\end{align}
\Cref{tab:SQOQ_cs} shows the explicit form of the RIXS cross section for magnetic excitations in a generic spin system with orbital order or with an orbital liquid state calculated via \cref{eq:KramersHeisenberg} using \cref{eq:RIXS-SQOQ,eq:RIXS-SQOL}. 

\section{Magnetic RIXS spectra and orbital order}
\label{sec:Magnetic-RIXS-orbital}
\subsection{RIXS spectra of two dimensional CuO systems}\label{sec:RIXSintensities}

The strong dependence of the RIXS cross section of magnetic excitations in \cref{tab:FMAFxFOAOOL_cs} and of the the local transition amplitudes in \cref{eq:LocalTransitionAmplitudes} on the orbital ground state is generic to any transition metal $L_{2,3}$ edge. 
This is because this dependence is merely due to the properties of the dipole transitions and to the spin-orbit coupling, and not to the particular electronic configuration of the transition metal ions. 

\begin{figure}
	\centering
	\subfigure[FM-FO]{\includegraphics[scale=.62]{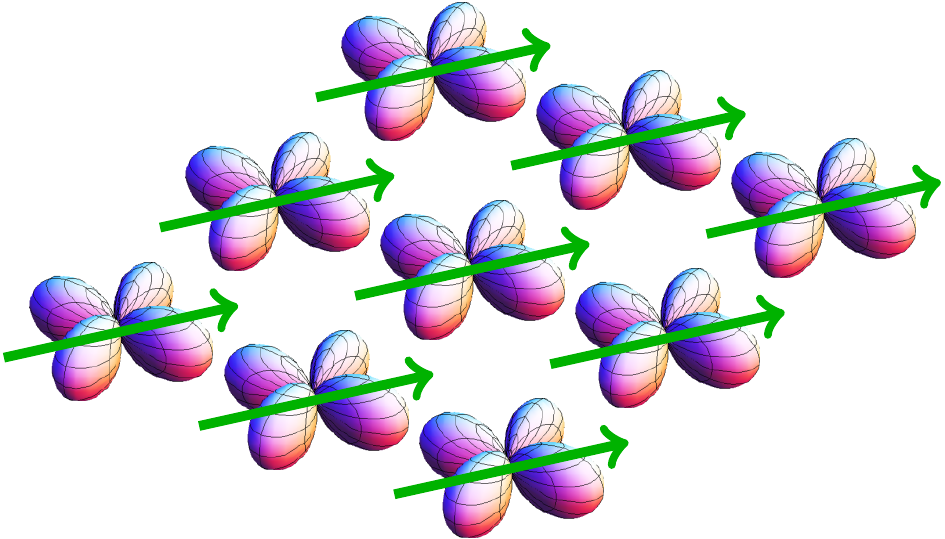}}\qquad\qquad
	\subfigure[AF-FO]{\includegraphics[scale=.62]{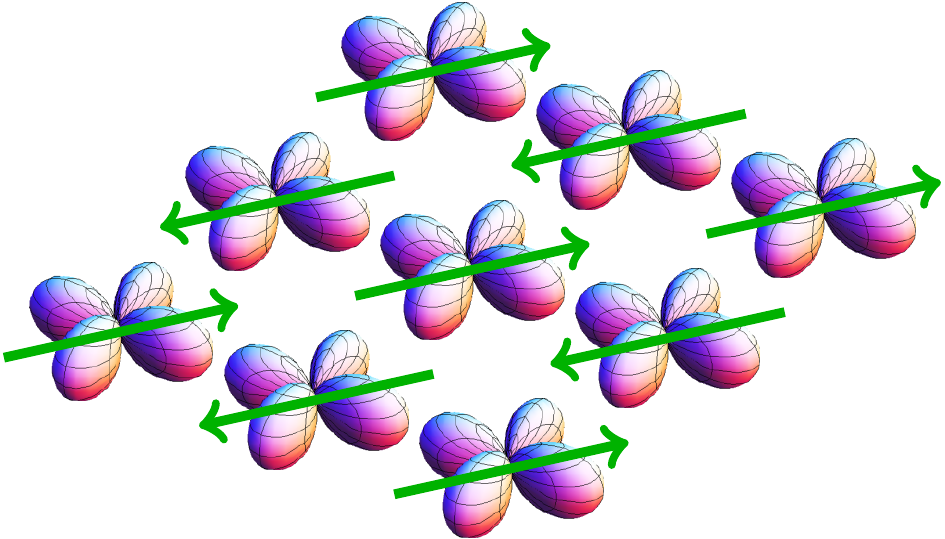}}\\[2mm]
	\subfigure[FM-AO]{\includegraphics[scale=.62]{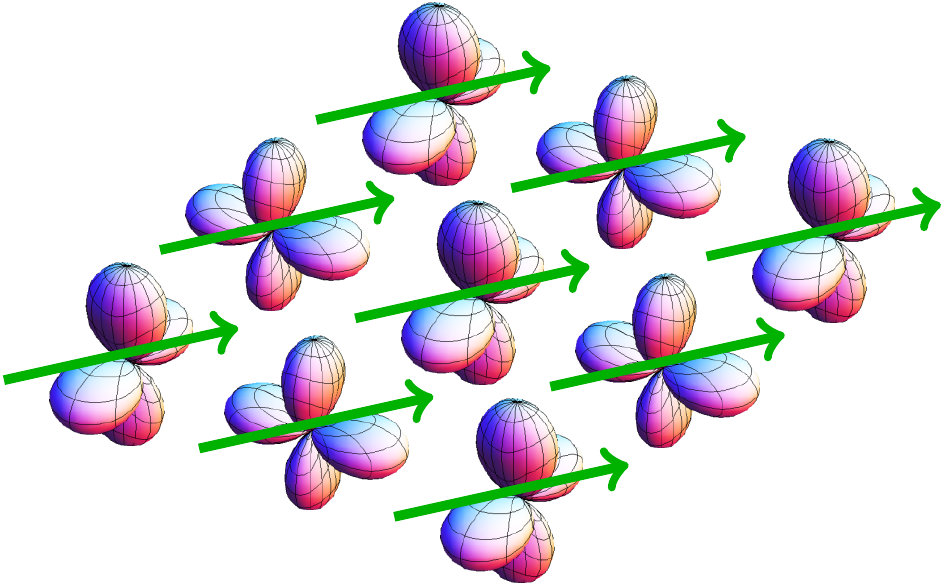}}\qquad\qquad
	\subfigure[AF-AO]{\includegraphics[scale=.62]{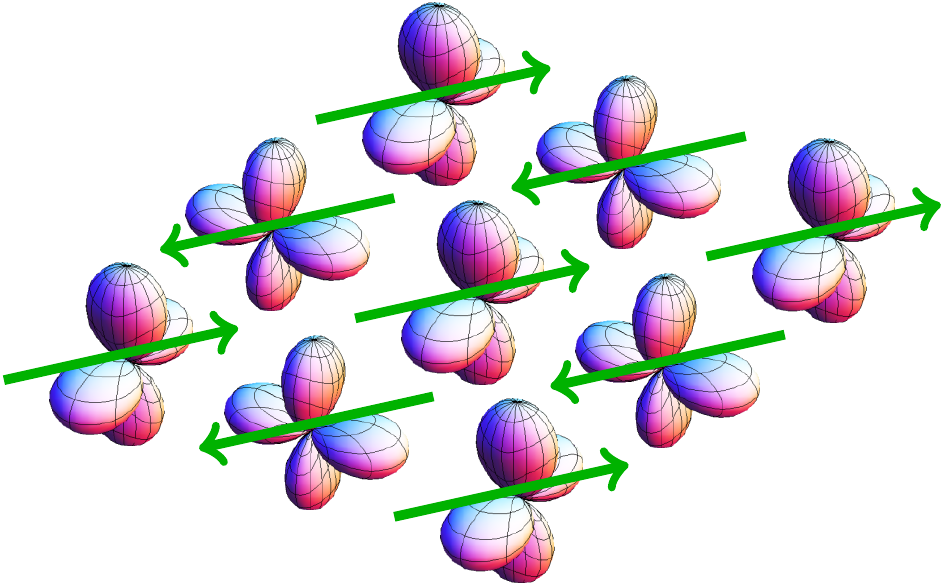}}
	\caption[Examples of systems with different magnetic and orbital orders]{
Examples of two dimensional systems with ferromagnetic and ferroorbital orders (FM-FO), antiferromagnetic and ferroorbital orders (AF-FO), ferromagnetic and alternating orbital orders (FM-AO), and antiferromagnetic and alternating orbital orders (AF-AO). 
}
	\label{fig:FMAFxFOAO}
\end{figure}

In this Section we will show how this dependence arises in a cuprate-like system, i.e., in a two dimensional layer of Cu$^{2+}$ ions, with one hole in the Cu $3d$ orbital. 
We will assume either a ferromagnetic or an antiferromagnetic order with the spin in the $xy$ plane.
On top of this,  we will consider three kinds of orbital ground state, i.e., the ferroorbital order formed by the hole state in the $d_{x^2-y^2}$ orbital at each transition metal ion; the alternating orbital order formed by the set of two alternating orbitals 
\begin{equation*}
\left\{d_{z^2-x^2},d_{y^2-z^2}\right\}=
\left\{\frac{\sqrt3}2d_{3z^2-r^2}-\frac12d_{x^2-y^2},-\frac{\sqrt3}2d_{3z^2-r^2}-\frac12d_{x^2-y^2}\right\}
\end{equation*}
at every other transition metal ion site; and the orbital liquid state where these two orbitals fluctuate with the same expectation value. 
The ferromagnetic and antiferromagnetic states with ferroorbital and alternating orbital orders considered are shown in \cref{fig:FMAFxFOAO,fig:CAOGAO}. 
The dispersion of magnetic excitations in such a two dimensional system are given by \cref{eq:MagnonDispersion,eq:AntiMagnonDispersion} respectively in the ferromagnetic and in the antiferromagnetic case, while the RIXS cross section of these excitations in the case of ferroorbital, alternating orbital order, and orbital liquid state are given in \cref{tab:FMAFxFOAOOL_cs}. 
The local transition amplitudes $W^{\pm}_{{\bf e}{\bf e}'}(d)$ are evaluated via \cref{eq:LocalTransitionAmplitudes} considering the relevant $t_{2g}$ orbitals and the spin in the $xy$ plane. 

\begin{figure}[t!]
\centering
\includegraphics[width=1\textwidth]{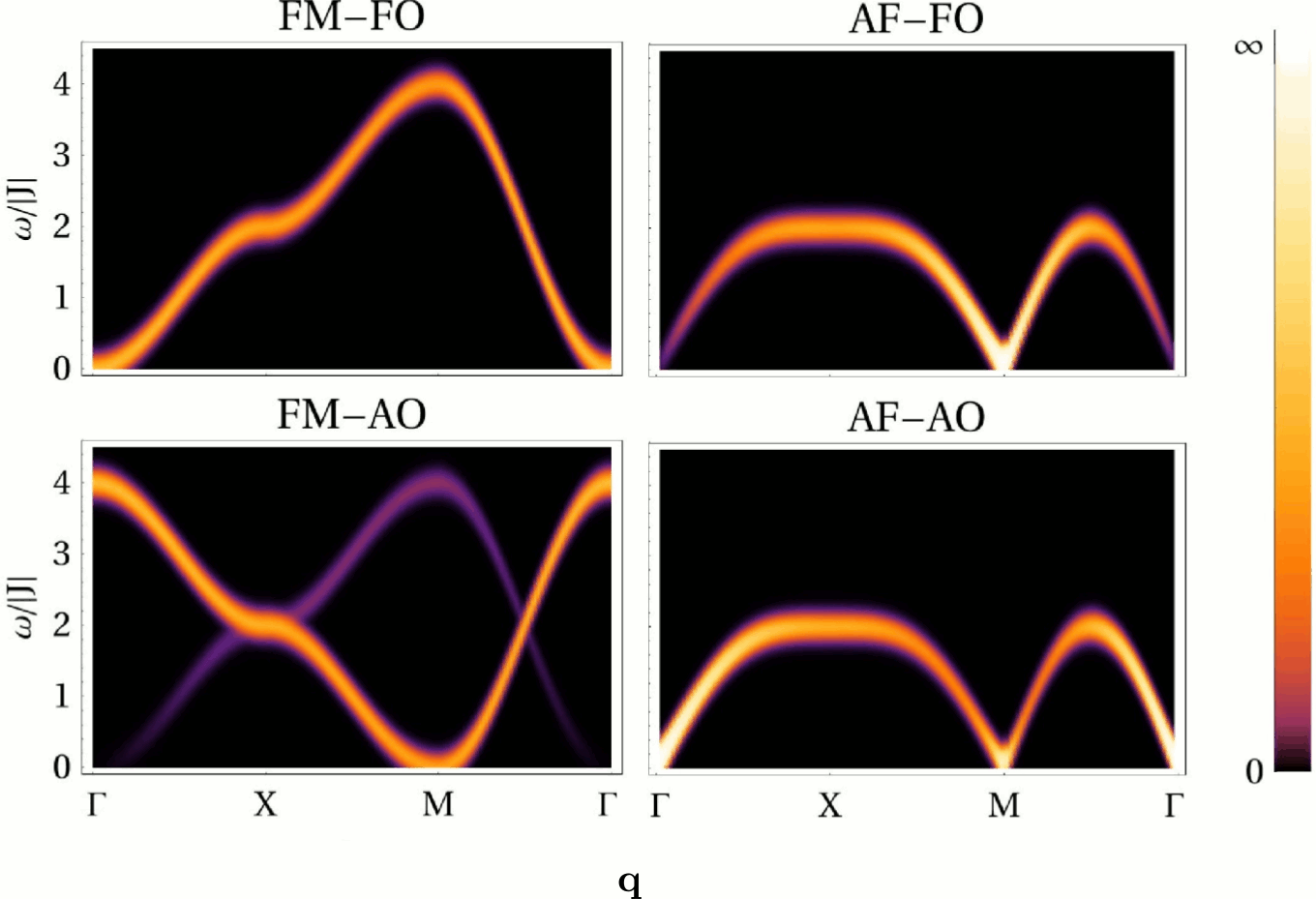}
\caption[Magnetic RIXS cross section in a ferromagnet and an antiferromagnet with ferroorbital and alternating orbital orders]{
Magnetic RIXS cross section $I_{{\bf e}{\bf e}'}({\bf q},\omega)$ for a ferromagnet (FM) and an antiferromagnet (AF) with ferroorbital (FO) and alternating orbital (AO) orders along a high symmetry path in the Brillouin zone [where $\Gamma=(0,0)$, ${\rm X}=(\pi,0)$, and ${\rm M}=(\pi,\pi)$], 
averaged over the polarizations of the incident and scattered radiation. 
The ferroorbital order consist of $d_{x^2-y^2}$ orbital at each transition metal ion, while the alternating orbital order of $d_{x^2-z^2}$ and $d_{y^2-z^2}$ orbital every other site.
In the ferromagnetic case, an optical mode signals the onset of the alternating orbital order, while in the antiferromagnetic case, spectral intensities at ${\bf q}=0$ vanish in the ferroorbital case, but diverge in the alternating orbital case. 
Spectra in the case of an orbital liquid state (not shown) differ only quantitatively from the ferroorbital case. 
}
\label{fig:FMAFxFOAO_cs}
\end{figure}

\Cref{fig:FMAFxFOAO_cs} shows the RIXS intensities as a function of transferred momentum and energy loss in a ferromagnet and an antiferromagnet, with ferroorbital and alternating orbital orders, averaged over all possible incident and scattered polarizations. 
In the ferromagnetic case, the additional optical branch of the magnetic excitation dispersion signals the onset of the alternating orbital order. 
The optical branch intensities are proportional to $\vert W^-_{{\bf e}{\bf e}'}(d_{z^2-x^2})-W^-_{{\bf e}{\bf e}'}(d_{y^2-z^2})\vert^2$ and therefore are finite since the local transition amplitudes relative to a spin flip transition in \cref{eq:LocalTransitionAmplitudes} differ for the two orbitals considered. 
The presence of the optical mode arises from breaking the translational symmetry into two physically inequivalent sublattices, which have a different orbital occupancy. 
Spectra in the case of the orbital liquid state differ only quantitatively from  the ferroorbital order case. 
In the antiferromagnetic case instead, no additional optical branch is present, since the translational symmetry is not further broken by the onset of an alternating orbital order. 
However, the magnetic RIXS spectra differ in the case of a ferroorbital (or orbital liquid state) and alternating orbital orbital orders. 
At the center of the Brillouin zone $\bf q=0$, spectral intensities vanish in the ferroorbital case while they diverge in the case of alternating orbital order. 

Moreover, in the antiferromagnetic case, RIXS spectra are strongly sensitive to the choice of the x-ray polarizations. 
In particular, one can define the circular dichroism of RIXS spectra as the normalized difference between RIXS intensities of orthogonal circular polarizations of the incident x-ray photon
\begin{equation}
	D_{LR}({\bf q},\omega)=
	\frac
	{I_{{\bf e}_L}({\bf q},\omega)-I_{{\bf e}_R}({\bf q},\omega)}
	{I_{{\bf e}_L}({\bf q},\omega)+I_{{\bf e}_R}({\bf q},\omega)}
	,
\end{equation}
where $I_{{\bf e}_L}({\bf q},\omega)$ and $I_{{\bf e}_R}({\bf q},\omega)$ are the spectral intensities corresponding respectively to the incident photon left and right circular polarizations ${\bf e}_L$ and ${\bf e}_R$. 
\Cref{fig:Dichroism} shows the RIXS circular dichroism for an antiferromagnet, as a function of transferred momentum ${\bf q}$ and at energy $\omega=\Omega_{\bf q}$, in the case of ferroorbital and alternating orbital orders. 
In the ferroorbital case, the circular dichroism vanishes, since left and circular polarizations are symmetric respect to the time reversal symmetry, as well as the antiferromagnetic system, up to lattice translations. 
In the alternating orbital case however, the time reversal symmetry of the antiferromagnet is not recovered by lattice translations. 
Therefore, the response of the system to left and right polarizations is no longer equivalent, and a finite circular dichroism appears. 

\begin{figure}[t!]\centering
\includegraphics[width=.7\textwidth]{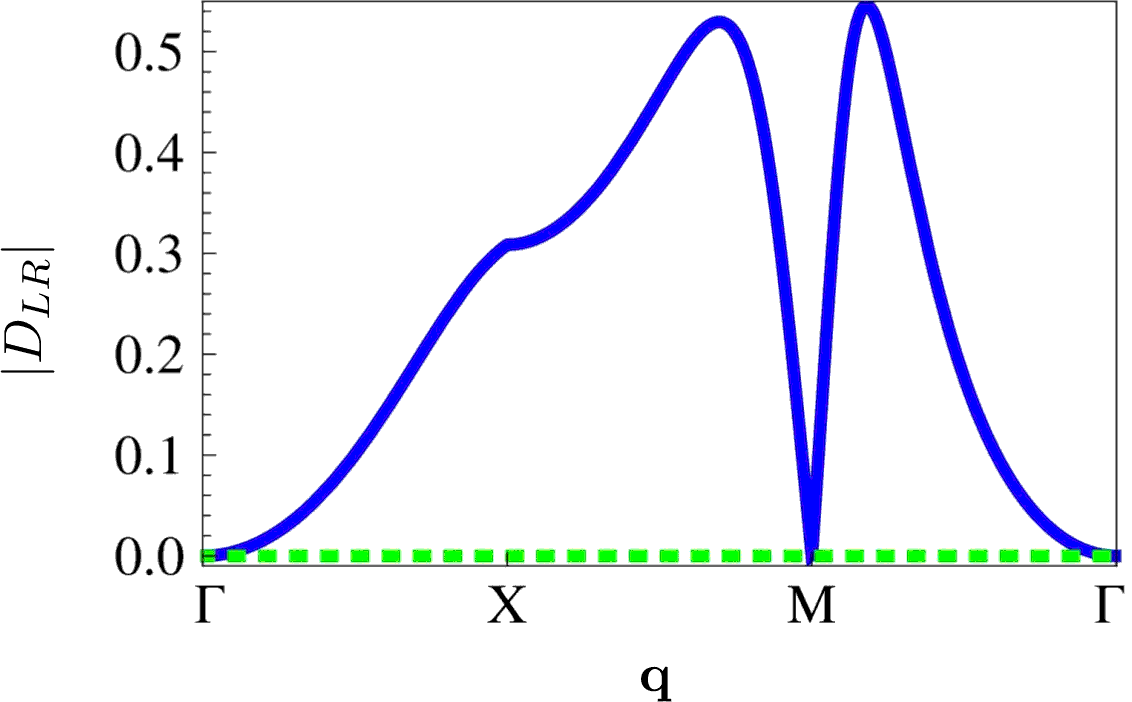}
\caption[RIXS circular dichroism of an antiferromagnet with different orbital orders]{
RIXS circular dichroism $|D_{LR}({\bf q},\omega)|$ at $\omega=\Omega_{\bf q}$ as a function of transferred momentum ${\bf q}$ of an antiferromagnet for ferroorbital (dashed line) and alternating orbitals (solid line) orders. 
}
\label{fig:Dichroism}
\end{figure}

\subsection[A case study: KCuF$_3$]{A case study: KCuF$_3$}

In this Section we will consider the RIXS spectra of magnetic excitations in the so-called $A$-AF state, i.e., ferromagnetic planes with antiferromagnetic coupling along the $z$ direction perpendicular to the planes, as realized in KCuF$_3$ below $T<T_N\sim38\unit{K}$~\cite{Satija1980, Lake2000}. 
The magnetic ordered state $A$-AF is characterized by an ordering vector ${\bf Q}=(0,0,\pi)$, and the corresponding magnetic excitations have dispersion given by \cref{eq:MagnonFFA}. 
Besides the magnetic ordering, we will consider here three different orbital ground states~\cite{Satija1980}: (i) $C$-AO state, i.e., alternating orbitals within the $xy$ planes with same orbitals stacked along the $z$ direction, realized in the (d)-type polytype of KCuF$_3$ below $T<T_S\sim800\unit{K}$~\cite{Oles2005}, (ii) $G$-AO state, i.e., isotropic three dimensional alternating orbital state, realized in the (a)-type polytype of KCuF$_3$ below $T<T_S\sim800\unit{K}$~\cite{Oles2005}, and (iii) orbital liquid (OL) state (not realized in KCuF$_3$ but included for comparison). 
The $C$-AO and the $G$-AO orders are described by the orbital ordering vector ${\bf\bar{Q}}=(\pi,\pi,0)$ and ${\bf\bar{Q}}=(\pi,\pi,\pi)$ respectively. 
Moreover, two different sets of alternating orbitals $\{a, b\}$ are considered: (i) $d_{x^2-y^2}$ and $d_{3z^2-r^2}$ orbitals, in the limit of vanishing orbital-lattice interactions (crystal field and Jahn-Teller interaction), and (ii) $d_{x^2-z^2}$ and $d_{y^2-z^2}$ orbitals, which are favored by the orbital-lattice interactions, which probably represent a realistic description of the actual orbital ground states realized in KCuF$_3$ polytypes~\cite{Oles2005}. 
The magnetic $A$-AF state with orbital $C$-AO and $G$-AO orders, for the two sets of alternating orbitals considered, are shown in \cref{fig:CAOGAO}. 

The spin wave excitations in the $A$-AF magnetic state are described by the Heisenberg Hamiltonian with nearest neighbor anisotropic interaction
\begin{equation}
	 H= J_{xy}	\sum_{\langle i, j \rangle || x,y} {\bf S}_i \cdot {\bf S}_j 
	  + J_z		\sum_{\langle i, j \rangle || z}   {\bf S}_i \cdot {\bf S}_j,  
\end{equation}
where the spin exchange constants $J_{xy}<0$ and $J_z>0 $ refer to nearest neighbors respectively within the planes $xy$ and along the $z$ direction. 
The anisotropic structure of this spin-only Hamiltonian stems from the full Kugel-Khomskii spin-orbital Hamiltonian~\cite{Oles2005}, when orbital degrees of freedom, which are responsible for the onset the $C$-AO or the $G$-AO orbital orders, are integrated out. 
As a consequence, the values of the spin exchange constants $J_{xy}$ and $J_z$ depend on the orbital ground state. 
The magnon dispersion in the $A$-AF magnetic state of KCuF$_3$ corresponds to the case of ferromagnetic coupling in the $xy$ plane and antiferromagnetic coupling along the perpendicular direction $z$ in \cref{eq:MagnonFFA}. 
The RIXS cross section of these excitations in the case of $C$-AO and $G$-AO orders and orbital liquid (OL) states are given in \cref{tab:SQOQ_cs}, considering the magnetic ordering vector ${\bf Q}=(0,0,\pi)$ of the $A$-AF magnetic state and the orbital ordering vector ${\bf\bar{Q}}=(\pi,\pi,0)$ and ${\bf\bar{Q}}=(\pi,\pi,\pi)$ respectively for the two orbital orders considered. 
The local transition amplitudes $W^{\pm}_{{\bf e}{\bf e}'}(d)$ for the two sets of $t_{2g}$ orbitals are evaluated via \cref{eq:LocalTransitionAmplitudes}. 

\begin{figure}
	\centering
	\subfigure[$C$-AO $\{d_{x^2-y^2},d_{3z^2-r^2}\}$]{\includegraphics[scale=.5]{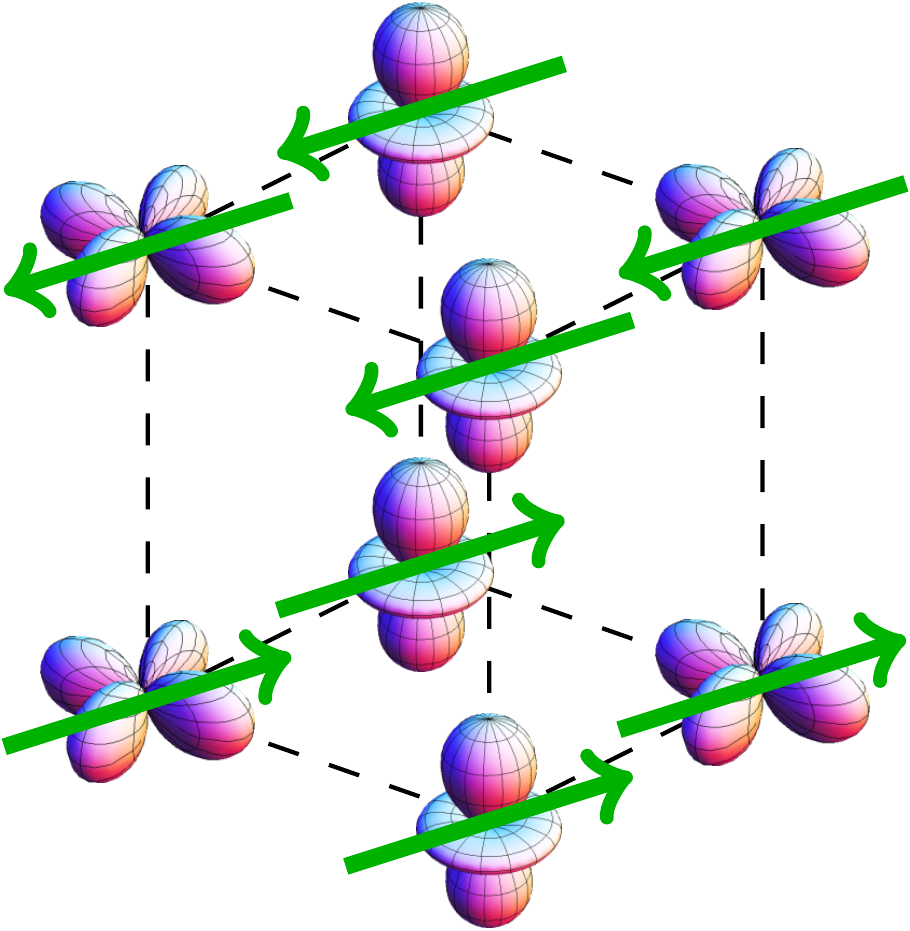}}\quad\quad\quad\quad
	\subfigure[$G$-AO $\{d_{x^2-y^2},d_{3z^2-r^2}\}$]{\includegraphics[scale=.5]{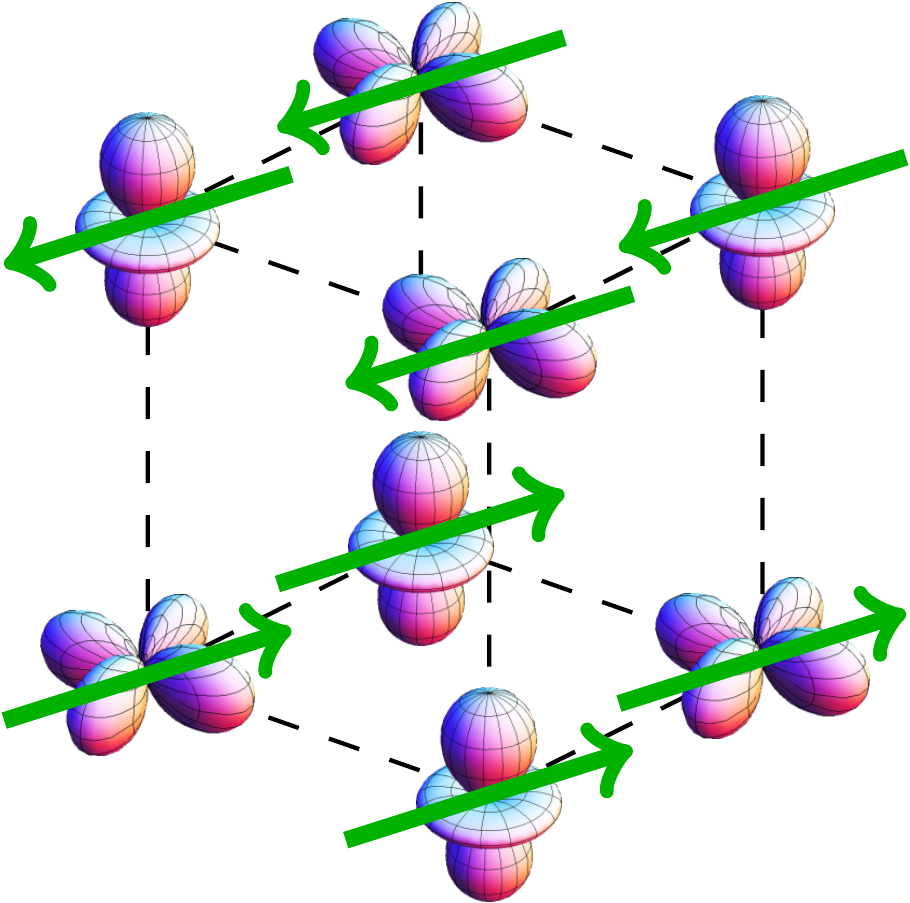}}
	\subfigure[$C$-AO $\{d_{x^2-z^2},d_{y^2-z^2}\}$]{\includegraphics[scale=.5]{A-AF-C-AO-b}}\quad\quad\quad\quad
	\subfigure[$G$-AO $\{d_{x^2-z^2},d_{y^2-z^2}\}$]{\includegraphics[scale=.5]{A-AF-G-AO-b}}
	\caption[The antiferromagnetic state with different orbital orders in KCuF$_3$]{
The magnetic $A$-AF state with orbital $C$-AO and $G$-AO orders, as realized respectively in the (d)-type and (a)-type KCuF$_3$ polytypes, for the two sets of alternating orbitals $\{d_{x^2-y^2},d_{3z^2-r^2}\}$ and $\{d_{x^2-z^2},d_{y^2-z^2}\}$ considered. 
}
\label{fig:CAOGAO}
\end{figure}

In \cref{fig:KCuF_cs} are shown the RIXS cross sections for the two different orbital orders and for the orbital liquid state, for two choices of the orbital occupancies, assuming that $|J_{xy}/J_z| = 0.06$ as in \onlinecite{Lake2000,Oles2005}.
While no clear signature in the RIXS spectra allows one to distinguish between the two sets of orbital occupancies $\{d_{x^2-y^2},d_{3z^2-r^2}\}$ and $\{d_{x^2-z^2},d_{y^2-z^2}\}$ (a subtle intensity shift between the optical and the acoustic branch is hardly visible), differences between the $C$-AO, $G$-AO orders and orbital liquid state show off strikingly (see \cref{fig:KCuF_cs}). 
In fact, in the case of an orbital liquid state, only the acoustic branch of the magnon dispersion is present, whereas in the case of alternating orbital orders ($C$-AO and $G$-AO) the additional optical branch appears (compare with the differences between the ferroorbital and alternating orbitals states in \cref{fig:FMAFxFOAO_cs}).
Moreover, spectral intensities of the optical branch at $\rm{M}=(\pi,\pi,0)$ discriminate between the two different alternating orbitals orders, vanishing in the $C$-AO, and diverging in the the $G$-AO case.

Therefore, magnetic RIXS cross section allows one to distinguish between the various orbitally ordered phases which are predicted to be stable in KCuF$_3$~\cite{Oles2005}. 
Furthermore, the spectrum of the orbital liquid phase (which is not stable in the magnetically ordered phase of KCuF$_3$~\cite{Oles2005}) is strikingly different from the one of the ordered phases.
This shows that the magnetic RIXS cross section strongly depends on orbital correlations --- as already discussed in simplest case of two dimensional magnetic systems in \cref{sec:RIXSintensities}. 

\begin{figure}[t!]\centering
\includegraphics[width=1\textwidth]{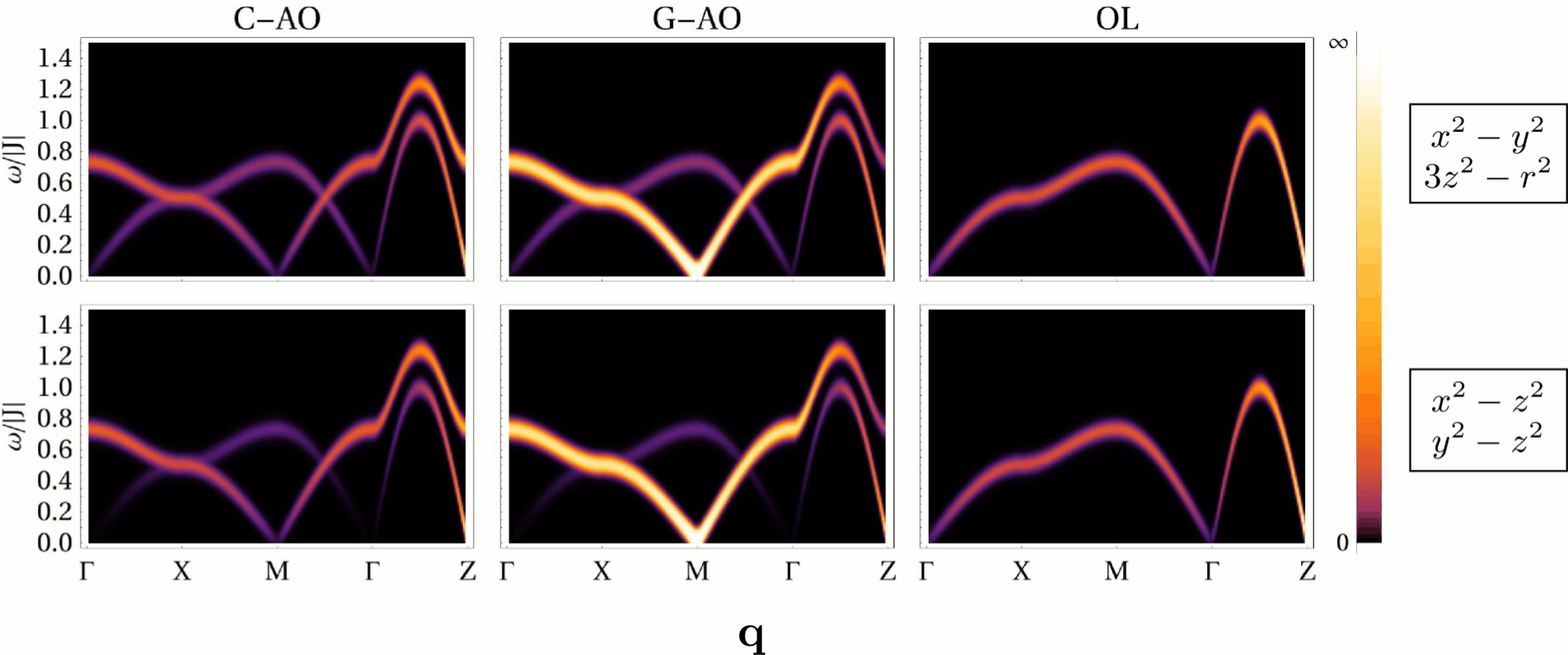}
\caption[Magnetic RIXS cross section for KCuF$_3$]{
Magnetic RIXS cross section $I_{{\bf e}{\bf e}'} ({\bf q}, \omega)$ for KCuF$_3$ along a high symmetry path in Brillouin zone for three different orbital ground states (from left to right, $C$-AO, $G$-AO and orbital liquid state), formed by alternating $d_{x^2-y^2}$  and $d_{3z^2-r^2}$ orbitals (top row), and alternating $d_{x^2-z^2}$  and $d_{y^2-z^2}$ orbitals (bottom row), and with the spin in the $xy$ plane. 
The cross section is averaged over incident and scattered photon polarizations and the exchange parameter are chose such that $|J_{xy}/J_z| = 0.06$. 
The high symmetry points in the Brillouin zone are defined as $\Gamma=(0,0,0)$, ${\rm X}=(\pi,0,0)$, ${\rm M}=(\pi,\pi,0)$, and ${\rm Z}=(0,0,\pi)$. 
The color scale is nonlinear, since intensities diverge at $\rm Z$ in every case, and at $\rm M$ in the $G$-AO case.
}
\label{fig:KCuF_cs}
\end{figure}

\subsection{Discriminating different orbital states}\label{sec:Discriminate}

As shown in \cref{fig:FMAFxFOAO}, RIXS spectra can discriminate the alternating orbital against ferroorbital order or orbital liquid ground state.
Whereas in the ferromagnetic case the optical magnon branch signals the onset of the alternating orbital order, in the antiferromagnetic case the intensity of magnons with momenta ${\bf q}\rightarrow0$ does not vanish in the alternating orbital case, contrarily to the ferroorbital and orbital liquid case.
This dependence is not due to distinct magnon dispersions for different orbital or electronic ground states~\cite{Wohlfeld2011,Tung2008,Ulrich2003}, but to the orbital dependency of magnetic RIXS amplitudes.

Furthermore, circular dichroism of \emph{magnetic} RIXS intensities allows one to distinguish between different \emph{orbital} ground states, as shown in \cref{fig:Dichroism}. 
While in ferromagnets the presence of a finite circular dichroism depends on the symmetry of the orbital occupied, in antiferromagnets it only depends on the system translational symmetry.
Specifically, in antiferromagnets with ferroorbital order (AF-FO) or with orbital liquid state (AF-OL) circular dichroism vanishes, while in the case of alternating orbital order (AF-AO) the circular dichroism is nonzero (provided that spin flip amplitudes are finite for both orbitals forming the alternating orbital ground state, cf.~\onlinecite{Ament2009}).

In fact, if there is an alternating orbital order in a magnetic system, translational symmetry is broken into two physically inequivalent sublattices, as shown in \cref{fig:Translations}. 
Consequently an optical branch in the magnon dispersion in a ferromagnet with alternating orbital order (FM-AO). 
On the other hand, while an antiferromagnet with ferroorbital order (AF-FO) or with a orbital liquid state (AF-OL) is symmetric under the combination of time reversal (which flips the spin directions) and a discrete translation~\cite{Dresselhaus2010}, in the case of an alternating orbital order (AF-AO) the latter is broken, as shown in \cref{fig:Translations}. 
Macroscopically~\cite{Birss1966}, that means that the system is no longer symmetric under the combination of time reversal and translation. 
As a consequence, a finite circular dichroism appears, i.e., RIXS intensities corresponding to left and right circular polarizations of the incident radiation are no longer equivalent. 

\begin{figure}
	\centering
	\subfigure[FM-FO: $\hat{T}$]{\includegraphics[scale=.6]{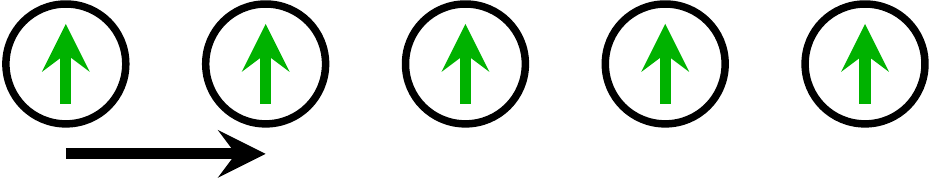}}\qquad\qquad
	\subfigure[FM-AO: $\hat{T}^2$]{\includegraphics[scale=.6]{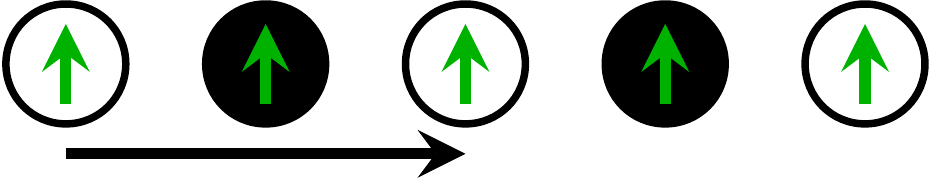}}
	\\[5mm]
	\subfigure[AF-FO: $\hat{\Theta}\hat{T}$,$\hat{T}^2$]{\includegraphics[scale=.6]{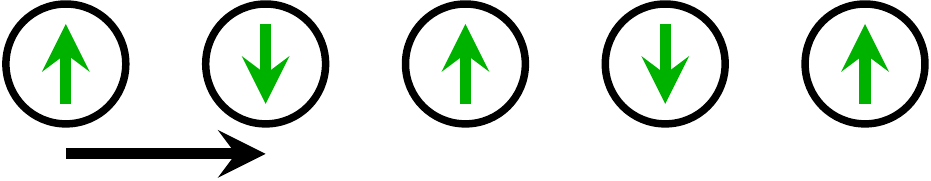}}\qquad\qquad
	\subfigure[AF-AO: $\hat{T}^2$]{\includegraphics[scale=.6]{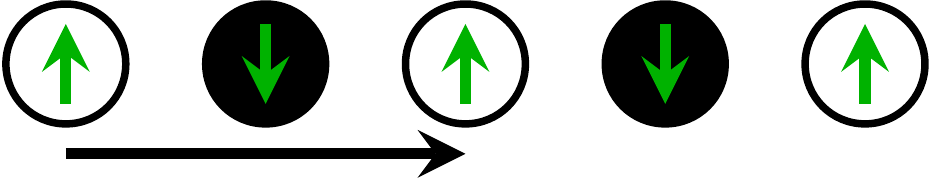}}
	\\
	\caption[Time reversal and lattice translations in magnetic systems with orbital order]{
(a) Translation vector $\hat{T}$ which transform the ferromagnetic system with ferroorbital order in itself. 
(b) In the case of an alternating orbital order, the ferromagnet is no longer invariant under a translation $\hat{T}$ of one lattice site, but it is still invariant under the translation $\hat{T}^2$ of two lattice sites. 
The translation symmetry is lowered by the onset of the alternating orbital order. 
(c) An antiferromaget with ferroorbital order is invariant under the combination of a lattice translation $\hat{T}$ and time reversal $\hat{\Theta}$ (which flips the spin directions), as well as the combination of two lattice translations $\hat{T}^2$. 
(d) If an alternating orbital order is introduced, the antiferromagnet is no longer invariant under the combination of lattice translation and time reversal, although the translation symmetry $\hat{T}^2$ is not further broken. 
}
	\label{fig:Translations}
\end{figure}

Although the actual values of the local transition amplitudes $W_{{\bf e}{\bf e}'}^\pm(d)$ depend on the orbital symmetry at each ion site,  differences in the RIXS spectra between different orbital ground states show up (cf. \cref{fig:FMAFxFOAO_cs} and \cref{fig:Dichroism}), as long as $W_{{\bf e}{\bf e}'}^\pm(a)\neq W_{{\bf e}{\bf e}'}^\pm(b)$. 
For this reason, the discrimination between different orbital states does not rely on the particular orbital occupancy on the single ion site, but on the breaking of the translational symmetry caused by the onset of the alternating orbital order orbital order. 

While other inelastic scattering methods have been theoretically proposed to detect orbital ordering~\cite{Ito1976,vanVeenendaal2008}, it should be stressed that, due to the onset of characteristic dispersion, the magnetic peaks in RIXS can, unlike, e.g., orbitons, be easily identified. 
Besides, as magnons typically interact weakly, quasiparticle peaks in RIXS spectra have sharp and well defined line shapes that will not be obliterated by other low energy excitations (cf. \onlinecite{Ament2011}) and therefore their dependence on the orbital ground state is very pronounced.

\section{Conclusions}

In this Chapter, we have shown in detail how orbital correlations in the ground state directly reflect themselves in magnetic RIXS intensities.
It follows that measuring the RIXS spectra at transition metal $L_{2,3}$ edges in correlated materials with orbital degrees of freedom and magnetic order allows one to distinguish between different orbital ground states.
Although it seems not possible to distinguish between ferroorbital order and orbital liquid state in magnetic RIXS, typically in orbital systems the main question is whether the orbital ground state has an alternating orbital order or is in a orbital liquid state~\cite{Khaliullin2000,Khaliullin2001,Pavarini2004,DeRaychaudhury2007} --- for which the proposed method is well suited.
This is possible because in magnetic RIXS the spin flip mechanism involves a strong spin-orbit coupling deep in the electronic core so that, unlike in inelastic neutron scattering, the magnetic scattering spectra strongly depend on the symmetry of the orbitals where the spins are in.

}
\chapter[RIXS as a probe of the superconducting order parameter]{RIXS as a probe of the superconducting order parameter\footnotemark{}}
\footnotetext{Part of this chapter has been published in \onlinecite{Marra2013}}
{
\newcommand{\kk}{{\bf k}}
\newcommand{\qq}{{\bf q}}
\newcommand{\vv}{{\bf v}}
\newcommand{\ee}{{\bf e}}
\newcommand{\kkA}{{\kk_A}}
\newcommand{\kkN}{{\kk_N}}
\newcommand{\kkF}{{\kk_F}}
\newcommand{\Ek}{E_\kk}
\newcommand{\uk}{u_\kk}
\newcommand{\vk}{v_\kk}
\newcommand{\vvk}{{\bf v_k}}
\newcommand{\vvkF}{{\bf v}_\kkF}
\newcommand{\eps}{\varepsilon}
\newcommand{\epsk}{\varepsilon_\kk}
\newcommand{\Sqomega}{S_\qq(\omega)}
\newcommand{\Deltak}{\Delta_\kk}
\newcommand{\DeltakF}{\Delta_\kkF}
\newcommand{\Fkq}{F_\kk(\qq)}
\newcommand{\gammadag}{\gamma^\dag}
\newcommand{\gammaks}{\gamma_{\kk\sigma}^{\nodag}}
\newcommand{\gammaksdag}{\gamma_{\kk\sigma}^{\dag}}
\newcommand{\BCSGround}{\vert{\rm BCS}\rangle}

\label{ch:BCS-Cuprates}

Probing the order parameter in unconventional superconductors is generally the first step for an investigation of the pairing mechanism and of the character of the superconducting state. 
In this Chapter we will develop the RIXS scattering theory of quasiparticle excitations in superconductors, calculating its momentum-dependent scattering amplitudes at zero temperature. 
In \cref{sec:Quasiparticle} we will introduce the concept of quasiparticle excitations in unconventional superconductors, and consequently in \cref{sec:RIXS-BCS-DSF} we will show how direct RIXS spectra are a direct probe of the charge and spin DSF\@. 
Finally, considering superconductors with different pairing symmetries, we will demostrate in \cref{sec:Phase-sensitivity} that the momentum dependence of RIXS spectra in the low energy range is intrinsically determined by the pairing symmetry, being sensitive not only to the magnitude of the superconducting gap and to the presence of nodes on the Fermi surface but also to the phase of the order parameter. 
This phase sensitivity is due to the appearance of coherence factors which, for instance, in STM determine to large extent the quasiparticle interference in the presence of impurities~\cite{Hoffman2002,McElroy2003,Hanaguri2007,Kohsaka2008,Hanaguri2009,Hanke2012,Sykora2011}.

\section{Quasiparticle excitations in a superconductor}
\label{sec:Quasiparticle}

The emergence of a superconducting state in a wide class of different materials, from metals to strongly correlated unconventional superconductors, is a consequence of the Cooper pairing mechanism~\cite{Cooper1956}, that is, an \emph{attractive} interaction between electrons which induces an instability in the Fermi gas with respect to the formation of Cooper pairs, i.e., a bound state of two electrons. 
In general, Cooper pairs can be formed by spin-singlet states (singlet pairing) with total spin $s=0$, or by spin-triplet states (triplet pairing) with total spin $s=1$, and can have a nonzero total momentum as in, e.g., the Fulde-Ferrell-Larkin-Ovchinnikov phase\cite{Fulde1964,Larkin1965}. 
However, in their simplest realization, Cooper pairs are spin-singlet states with zero total momentum and spin, i.e., the pairing mechanism couples electrons on the Fermi surface which have opposite momenta and spins. 
In conventional superconductors, the superconducting state can be described by the BCS theory~\cite{Bardeen1957a,Bardeen1957b,Tinkham2004}, where the pairing mechanism is assumed to be a phonon-mediated attractive interaction of the conduction electrons in a metal. 
Although the origin of the electron pairing is still not clear in high temperature superconductors (e.g., cuprates, and more recently, pnictides), the superconducting state in this unconventional superconductors can be still, in general, described by the BCS theory or by one of its generalizations~\cite{Eliashberg1963,DeGennes1999}. 
In fact, BCS and BCS-like theories do not require any assumptions on the origin of the pairing mechanism and, moreover, can be extended to cases where the attractive interaction depends in a non trivial way on the spin and on the momentum of the interacting electrons. 
In this Section we will concentrate on singlet pairing superconductors, making no assumptions on the origin of the pairing mechanism. 
However, we will assume that the normal state of the superconductor is a non-correlated electron system, i.e., either a non-interacting electron gas or a Fermi liquid with renormalized quasiparticle interactions. 

In a single band Fermi gas of non-interacting electrons, the singlet pairing mechanism is described by an effective Hamiltonian in the form
\begin{emphequation}\label{eq:BCSPairingHamiltonian}
	\hat{\cal H}-\eps_FN=
	\sum\limits_{\kk\sigma}	
	\eps_\kk
	c^\dag_{\kk\sigma}c^\nod_{\kk\sigma}
	+
	\sum\limits_{\kk\kk'}
	V_{\kk\kk'}
	c^\dag_{\kk'\uparrow}c^\dag_{-\kk'\downarrow}
	c^\nod_{-\kk\downarrow}c^\nod_{\kk\uparrow}
	,
\end{emphequation}
where the first term describes the kinetic energy of tight-binding electrons with bare electron dispersion $\eps_\kk$, whereas the second contains the two electron attractive interaction, and where $V_{{\bf k}{\bf k}'}$ is the pairing potential which depends on the momenta of the two interacting electrons. 
The common and most straightforward approach to deal with two particle interactions is to introduce a mean field approximation. 
Expanding the two particle operator $c^\nod_{-\kk\downarrow}c^\nod_{\kk\uparrow}$ in terms of fluctuations around its mean value and consequently neglecting higher order terms, one obtains the BCS mean field Hamiltonian which reads, up to a constant term
\begin{align}\label{eq:BCSHamiltonian}
	\hat{\cal H}=
	\sum\limits_{\kk\sigma}	
	\eps_\kk
	c^\dag_{\kk\sigma}c^\nod_{\kk\sigma}
	-\sum\limits_{\kk}
	\left(
	\Delta_\kk^*	c^\nod_{-\kk\downarrow}c^\nod_{\kk\uparrow}
	+
	\Delta_\kk	c^\dag_{\kk\uparrow}c^\dag_{-\kk\downarrow}
	\right)
	,
\end{align}
where the order parameter $\Delta_\kk$ is related to the mean value of the two electron operators via
\begin{equation}
	\Delta_\kk=-\sum_{\kk'}V_{\kk\kk'}\langle c^\nod_{-\kk'\downarrow}c^\nod_{\kk'\uparrow} \rangle
	,
\end{equation}

In order to find the ground state and the excitations of the BCS Hamiltonian \ref{eq:BCSHamiltonian}, one can introduce a new set of fermionic quasiparticle operators $\gamma^\dag_{\kk\sigma}$ and $\gamma^\nod_{\kk\sigma}$ defined by the Bogoliubov transformation
\begin{align}\label{eq:BogoliubovBCS}
	c^\nod_{\kk\uparrow}&=u_{\kk}\gamma^\nod_{\kk\uparrow} + v_{\kk}\gamma^\dag_{-\kk\downarrow}
	\nonumber\\
	c^\dag_{\kk\downarrow}&=u_{\kk}\gamma^\dag_{\kk\downarrow} - v_{\kk}\gamma^\nod_{-\kk\uparrow}
	,
\end{align}
with the conditions $u_\kk v_{-\kk}=u_{-\kk}v_\kk$ and $|u_{\bf q}|^2+|v_{\bf q}|^2=1$, which ensure that the new operators satisfy the canonical commutation relations $\{\gamma^\nod_{\kk\sigma},\gamma^\dag_{\kk'\sigma'}\}=\delta_{\kk\kk'}\delta_{\sigma\sigma'}$. 
The Hamiltonian \ref{eq:BCSHamiltonian} is diagonal in terms of the Bogoliubov fermions, in the form
\begin{emphequation}\label{eq:BCSHamiltonianDiagonalized}
	\hat{\cal H}= \sum_{\kk\sigma} E_{\kk} \gamma^\dag_{\kk\sigma}\gamma^\nod_{\kk\sigma}
	= \sum_{\kk\sigma} \sqrt{\eps_{\kk}^2+|\Delta_\kk|^2} \gamma^\dag_{\kk\sigma}\gamma^\nod_{\kk\sigma}
	.
\end{emphequation}
where $E_{\kk}=\sqrt{\eps_{\kk}^2+|\Delta_\kk|^2}$ is the quasiparticle dispersion and with the Bogolubov factors in \cref{eq:BogoliubovBCS} which satisfy the conditions
\begin{equation}
	E_\kk\left(|u_\kk|^2-|v_\kk|^2\right)=\eps_\kk,	\qquad u_\kk v_\kk=\frac{\Delta_\kk}{2E_\kk},
\end{equation}
which leads to 
\begin{align}
	|u_\kk|^2=\frac12\left(1+\frac{\eps_\kk}{E_\kk}\right),\qquad
	|v_\kk|^2=\frac12\left(1-\frac{\eps_\kk}{E_\kk}\right),
\end{align}
with the reciprocal phase determined by the previous equation. 
The BCS ground state is defined by the quasiparticle vacuum $\gamma_{\kk\sigma}\ket{\text{BCS}}=0$, i.e, the state which is annihilated by any quasiparticle operator $\gamma_{\kk\sigma}$, which is given by
\begin{equation}\label{eq:BCSGroundState}
	\ket{\text{BCS}}\propto\prod\limits_{\kk\sigma}\gamma_{\kk\sigma}\ket{0}
	,
\end{equation}
up to a normalization factor, where $\ket{0}$ is the bare electron vacuum. 
The excited states of the BCS Hamiltonian in \cref{eq:BCSHamiltonianDiagonalized}, known as quasiparticle excitations, are a linear combination of electron and hole states in the form 
\begin{align}\label{eq:BCSExcitations}
	\gamma^\dag_{\kk\uparrow}\ket{\text{BCS}}=\left(u_\kk c^\dag_{\kk\uparrow}-v_\kk c^\nod_{-\kk\downarrow}\right) \ket{\text{BCS}},
	\nonumber\\
	\gamma^\dag_{\kk\downarrow}\ket{\text{BCS}}=\left(u_\kk^* c^\dag_{\kk\downarrow}+v_\kk^* c^\nod_{-\kk\uparrow}\right) \ket{\text{BCS}},
\end{align}
and have dispersion $E_\kk$.
In the normal state, where $\Delta_\kk=0$, the BCS Hamiltonian in \cref{eq:BCSHamiltonianDiagonalized} reduces to a tight-binding Hamiltonian and the quasiparticle excitations in \cref{eq:BCSExcitations} reduce to single electron excitations with bare dispersion $|\eps_\kk|$. 

\section{RIXS as a probe of the DSF of a superconductor}
\label{sec:RIXS-BCS-DSF}

\subsection{Direct RIXS and DSF}

In direct RIXS scattering at transition metal ion $L_{2,3}$ edges, the incident photon resonantly excites the core shell $2p$ electron into the $3d$ shell which consequently decays into a scattered photon and a charge, spin, or orbital excitation in the electronic system.  
In this case the RIXS cross section can be decomposed into a sum of the charge DSF the spin DSF as in \cref{eq:DSF_W} with prefactors $W^0_{{\bf e}{\bf e}'}$ and ${\bf W}_{{\bf e}{\bf e}'}$ which correspond to the local transition amplitudes of the RIXS operator $\hat{O}_i$ defined in \cref{eq:RIXSOperatorLocal}. 
These local transition amplitudes depend on the specific transition metal ion, the spin and the orbitals occupied, and on the polarizations of the incident and scattered photon. 
In the case that the spin DSF has the same momentum and energy dependence for all the three different components of the spin operator --- as is the case of unconventional superconductors~\cite{Andersen2005} --- the RIXS cross section in \cref{eq:KramersHeisenberg} can be rewritten as in \cref{eq:KramersHeisenberg_DSF} in \cref{sec:DSF_W}, in terms of charge and spin DSF as
\begin{emphalign}\label{eq:BCS_cs}
	\frac{d^2 \sigma}{d\omega d\Omega}\propto
	|W^0_{{\bf e}{\bf e}'}|^2 \chi^0 ({\bf q}, \omega)
	+
	|W^z_{{\bf e}{\bf e}'}|^2 \chi^z ({\bf q}, \omega)
	,
\end{emphalign}
where one assumes that the local transition amplitudes can be tuned in such a way that the terms in \cref{eq:KramersHeisenberg_DSF} which mix charge and spin excitations can be neglected.
The charge DSF and the $z$ component of the spin DSF are defined respectively as
\begin{align}
	\chi^0(\qq,\omega)&=\sum_f | \langle f | \hat{\rho}_{\bf q} | i \rangle|^2 \delta (\hbar \omega - E_f + E_i)
	\nonumber\\
	\chi^z(\qq,\omega)&=\sum_f | \langle f | \hat{S}^z_{\bf q}  | i \rangle|^2 \delta (\hbar \omega - E_f + E_i)
	,
\end{align}
where $\rho_\qq=\sum_{\kk\tau}c^\dag_{\kk+\qq\tau}c^\nod_{\kk\tau}$ and $\hat{S}^z_\qq=\sum_{\kk\tau\tau'}\sigma^z_{\tau\tau'}c^\dag_{\kk+\qq\tau}c^\nod_{\kk\tau'}$ are the density and the spin operators of conduction electrons, with $\sigma^z_{\tau\tau'}$ the third Pauli matrix. 
Since the local transition amplitudes $W^0_{{\bf e}{\bf e}'}$ and $W^z_{{\bf e}{\bf e}'}$ depends on the polarizations of the incident and scattered photon, they can be tuned by properly adjusting the scattering geometry of the RIXS experiment. 
This implies that direct RIXS at $L_{2,3}$ edges can measure either spin or charge DSF depending on the chosen polarization, which is a unique feature of RIXS spectroscopy. 

In what follows we will derive an explicit expression of the charge and spin DSF of a superconductor in the framework of the BCS theory, and we will show how the DSF, and therefore RIXS spectra, can probe the quasiparticle excitation spectra of a superconductor, and allow one to disentangle the nature of these excitations and of the underlaying pairing mechanism. 
The Bogoliubov transformation allows one to evaluate the DSF of a superconductor via the calculation of the transition amplitudes $\bra{f}\rho_\qq\ket{g}$ and $\bra{f}\hat{S}^z_\qq\ket{g}$ between the ground state $\ket{\text{BCS}}$ in \cref{eq:BCSGroundState} and any excited state of the Hamiltonian. 
At zero temperature, the excited states which contribute to DSF have the form  $\gammadag_{\kk+\qq,\sigma}\gammadag_{-\kk,-\sigma}\ket{\text{BCS}}$, i.e., corresponding to two-quasiparticle excitations with dispersion $E_{\kk,\qq}=E_{\kk+\qq}+E_{\kk}$. 
Using the Bogoliubov transformation in \cref{eq:BogoliubovBCS} one then finds that the DSF for a superconductor reads
\begin{emphalign}\label{eq:DSF_BCS}
	\chi^{0,z} ({\bf q}, \omega) = \sum_\kk & 
	\left[ 1 \pm \frac{\Re{\left(\Deltak^{}\Delta_{\kk+\qq}^*\right)}\mp\epsk\eps_{\kk+\qq}}{E_\kk E_{\kk+\qq}}\right]
	\delta(\hbar\omega-E_{\kk}-E_{\kk+\qq})
	,
\end{emphalign}
where $\pm$ sign corresponds to the charge and spin DSF respectively~\cite{Kee1998,Kee1999,Voo2000}. 
Note that the transition amplitudes appearing in the sum in \cref{eq:DSF_BCS} have a similar structure as the coherence factors which are known to determine the quasiparticle interference in the presence of impurities~\cite{Tinkham2004}. 

Thus, the DSF can be evaluated as a sum over all momenta within the Brillouin zone, where the transition amplitudes are strongly influenced by the character of the superconducting state. 
Although quasiparticle interactions substantially affect the DSF, they do not alter its intrinsic sensitivity to the character of the superconducting state (in particular to the symmetry of its gap function), as can be seen, e.g., at the random-phase approximation level~\cite{Kao2005}, or by considering a strongly correlated system with Hubbard interactions, as shown in \cref{sec:DSF_BCS_correlations}. 

The dependence of the DSF of quasiparticle excitations upon the superconducting order parameter $\Delta_\kk$ has to be sought in the low energy regime, i.e., for excitation energies which are comparable to the superconducting gap. 
For this reason, we will focus hereafter on low energy quasiparticle excitations with momenta close to the Fermi surface, i.e., with $\eps_\kk\approx\eps_{\kk+\qq}\approx0$. 

\subsection{DSF for small momenta}

\begin{figure}[t!]
\centering
\subfigure[normal state]{\includegraphics[width=.33\textwidth]{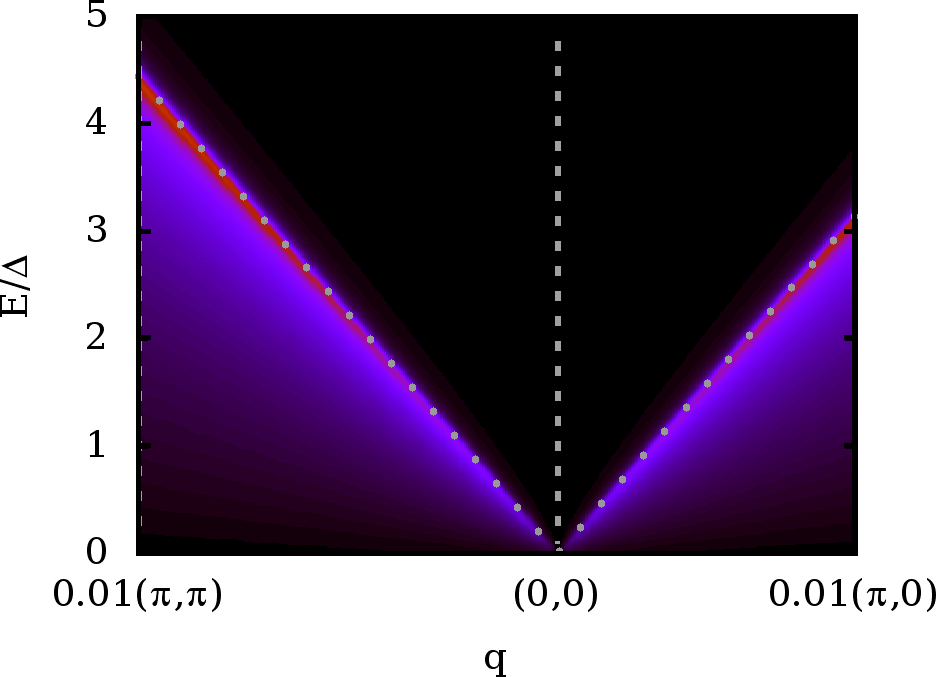}}\!
\subfigure[$s$-wave]	{\includegraphics[width=.33\textwidth]{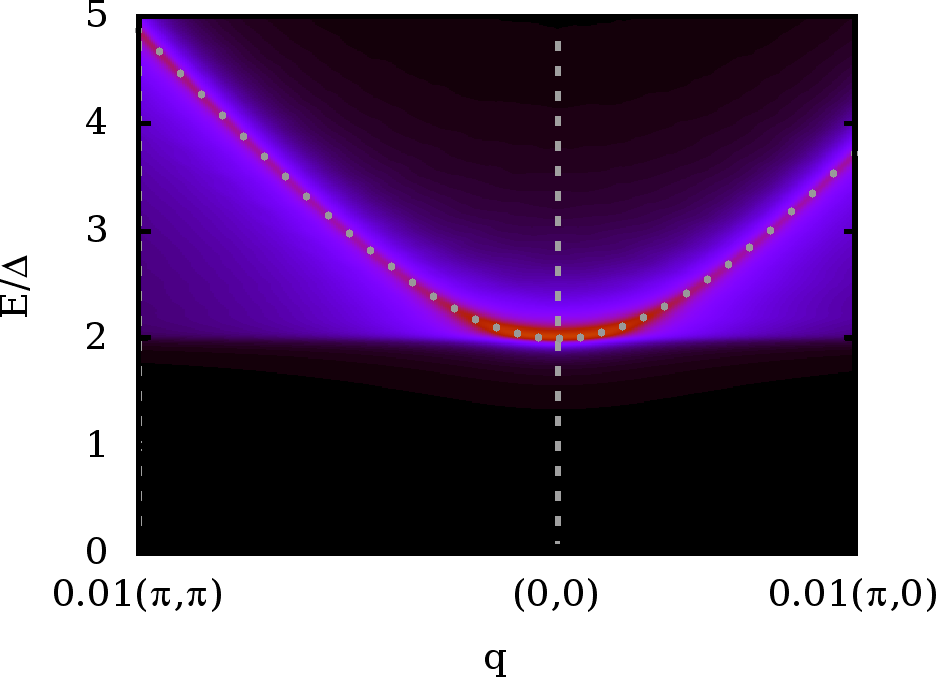}}\!
\subfigure[$d$-wave]	{\includegraphics[width=.33\textwidth]{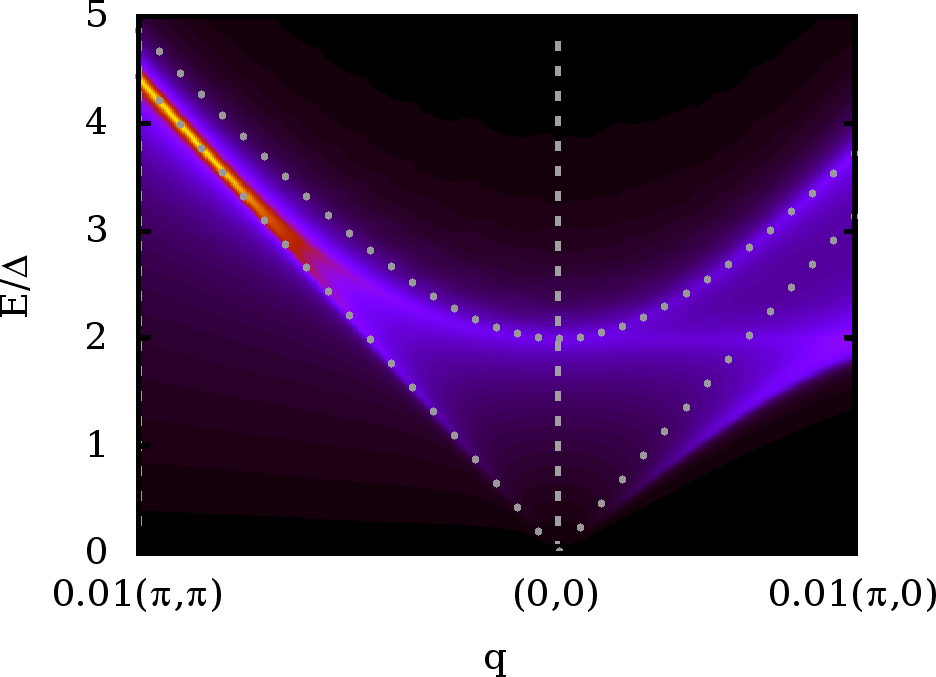}}\!
\caption[Charge DSF of an isotropic Fermi surface]{
Charge DSF of an isotropic Fermi surface with parabolic dispersion as a function of transferred momentum $\qq\rightarrow0$, calculated via \cref{eq:DSF_BCS} for a metal (a), an $s$-wave (b) and a $d$-wave (c) superconductor. 
The energy dispersion is drawn (dotted lines) for gapless excitations (a) in a metal and for gapped  excitations (b) in the conventional superconducting state. 
In the case of an unconventional superconductor (c) the energy dispersions (dotted lines) is drawn for excitations with momenta on nodal (gapless) and antinodal points (gapped) on the Fermi surface. 
}
\label{fig:IsotropicFS}
\end{figure}

In this Section we will discuss the effect of the momentum dependence of the DSF on the superconducting order parameter in the case where the transferred momentum $\qq$ is small compared with momenta $\kk_F$ on the Fermi surface. 
For small transferred momentum and for low energy quasiparticle excitations, the main contributions to the DSF in \cref{eq:DSF_BCS} come from two-quasiparticle excitations corresponding to electron states with momenta $\kk$ and $\kk+\qq$ on the Fermi surface. 
As a consequence, the spin DSF in \cref{eq:DSF_BCS} is strongly suppressed, while the charge DSF reduces, in the case of a finite superconducting order parameter $|\Delta_\kk|\gg|\eps_\kk|$, to the spectra of the two-quasiparticle excitations
\begin{align}\label{eq:DSF_BCS_smallq}
	\chi^{0} (\qq, \omega) = 2\sum_\kk & 
	\delta(\hbar\omega-E)
	,
\end{align}
where the quasiparticle dispersion $E_{\kk,\qq}=E_{\kk+\qq}+E_\kk$ in the case of small transferred momenta $\qq\rightarrow0$ is given by
\begin{equation}\label{eq:PHDispersion}
	E_{\kk,\qq}\approx
	2\sqrt{|\Delta_\kk|^2+\left(\frac{\hbar \vv_\kk\cdot\qq}{2}\right)^2}
	,
\end{equation}
where the order parameter magnitude $|\Delta_\kk|$ and the electron group velocity $\vv_\kk=\rm{d}\eps_\kk/\hbar \rm{d}\kk$ depend on the the momentum $\kk$ on the Fermi surface. 
As one can see, in this case the quasiparticle dispersion inherently depends both on the order parameter magnitude and on the bare electron velocity. 
Therefore, in this regime the DSF spectral intensities depend strongly on the superconducting gap magnitude and on the presence of nodes of the order parameter on the Fermi surface, albeit they are rather insensitive with respect to the order parameter phase. 
In what follows we will consider the DSF in the normal state (vanishing order parameter $\Delta_\kk=0$), in a conventional $s$-wave superconductor, i.e., with an isotropic order parameter $\Delta_\kk=\Delta_0$, and an unconventional superconductor with $d$-wave order parameter, i.e., with nodes on the Fermi surface, given by $\Delta_\kk=\frac12\Delta_0\left(\cos{k_y}+\cos{k_y}\right)$. 

In the normal state the quasiparticle spectra reduces to the particle-hole continuum, with particle-hole excitations which disperse according to the bare electron dispersion. 
In this case the charge DSF in \cref{eq:DSF_BCS} vanishes at $\qq=0$, while for small, but finite, transferred momenta $q\ll k_F$ instead, transition amplitudes are finite for two-quasiparticle excitations corresponding to momenta $\kk$ and $\kk+\qq$ close to the Fermi surface, having a gapless dispersion which is linear with respect to small transferred momenta $\qq\rightarrow0$, and which reads
\begin{equation}\label{eq:PHDispersion_gapless}
	E\approx \hbar |\vv_\kk\cdot\qq|
	,
\end{equation}
with the momentum $\kk$ on the Fermi surface. 
In the superconducting state of a conventional $s$-wave superconductor instead, the charge DSF in \cref{eq:DSF_BCS,eq:DSF_BCS_smallq} at $\qq=0$ does not vanish for excitation energies above the gap energy $2|\Delta_0|$. 
Hence, the quasiparticle spectra are gapped and two-quasiparticle excitations have dispersion which reads
\begin{equation}\label{eq:PHDispersion_gapped}
	E\approx
	2\sqrt{|\Delta_0|^2+\left(\frac{\hbar \vv_\kk\cdot\qq}{2}\right)^2}
	,
\end{equation}
for small transferred momenta $\qq\rightarrow0$ and for momenta $\kk$ on the Fermi surface. 
As a consequence, the DSF shows a coherence peak at $\qq=0$ at at energy $E=2|\Delta_0|$, corresponding to gapped excitations with momentum on the Fermi surface. 

In the case of an unconventional superconductors with nodes on the Fermi surface (e.g., a $d$-wave superconductor) both gapless and gapped excitations contribute to the low energy spectrum, corresponding to excitations with different momenta $\kk$ in \cref{eq:DSF_BCS}. 
In this case, the quasiparticle dispersion is given by \cref{eq:PHDispersion} where the order parameter magnitude $|\Delta_\kk|$ depends on the the momentum $\kk$ on the Fermi surface. 
If the momentum $\kk$ coincides or it is close to the nodal point $\kk_N$ of the Fermi surface (i.e., $\Delta_\kkN=0$), the quasiparticle dispersion reduces to the case described in \cref{eq:PHDispersion_gapless}, i.e., gapless excitations with linear dispersion in the transferred momentum $\qq$. 
If, otherwise, the momentum is away from any nodal points (i.e. $\Delta_\kk>0$), the quasiparticle excitations are gapped with an energy gap given by $2|\Delta_\kk|$ and depending on the quasiparticle momentum on the Fermi surface. 
In particular, if the momentum coincide or it is close to an antinodal point $\kk_A$ of the Fermi surface (i.e., where the order parameter magnitude $|\Delta_{\kk_A}|$ assumes the maximum value) the excitation gap assumes its largest value $2|\Delta_\kkA|$. 
The coherence peak at $\qq=0$ is now broadened with a low energy tail down to gapless excitations with momentum close to a nodal point. 

\Cref{fig:IsotropicFS} shows the charge DSF for a normal metal (a), for a conventional $s$-wave (b) and for an unconventional $d$-wave (c) superconductor, calculated using \cref{eq:DSF_BCS} for an isotropic Fermi surface as a function of the transferred momentum $q\ll k_F$. 
In the normal state (a) the charge DSF vanishes at $\qq=0$ while for finite, but small, momenta, spectral intensities are enhanced for gapless two-quasiparticle excitations with linear dispersion as in \cref{eq:PHDispersion_gapless} (dotted line). 
In the conventional case (b) spectral intensities are enhanced for two-quasiparticle excitations with dispersion as in \cref{eq:PHDispersion_gapped} (dotted line), whereas they vanish for $E<2|\Delta_0|$. 
In particular, the quasiparticle spectra show a coherence peak at $\qq=0$ at at energy $E=2|\Delta_0|$, corresponding to gapped excitations with momentum on the Fermi surface. 
In the unconventional case (c) instead the coherence peak at $\qq=0$ is broadened, and both gapless and gapped excitations contribute, in particular those with momentum which coincides respectively with a nodal or an antinodal points on the Fermi surface (dotted lines). 
Therefore, the presence of a two-quasiparticle excitation with linear dispersion for $\qq\rightarrow0$ indicates the presence of nodes in the superconducting order parameter. 
Hence quasiparticle excitations spectra are affected by the order parameter symmetry and, in particular, gapless excitations allows one to unveil the presence of nodes on the Fermi surface.

\subsection{DSF for large momenta: phase sensitivity}

In order to characterize the symmetry of the order parameter in the superconducting state, we will focus hereafter on the DSF for finite transferred momenta ($q\approx k_F$) and at energies close to the superconducting gap, i.e, with both momenta $\kk$ and $\kk+\qq$ on the Fermi surface. 
Assuming an unconventional superconductor with a pairing governed by a phase dependent order parameter $\Delta_\kk=|\Delta_\kk|e^{i\phi_{\bf k}}$ the DSF in \cref{eq:DSF_BCS} for excitations near the Fermi surface becomes
\begin{emphalign}\label{eq:DSF_BCS_largeq}
	\chi^{0,z} (\qq,\omega)
	 \approx \sum_{\kk,\kk+\qq\in FS} & 
	\left[1 \pm \cos(\phi_\kk - \phi_{\kk+\qq}) \right]
	\delta(\hbar\omega-E)
	,
\end{emphalign}
where the two-quasiparticle dispersion reduces to $E\approx|\Delta_{\kk+\qq}|+|\Delta_\kk|$. 
As a consequence, the charge and spin DSF strongly depend on the relative phase $\Delta\phi=\phi_\kk-\phi_{\kk+\qq}$ of the order parameter corresponding to points on the Fermi surface connected by the transferred momentum $\qq$. 
In particular, one can distinguish between sign-reversing and sign-preserving excitations, i.e, between two-quasiparticle excitations where the transferred momentum $\qq$ connects points respectively with opposite phase ($\Delta\phi=\pi$) or the same phase ($\Delta\phi=0$) of the order parameter on the Fermi surface. 
As one can see from \cref{eq:DSF_BCS_largeq}, charge DSF vanishes for sign-reversing, while it is enhanced for sign-preserving excitations. 
On the other hand, spin DSF vanishes for sign-preserving, while it is enhanced for sign-reversing excitations. 
Thus, the momentum-dependent intensity distribution of the low energy DSF represents the variation of the superconducting order parameter phase along the Fermi surface. 

\section{Phase sensitivity in cuprates}
\label{sec:Phase-sensitivity}

\subsection{DSF in cuprates}

In this Section we will concentrate on determining the properties of DSF for different types of singlet-pairing superconductors, and to be even more specific we will consider the case of a high $T_c$ cuprate superconductor. 
The main aim in this context is to establish how a variation of the phase of the superconducting order parameter is reflected in the spin and charge DSF\@. 
Following the most direct theoretical inroad and avoiding model-specific technical details, we will consider a singlet-pairing superconductor described by the BCS Hamiltonian in \cref{eq:BCSHamiltonianDiagonalized} with a superconducting order parameter varying along the Fermi surface. 
Even if electron correlations are not fully taken into account, this approach is commonly used --- and is very successful to calculate quasiparticle interference in cuprates~\cite{Hanaguri2007,Hanaguri2009,Kohsaka2008,Fischer2007}. 
In addition, in \cref{sec:DSF_BCS_correlations} we will show that it is actually possible to introduce the effect of correlations into the calculations, and that doing so does not affect the main results presented below. 

\begin{figure}[t!]
\centering
\includegraphics[width=1\textwidth]{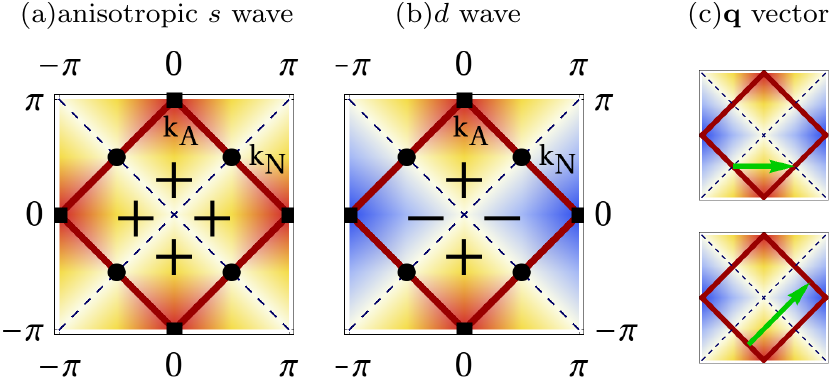}
\caption[Order parameters of an anisotropic $s$-wave and $d$-wave superconductor with nested Fermi surface]{
Order parameters of an anisotropic $s$-wave (a) and $d$-wave (b) superconductor with a perfectly nested Fermi surface (solid line). 
The order parameter vanishes (has maxima) at the nodal points $\kkN$ (antinodal points $\kkA$). 
(c) Particle-hole excitations with and without sign reversal in case of $d$-wave pairing. 
}
\label{fig:OrderParameter}
\end{figure}

To determine in detail how the RIXS spectra of unconventional superconductors reflect the phase of the order parameter, we model the bare electron dispersion of the cuprate superconductor as
\begin{equation}\label{eq:TBCuprates}
	\eps_\kk=-2t\left(\cos{k_x}+\cos{k_y}\right)
	,
\end{equation}
where $t$ is the tight-binding parameter. 
This bare electron dispersion follows from a single-band tight-binding model of Cu ions in a two dimensional lattice with tetragonal symmetry~\cite{Tohyama2012}, i.e., the one which has direct relevance to the high $T_c$ superconductors, and give rise to a perfectly nested Fermi surface, as in \cref{fig:OrderParameter}. 
Moreover, we consider two different pairing symmetries which differ from each other only in the superconducting order parameter phase, that is, a $d$-wave pairing and an anisotropic $s$-wave pairing defined respectively as
\begin{align}
	\Delta_\kk&=\frac12\Delta_0 \left(\cos{k_x}-\cos{k_y}\right) \qquad\text{$d$-wave},
	\nonumber\\
	\Delta_\kk&=\frac12\Delta_0 \vert\cos{k_x}-\cos{k_y}\vert \,\,\qquad\text{anisotropic $s$-wave}
	.
\end{align}
The gap functions considered here along with the nested Fermi surface are shown in \cref{fig:OrderParameter}. 
In the anisotropic $s$-wave case, two-quasiparticle excitations are sign-preserving all over the Brillouin zone, and therefore the spin DSF is strongly suppressed at any transferred momenta.
In the $d$-wave case instead, the charge DSF is suppressed for sign-reversing excitations, i.e., with a transferred momentum which connects points on the Fermi surface with opposite phases of the order parameter, as one can see from \cref{eq:DSF_BCS,eq:DSF_BCS_largeq}. 

\begin{figure}[t!]
\centering
\includegraphics[width=.8\textwidth]{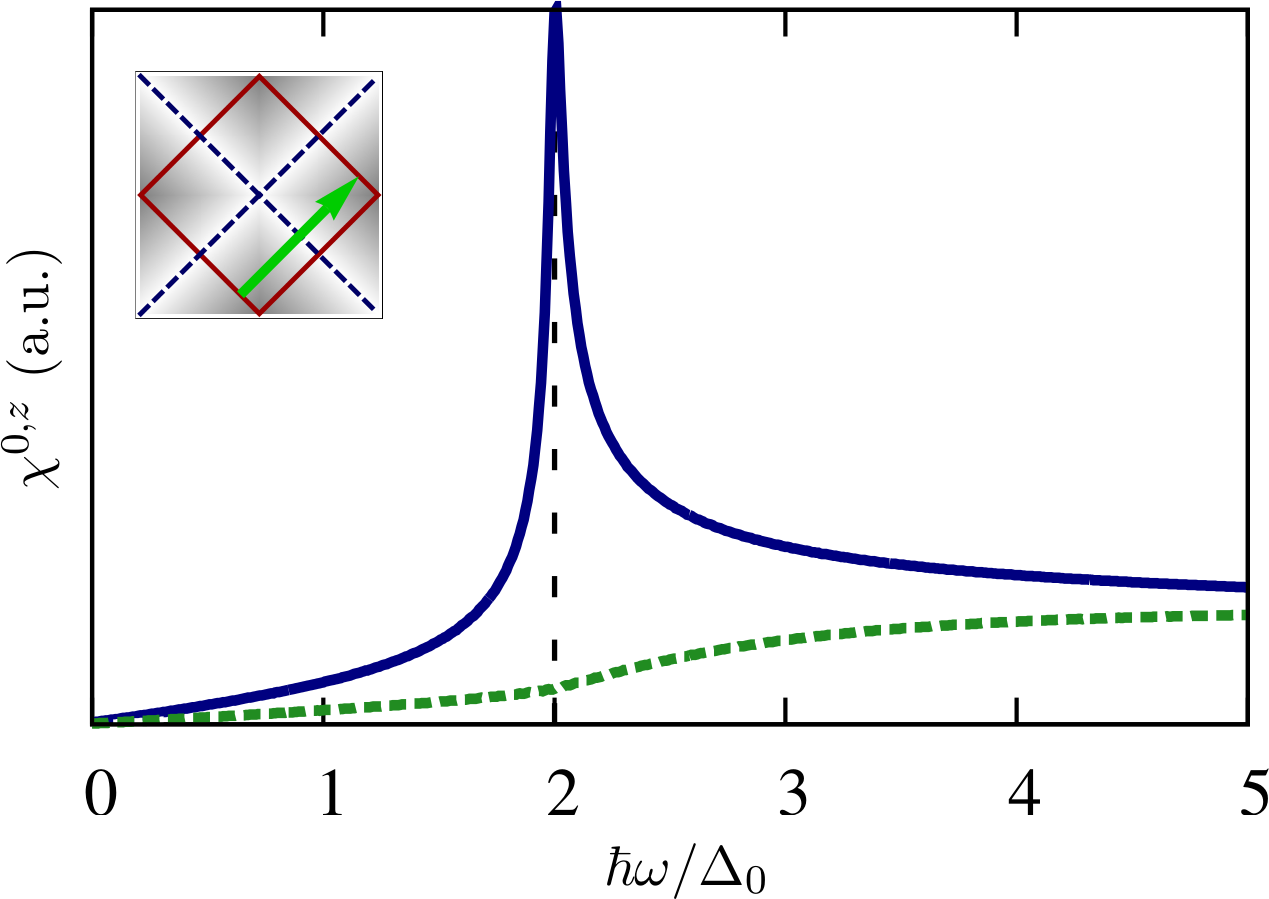}
\caption[Charge and spin DSF of a superconductor with a nested Fermi surface at fixed momentum]{
DSF of a superconductor with a nested Fermi surface with transferred momentum $\qq=(\pi,\pi)$, for charge excitations ($\chi^0$) with anisotropic $s$-wave (solid line) and $d$-wave (dashed line), or equivalently for spin excitations ($\chi^z$) with $d$-wave (solid line) and anisotropic $s$-wave (dashed line) order parameter, as a function of the transferred energy $\hbar\omega$. 
}
\label{fig:PhaseCoherence}
\end{figure}

In unconventional superconductors, where the pairing is generally considered as mediated by antiferromagnetic spin fluctuations~\cite{Schrieffer1989}, the superconducting gap function is expected to exhibit a sign reversal between the Fermi momenta connected by characteristic wave vector ${\bf Q}=(\pi,\pi)$ of the spin fluctuations~\cite{Kuroki2001}. 
As a consequence, the conduction electrons of such superconductors show a tendency to Fermi surface nesting with a typical nesting vector ${\bf Q}$. 
The scattering intensities as a function of energy calculated using \cref{eq:DSF_BCS} for the two pairing symmetries considered above are shown in \cref{fig:PhaseCoherence}, for a perfectly nested cuprate-like Fermi surface (see inset), for a fixed transferred momentum equal to the nesting vector. 
Note that direct RIXS at the Cu $L_{2,3}$ edge in two dimensional cuprates allows momentum transfers $q\lesssim 0.87 \pi$~\cite{Braicovich2010b} and therefore one is able to access momentum transfers of the order of the nesting vector $\bf Q$. 
For such excitations, the sign of the order parameter in the $s$-wave case is preserved whereas in the $d$-wave case is reversed. 
The solid line in \cref{fig:PhaseCoherence} corresponds to the charge DSF in the anisotropic $s$-wave and to the spin DSF in the $d$-wave case, and shows a coherence peak at $\hbar\omega=2|\Delta|$ which is strongly enhanced due to the nesting effect. 
However, sign-reversing excitations occurring in the $d$-wave are strongly suppressed in the charge DSF, as well as sign-preserving excitations in the anisotropic $s$-wave case in the spin DSF, according to \cref{eq:DSF_BCS,eq:DSF_BCS_largeq}. 
On the other hand, the dashed line in \cref{fig:PhaseCoherence} corresponds to the charge DSF in the $d$-wave and to the spin DSF in the anisotropic $s$-wave case, and is strongly suppressed due to the sign-reversing and the sign-preserving excitations occurring respectively in the case of $d$-wave and anisotropic $s$-wave. 
The symmetry between the charge and the spin DSF with respect to the two choices of the order parameter is due to the $\pm$ sign in \cref{eq:DSF_BCS}. 
Note that since the order parameter magnitude is equal in both cases, the obtained effect is entirely due to phase changes of the superconducting order parameter along the Fermi surface.

To highlight its strong dependence on the order parameter phase, the DSF is shown in \cref{fig:PhaseBZ} at a fixed energy $\hbar \omega=2|\Delta|$ as a function of momentum $\qq$ in the entire Brillouin zone, both for the anisotropic $s$-wave and the $d$-wave pairing, for a perfectly nested Fermi surface. 
Because of the nesting effect, coherence peaks are clearly visible in the charge (spin) DSF in the anisotropic $s$-wave ($d$-wave) case if the transferred momenta coincide with the nesting vector $\qq=(\pi,\pi)$, while they are strongly suppressed in the $d$-wave (anisotropic $s$-wave) case (\cref{fig:PhaseBZ}). 
Clearly, the symmetry of the order parameter is reflected by the symmetry of the charge and spin DSF spectrum. 
Since the charge and spin DSF are complementary with respect to the spectral suppression of the sign-reversing and sign-preserving excitations, the phase sensitivity is enhanced when these two components are fully disentangled. 
This can be done by tuning the polarization dependence in the form factors $W_{{\bf e}{\bf e}'}^0$ and ${\bf W}_{{\bf e}{\bf e}'}^z$ in \cref{eq:BCS_cs}. 

\begin{figure}[t!]
\centering
\includegraphics[width=.8\textwidth]{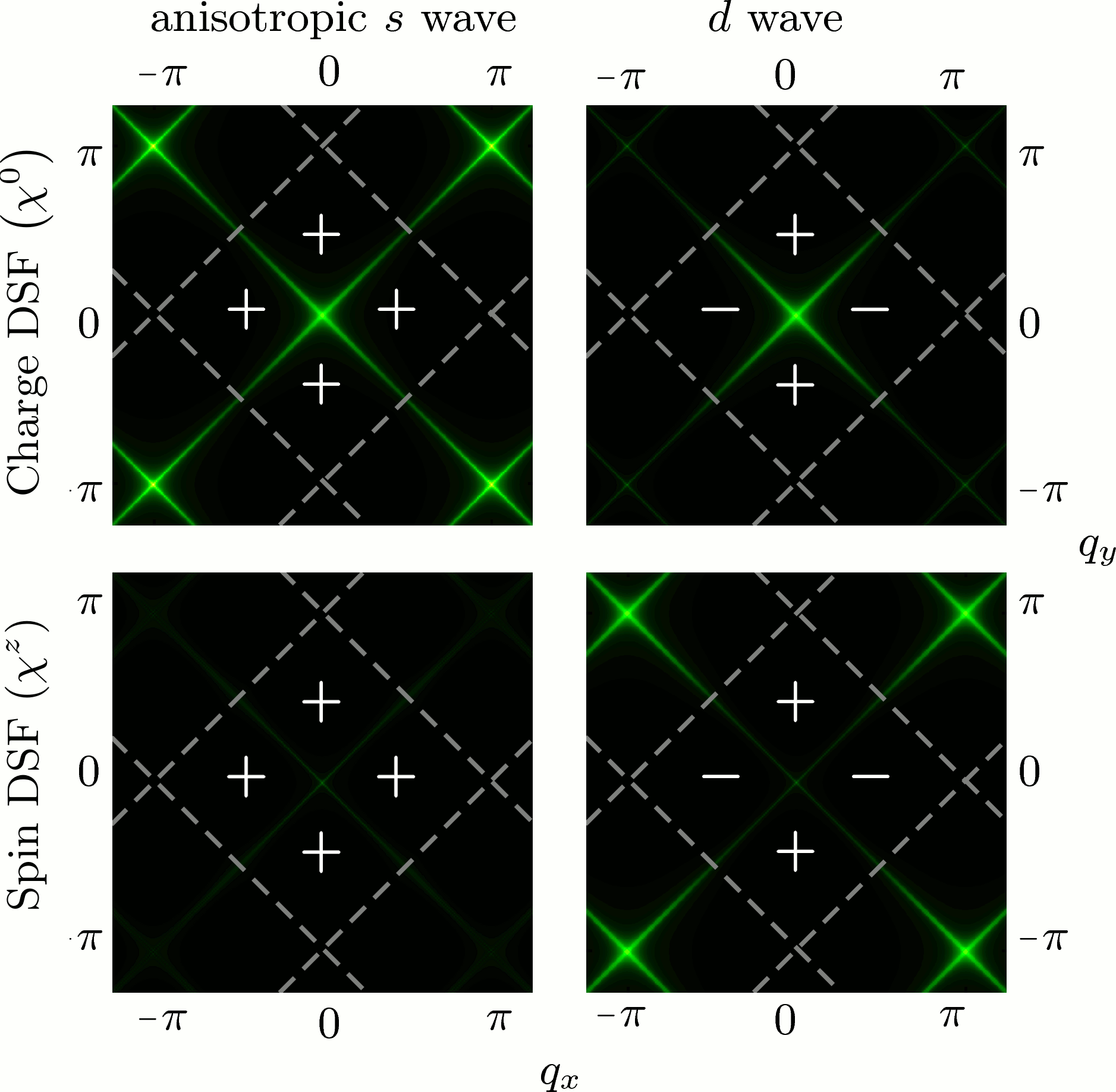}
\caption[Charge and spin DSF at fixed energy of a superconductor with nested Fermi surface]{
Charge ($\chi^0$) and spin ($\chi^z$) DSF at fixed energy $\hbar \omega=2|\Delta|$, as a function of transferred momentum of an anisotropic $s$-wave and a $d$-wave superconductor with a nested Fermi surface.
}
\label{fig:PhaseBZ}
\end{figure}

\subsection{DSF in the strongly correlated limit}\label{sec:DSF_BCS_correlations}

In this Section we will show how the sensitivity of the DSF to the order parameter phase is not obliterated in the 
presence of strong Coulomb repulsion of the conduction electrons. 
To simplify the presentation of our arguments, we will focus on the limit of infinitely strong correlations $U\rightarrow \infty$ where the double occupancy of a single transition metal ion site is strictly forbidden~\cite{Trugman1990,Becca2000}. 
As is well-known~\cite{Chao1978} using a Schrieffer-Wolf transformation the Hubbard model can then be replaced with a $t$-$J$ model where the superexchange $J\rightarrow 0$. 
Such regime of very strong correlations may be still regarded as rather realistic in describing many basic properties of the strongly correlated transition metal oxides such as cuprates in the overdoped limit~\cite{Putikka1992,Tandon1999,Cosentini1998}.
In this case, the superconducting phase is well described by the Hamiltonian
\begin{equation}\label{eq:BCSUHamiltonian}
	\mathcal H =
	\sum_{\kk\sigma}\varepsilon_\kk	d_{\kk \sigma}^\dag  d_{\kk\sigma}  
	-\sum_\kk 	\Delta_\kk 	d_{\kk \uparrow}^\dag  d_{-\kk \downarrow}^\dag
	+ \text{h.c.}
	,
\end{equation}
where the operators $d_{i\sigma}^\dag=c_{i\sigma}^\dag(1- n_{i,-\sigma})$, and $d_{i\sigma}=c_{i\sigma}(1- n_{i,-\sigma})$ are the Hubbard creation and annihilation operators.
These transformed operators are introduced since double occupancy is strictly forbidden, due to the presence of strong electronic correlations. 
They obey unusual anticommutation relations, e.g., $[d_{i\sigma}, d_{j\sigma}^\dag ]_+ = \delta_{ij}{\mathcal D}_{\sigma}(i)$, with ${\mathcal D}_{\sigma}(i)= 1 - n_{i, -\sigma}$. 

For the sake of simplicity, let us consider the spin DSF $\chi^z(\qq,\omega)$, which can be written in terms of the spin correlation function as
\begin{equation}\label{eq:SDSFCorrelation}
	\chi^z(\qq,\omega)=
	\imath \int_0^\infty {\rm d}t \, \langle S_{-{\mathbf q}}^z(t) 
	S_{\mathbf q}^z \rangle_{\mathcal H} \ e^{\imath(\omega +\imath\eta)t}   .
\end{equation}
Here, the time dependence and the expectation are evaluated via the Hamil-tonian \ref{eq:BCSUHamiltonian}. 
To calculate the expectation value and the dynamical behavior, one can diagonalize the Hamiltonian using new approximate quasiparticle operators $\gamma^\nod_{\kk\sigma}$ and $\gamma^\dag_{\kk\sigma}$, which are related to the original correlated electron operators via the Bogoliubov transformation
\begin{align}\label{eq:BogoliubovBCSU}
	d_{\kk\uparrow}=\uk^* \gamma^\nod_{\kk\uparrow}-\vk \gamma^\dag_{-\kk\downarrow},
	\nonumber\\
	d_{\kk\downarrow}=\uk^* \gamma^\nod_{\kk\downarrow}+\vk \gamma^\dag_{-\kk\uparrow}. 
\end{align}
In the case of a sufficiently large hole concentration the operator ${\mathcal D}_{\sigma}(i)$ can approximately be replaced by its expectation value $D$ and the Bogoliubov quasiparticle operators fulfill the following relations~\cite{Sykora2009}:  $[\mathcal H,\gamma_{\kk\sigma}^\dag] = E_\kk \gamma_{\kk\sigma}^\dag$, where $E_\kk=\sqrt{\varepsilon_\kk^2 + D^2 \Delta_\kk^2}$ and $D = 1 - n/2$. 
Replacing all the Hubbard operators $d_{\bf k\sigma}^{(\dag)}$ with the quasiparticle operators $\gamma_{\bf k\sigma}^{(\dagger)}$ via the Bogoliubov transformation in \cref{eq:BogoliubovBCSU}, the time dependence in \cref{eq:SDSFCorrelation} can be easily evaluated. 
Each of the remaining expectation values contain a product of four quasiparticle operators. 
A final factorization leads to the following expectation values of the two-quasiparticle operators
\begin{align}\label{eq:BCSUExpectationValues}
	\langle \gamma_{\kk\sigma}^\dag \gamma_{\kk\sigma} \rangle &=
	\frac12\left(1 + \frac{\varepsilon_\kk}{E_\kk}  \right) n_\kk +
	\frac12\left(1 - \frac{\varepsilon_\kk}{E_\kk}  \right) m_\kk
	-
	\frac{D^3 | \Delta_\kk|^2}{2 E_\kk^2}, \nonumber \\
	\langle \gamma_{\kk\sigma} \gamma_{\kk\sigma}^\dag \rangle &=
	D - \langle \gamma_{\kk\sigma}^\dag \gamma_{\kk\sigma} \rangle,
\end{align}
where $D= \langle {\mathcal D_\sigma (i)} \rangle = 1 -n/2$ and $n_\kk$ and $m_\kk$ are defined by $n_\kk= \langle  d_{\kk\sigma}^\dag d_{\kk\sigma}\rangle$ and  $m_\kk= \langle d_{\kk\sigma} d_{\kk\sigma}^\dagger\rangle$. 
They are evaluated using the Gutzwiller approximation (cf. \onlinecite{Fazekas1999}) as $n_{\kk} = (D-q)  + q  \, f( \varepsilon_\kk)$ and $m_\kk = D - n_{\kk}$, with $q=(1-n)/(1-n/2)$ and where $f( \varepsilon_\kk)$ is the Fermi function at $T=0$. 
Finally, one obtains the spin and charge DSF in the strongly correlated case
\begin{align}\label{eq:DSF_BCSU}
	\chi^{0,z}(\qq,\omega)=
	\sum_\kk
	&\left[ \frac{A^\pm(\kk,\qq)}{E_\kk+E_{\kk+\qq}- (\omega + \imath \eta)} 
	\langle \gamma_{\kk+\qq,\downarrow} \gamma_{\kk+\qq,\downarrow}^\dag \rangle
	\langle \gamma_{\kk\uparrow} \gamma_{\kk\uparrow}^\dag \rangle
	\right. +
	\nonumber\\&\left.
	\frac{A^\pm(\kk,\qq)}{-E_\kk- E_{\kk+\qq}- (\omega + \imath\eta)} 
	\langle \gamma_{\kk+\qq,\uparrow}^\dag \gamma_{\kk+\qq,\uparrow} \rangle
	\langle \gamma_{\kk\downarrow}^\dag \gamma_{\kk\downarrow} \rangle
	\right. +
	\nonumber\\&\left.
	\frac{2 A^\mp(\kk, \qq)}{E_\kk- E_{\kk+\qq}- (\omega + \imath\eta)} 
	\langle \gamma_{\kk+\qq,\uparrow}^\dag \gamma_{\kk+\qq,\uparrow} \rangle
	\langle \gamma_{\kk\uparrow} \gamma_{\kk\uparrow}^\dag \rangle
	\right],
\end{align}
where the transition amplitudes $A^\pm(\kk,\qq)$ are defined by
\begin{equation}\label{eq:CoherenceFactors_BCSU}
	A^\pm(\kk, \qq) =
	1 \pm \frac{ D^2 \Re{(\Delta_\kk \Delta_{\kk+\qq}^*)}  \mp \varepsilon_\kk \varepsilon_{\kk+\qq}	}
	{E_\kk E_{\kk+\qq}  }
	.
\end{equation}

\begin{figure}[t!]
\centering
\includegraphics[width=1\textwidth]{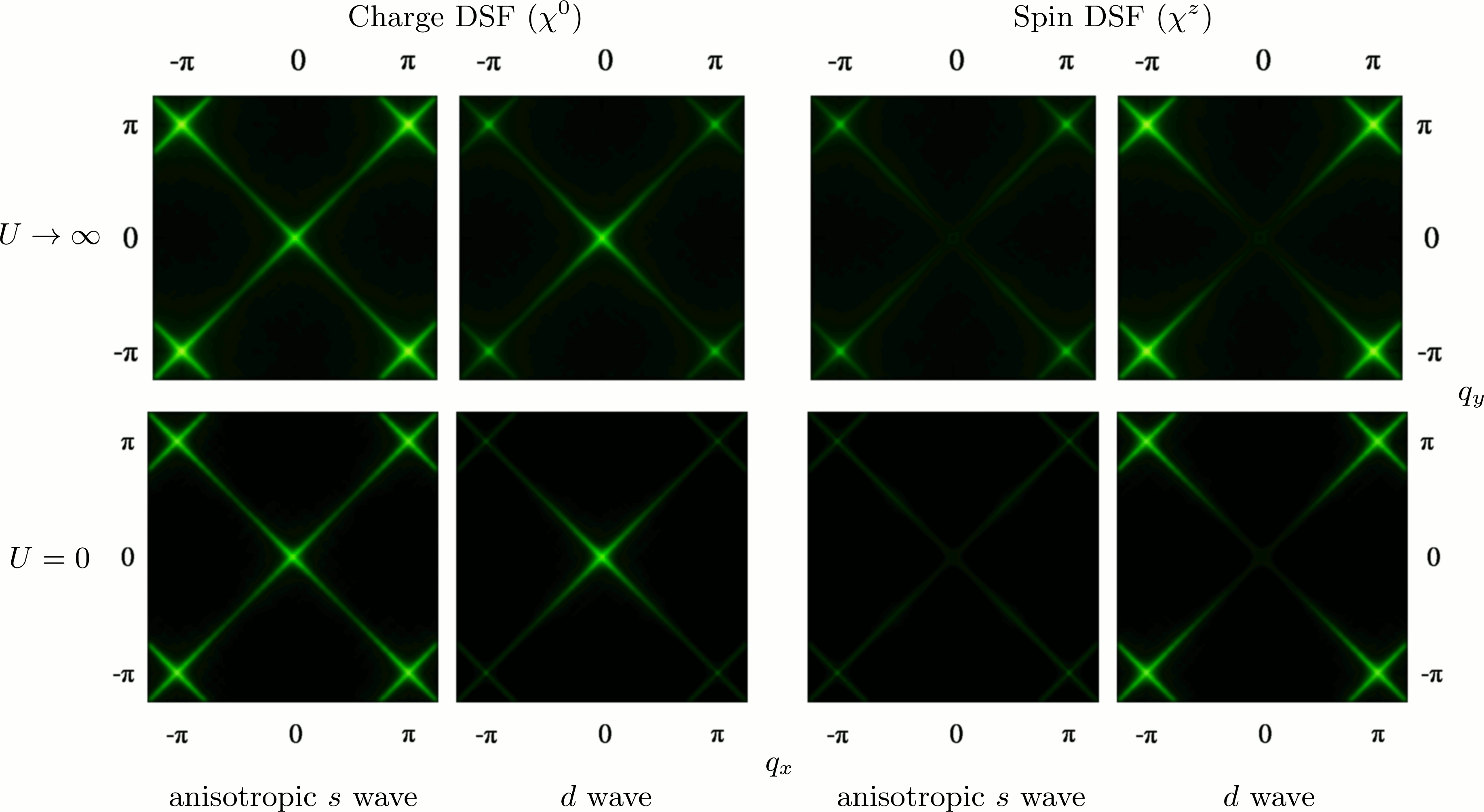}
\caption[Charge and spin DSF at fixed energy of a superconductor in the infinitely strongly correlated and the uncorrelated limits]
{ 
Charge ($\chi^0$) and spin ($\chi^z$) DSF at the quasiparticle gap level, as a function of transferred momentum for an anisotropic $s$-wave (first and third columns) and a $d$-wave (second and fourth columns) superconductor, for a cuprate-like system with a nested Fermi surface,  in the infinitely strongly correlated ($U\rightarrow\infty$, first row) and the uncorrelated ($U=0$, second row) limits.
In the infinite repulsion case, the Gutzwiller approximation is used, assuming $D=\langle{\mathcal D}_{\sigma}(i)\rangle=0.8$ (hole doped system), and the gap level is renormalized as $2|\Delta|\rightarrow 2D|\Delta|$. 
In the $d$-wave (anisotropic $s$-wave) superconductor, the charge (spin) DSF peak at $\qq=(\pi,\pi)$ is suppressed, for both the infinitely strongly correlated and the uncorrelated limits. 
}
\label{fig:U}
\end{figure}

To summarize, the main effect of correlations in the limit of infinitely strong correlations $U\rightarrow\infty$ is to rescale the magnitude of the order parameter. 
In fact, up to a renormalized gap function $\Delta_\kk\rightarrow D\Delta_\kk$, the transition amplitudes in \cref{eq:CoherenceFactors_BCSU} have the same form as in the uncorrelated case in \cref{eq:DSF_BCS}. 
Due to this renormalization, the quasiparticle excitations gap is lowered in energy by a factor $D$. 
Moreover, the phase sensitivity of the DSF is reduced by the presence of the third term of \cref{eq:DSF_BCSU}. 
However, when the hole doping is rather large ($D\approx1$, $n\ll1$), the order parameter decrease is negligible, while the main contributions to the quasiparticle spectrum are given by the first term in \cref{eq:DSF_BCSU}. 
As a consequence, the phase sensitivity of the DSF is not affected. 
In \cref{fig:U} we compare the spin and charge DSF of a strongly correlated hole doped system with those of an uncorrelated one (cf. \cref{fig:PhaseBZ}), for different order parameter symmetries ($d$-wave and anisotropic $s$-wave). 
As one can see, the presence of electronic correlations does not change RIXS spectra qualitatively.
Hence the charge and spin DSF in a strongly correlated electron system are governed, as well as in an uncorrelated one, by coherence factors which are responsible for the sensitivity of RIXS spectra to the order parameter phase. 

\section{Conclusions}

In principle, other two-particle spectroscopies (see, e.g., \onlinecite{Ament2011} for an overview) can also be \emph{directly} sensitive to the DSF of superconductors. 
Even if none can match RIXS in measuring both spin and charge DSF of superconductors, already probing either of the two is in general challenging. 
For example, electron energy loss spectroscopy cannot reliably measure spectra with high momentum transfers, while inelastic neutron scattering does not directly probe the charge DSF, and non-resonant inelastic x-ray scattering is extremely photon hungry. 
Nevertheless, transition amplitudes of the same type as in \cref{eq:DSF_BCS,eq:DSF_BCS_largeq} are also encountered when determining the scattering rate of conduction electrons in the presence of impurities, as observed in the surface-sensitive STM~\cite{Hoffman2002,McElroy2003,Hanaguri2007,Kohsaka2008,Hanaguri2009,Hanke2012,Sykora2011}. 
This is because these transition amplitudes have a similar structure as the ones which are known to govern the quasiparticle interference [in which case the transition amplitudes, whose sum over the momentum ${\bf k}$ contribute to the DSF in \cref{eq:DSF_BCS}, are termed ``coherence factors''] in the presence of impurities. 
Since the quasiparticle interference patterns explored by STM have turned out to be very successful to uncover the pairing symmetries of the unconventional superconductors~\cite{Hoffman2002,McElroy2003,Hanaguri2007,Kohsaka2008,Hanaguri2009,Hanke2012}, this gauges the potential of RIXS to observe and unravel symmetries of superconducting pairing and pairing mediators. 
Compared to STM however, RIXS has the advantage of simplicity. 
Whereas the theoretical interpretation of STM in the framework of quasiparticle interference relies crucially on the form of the underlying impurity system showing various components of scattering~\cite{Sykora2011}, in the case of RIXS the interpretation of spectroscopic features relies neither on the presence of impurities in the superconductor nor on the modeling thereof. 

In this Chapter, we have shown that RIXS, in contrast to other two-particle spectroscopies, is directly sensitive to the spin \emph{and} to the charge DSF of a superconductor and, in particular, that the DSF of a superconductor observed in RIXS is sensitive to the symmetry of the order parameter. 
This is rooted in the quasiparticle spectra reflecting sign-reversing excitations at large transferred momenta which arise for order parameters with a phase that varies over the Fermi surface. 
This, together with the recent experimental successes of RIXS, including, in particular, the major enhancements in resolution and pioneering study of hole doped cuprates~\cite{LeTacon2011}, establishes the potential of RIXS as a versatile and practical spectroscopic technique to probe the elementary excitations, disentangle the pairing symmetry, and to investigate the fundamental properties of unconventional superconductors.

}

\chapter[RIXS on iron based superconductors]{RIXS on iron based superconductors\footnotemark{}}
\footnotetext{Part of this chapter has been published in \onlinecite{Marra2016}}
{
\newcommand{\meV}{{~\mathrm{meV}}}

\label{ch:BCS-Pnictides}

In this Chapter, we will develop a phenomenological theory to predict the characteristic features of the momentum-dependent scattering amplitudes in RIXS spectra at the energy scale of the superconducting gap of iron based superconductors. 
In particular in \cref{sec:theo} we will develop a theoretical approach to calculate the spectra of quasiparticle excitations at zero temperature in unconventional superconductors with more than one relevant orbitals close to the Fermi level. 
Afterwards, taking into account all \mbox{relevant} orbital states as well as their specific content along the Fermi surface, we will evaluate in \cref{sec:laofeas,sec:lifeas} the charge and spin DSF for the compounds LaOFeAs and LiFeAs, using simple tight-binding models which are fully consistent with recent ARPES data. 
While the orbital content affects the momentum-dependence of RIXS intensities, the intensity of sign-reversing excitations are considerably enhanced in the spin structure factor. 
The calculated RIXS spectra for different types of superconducting pairing ($s_\pm$, $s_{++}$, $p$~wave) show a characteristic intensity redistribution between charge and spin dynamical structure factors which discriminates between sign-reversing and sign-preserving quasiparticle excitations.
Consequently, RIXS spectra can discriminate between $s_\pm$ and $s_{++}$~wave gap functions in the singlet pairing case. 
In addition, an analogous intensity redistribution at small momenta allows one to distinguish between triplet pairing (chiral $p$ wave) and singlet pairing ($s_\pm$ or $s_{++}$~wave) superconductivity.

\section{The challenge of RIXS in iron based superconductors}
\label{sec:challenge}

The Fe $L_3$-edge RIXS response of iron-bases superconductors has been already studied theoretically~\cite{Kaneshita2011} at energies higher than the superconducting gap, suggesting that RIXS can in principle probe magnon excitations in these systems.
This theoretical work stimulated a subsequent experimental study~\cite{Zhou2013}, which not only confirmed the presence of the magnon excitations in an energy range up to an energy of 200~meV in magnetically ordered iron-based superconductors~\cite{Harriger2011}, but also revealed the persistence of magnon-like modes in this energy range as the material is doped and becomes superconducting.
It also motivates the question whether and how RIXS can pick up the signatures of the superconducting gap in these materials.

In fact, the relative sign change of the superconducting in iron-based pnictides gap can be determined, in principle, by phase-sensitive experiments such as Josephson junctions experiments~\cite{Tsuei2000}, composite superconducting loops~\cite{Chen2010}, scanning tunneling microscopy~\cite{Hoffman2002,McElroy2003,Hanaguri2007,Kohsaka2008,Hanaguri2009}, and inelastic neutron scattering~\cite{Maier2008,Korshunov2008,Christianson2008,Maier2011,Qiu2008,Inosov2010,Knolle2012}.
Of particular interest in this context is the material LiFeAs~\cite{Wang2008,Tapp2008}.
Contrarily to the expected scenario in fact, ARPES have proven the absence of Fermi surface nesting in this compound~\cite{Borisenko2010,Borisenko2012}.
Furthermore, the presence of a sign-reversal of the order parameter and the nature of the superconducting pairing in LiFeAs is still debated.
In fact, theoretical works have given contradictory results, suggesting the presence of either $s$-wave singlet~\cite{Platt2011,Wang2013,Ahn2014} (with or without sign-reversal~\cite{Kontani2010}) or $p$-wave triplet pairing~\cite{Brydon2011}.
On the experimental side, neutron scattering experiments~\cite{Taylor2011} seem consistent with the presence of spin-singlet pairing with $s$-wave symmetry, whereas STM experiments of the quasiparticle interference~\cite{Hanke2012} indicate a $p$-wave spin-triplet state or a singlet pairing mechanism with a more complex order parameter ($s+\imath d$~wave).

Probing the superconducting order parameter with RIXS requires an energy resolution roughly comparable with twice the magnitude of the gap, i.e., in the order of 10~meV in the iron-based superconductor LiFeAs, which is below the energy resolution of present RIXS facilities~\cite{Hancock2010,Zhou2013} at the Fe $L_3$ edge, but in the targeted range of, e.g., the Soft Inelastic X-ray Scattering (SIX) beamline at NSLS-II which is presently under construction.
Nevertheless, the energy resolution of RIXS experiments has improved in the last decade at a very fast pace~\cite{Ament2011}, such that one can realistically expect to reach the energy scale of the superconducting gap in the near future.
On top of that, RIXS spectra of iron-based superconductors, as well as any other metallic compound, is strongly influenced by fluorescence.
However, it has been shown that, in order to reliably uncover electronic excitations, the fluorescence background can be subtracted from the spectra~\cite{Zhou2013,Hancock2010,Yang2009}.
Moreover, since any spectral feature related to the superconducting order parameter is obviously absent in the normal state, one can in principle obtain the spectra of quasiparticle excitations as the difference between the total inelastic scattering in the normal and in the superconducting state.
Such differential spectra can be obtained by probing the inelastic response slightly above and sightly below the $T_c$ of the material.
This would have the additional advantage of canceling out not only the fluorescence background, but also any other contribution from excitations which are not related, and thus not affected, by the onset of the superconducting state.

\section{RIXS cross section in iron based superconductors}
\label{sec:theo}

In a \emph{direct} RIXS process at a transition-metal ion $L_{2,3}$ edge, the incident photon excites a core shell 2$p$ electron into the 3$d$ shell, which consequently decays into a scattered photon and a charge, spin, or orbital excitation in the electronic system~\cite{Ament2011}.
In the case of iron-based superconductors considered in this work, one has to take into account more than one relevant orbital states of the 3$d$ shell.
Note that this multi-orbital structure leads to the characteristic disconnected Fermi surface branches dominating the low-energy properties in these materials.
As in \cref{sec:RIXS-DSF}, using the fast collision approximation the RIXS cross section can be decomposed into a combination of the charge and the spin DSF of $3d$ electrons as~\cite{Kaneshita2011,Haverkort2010}
\begin{equation}\label{eq:CrossSection}
	I({\bf e},{\bf q},\omega) = \sum\limits_{\alpha\beta}
	|W^0_{\alpha\beta}({\bf e})|^2 \chi^0_{\alpha\beta}({\bf q},\omega) + 
	|W^z_{\alpha\beta}({\bf e})|^2 \chi^z_{\alpha\beta}({\bf q},\omega)
	,
\end{equation}
where the charge and spin DSF corresponding to the orbitals $\alpha,\beta$ are defined as
\begin{eqnarray}
	\chi^0_{\alpha\beta}({\bf q},\omega) = \sum\limits_f|\langle f|
	\rho_{\alpha\beta}({\bf q})|i\rangle|^2\delta(\hbar\omega+E_i-E_f), 	
\nonumber
\\
	\chi^z_{\alpha\beta}({\bf q},\omega) = \sum\limits_f|\langle f| 
	S^z_{\alpha\beta}({\bf q})
	|i\rangle|^2\delta(\hbar\omega+E_i-E_f),
\label{eq:ChargeSpinDSF}
\end{eqnarray}
being $|i\rangle$ and $|f\rangle$ the initial and final states of the RIXS process with energy $E_i$ and $E_f$, and with $\hbar\omega$ and ${\bf q}$ the transferred photon energy and momentum.
Note that the spin DSF is assumed to have the same momentum and energy dependence for any direction of the spin~\cite{Andersen2005}.

Here, the density and the spin of $3d$ electrons $\rho_{\alpha\beta}({\bf q})=\sum_{{\bf k}\tau} d^{\dagger}_{\alpha\tau{{\bf k}+{\bf q}}} d^{}_{\beta\tau{\bf k}}$ and $S^z_{\alpha\beta}({\bf q})=\sum_{{\bf k}\tau\tau'} d^{\dagger}_{\alpha\tau{{\bf k}+{\bf q}}} \sigma^z_{\tau\tau'} d^{}_{\beta\tau'{\bf k}}$ are defined in terms of the orbital operators $d^{\dag}_{\alpha\tau{\bf k}}$ and $d^{}_{\alpha\tau{\bf k}}$, which respectively creates and annihilates an electron in the orbital $\alpha$ with spin $\tau$ and momentum $\bf k$.
The RIXS form factors $W^0_{\alpha\beta}({\bf e})$ and $W^z_{\alpha\beta}({\bf e})$ in \cref{eq:CrossSection} depend on the transition-metal ion, the orbital symmetry of the system, the specific geometry of the experiment, and on the polarization ${\bf e}$ of the incident and scattered x-ray beams~\cite{Ament2009,Haverkort2010}, as shown in \cref{ch:RIXS}.
Thus, these parameters can be adjusted in the RIXS experiment, and therefore, under construction of a particular experimental setup, the cross section will be solely determined either by the charge or by the spin DSF\@. 
As it has been shown in the previous Chapter, this property can be used to reveal the character of the pairing mechanism in unconventional superconductors. 
Hereafter we will study the charge and spin DSF for iron-based superconductors using the band structure of tight-binding models, and comparing different pairing mechanisms and order parameter symmetries.

In order to reproduce correctly the characteristic disconnected Fermi surface of iron-based superconductors, a minimal model for these systems must include more than one $3d$ orbital state on the Fermi surface. 
Therefore, a phenomenological description of the unconventional superconducting state in iron based pnictides can be achieved considering a generalized multi-band mean-field Hamiltonian in the form~\cite{Tinkham2004,Sigrist2005} 
\begin{equation}\label{eq:Hamiltonian}
	{\cal H} = \sum\limits_{i\tau{\bf k}} \varepsilon_{i{\bf k}} \, 
	c^\dag_{i\tau{\bf k}}c^{}_{i\tau{\bf k}} 
	- \frac12 \sum\limits_{i\tau{\bf k}} \xi_\tau 
	\left( \Delta_{\bf k} c^\dag_{i\tau{\bf k}}c^\dag_{i-\tau-{\bf k}} + \Delta_{\bf k}^* 
	c^{}_{i-\tau-{\bf k}} c^{}_{i\tau{\bf k}} \right),
\end{equation}
where the operators $c^\dag_{i\tau{\bf k}}$ and $c^{}_{i\tau{\bf k}}$ respectively create and annihilate an electron with spin $\tau$ in the energy band $i$, which is described by the bare electron dispersion $\varepsilon_{i{\bf k}}$, and with $\Delta_{\bf k}$ the momentum-dependent superconducting order parameter. 
The second term in \cref{eq:Hamiltonian} is responsible for the superconducting state, with the pairing character determined by $\xi_\tau$. 
The case of $\xi_\tau=\pm1$ for up and down spin describes the spin-singlet pairing, whereas the case $\xi_\tau=1$ for both spin directions leads to a special type of spin-triplet state. 
In general, the triplet pairing term is given by $-\frac12\Delta_{{\bf k}\tau\tau'} c^\dag_{i\tau{\bf k}}c^\dag_{i\tau'-{\bf k}} + \mbox{h.c.}$, with a multi-component superconducting order parameter of the form $\Delta_{{\bf k}\tau\tau'}=\imath\left[{\bf d}({\bf k})\cdot{\boldsymbol \sigma}\right]\sigma_y$ (see, e.g., \onlinecite{Sigrist2005}). 
However, in this Chapter only the simplest case $d_x({\bf k})=d_y({\bf k})=0$ and $d_z({\bf k})=\Delta_{\bf k}$ is considered, and therefore the gap function simplifies to $\Delta_{{\bf k}\uparrow\uparrow}=\Delta_{{\bf k}\downarrow\downarrow}=0$ and $\Delta_{{\bf k}\uparrow\downarrow}=\Delta_{{\bf k}\downarrow\uparrow}=\Delta_{{\bf k}}$. 

To investigate the RIXS cross section given by \cref{eq:CrossSection} one can calculate the DSF $\chi^{0,z}_\alpha({\bf q},\omega)$ on the basis of the model Hamiltonian~\eqref{eq:Hamiltonian} separately for each of the relevant orbitals. 
Using the unitary transformation between orbital and energy band representation defined as
\begin{equation}\label{eq:Transformation}
	c_{i\tau{\bf k}}=\sum\limits_\alpha \lambda_{i\alpha,{\bf k}} \,
	d_{\alpha\tau{\bf k}}
	,
\end{equation}
one can rewrite the density and spin operators $\rho_{\alpha\beta}({\bf q})$ and $S^z_{\alpha\beta}({\bf q})$ in \cref{eq:ChargeSpinDSF} in terms of the operators $c^{}_{i\tau{\bf k}}$ and $c_{i\tau{\bf k}}^\dag$ in the band representation.
This step is necessary for the calculation of the matrix elements and excitation energies in \cref{eq:ChargeSpinDSF}. 
Note that, in general, the Hamiltonian is not diagonal with respect to the orbital states because the different orbitals can hybridize with each other. 
The transformation matrix elements $\lambda_{i\alpha,{\bf k}}$, which describe the orbital content of conduction bands, are obtained diagonalizing the low-energy tight-binding Hamiltonian of the system. 
For this purpose, in \cref{sec:laofeas,sec:lifeas} the two iron superconductors LaOFeAs and LiFeAs will be described respectively by the tight-binding model in \onlinecite{Raghu2008} and in \onlinecite{Brydon2011}.

Having expressed the density and spin operators in the DSF in terms of the one-particle operators in the band representation, the next step is to diagonalize the Hamiltonian~\eqref{eq:Hamiltonian} by the Bogoliubov transformation $c_{i\uparrow{\bf k}} = u_{i{\bf k}}^* \gamma_{i\uparrow{\bf k}} - v_{i{\bf k}} \gamma_{i\downarrow -{\bf k}}^\dag$ and $c_{i\downarrow{\bf k}} = u_{i{\bf k}}^*\gamma_{i\downarrow{\bf k}} + v_{i{\bf k}} \gamma_{i\uparrow -{\bf k}}^\dag$, with $|u_{i{\bf k}}|^2=\frac12\left(1+\varepsilon_{i{\bf k}} / E_{i{\bf k}}\right)$, $|v_{i{\bf k}}|^2=\frac12\left(1-\varepsilon_{i{\bf k}} / E_{i{\bf k}}\right)$, and $u_{i{\bf k}}^* v_{i{\bf k}} =\frac12 \Delta_{\bf k} / E_{i{\bf k}}$ for each of the different bands. 
This allows one to determine the ground state $|{\rm BCS}\rangle$ and the excitations of the system, in terms of the quasiparticle operators $\gamma_{i\tau{\bf k}}$ and of the quasiparticle dispersion $E_{i{\bf k}}=\sqrt{\varepsilon_{i{\bf k}}^2+\vert\Delta_{\bf k}\vert^2}$. 
In a centrosymmetric superconductor at zero temperature, the excited states contributing to DSF have the form $\gamma^\dag_{j\tau{\bf k}+{\bf q}}\gamma^\dag_{i-\tau-{\bf k}}|{\rm BCS}\rangle$ with energy $E_{i{\bf k}}+E_{j{\bf k}+{\bf q}}$.
It follows that the charge and spin DSF $\hat{\chi}^{0,z}({\bf q},\omega)$ of quasiparticle excitations is described by a matrix of intra-orbital ($\alpha=\beta$) and inter-orbital ($\alpha\neq\beta$) components given by
\begin{align}\label{eq:StructureFactor}
	&\chi^{0,z}_{\alpha\beta}({\bf q},\omega) = 
	\sum\limits_{i j {\bf k}} 
	\delta\left(\hbar\omega-E_{i{\bf k}}-E_{j{{\bf k}+{\bf q}}}\right)
	\times\nonumber\\&
	|\lambda_{i\alpha,{\bf k}}
	\lambda_{j\beta,{\bf k+q}}|^2
	\left[ 
	1 \pm
	\frac{{\rm Re}(\Delta_{\bf k}^{}\Delta_{{\bf k}+{\bf q}}^*)
	\mp\varepsilon_{i{\bf k}}\varepsilon_{j{{\bf k}+{\bf q}}}}
	{E_{i{\bf k}} E_{j{{\bf k}+{\bf q}}}} 
	\right]
	,
\end{align}
where $\alpha,\beta$ span the relevant orbitals of the system and the $\pm$ sign distinguishes between charge and spin DSF~\cite{Kee1998,Kee1999,Voo2000}.
This result shows that the momentum-dependent DSF of low-energy quasiparticle excitations is strongly affected by the orbital content of bare electrons and the structure of the superconducting order parameter.

The character of the superconducting pairing, which is described by the gap function $\Delta_{\bf k}$, arises at energies close to the Fermi level $\hbar\omega \approx \varepsilon_F$. 
There, the main contributions to the DSF correspond to excitations close to the Fermi surface, i.e., those which fulfill the condition $\varepsilon_{i{\bf k}}\varepsilon_{j{\bf k + q}} \ll |\Delta_{\bf k}\Delta_{{\bf k} + {\bf q}}|$. 
Assuming a phase dependent order parameter in the form $\Delta_{\bf k} =|\Delta_{\bf k}|e^{\imath\phi_{\bf k}}$, the DSF in \cref{eq:StructureFactor} for low-energy excitations becomes approximately
\begin{align}\label{eq:CoherenceFactors}
	&\chi^{0,z}_{\alpha\beta}({\bf q},\omega)
	\approx 
	\sum\limits_{i j {\bf k}}
	\delta(\varepsilon_{i{\bf k}}\varepsilon_{j{\bf k+q}})
	\delta(\hbar\omega-|\Delta_{{\bf k}+{\bf q}}| - |\Delta_{\bf k}|)
	\times\nonumber\\&\qquad\qquad
	|\lambda_{i\alpha,{\bf k}}
	\lambda_{j\beta,{\bf k+q}}|^2
	\left[
	1 \pm \cos(\phi_{\bf k} - \phi_{{\bf k}+{\bf q}}) 
	\right]
	.
\end{align}
Hence the DSF is influenced significantly by the order parameter phase $\phi_{\bf k}$ on the Fermi surface. 
In particular, the charge DSF is suppressed for sign-reversing ($\phi_{\bf k}-\phi_{\bf k+q}=\pi $), whereas the spin DSF is suppressed for sign-preserving excitations ($\phi_{\bf k}-\phi_{\bf k+q}=0$). 

Interactions between the conduction electrons are taken into account within the RPA\@.
The matrix function of the DSF $\hat{\chi}^{0,z}_{\rm RPA}({\bf q},\omega)$ with interactions is related to the matrix function of the bare DSF $\hat{\chi}^{0,z}({\bf q},\omega)$ from \cref{eq:StructureFactor} via the equation
\begin{equation}
\label{eq:RPA}
\hat{\chi}^{0,z}_{\rm RPA}({\bf q},\omega) = \hat{\chi}^{0,z}({\bf q},\omega) \left[\mathbbm{1} - \hat{\Gamma}\hat{\chi}^{0,z}({\bf q},\omega) \right]^{-1},
\end{equation}
where $\mathbbm{1}$ is the identity matrix and $\hat{\Gamma}=U\mathbbm{1}$ the interaction matrix.
Following \onlinecite{Raghu2008}, the intra-orbital interaction is taken as $U=W/4$, where $W$ is the bandwidth of the relevant bands, and neglected the inter-orbital interaction ($J=0$) in the actual calculations of the DSF of LaOFeAs and LiFeAs.

Since the RIXS form factors $W^0_{\alpha\beta}({\bf e})$ and $W^z_{\alpha\beta}({\bf e})$ in \cref{eq:CrossSection} can be tuned by properly choosing the experimental setup, RIXS can probe both charge and spin DSF, which is a unique feature among other spectroscopies.
A comparison between the charge and the spin DSF of quasiparticle excitations allows one to disclose the momentum dependence of the magnitude and of the phase of the superconducting order parameter and, therefore, the underlying symmetry of the pairing mechanism. 
In the next Section we will show the predicted RIXS spectra for the LaOFeAs and LiFeAs iron based superconductors obtained numerically using the theoretical framework described above.

\section{Phase and orbital sensitivity in iron based superconductors}
\subsection{General model}
\label{sec:laofeas}

\begin{figure}[t]
\centering\includegraphics[scale=1.33]{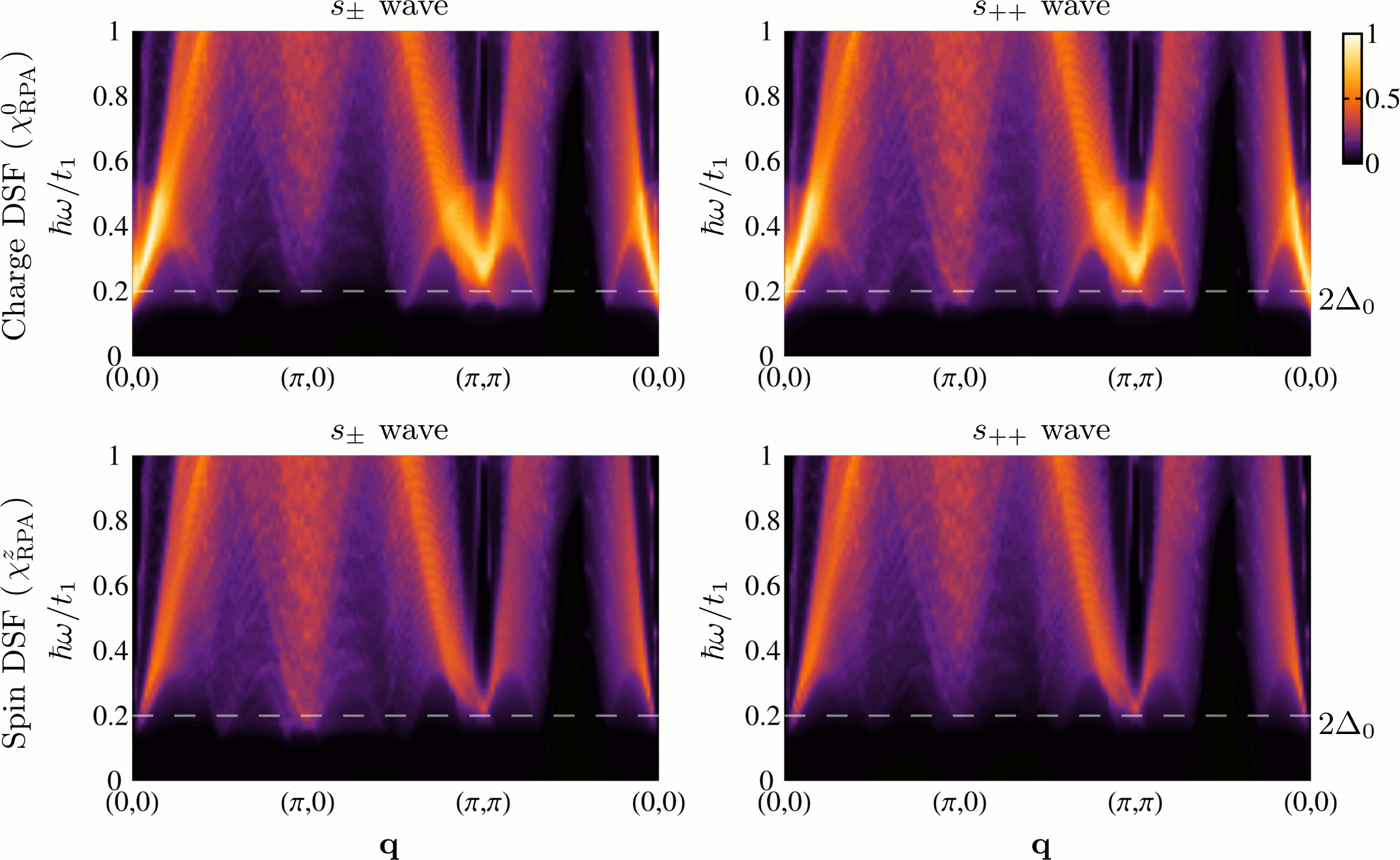}
\caption[RIXS intensities as a function of energy loss $\hbar\omega$ and transferred momentum $\bf q$ for the charge and spin DSF in LaOFeAs, with different order parameter symmetries]{
RIXS intensities as a function of energy loss $\hbar\omega$ and transferred momentum $\bf q$ along the high symmetry path in the Brillouin zone, for the charge and spin DSF in LaOFeAs, assuming respectively an $s_\pm$~wave and an $s_{++}$~wave order parameter, calculated via \cref{eq:RPA}, and assuming the bare electron dispersion of the tight-binding model in \onlinecite{Raghu2008}.
The resonance peak appears at relatively small energy values close to $(0,0)$ and is dispersive.
Spectral intensities at $(0,0)$ and $(\pi,\pi)$ are suppressed in the spin DSF for both order parameter choices, while spectral intensities at ${\bf Q}_{\rm AF}=(\pi,0)$ are suppressed in the charge (spin) spectra in the $s_\pm$ ($s_{++}$) wave state.
Intensities are in arbitrary units.
}
\label{fig:raghu-confront}
\end{figure}

\begin{figure}[t]
\centering\includegraphics[scale=1.33]{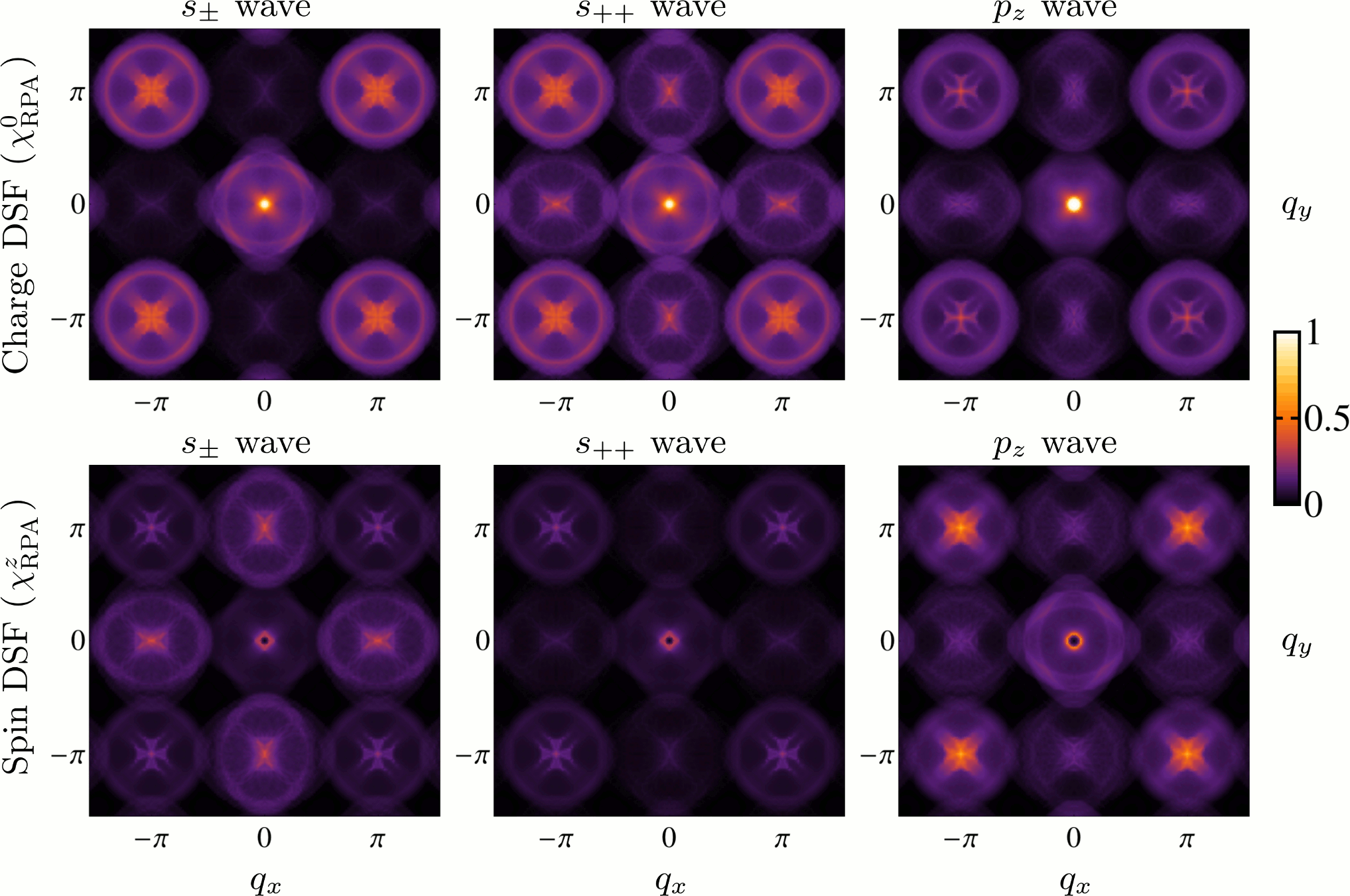}
\caption[RIXS intensities at a fixed energy loss for the charge and spin DSF in LaOFeAs, with different order parameter symmetries]{
Charge and spin DSF $\chi^{0,z}_{\rm RPA}$ of quasiparticle excitations in LaOFeAs at a fixed energy loss $\hbar\omega=2\Delta_0$ as a function of the transferred momentum $\bf q$, with $s_\pm$, $s_{++}$, and $p_z$~wave order parameter ($\Delta_0=0.1t_1$, see main text), calculated using \cref{eq:StructureFactor,eq:RPA} assuming the bare electron dispersion and the orbital symmetry of the tight-binding model in \onlinecite{Raghu2008}, summing over inter-orbital and intra-orbital contributions.
The coherence peak from \cref{fig:raghu-confront} which appears here close to the point $(0,0)$ is largely dominant in the charge DSF spectra, while intensity distributions around ${\bf Q}_{\rm AF}=(\pi,0)$ and at $|{\bf q}|\approx\pi/2$ are sensitive to the differences in the order parameter phase along the Fermi surface.
Spectral intensities at ${\bf Q}_{\rm AF}$ are strongly suppressed in the charge (spin) spectra in the $s_\pm$ ($s_{++}$) wave state, while intensities in the region $|{\bf q}|\approx\pi/2$ around the point $(0,0)$ are suppressed in the charge (spin) DSF in the $p_z$ ($s_\pm$ or $s_{++}$) wave state.
Intensities are in arbitrary units.
}
\label{fig:raghu}
\end{figure}

In this Section, in order to describe the LaOFeAs superconductor, we will employ the effective two-band tight-binding model proposed in \onlinecite{Raghu2008}, which is regarded as a minimal model for conduction electrons in iron based superconductors.
This model takes into account the effective hoppings between the two orbitals $d_{xz}$ and $d_{yz}$ of the iron ions, and correctly reproduces the band structure of the compound LaOFeAs, which consists of disconnected hole-like Fermi surface branches around $(0,0)$ and $(\pi,\pi)$ and separate electron pockets of similar size around $(0,\pm\pi)$ and $(\pm\pi,0)$ in the Brillouin zone [compare Fig.~\ref{fig:raghu-confront}~(c)]. 
We assume three different symmetries for the superconducting gap, i.e., $s_\pm$ wave~\cite{Mazin2008}, $s_{++}$ wave~\cite{Kontani2010}, and a spin-triplet $p_z$ wave~\cite{Brydon2011}, where the momentum dependence is modeled respectively by $\Delta^{s_\pm}_{\bf k}=\Delta_0 \cos{k_x}\cos{k_y}$, $\Delta^{s_{++}}_{\bf k}=\vert\Delta^{s_\pm}_{\bf k}\vert$, and $\Delta^{p_z}_{\bf k}=\Delta_0\left(\sin{k_x}-\imath \sin{k_y}\right)$, with $\Delta_0=0.1|t_1|$, where $t_1$ is the magnitude of the dominant nearest-neighbor hopping (cf.~\onlinecite{Raghu2008}). 
For these choices of the order parameter, the gap magnitude in the spin-singlet case varies around $\approx0.75\Delta_0$ along the electron pockets and the inner hole pocket, and around $\approx0.6\Delta_0$ along the outer hole pocket, with opposite sign in the case of the $s_\pm$~wave symmetry. 
In the spin-triplet case instead, the gap magnitude varies around $\approx0.65\Delta_0$ along the electron pockets and the inner hole pocket, and around $\approx0.83\Delta_0$ along the outer hole pocket. 
The inter-band interaction in the RPA is fixed to the value $U=W/4=3t_1$.

At first we study the general behavior of the RIXS spectra in a large energy range.
\Cref{fig:raghu-confront} shows the charge and spin DSF as a function of the transferred momentum $\bf q$ in an energy range up to $\hbar\omega\in[0,t]$, calculated in the RPA using \cref{eq:RPA}.
We consider the two cases of $s_\pm$ and $s_{++}$~wave symmetry.
A dispersive resonance peak is clearly visible at rather small momentum transfer, which arises due to the interaction processes.
Significant differences between charge and spin DSF are obtained at low-energy values in the order of the superconducting gap.
In particular, the spectral weight of the spin DSF at momentum vectors close to the points $(0,0)$ and $(\pi,\pi)$ is strongly suppressed, whereas spectral intensities at ${\bf Q}_{\rm AF}=(\pi,0)$ are suppressed in the charge (spin) spectra in the $s_\pm$ ($s_{++}$) wave state.
This spectral redistribution is indeed sensitive to the symmetry of the superconducting order parameter, as seen in \cref{ch:BCS-Cuprates}.
In order to investigate this feature in more detail, we will focus hereafter on the differences between charge and spin DSF by fixing the transferred energy $\hbar\omega$ to a value which is comparable with the energy scale of the superconducting gap.

For these purposes, we show in \cref{fig:raghu} the charge and spin DSF at a fixed energy loss $\hbar\omega=2\Delta_0$ as a function of the transferred momentum $\bf q$, for the three choices of the order parameter defined above.
As one can see, low-energy excitations which are sign-reversing, (opposite phase of the order parameter), suppress the charge component of the DSF, whereas sign-preserving excitations (same phase of the order parameter) suppress the spin component in the low-energy quasiparticle spectra.
For this reason, spectral intensities at ${\bf Q}_{\rm AF}=(\pi,0)$ in \cref{fig:raghu} are suppressed in the charge and in the spin DSF respectively in $s_\pm$~wave and in the $s_{++}$~wave superconducting states.
Such transferred momentum, which corresponds to the ordering vector of the antiferromagnetic phase, is in fact a nesting vector between the hole pockets and the electron pockets in the Brillouin zone, which have an opposite sign or the same sign of the order parameter alternatively in the $s_\pm$~wave and in the $s_{++}$~wave states.
Note that the enhancement of spectral intensity in the spin response functions at momentum ${\bf Q}_{\rm AF}$, as it appears in Fig.~\ref{fig:raghu} for the $s_\pm$ case, has been found also in neutron scattering experiments~\cite{Qiu2008,Inosov2010}.

On the other hand, based on the result in \cref{fig:raghu} we propose that RIXS will be able to detect a characteristic signature of the $p$-wave order parameter.
Namely, the odd-symmetry in momentum space $\Delta_{\bf -k}=-\Delta_{\bf k}$ should produce signatures in the spectral intensities of excitations with transferred momentum $|{\bf q}|\approx\pi/2$ (see \cref{fig:raghu}), corresponding to a \emph{self-nesting} of the hole pockets.
This type of excitations, which lead to characteristic intensity features also in LiFeAs (see next Section), refer to \emph{intra-band} contributions located in a narrow momentum range similar to the conventional nesting scenario between the electron and hole pockets.
In the $s$-wave case these excitations preserve the sign of the order parameter ($\Delta_{\bf k+q}=\Delta_{\bf -k}=\Delta_{\bf k}$), leading to a suppression of spectral intensities in the spin DSF\@.
In the $p$-wave case instead, these excitations are sign-reversing ($\Delta_{\bf k+q}=\Delta_{\bf -k}=-\Delta_{\bf k}$), with a consequent suppression in the charge DSF\@.

\subsection{LiFeAs}
\label{sec:lifeas}

\begin{table}[t!]
	\centering\footnotesize
\renewcommand{\tabcolsep}{.65mm}
\begin{tabular*}{\textwidth}{c @{\extracolsep{\fill}} cccccccccccccc}
		\hline\hline
		& $t_1$ & $t_2$ & $t_3$ & $t_4$ & $t_5$ & $t_6$ & $t_7$ 
		& $t_8$ & $t_9$ & $t_{10}$ & $t_{11}$ &\!\!\!\! $\Delta_{xy}$ \\
		\hline
		fit & 
		$0.019$ & $0.123$ & $0.014$ & $-0.055$ & $0.217$ & $0.264$ & $-0.137$ 
		& $-t_7/2$ & $-0.060$ & $-0.057$ & $0.016$ & \!\!\!\! $1$\\
		\onlinecite{Brydon2011}\,\,\,&
		$0.020$ & $0.120$ & $0.020$ & $-0.046$ & $0.200$ & $0.300$ & $-0.150$ 
		& $-t_7/2$ & $-0.060$ & $-0.030$ & $0.014$ & \!\!\!\! $1$\\
	\end{tabular*}
\caption[Hopping parameters for the three-band effective tight-binding model compared with those fitted with the experimental Fermi surface of LiFeAs]{
Hopping parameters for the three-band effective tight-binding model (see \onlinecite{Brydon2011} for the definitions of the parameters and for details) compared with those fitted with the experimental Fermi surface of LiFeAs~\cite{Borisenko2010,Knolle2012}. 
The chemical potential is $\mu=0.338$, that corresponds to a filling of four electrons per site. 
Energy units are in electron volts. 
}
\label{tab:tb}
\end{table}

\begin{figure}[t!]
\centering\includegraphics[scale=.2]{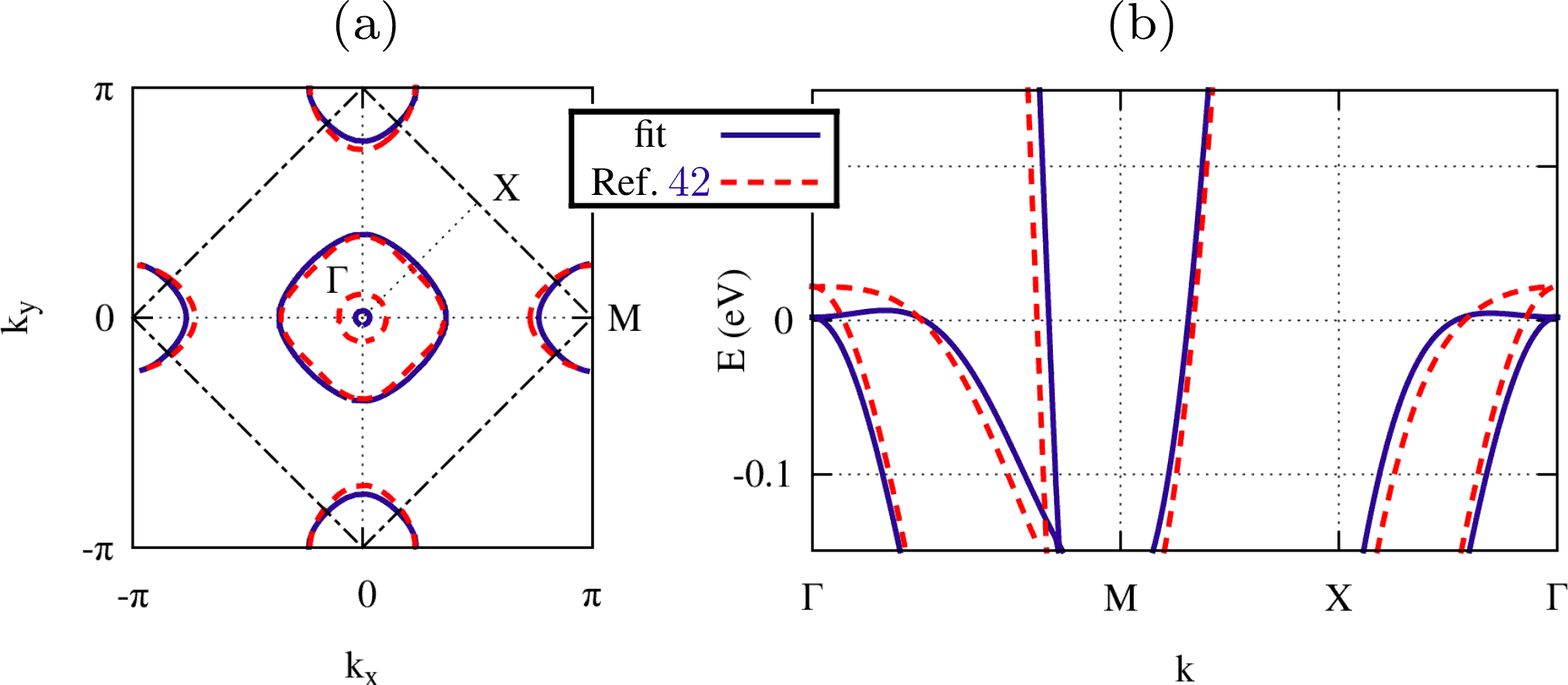}
\caption[Fermi surface and electronic dispersion for the tight-binding model of LiFeAs]{
Fermi surface (a) and electronic dispersion (b) for the tight-binding model in \onlinecite{Brydon2011} (dashed line) and for the same model with hopping parameters (see \cref{tab:tb}) fitted with the experimental Fermi surface of LiFeAs~\cite{Borisenko2010,Knolle2012} (solid line). 
}
\label{fig:bands}
\end{figure}

In contrast to other iron based superconductors, there is no general agreement about the nature of the superconducting state in LiFeAs, and in particular about the pairing mechanism. 
Along with the $s_\pm$~wave pairing in fact, different scenarios have been proposed, e.g., an $s_{++}$~wave state, driven by the critical $3d$-orbital fluctuations induced by moderate electron-phonon interactions~\cite{Kontani2010}, or even a spin-triplet pairing driven by ferromagnetic fluctuations~\cite{Brydon2011}. 
While singlet pairing is supported by some neutron scattering experiments~\cite{Taylor2011}, the unusual shape of the Fermi surface and the momentum dependency of the superconducting gap measured by ARPES~\cite{Borisenko2012} is in conflict with the $s_\pm$~wave symmetry. 
Moreover, quasiparticle interference probed by STM experiments~\cite{Hanke2012} are consistent with a $p$~wave spin-triplet state or with a singlet pairing mechanism with a more complex order parameter ($s+\imath d$~wave). 
Whereas ARPES has been proven to be powerful in measuring the momentum dependence of the superconducting gap on the Fermi surface~\cite{Borisenko2010,Borisenko2012}, it should be noted here that ARPES, since not sensitive to the order parameter phase, cannot distinguish between singlet and triplet pairing, i.e., between even ($\Delta_{\bf k}=\Delta_{-\bf k}$) and odd ($\Delta_{\bf k}=-\Delta_{-\bf k}$) symmetry of the order parameter. 
In fact, the experimental momentum dependence of the superconducting gap measured by ARPES~\cite{Borisenko2012} is consistent, in principle, with a spin-singlet as well as with a spin-triplet state. 

An appropriate RIXS experiment may help to clarify this complicated and controversial situation in LiFeAs. 
In particular, to highlight the characteristic features of RIXS spectra which allow one to discriminate between different pairing mechanisms, we will present some theoretical RIXS spectra for different order parameter symmetries, corresponding to spin-singlet and spin-triplet pairing. 
To achieve this goal, we consider different order parameter symmetries, corresponding to spin-singlet and spin-triplet pairing. 
In order to properly take into account the orbital degrees of freedom of the system, we construct our model on the basis of the effective three-band tight-binding model proposed in \onlinecite{Brydon2011}, which includes the effective hoppings between the $t_{2g}$ orbitals of the iron ions, within a single Fe-ion unit cell. 
Nevertheless, a comparison with ARPES measurements~\cite{Borisenko2010,Knolle2012} shows that the inner hole pocket in LiFeAs is much smaller than the one produced by the tight-binding model in \onlinecite{Brydon2011}. 
Furthermore, the superconducting gap is significantly larger~\cite{Borisenko2012} on the inner hole pocket than on the outer one. 
For this reason, one can redefine the hopping parameters in order to fit the experimental Fermi surface~\cite{Knolle2012,Borisenko2010}. 
These parameters are given in \cref{tab:tb}, while in \cref{fig:bands} the fitted Fermi surface (a) and bare electron dispersion (b) is compared with the original model (cf.~\onlinecite{Brydon2011}). 

Besides the $s_\pm$, $s_{++}$, and $p_z$~wave defined above, we consider here also a triplet pairing order parameter $\widetilde{p}_z$, defined as $\Delta^{\widetilde{p}_z}_{\bf k}=|\Delta^{s_{\pm}}_{\bf k}|e^{\imath\phi_{\bf k}}$, i.e., having the same magnitude of the $s_{\pm}$ (or $s_{++}$) wave and the same phase $\phi_{\bf k}=\arg{\Delta^{p_z}_{\bf k}}$ of the $p_z$~wave order parameter.
This superconducting order parameter is considered here for comparison, in order to have an example of a spin-triplet pairing which reproduces the experimental gap magnitude on the different branches of the Fermi surface in LiFeAs.
Moreover, the equal gap structure in comparison to the singlet pairing models allows us to study those features of spectra which are solely attributed to the phase variation.
The order parameter magnitude is $\Delta_0=6\meV$, in order to be consistent with the measured value of the superconducting gap in LiFeAs~\cite{Borisenko2012}.
We consider the inter-band interaction in the RPA as $U=W/4\approx0.7$~eV.
Therefore, in the case of the $s$-wave states ($s_\pm$ and $s_{++}$), and of the $\widetilde{p}_z$~wave state, the gap magnitude varies around $\approx4.6\meV$ along the electron pockets, around $\approx6\meV$ along the inner hole pocket, and around $\approx3\meV$ along the outer hole pocket, with opposite sign in the case of the $s_\pm$~wave symmetry, and with the phase continuously varying on the Fermi surface in the case of the $\widetilde{p}_z$~wave state.
In the $p_z$~wave case instead, the gap magnitude varies around $\approx4\meV$ along the electron pockets, around $\approx0.6\meV$ along the inner hole pocket, and around $\approx5.6\meV$ along the outer hole pocket.
In any of the case considered, the low-energy quasiparticle excitations contribute to coherence peaks at $(0,0)$ with energy in the range $6\meV<E<12\meV$ ($\Delta_0<E<2\Delta_0$).

\begin{figure}[t]
\centering\includegraphics[scale=1.33]{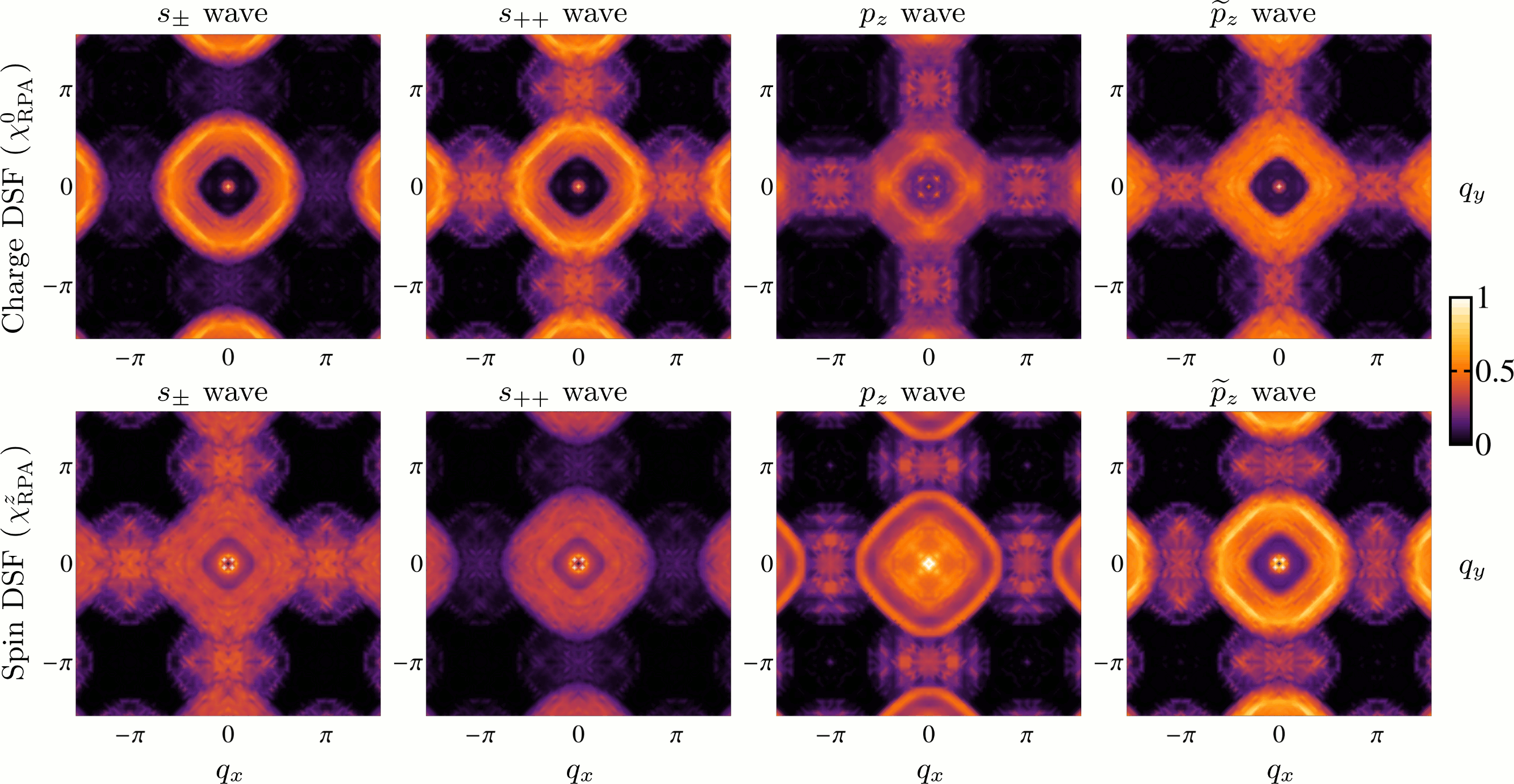}
\caption[RIXS intensities at a fixed energy loss for the charge and spin DSF in LiFeAs]{
Charge and spin DSF $\chi^{0,z}_{\rm RPA}$ of quasiparticle excitations in LiFeAs at a fixed energy loss $\hbar\omega=2\Delta_0=12\meV$ as a function of the transferred momentum $\bf q$, with the $s_\pm$, $s_{++}$, $p_z$, and $\widetilde{p}_z$~wave superconducting order parameter, calculated using \cref{eq:StructureFactor,eq:RPA} assuming the bare electron dispersion and the orbital symmetry of the tight-binding mode in \onlinecite{Brydon2011} fitted with ARPES data~\cite{Knolle2012,Borisenko2010}, and summing over inter-orbital and intra-orbital contributions.
Spectral intensities at ${\bf Q}_{\rm AF}=(\pi,0)$ are suppressed in the charge (spin) spectra in the $s_\pm$ ($s_{++}$) wave state, while intensities in the region $|{\bf q}|\approx\pi/2$ around the point $(0,0)$ are suppressed in the charge (spin) DSF in the $p_z$ and $\widetilde{p}_z$ ($s_\pm$ and $s_{++}$) state.
Intensities are in arbitrary units.
}
\label{fig:lifeas}
\end{figure}

In \cref{fig:lifeas} are shown the RIXS intensities for the charge and spin DSF at a fixed energy loss $\hbar\omega=2\Delta_0=12 \meV$ as a function of the transferred momentum $\bf q$, for different choices of the supeconducting order parameter symmetry, calculated using \cref{eq:StructureFactor}.
The resonant peaks at ${\bf Q}_{\rm AF}$ and at $|{\bf q}|\approx\pi/2$ are clearly visible in the calculated spectra, and are consistent with recent neutron scattering experiments~\cite{Knolle2012,Qureshi2012}.
However, in LiFeAs, no nesting occurs between the hole and the electron pockets~\cite{Borisenko2010}, and therefore the peak at ${\bf Q}_{\rm AF}$ in the quasiparticle spectra, which corresponds to the scattering between hole and electron pockets, is much weaker and broader than in the LaOFeAs case.
The square-like intensity distribution at small momenta for all the considered pairing symmetries is a typical feature of the low-energy spectrum in LiFeAs arising from inter-band scattering processes between the two hole pockets of the Fermi surface~\cite{Hanke2012,Hess2013}.
Indeed, as in the previous case, RIXS spectra in LiFeAs are strongly sensitive to the symmetry of the supeconducting order parameter and on its relative phase differences along the Fermi surface.
In fact, spectral intensities at ${\bf Q}_{\rm AF}$ are further suppressed in the charge and in the spin DSF respectively in the $s_\pm$~wave and in the $s_{++}$~wave states.
This is because one has $\Delta_{{\bf k}+{\bf Q}_{\rm AF}}=\pm\Delta_{\bf k}$, with the $\pm$ sign corresponding to the $s_{++}$ and $s_\pm$~wave, resulting in sign-preserving and sign-reversing excitations respectively.
In the $p$-wave states no suppression occurs, being $\Delta_{{\bf k}+{\bf Q}_{\rm AF}}=\Delta_{\bf k}^*$, i.e., with a phase difference given by $2\phi_{\bf k}$, resulting in charge and spin coherence factors [see \cref{eq:CoherenceFactors}] which continuously vary on the Fermi surface.
Again, the signature of the $p$-wave odd-symmetry is in the spectral intensities of excitations with transferred momentum $|{\bf q}|\approx\pi/2$, corresponding to a self-nesting of the larger hole pocket (see \cref{fig:lifeas}).
While in the $s$-wave case excitations with $|{\bf q}|\approx\pi/2$ are sign-preserving, with a consequent suppression of spectral intensities in the spin DSF, in the $p$-wave case they are sign-reversing, resulting instead in an enhancement in the spin DSF\@.

\begin{figure}[t]
\centering\includegraphics[width=\textwidth]{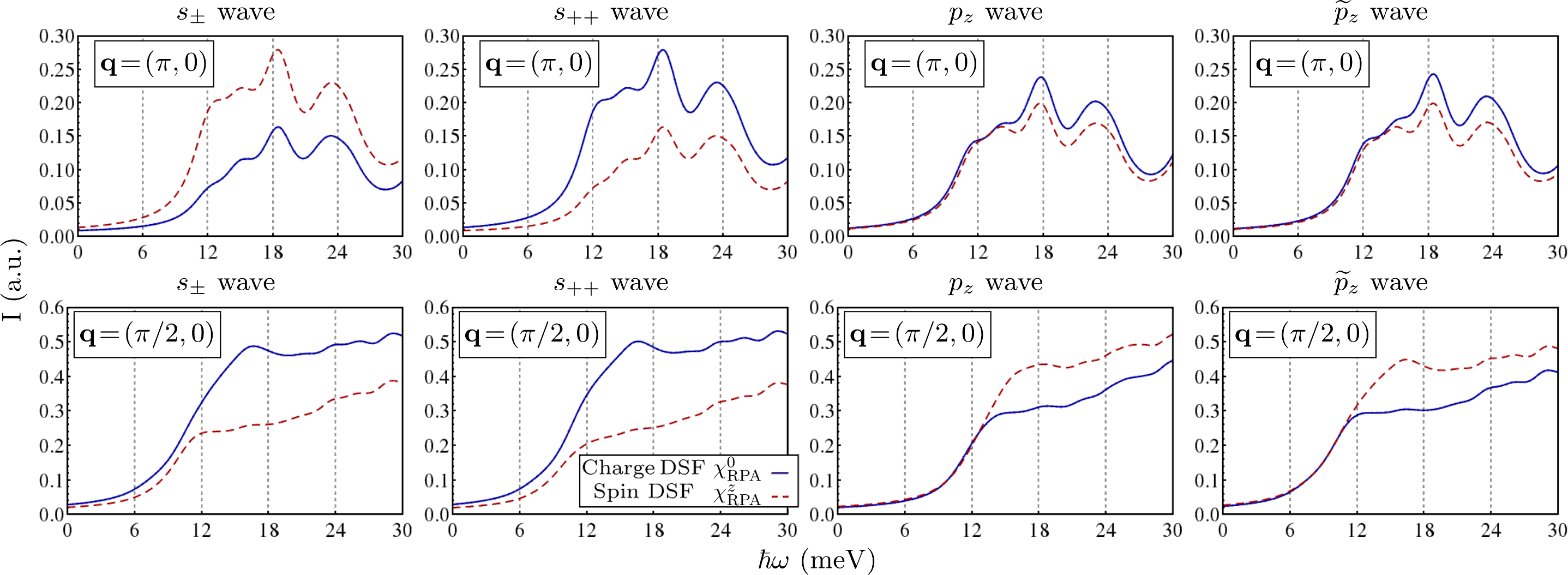}
\caption[RIXS intensities at a fixed energy loss for the charge and spin DSF in LiFeAs at fixed momenta]{
Charge (solid line) and spin (dashed line) DSF $\chi^{0,z}_{\rm RPA}$ of quasiparticle excitations in LiFeAs at ${\bf Q}_{\rm AF}=(\pi,0)$ and at $(\pi/2,0)$ as a function of the energy loss, calculated using \cref{eq:StructureFactor,eq:RPA}, respectively with $s_\pm$, $s_{++}$, $p_z$, and $\widetilde{p}_z$~wave superconducting order parameter ($\Delta_0=6\meV$).
Spectral intensities at ${\bf Q}_{\rm AF}$ are larger for the spin (charge) DSF spectra in the $s_\pm$ ($s_{++}$) wave state, while intensities at $(\pi/2,0)$ are larger in the spin (charge) DSF in the $p_z$ and $\widetilde{p}_z$ ($s_\pm$ and $s_{++}$) wave state.
Intensities are in arbitrary units.
}
\label{fig:lifeas-hsp}
\end{figure}

In order to present in the most clear way how to distinguish between the different pairing scenarios in LiFeAs, we show in \cref{fig:lifeas-hsp} the RIXS spectra as a function of the energy loss for the charge and spin DSF of quasiparticle excitations at ${\bf Q}_{\rm AF}$ and at $(\pi/2)$, again for different choices of the superconducting order parameter symmetry.
As we have seen, these particular momenta are those where the sensitivity to the order parameter phase is more pronounced.
In particular, spectral intensities corresponding to the transferred momentum ${\bf Q}_{\rm AF}$ are sensitive to sign changes of the order parameter between hole and electron pockets.
Indeed, as one can see in \cref{fig:lifeas-hsp}, the charge (spin) DSF is suppressed in the $s_\pm$ ($s_{++}$) wave state.
Therefore, a comparison between charge and spin DSF can be revealing of a sign-reversal in the order parameter between disconnected branches of the Fermi surface.
On the other hand, the spectral contributions of the intra-band scattering within the hole pockets, which correspond to a transferred momentum $|{\bf q}|\approx\pi/2$, are strongly affected by the parity of the order parameter, and therefore can discriminate between spin-singlet (e.g., $s$-wave) and spin-triplet pairing (e.g., $p$-wave).
In fact, spectral intensities at $(\pi/2,0)$ in \cref{fig:lifeas-hsp} are suppressed in the charge and spin DSF respectively in the spin-triplet ($p_z$ and $\widetilde{p}_z$~wave) and in the spin-singlet ($s_\pm$ and $s_{++}$~wave) cases.
It should be noticed here that this result is general, and does not depend on the gap magnitude dependence along the Fermi surface, but only on its phase variations, and therefore is a mere consequence of the odd parity of the order parameter.
This aspect is clearly displayed by the two panels of \cref{fig:lifeas-hsp} referring to the $p$-wave state at $(\pi/2,0)$.
The suppression of the charge DSF occurs for both $p_z$ and $\widetilde{p}_z$~wave, which have a different gap magnitude dependence, but nevertheless the same phase variations and the same parity.

\section{Conclusions}
\label{sec:con}

In this Chapter we have shown how RIXS spectra of quasiparticle excitations are sensitive to phase differences of the superconducting order parameter along the Fermi surface, and hence allow one to distinguish among different superconducting states, in particular between spin-singlet and spin-triplet pairing and between sign-preserving and sign-reversing $s$-wave states in iron based superconductors. 
In particular, RIXS spectral intensities corresponding to a self-nesting of the hole pockets can discriminate between singlet and triplet pairing, while RIXS spectra corresponding to a scattering between hole and electron pockets [${\bf Q}_{\rm AF}=(\pi,0)$] can discriminate between an $s_\pm$~wave and an $s_{++}$~wave order parameter. 
The contribution of quasiparticle excitations can be separated from other effects (e.g., fluorescence) by considering the difference between the total inelastic scattering measured slightly above and slightly below the critical temperature.
Therefore RIXS has the potential to serve as a tool to probe the symmetry of the superconducting order parameter in iron-based superconductors, as soon as the energy resolution will reach the energy scale of the superconducting gap.

}

\chapter{Conclusions}
\label{ch:Conclusions}

The interplay between superconductivity and orbital degree of freedom seems to be of fundamental importance to understand electronic correlations, the superconducting pairing mechanism, and in general the low energy properties of unconventional superconductors. 
In this work we presented some theoretical predictions on RIXS spectra of magnetic and quasiparticle excitations, which may help to probe the spin and orbital correlations and the superconducting order parameter in cuprate and pnictide superconductors. 

In \cref{ch:Orbitals}, we proposed a new method to reveal orbital ordered states in transition metal compounds via the analysis of the RIXS spectra of magnetic excitations. 
In fact, the momentum and polarization dependence of the RIXS cross section of magnetic excitations is strongly influenced by the presence, or the lack, of a long-range orbital order, and therefore allow one to distinguish between different orbital ground states in the material under study. 
Moreover, in \cref{ch:BCS-Cuprates} we have shown how RIXS spectra of quasiparticle excitations in superconductors not only can measure the superconducting gap magnitude and reveal the presence of nodal points but, more importantly, is also sensitive to phase differences of the order parameter on the Fermi surface. 
This can allow one to get an insight on the pairing mechanism in unconventional superconductors, discriminating between different superconducting parameter symmetries, such as $s$-wave or $d$-wave (singlet pairing) and $p$-wave (triplet pairing). 
Finally in \cref{ch:BCS-Pnictides} we have studied the interplay between the orbital content on the Fermi surface and the order parameter in RIXS spectra of pnictide superconductors, and proposed a way to distinguish between singlet and triplet pairing. 
In fact, such a direct experimental probe of the order parameter by RIXS spectroscopy may be helpful to shed a light on the pairing mechanism in some pnictide compounds, in particular in LiFeAs, in which the nature of the superconducting state is still debated.


\begin{appendices}

\chapter{Orbitals}\label{sec:Orbitals}
\section*{Broken rotational symmetry in crystals}

A free ion has full rotational symmetry. 
That is, all the rotations about any axis, which form the group of rotations $SO(3)$ in three dimensions, are symmetry operations of the system, i.e., they commute with the Hamiltonian. 
Since the total angular momentum operator is the generator of the rotation group $SO(3)$, in the case where the spin-orbit coupling is negligible, the full rotational symmetry implies that all the $2l+1$ electron states with azimuthal quantum number $l$ have the same energy, i.e., they are degenerate.\footnote{
This is generally the case of transition metal ions. 
In the case of heavy ions instead, e.g, rare earth ions, the spin-orbit coupling in the valence shell is not negligible, and therefore the rotational symmetry implies that all the electron states with the same total angular momentum quantum number $j$ are degenerate. 
}.
In this case, the eigenstates of the orbital angular momentum $\ket{nlm}$ are the natural choice to describe the basis of single electron states. 
In a crystal environment instead, the periodic arrangement of ions in the crystal lattice does not have the full rotational symmetry. 
In a crystal, this full symmetry is reduced to the so called the point group $\cal{G}$ which includes, in general, symmetry operations like rotations about finite angles and around a specific axes, inversions and reflections, and which define the symmetry of the crystal. 
The full rotation group contains all these symmetry operations, i.e., the point group of a crystal $\cal{G}$ is a subset of the full rotational group of a free ion. 
As a consequence of the lower symmetry of the crystal, the $(2l+1)$-fold degeneracy of electron states may be totally or partially lifted. 
In general, the influence of the crystal environment on the single ion is described as a perturbation to the free ion Hamiltonian $\hat{\cal H}_0$ in terms of a crystal field potential\cite{Yosida1996} $V_{cf}$, or any other more complex term $\hat{U}_{c}$ which describes the interaction between the ion and the sourronding crystal environment. 
In its simplest form, the crystal field potential $V_{cf}$ is given by the electrostatic interaction with the charge distribution surrounding the lattice ion. 
Whereas the unperturbed Hamiltonian commute with any symmetry operation of the full space rotational group, these additional terms in the Hamiltonian commute only with the symmetry operation of the crystal point group, and therefore lowers the full rotational symmetry of the system. 
The splitting of the energy levels and the lifting of the degeneracy is therefore a direct consequence of the interaction with the crystal environment and of the broken rotational symmetry of the crystal. 
The actual energies of the different levels depend on the strength of the crystal field and on the details of the crystal field potential $V_{cf}$ or, more generally, of the interaction term $\hat{U}_c$.
However, it can be shown that the energy level degeneracy depends only on the crystal symmetry\cite{Dresselhaus2010}. 

\section*{Tesseral harmonics}

\begin{figure}[t!]
	\centering
	\includegraphics[width=1\textwidth]{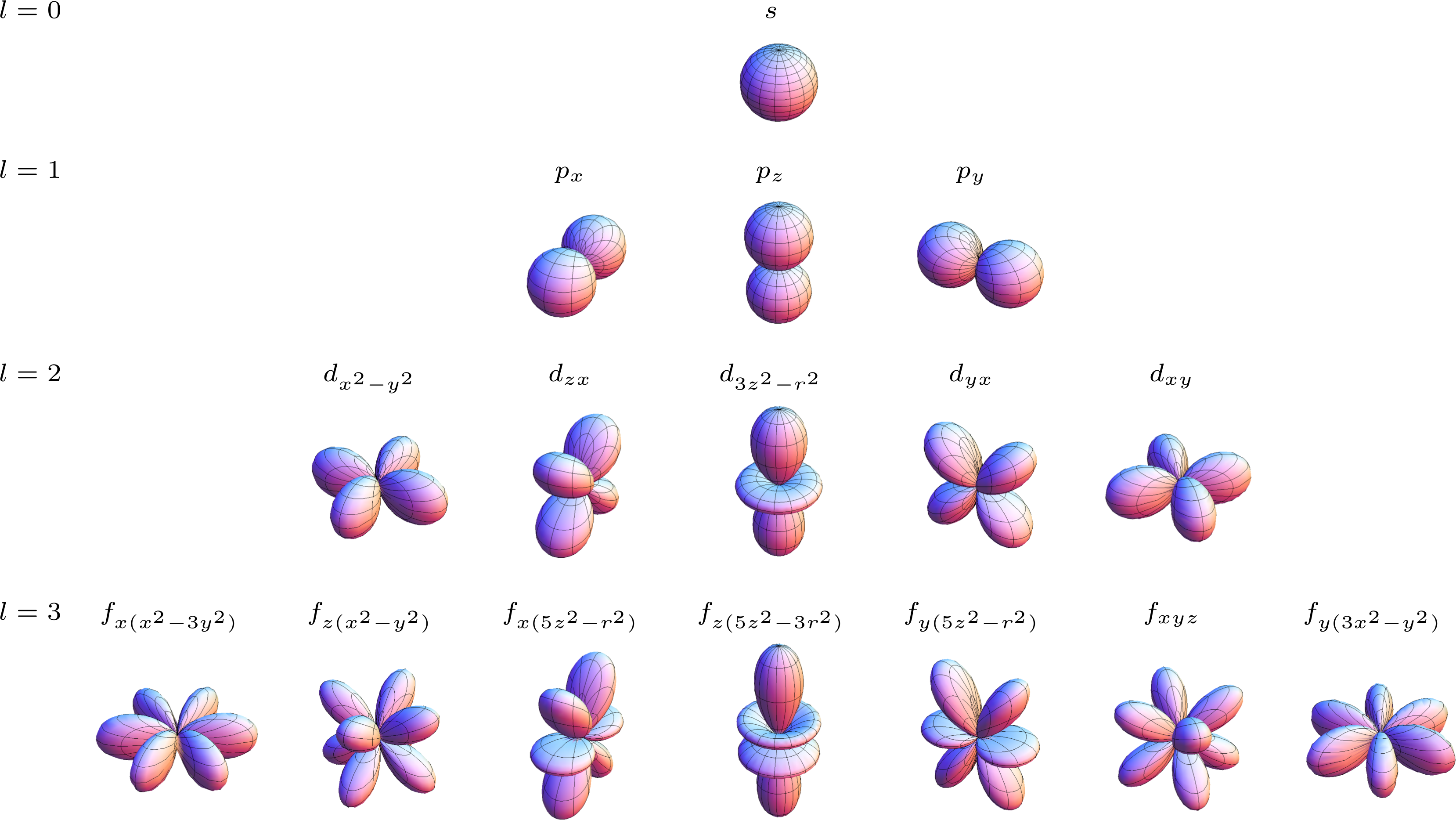}
\caption[Orbitals]{
Angular dependence of the $s$, $p$, $d$, and $f$ orbitals with $l=0,1,2,3$. 
The subscript describes the angular dependence of the orbital (and of the corresponding tesseral harmonic) in Cartesian coordinates (cf.~\cref{tab:TesseralBasis}).
}
\label{fig:Orbitals}
\end{figure}

In the free ion case, the electron states corresponding to the azimuthal quantum number $l$ are in general described in terms of spherical harmonics, i.e., as a linear combination of states $\ket{R_{nl}Y_l^m}$, with the radial and angular parts of the wavefunction described respectively by a radial function $R_{nl}(r)$ and a spherical harmonic $Y_l^m(\theta,\phi)$. 
However, in the case of an ion in a crystal with orthorhombic, tetragonal, or cubic symmetry, the point group contains rotations about finite angles around three orthogonal axis, and the energy levels of the crystal field Hamiltonian correspond in general to electron states which are better described in terms of tesseral harmonics $Y_{lm}$, which can be defined as
\begin{equation}\label{eq:TesseralHarmonics}
	Y_{lm}(\theta,\phi)\equiv
	\begin{cases}
	{\frac1{\sqrt2}}\: 	\left[ (-1)^m Y_l^m(\theta,\phi)  + Y_l^{-m}(\theta,\phi) \right] & \text{if } m>0 \\
	Y_l^0(\theta,\phi)  & \text{if } m=0 \\
	{\frac1{\imath\sqrt2}}	\left[ (-1)^m Y_l^{-m}(\theta,\phi) - Y_l^m(\theta,\phi)  \right] & \text{if } m<0 \\
	\end{cases}
	.
\end{equation}
Tesseral harmonics represent a natural choice to describe electrons in a non-spherical potential. 
The angular part of the wavefunction of an electron in a solid is usually described conveniently in terms of tesseral harmonics.
In particular, if the crystal has an orthorhombic, tetragonal, or cubic symmetry, the Hamiltonian of the single ion is diagonalized by the tesseral harmonic basis. 
These different single-electron states, which correspond to the different tesseral harmonic in each of the electronic shell of the atom, are called orbitals. 
Orbitals are commonly referred as $s$, $p$, $d$, and $f$ orbitals for $l=0,1,2,3$, and are usually labelled by a subscript which describes the functional dependence of the corresponding tesseral harmonic in Cartesian coordinates. 
A list of tesseral harmonics in terms of the usual spherical harmonics for $l\le3$ and their explicit angular dependence in Cartesian coordinates is given in \cref{tab:TesseralBasis}. 
Their angular dependence is shown in \cref{fig:Orbitals}. 

\begin{table}[t]
	\centering
\renewcommand{\arraycolsep}{.4mm}
\newcommand{\pmath}{\phantom{\imath}}
$\begin{array}{l|rccclclc}
l=0 \qquad\qquad&\qquad\qquad\quad
s &=& Y_{00} &=& \quad Y_0^0 &=& \frac{1}{2} \sqrt{\frac{1}{\pi}}\\[4mm]\hline\\[-.5mm]
&
p_x &=& Y_{11} &=& \pmath \sqrt{\frac{1}{2}} \left( Y_1^{- 1} - Y_1^1 \right) &=& \sqrt{\frac{3}{4 \pi}} \cdot \frac{x}{r}\\
l=1&
p_z &=& Y_{10} &=& \quad Y_1^0 &=& \sqrt{\frac{3}{4 \pi}} \cdot \frac{z}{r} \\
&
p_y &=& Y_{1,-1} &=& \imath \sqrt{\frac{1}{2}} \left( Y_1^{- 1} + Y_1^1 \right) &=& \sqrt{\frac{3}{4 \pi}} \cdot \frac{y}{r}  \\[4mm]\hline\\[-.5mm]
&
d_{x^2-y^2} &=& Y_{22} &=& \pmath\sqrt{\frac{1}{2}} \left( Y_2^{- 2} + Y_2^2 \right) &=& \frac{1}{4} \sqrt{\frac{15}{\pi}} \cdot \frac{x^2 - y^2 }{r^2}\\
&
d_{xz} &=& Y_{21} &=& \pmath\sqrt{\frac{1}{2}} \left( Y_2^{- 1} - Y_2^1 \right) &=& \frac{1}{2} \sqrt{\frac{15}{\pi}} \cdot \frac{z x}{r^2} \\
l=2&
d_{z^2} &=& Y_{20} &=& \quad Y_2^0 &=& \frac{1}{4} \sqrt{\frac{5}{\pi}} \cdot \frac{- x^2 - y^2 + 2 z^2}{r^2}  \\
&
d_{yz} &=& Y_{2,-1} &=& \imath \sqrt{\frac{1}{2}} \left( Y_2^{- 1} + Y_2^1 \right) &=& \frac{1}{2} \sqrt{\frac{15}{\pi}} \cdot \frac{y z}{r^2} \\
&
d_{xy} &=& Y_{2,-2} &=& \imath \sqrt{\frac{1}{2}} \left( Y_2^{- 2} - Y_2^2\right) &=& \frac{1}{2} \sqrt{\frac{15}{\pi}} \cdot \frac{x y}{r^2}  \\[4mm]\hline\\[-.5mm]
&
f_{x(x^2-3y^2)} &=& Y_{33} &=& \pmath\sqrt{\frac{1}{2}} \left( Y_3^{- 3} - Y_3^3 \right) &=& \frac{1}{4} \sqrt{\frac{35}{2 \pi}} \cdot \frac{\left( x^2 - 3 y^2 \right) x}{r^3}\\
&
f_{z(x^2-y^2)} &=& Y_{32} &=& \pmath\sqrt{\frac{1}{2}} \left( Y_3^{- 2} + Y_3^2 \right) &=& \frac{1}{4} \sqrt{\frac{105}{\pi}} \cdot \frac{\left( x^2 - y^2 \right) z}{r^3} \\
&
f_{x(5z^2-r^2)} &=& Y_{31} &=& \pmath\sqrt{\frac{1}{2}} \left( Y_3^{- 1} - Y_3^1 \right) &=& \frac{1}{4} \sqrt{\frac{21}{2 \pi}} \cdot \frac{x (4 z^2 - x^2 - y^2)}{r^3} \\
l=3&
f_{z(5z^2-3r^2)} &=& Y_{30} &=& \quad Y_3^0 &=& \frac{1}{4} \sqrt{\frac{7}{\pi}} \cdot \frac{z (2 z^2 - 3 x^2 - 3 y^2)}{r^3} \\
&
f_{y(5z^2-r^2)} &=& Y_{3,-1} &=& \imath \sqrt{\frac{1}{2}} \left( Y_3^{- 1} + Y_3^1 \right) &=& \frac{1}{4} \sqrt{\frac{21}{2 \pi}} \cdot \frac{y (4 z^2 - x^2 - y^2)}{r^3} \\
&
f_{xyz} &=& Y_{3,-2} &=& \imath \sqrt{\frac{1}{2}} \left( Y_3^{- 2} - Y_3^2 \right) &=& \frac{1}{2} \sqrt{\frac{105}{\pi}} \cdot \frac{xy z}{r^3} \\
&
f_{y(3x^2-y^2)} &=& Y_{3,-3} &=& \imath \sqrt{\frac{1}{2}} \left( Y_3^{- 3} + Y_3^3 \right) &=& \frac{1}{4} \sqrt{\frac{35}{2 \pi}} \cdot \frac{\left( 3 x^2 - y^2 \right) y}{r^3} \\
\end{array}$
\\[1cm]
	\caption[Tesseral harmonics basis in terms of orbital angular momentum eigenstates]{
Tesseral harmonics basis in terms of orbital angular momentum eigenstates and their explicit angular dependence in Cartesian coordinates, where 
$x  = r \sin\theta\cos\varphi$, $y = r \sin\theta\sin\varphi$, and $z = r \cos\theta$.
}
	\label{tab:TesseralBasis}
\end{table}

\chapter{Wigner-Eckart theorem and dipole operator}\label{sec:WignerEckartAppendix}
\section*{Wigner-Eckart theorem}\label{sec:WignerEckartTheorem}

A spherical vector operator\cite{Sakurai2006} $\hat{\bf V}$ is defined as a vector operator whose components transforms under rotations in the same way as the components of the orbital angular momentum $\hat{\bf L}$. 
For a spherical vector operator, the following canonical commutation relations must hold
\begin{equation}
	\label{eq:Spherical}
	[\hat{L}_\lambda,\hat{V}_\mu]=\imath\hbar\epsilon_{\lambda\mu\nu}\hat{V}_\nu
	,
\end{equation}
where $\epsilon_{\lambda\mu\nu}$ is the Levi-Civita symbol. 
It is useful to introduce the spherical basis representation, defining the spherical components of the vector as
\begin{equation}\label{eq:SphericalBasis}
	\hat{V}_0 =\hat{V}_z, \qquad
	\hat{V}_{\pm1} =\mp\frac1{\sqrt{2}}(\hat{V}_x\pm\imath \hat{V}_y)
	,
\end{equation}
where the components $\hat{V}_{\pm1}$ are proportional to the ladder operators in the case of angular momenta or spin. 
If one assumes that the vector operator is hermitian, one has $\hat{V}_0^\dag=\hat{V}_0$ and $\hat{V}_{\pm1}^\dag=-\hat{V}_{\mp1}$. 

The Wigner-Eckart theorem\cite{Eckart1930,Wigner1959} states that matrix elements of a spherical vector operator $\hat{\bf V}$ on the basis of the orbital angular momentum eigenstates can be expressed as a combination of Clebsch-Gordan coefficients and a set of \emph{reduced matrix elements}, in the form
\begin{align}\label{eq:WignerEckart1st}
	\bra{n'l'm'}\hat{V}_q\ket{nlm}&=
	\frac{\ClebschGordan{l}{m}{1}{q}{l'}{m'}}{\sqrt{2l'+1}}
	{\rbra{n'l'}V\rket{nl}}
	\qquad\text{with } q=0,\pm1
	,
\end{align}
where $\ClebschGordan{l}{m}{1}{q}{l'}{m'}$ are the Clebsch-Gordan coefficients, $\rbra{n'l'}V\rket{nl}$ are the reduced matrix elements of the spherical operator, which do not depend on the magnetic quantum number $m$, and $\ket{nlm}$ are the eigenstates of the angular momentum with $l$ and $m$ the azimuthal and magnetic quantum numbers, where $n$ represents all the other quantum numbers which define the state of the system. 
In the second quantization language, the Wigner-Eckart representation of a spherical vector operator reads
\begin{equation}\label{eq:WignerEckart2nd}
	\hat{V}_q=\sum\limits_{nlmn'l'm'}
	\frac{\ClebschGordan{l}{m}{1}{q}{l'}{m'}}{\sqrt{2l'+1}}
	{\rbra{n'l'}\hat{V}\rket{nl}} c^\dag_{n'l'm'} c^{\nodag}_{nlm} 
	\qquad\text{with } q=0,\pm1
	,
\end{equation}
where $c^\dag_{nlm}$ and $c^{\nodag}_{nlm}$ creates and annihilates respectively an electron state with quantum numbers $n$, $l$, and $m$. 

The Clebsch-Gordan coefficients $\ClebschGordan{l}{m}{1}{q}{l'}{m'}$ can be written in terms of Wigner $3j$ symbols as
\begin{equation}\label{eq:ClebschGordan}
	\ClebschGordan{l}{m}{1}{q}{l'}{m'}= (-1)^{l+m'}\sqrt{2l'+1}\ThreeJSymbol{l'}{-m'}{1}{q}{l}{m}
	.
\end{equation}
The general definition, symmetry properties, and selection rules of the Wigner $3j$ symbols can be found in \onlinecite{Messiah1981}. 
Since nonzero Clebsch-Gordan coefficients and Wigner $3j$ symbols in \cref{eq:ClebschGordan} satisfy the triangular inequality $|l'-l|\le1\le l'+l$ and the condition $m+q-m'=0$, the matrix elements of the spherical vector operator $\hat{\bf V}$ vanish for $l'=l=0$ while are nonzero if the following selection rules are satisfied
\begin{equation}\label{eq:SelectionRulesSphericalVectors}
	\bra{n'l'm'}\hat{V}_q\ket{nlm}\ne0	\quad\Rightarrow\quad	\Delta l=0,\pm1, \text{ and } \Delta m=q
	,
\end{equation}
where $\Delta l=l'-l$ and $\Delta m=m'-m$. 
Therefore, the nonzero matrix elements of a spherical vector operator connect eigenstates with azimuthal quantum numbers $l$ which differs at most by 1. 
In particular, the component $\hat{V}_z=\hat{V}_0$ allows nonzero matrix elements between eigenstates with the same magnetic quantum number $m=m'$, while the spherical components $\hat{V}_{\pm1}$ allows nonzero matrix elements for $m'=m\pm1$. 
Expanding the Clebsch-Gordan coefficients in \cref{eq:WignerEckart2nd} via \cref{eq:ClebschGordan}, and writing down explicitly the relevant Wigner $3j$ symbols, the components of the spherical vector $\hat{\bf V}$ become\cite{Landau1981}
\begin{align}
	\label{eq:WignerEckart}
	\hat{V}_0=\sum\limits_{nn'lm}&
	\tfrac{m}{\sqrt{l(l+1)(2l+1)}}
	\rbra{n'l}V\rket{nl}    c^\dag_{n'lm}   c^\nod_{nlm}
	+\nonumber\\
	+&\sqrt{\tfrac{l^2-m^2}{l(2l-1)(2l+1)}}
	\rbra{n'l-1}V\rket{nl}  c^\dag_{n'l-1m} c^\nod_{nlm} 
	+\nonumber\\
	-&\sqrt{\tfrac{l^2-m^2}{l(2l-1)(2l+1)}}
	\rbra{nl}V\rket{n'l-1}  c^\dag_{nlm}   c^\nod_{n'l-1m}
	,
	\nonumber\\[1em]
	\hat{V}_{-1}=-\hat{V}_{+1}^\dag=\sum\limits_{nn'lm}&
	\sqrt{\tfrac{(l+m)(l-m+1)}{2l(l+1)(2l+1)}}	\rbra{n'l}V\rket{nl}    c^\dag_{n'lm-1}		c^\nod_{nlm}
	+\nonumber\\
	-&\sqrt{\tfrac{(l+m)(l+m-1)}{2l(2l-1)(2l+1)}}	\rbra{n'l-1}V\rket{nl}  c^\dag_{n'l-1m-1}	c^\nod_{nlm}
	+\nonumber\\
	-&\sqrt{\tfrac{(l-m)(l-m+1)}{2l(2l-1)(2l+1)}}	\rbra{nl}V\rket{n'l-1}	c^\dag_{nlm-1}		c^\nod_{n'l-1m}
	.
\end{align}
Note that the reduced matrix element are not independent, since for the hermiticity of the component $\hat{V}_0$ one has that 
\begin{align}
	\rbra{nl}V\rket{n'l}&=\rbra{n'l}V\rket{nl}^*
	,
	\nonumber\\
	\rbra{nl}V\rket{n'l-1}&=-\rbra{n'l-1}V\rket{nl}^*
	.
\end{align}

\section*{Angular momentum}

The Wigner-Eckart theorem prescribes the general structure of a spherical vector operator, but does not define the actual representation of the vector in the momentum operator basis. 
In particular, it says nothing about the reduced matrix elements, which are in fact implicitly defined and have to be calculated specifically for any given operator. 
This is usually done by a comparison of the Wigner-Eckart representation with a direct calculation of the matrix element of the vector operator. 

As an example, one can consider the orbital angular momentum $\hat{\bf L}$, which is a spherical vector by definition. 
Comparing \cref{eq:WignerEckart} with the matrix element of the orbital angular momentum on its eigenvectors basis, it follows that its reduced matrix elements are 
\begin{equation}
\rbra{n'l'}L\rket{nl}=\delta_{nn'}\delta_{ll'}\hbar\sqrt{l(l+1)(2l+1)}
\end{equation}
which leads to the representation of the orbital angular momentum in spherical components as
\begin{align}\label{eq:Ls}
	\hat{L}_0&=\sum\limits_{nn'lm} 
	\hbar m\	c^\dag_{n'lm}	c^{\nodag}_{nlm}
	\nonumber\\
	\hat{L}_{-1}=-\hat{L}_{+1}^\dag&=\frac1{\sqrt2}\sum\limits_{nn'lm}
	\hbar\sqrt{(l+m)(l-m+1)}\	c^\dag_{n'lm-1}   c^{\nodag}_{nlm}
	,
\end{align}
or, alternatively, in Cartesian coordinates as
\begin{align}\label{eq:Lc}
	\hat{L}_{x}&=\frac12\sum\limits_{nn'lm}
	\hbar\sqrt{(l+m)(l-m+1)}\	
	\left(
	c^\dag_{n'lm-1}	c^{\nodag}_{nlm} +
	c^{\dag}_{nlm}	c^\nodag_{n'lm-1}
	\right)
	\nonumber\\
	\hat{L}_{y}&=\frac\imath2\sum\limits_{nn'lm}
	\hbar\sqrt{(l+m)(l-m+1)}\	
	\left(
	c^\dag_{n'lm-1}	c^{\nodag}_{nlm}
	-c^{\dag}_{nlm}	c^\nodag_{n'lm-1}
	\right)
	\nonumber\\
	\hat{L}_z&=\sum\limits_{nn'lm} 
	\hbar m\	c^\dag_{n'lm}	c^{\nodag}_{nlm}
	.
\end{align}
Note that the orbital angular momentum, being diagonal respect to the azimuthal quantum number $l$, does not allow transitions with $\Delta l=\pm1$. 

\section*{Position and momentum operators}

The position $\hat{\bf r}$ and linear momentum $\hat{\bf p}$ operators are spherical also spherical operators, since they both satisfy \cref{eq:Spherical}. 
In order to calculate the reduced matrix elements of the position and momentum operators, one can directly calculate the matrix elements of one component in spherical coordinates using the spherical harmonics representation of the orbital angular momentum eigenstates $\ket{nlm}=\ket{R_{nl}Y_l^m}$. 
In this way, the calculations reduce to the evaluation of spherical harmonics integrals in the form
\begin{align}\label{eq:SphericalHarmonicsIntegrals}
	&\sbra{Y_{l_1}^{m_1}}Y_{l_2}^{m_2}\sket{Y_{l_3}^{m_3}}=
	\int Y_{l_1}^{m_1*}(\theta,\varphi)Y_{l_2}^{m_2}(\theta,\varphi)Y_{l_3}^{m_3}(\theta,\varphi)\,\sin\theta\,\mathrm{d}\theta\,\mathrm{d}\varphi 
	\nonumber\\&=
	(-1)^{m_1}\sqrt{\tfrac{(2l_1+1)(2l_2+1)(2l_3+1)}{4\pi}}
	\ThreeJSymbol{l_1}{0}{l_2}{0}{l_3}{0}
	\ThreeJSymbol{l_1}{-m_1}{l_2}{m_2}{l_3}{m_3}
	.
\end{align}
Since the reduced matrix elements do not depend on the magnetic quantum number, one can restrict the calculation, for the sake of simplicity, to matrix elements between eigenstates with $m=m'$. 

The matrix elements of the component $\hat{r}_0=\hat{r}_z=r \cos{\theta}$ of the position operator between orbital angular momentum eigenstates can be calculated using the fact that $Y_1^0=\frac12\sqrt{3/\pi}\cos{\theta}$, which leads to
\begin{align*}
	\bra{n'l'm} \hat{r}_0 \ket{nlm}&= 
	\sbra{R_{n'l'}Y_{l'}^m} r \cos\theta \sket{R_{nl}Y_l^m}
	\\&=
	\sqrt{\tfrac{4\pi}{3}}
	\bra{R_{n'l'}} r \ket{R_{nl}}
	\sbra{Y_{l'}^m} Y_1^0 \ket{Y_l^m}
	,
\end{align*}
where the spherical harmonics integral can be calculated via \cref{eq:SphericalHarmonicsIntegrals}. 
The three spherical harmonics integral vanishes for $l'=l$, while it is nonzero for $\Delta l=\pm1$ and therefore, by a comparison with \cref{eq:WignerEckart}, one has
\begin{align}\label{RMEPosition}
	\rbra{n'l}r\rket{nl}&=0
	,\nonumber\\
	\rbra{n'l-1}r\rket{nl}&=\sqrt{l}\bra{R_{n'l-1}} r \ket{R_{nl}}
	.
\end{align}

To evaluate the reduced matrix elements of the momentum operator, one can calculate the matrix elements of the component $\hat{p}_0=\hat{p}_z$, which in spherical coordinates reads
\begin{equation}
	\hat{p}_0=\hat{p}_z=-\imath \hbar \frac{\partial}{\partial z}=
	-\imath \hbar \left(
	\cos{\theta}\frac{\partial}{\partial r}+\sin{\theta}\frac{1}{r}\frac{\partial}{\partial\theta}
	\right)
	.
\end{equation}
Using the differentiation rules for spherical harmonics, and the fact that $Y_1^0=\frac12\sqrt{3/\pi}\cos{\theta}$ and $Y_1^{-1}=\frac12\sqrt{3/2\pi}\sin\theta e^{-\imath\phi}$, one has 
\begin{align*}
	&\bra{n'l'm} \hat{p}_0 \ket{nlm}=
	-\imath \hbar 
	\sbra{R_{n'l'}Y_{l'}^m} 
	\left(\cos{\theta}\frac{\partial}{\partial r}+\sin{\theta}\frac{1}{r}\frac{\partial}{\partial\theta}\right)
	 \sket{R_{nl}Y_l^m}
	\\=&
	-\imath \hbar 
	\sbra{R_{n'l'}}\frac{\partial}{\partial r}\sket{R_{nl}}
	\sbra{Y_{l'}^m}	\cos{\theta}\sket{Y_l^m}
	-\imath \hbar 
	\sbra{R_{n'l'}}\frac{1}{r}\sket{R_{nl}}
	\\
	&\times
	\left(
	m\sbra{Y_{l'}^m}\cos{\theta}\sket{Y_l^m}+
	\sqrt{(l-m)(l+m+1)}\sbra{Y_{l'}^m}\sin{\theta}e^{-\imath\phi}\sket{Y_l^{m+1}}
	\right)
	\\=&
	-\imath \hbar \sqrt{\tfrac{4\pi}{3}}
	\left(
	\sbra{R_{n'l'}}\frac{\partial}{\partial r}\sket{R_{nl}}
	+m\sbra{R_{n'l'}}\frac{1}{r}\sket{R_{nl}}
	\right)
	\sbra{Y_{l'}^m}	Y_1^0	\sket{Y_l^m}
	\\&
	-\imath \hbar \sqrt{\tfrac{4\pi}{3}}
	\sbra{R_{n'l'}}\frac{1}{r}\sket{R_{nl}}
	\sqrt{2(l-m)(l+m+1)}
	\sbra{Y_{l'}^m}Y_1^{-1}\sket{Y_l^{m+1}}
	,
\end{align*}
where the integrals in the last line can be again calculated using \cref{eq:SphericalHarmonicsIntegrals}. 
The three spherical harmonics integrals vanish for $l=l'$ while they are nonzero for $\Delta l=\pm1$, and therefore by a comparison with \cref{eq:WignerEckart} one obtains the reduced matrix elements of the momentum operator, which read
\begin{align}\label{RMEMomentum}
	\rbra{n'l}p\rket{nl}&=0
	,\nonumber\\
	\rbra{n'l-1}p\rket{nl}&=
	-\imath \hbar\sqrt{l}
	\bra{R_{n'l-1}}\frac{\partial}{\partial r}-(l+1)\frac{1}{r}\ket{R_{nl}}
	.
\end{align}

Eventually, since the reduced matrix elements in \cref{RMEPosition,RMEMomentum} corresponding to transitions $l'=l$ are zero in both cases, the spherical components of the position and of the momentum operators are given by
\begin{align}
	\label{eq:DipoleOperatorSpherical}
	\hat{D}_0=\sum\limits_{nn'lm}&
	\sqrt{\tfrac{l^2-m^2}{(2l-1)(2l+1)}}		\delta_{nn'll-1}	c^\dag_{n'l-1m}	c^{\nodag}_{nlm}
	+\text{h.c.}
	,\nonumber\\
	\hat{D}_{-1}=-\hat{D}_{+1}^\dag=\frac1{\sqrt2}\sum\limits_{nn'lm}
	&
	\sqrt{\tfrac{(l-m)(l-m+1)}{(2l+1)(2l-1)}}	\delta_{nn'll-1}^*	c^\dag_{nlm-1}	c^{\nodag}_{n'l-1m}
	+
	\nonumber\\
	-&
	\sqrt{\tfrac{(l+m)(l+m-1)}{(2l+1)(2l-1)}}	\delta_{nn'll-1}	c^\dag_{n'l-1m-1}   c^{\nodag}_{nlm}
	,
\end{align}
where $\hat{\bf D}$ is either the position $\hat{\bf r}$ or the momentum $\hat{\bf p}$ operator, and where $\delta_{nn'll-1}$ represents respectively the radial integrals $\rho_{nn'll-1}$ and $\pi_{nn'll-1}$ defined by
\begin{align}\label{eq:RadialIntegrals}
	\rho_{nn'll-1}&=\bra{R_{n'l-1}} r \ket{R_{nl}}=\int_0^\infty R_{n'l-1}^*(r) r R_{nl}(r) r^2 {\rm d} r
	,\nonumber\\
	\pi_{nn'll-1}&=-\imath\hbar\bra{R_{n'l-1}} \frac{\partial}{\partial r}-(l+1)\frac{1}{r} \ket{R_{nl}}
	\nonumber\\&=
	-\imath\hbar\int_0^\infty R_{n'l-1}^*(r) \left(\frac{\partial}{\partial r}-(l+1)\frac{1}{r}\right) R_{nl}(r) r^2 {\rm d} r
	,
\end{align}
which inherently depends on the eigenstates of the system. 
Since no hypothesis have been done on the system eigenstates, the radial functions $R_{nl}(r)$ do not necessary correspond to the eigenstates of the hydrogen-like ion. 
Using the definition of spherical vector in \cref{eq:SphericalBasis}, one can obtain the Cartesian components of the position and momentum operators, which read
\begin{align}\label{eq:DipoleOperatorCartesian}
		\hat{D}_x=\frac12\sum\limits_{nn'lm}&
		\sqrt{\tfrac{(l-m)(l-m+1)}{(2l+1)(2l-1)}}	\delta_{nn'll-1}
		c^{\dag}_{n'l-1m} c^\nodag_{nlm-1} +
		\nonumber\\
		-&
		\sqrt{\tfrac{(l+m)(l+m-1)}{(2l+1)(2l-1)}}	\delta_{nn'll-1}
		c^\dag_{n'l-1m-1}   c^{\nodag}_{nlm} +
		\text{h.c.}
		,\nonumber\\
		\hat{D}_y=-\frac\imath2\sum\limits_{nn'lm}&
		\sqrt{\tfrac{(l-m)(l-m+1)}{(2l+1)(2l-1)}}	\delta_{nn'll-1}
		c^{\dag}_{n'l-1m} c^\nodag_{nlm-1} +
		\nonumber\\
		+&
		\sqrt{\tfrac{(l+m)(l+m-1)}{(2l+1)(2l-1)}}	\delta_{nn'll-1}
		c^\dag_{n'l-1m-1}   c^{\nodag}_{nlm} +
		\text{h.c.}
		,\nonumber\\
		\hat{D}_z=\sum\limits_{nn'lm}&
		\sqrt{\tfrac{l^2-m^2}{(2l-1)(2l+1)}}		\delta_{nn'll-1}
		c^\dag_{n'l-1m} c^{\nodag}_{nlm} +
		\text{h.c.}
		.
\end{align}
The position and the momentum operators allow nonzero matrix elements with $\Delta l=\pm1$, while transitions with $\Delta l=0$ are forbidden. 
Moreover, the components $\hat{r}_z$ and $\hat{p}_z$ preserve the magnetic quantum number $m$, while the components $\hat{r}_x$, $\hat{r}_y$ and $\hat{p}_x$, $\hat{p}_y$ allow transitions to states which are superpositions of eigenstates with $m\pm1$. 
The representation of the position $\hat{\bf r}$ and of the momentum operator $\hat{\bf p}$ in the basis of orbital angular momentum eigenvalues are similar, and that they indeed differ only with respect to the radial integrals $\rho_{nn'll-1}$ and $\pi_{nn'll-1}$. 
Nevertheless, since they satisfy the canonical commutation relation $[\hat{r}_\lambda,\hat{p}_\lambda]=\imath\hbar$, position and momentum operators are not proportional to each another.

\section*{Tesseral harmonics representation}
\label{sec:TesseralHarmonics}

The orbital angular momentum operator allows nonzero matrix elements only if $\Delta l=0$, i.e., it is diagonal respect to the azimuthal quantum number. It is natural therefore to decompose the orbital angular momentum $\hat{\bf L}$ into a sum of operators which has nonzero matrix elements only between eigenstates with fixed quantum number $l$.
On the other hand, the position and momentum operators allow nonzero matrix elements only if $\Delta l=\pm1$. 
Hence, these operators can be decomposed in the form
\begin{equation}
	\hat{\bf L}=\sum\limits_{nn'l} \hat{\bf L}^{nn'll}
	,
	\qquad
	\hat{\bf D}=\sum\limits_{nn'l} \hat{\bf D}^{nn'll-1}
	,
\end{equation}
where the operator $\hat{\bf D}$ is either the position $\hat{\bf r}$ or the momentum $\hat{\bf p}$ operator, and $\hat{\bf L}^{nn'll}$ acts only between eigenstates with quantum numbers $n,l$ and $n',l$, while $\hat{\bf D}^{nn'll-1}$ acts only between eigenstates with quantum numbers $n,l$ and $n',l-1$.

In solids, the full rotational symmetry $SO(3)$ of the single ion is broken by the presence of the surrounding crystal lattice, which corresponds to a lifting of the atomic states degeneracy. 
In this case, the single electron states, i.e., the orbitals, can be described in terms of tesseral harmonics $Y_{lm}$, as shown in \cref{sec:Orbitals}. 
It is therefore reasonable to expand the operators $\hat{\bf L}^{nn'll}$ and $\hat{\bf D}^{nn'll-1}$ in terms of the tesseral harmonics basis. 
In \cref{tab:MomentumDipole} are listed the explicit form of the operators $\hat{\bf L}^{nn'll}$ and $\hat{\bf D}^{nn'll-1}$ in terms of the tesseral harmonics basis defined in \cref{eq:TesseralHarmonics} and listed in \cref{tab:TesseralBasis}. 

{
\newcommand{\ppx}{
		\hat{L}^{p}_x=
		-\imath\hbar\	p_y^\dag p_z^\nodag
		+\text{h.c.}
		}
\newcommand{\ppy}{
		\hat{L}^{p}_y=
		-\imath\hbar\	p_z^\dag p_x^\nodag
		+\text{h.c.}
		}
\newcommand{\ppz}{
		\hat{L}^{p}_z=
		-\imath\hbar\	p_x^\dag p_y^\nodag
		+\text{h.c.}
		}
\newcommand{\ddx}{
		\hat{L}^{d}_x=
		-\imath\hbar
		\left(
		d_{yz}^\dag d_{x^2-y^2}
		-d^\dag_{xz} d_{xy}
		+\sqrt3 d_{yz}^\dag d_{3z^2-r^2}
		\right)
		+\text{h.c.}
		}
\newcommand{\ddy}{
		\hat{L}^{d}_y=
		-\imath\hbar
		\left(
		d_{xz}^\dag d_{x^2-y^2}
		+d_{yz}^\dag d_{xy}
		-\sqrt3 d_{xz}^\dag d_{3z^2-r^2}
		\right)
		+\text{h.c.}
		}
\newcommand{\ddz}{
		\hat{L}^{d}_z=
		-\imath\hbar
		\left(
		d^\dag_{xz} d_{yz}
		+2d^\dag_{x^2-y^2} d_{xy}
		\right)
		+\text{h.c.}
		}
\newcommand{\spx}{
		\hat{D}^{ps}_x=
		\tfrac{1}{\sqrt{3}}
		\delta_{ps}\		s^\dag p_x
		+
		\text{h.c.}
		}
\newcommand{\spy}{
		\hat{D}^{ps}_y=
		\tfrac{1}{\sqrt{3}}
		\delta_{ps}\		s^\dag p_y
		+
		\text{h.c.}
		}
\newcommand{\spz}{
		\hat{D}^{ps}_z=
		\tfrac{1}{\sqrt{3}}
		\delta_{ps}\		s^\dag p_z
		+
		\text{h.c.}
		}
\newcommand{\pdx}{
		\hat{D}^{dp}_x=
		\tfrac{1}{\sqrt{5}}
		\delta_{dp}
		\left(
		p_y^\dag d_{xy}
		+p_x^\dag d_{x^2-y^2}
		+p_z^\dag d_{xz}
		-\tfrac{1}{\sqrt3} p_x^\dag d_{3z^2-r^2}
		\right)
		+
		\text{h.c.}
		}
\newcommand{\pdy}{
		\hat{D}^{dp}_y=
		\tfrac{1}{\sqrt{5}}
		\delta_{dp}
		\left(
		p_x^\dag d_{xy}
		-p_y^\dag d_{x^2-y^2}
		+p_z^\dag d_{yz}
		-\tfrac{1}{\sqrt3} p_y^\dag d_{3z^2-r^2}
		\right)
		+
		\text{h.c.}
		}
\newcommand{\pdz}{
		\hat{D}^{dp}_z=
		\tfrac{1}{\sqrt{5}}
		\delta_{dp}
		\left(
		p_x^\dag d_{xz}+
		p_y^\dag d_{yz}+
		\tfrac{2}{\sqrt3}p_z^\dag d_{3z^2-r^2}
		\right)
		+
		\text{h.c.}
		}
\newcommand{\dfx}{
		\hat{D}^{fd}_x=
		\tfrac{1}{\sqrt{7}}
		\delta_{fd}
		\left(
		\sqrt{\tfrac{3}{2}}  d_{x^2-y^2}^\dag f_{x(x^2-3y^2)}
		-\tfrac{1}{\sqrt10}  d_{x^2-y^2}^\dag f_{x(5z^2-r^2)}
		+\sqrt{\tfrac{3}{2}} d_{xy}^\dag      f_{y(3x^2-y^2)}
		-\tfrac{1}{\sqrt10}  d_{xy}^\dag      f_{y(5z^2-r^2)}
+\right.\\&\qquad\qquad\qquad\left.
		+                    d_{yz}^\dag      f_{xyz}
		+                    d_{xz}^\dag      f_{z(x^2-y^2)}
		-\sqrt{\tfrac{3}{5}} d_{xz}^\dag      f_{z(5z^2-3r^2)}
		+\sqrt{\tfrac{6}{5}} d_{3z^2-r^2}^\dag     f_{x(5z^2-r^2)}
		\right)
		+
		\text{h.c.}
		}
\newcommand{\dfy}{
		\hat{D}^{fd}_y=
		\tfrac{1}{\sqrt{7}}
		\delta_{fd}
		\left(
		\sqrt{\tfrac{3}{2}}  d_{x^2-y^2}^\dag f_{y(3x^2-y^2)}
		+\tfrac{1}{\sqrt10}  d_{x^2-y^2}^\dag f_{y(5z^2-r^2)}
		-\sqrt{\tfrac{3}{2}} d_{xy}^\dag      f_{x(x^2-3y^2)}
		-\tfrac{1}{\sqrt10}  d_{xy}^\dag      f_{x(5z^2-r^2)}
+\right.\\&\qquad\qquad\qquad\left.
		+                    d_{xz}^\dag      f_{xyz}
		-                    d_{yz}^\dag      f_{z(x^2-y^2)}
		-\sqrt{\tfrac{3}{5}} d_{yz}^\dag      f_{z(5z^2-3r^2)}
		+\sqrt{\tfrac{6}{5}} d_{3z^2-r^2}^\dag     f_{y(5z^2-r^2)}
		\right)
		+
		\text{h.c.}
		}
\newcommand{\dfz}{
		\hat{D}^{fd}_z=
		\tfrac{1}{\sqrt{7}}
		\delta_{fd}
		\left(
		                      d_{x^2-y^2}^\dag f_{z(x^2-y^2)}
		+                     d_{xy}^\dag      f_{xyz}
		+2\sqrt{\tfrac{2}{5}} d_{yz}^\dag      f_{y(5z^2-r^2)}
		+2\sqrt{\tfrac{2}{5}} d_{xz}^\dag      f_{x(5z^2-r^2)}
		+\tfrac{3}{\sqrt5}    d_{3z^2-r^2}^\dag     f_{z(5z^2-3r^2)}
		\right)
		+
		\text{h.c.}
		}
\begin{landscape}
\begin{table}[ht]
	\scriptsize
	\flushleft
	\begin{tabular}{lll}
	{\normalsize Position and momentum}
	&\qquad\qquad\qquad\qquad\qquad\qquad\qquad&
	{\normalsize Orbital angular momentum}	\\
	&\\
	$\begin{aligned}
	&s\leftrightarrow p\\\midrule	&\spx\\	&\spy\\	&\spz\\
	\end{aligned}$
	&&
	$\begin{aligned}
	&p\leftrightarrow p\\\midrule	&\ppx\\	&\ppy\\	&\ppz\\
	\end{aligned}$
	\\&&\\
	$\begin{aligned}
	&p\leftrightarrow d\\\midrule	&\pdx\\	&\pdy\\	&\pdz\\
	\end{aligned}$
	&&
	$\begin{aligned}
	&d\leftrightarrow d\\\midrule	&\ddx\\	&\ddy\\	&\ddz\\
	\end{aligned}$
	\\&&\\
	\multicolumn{3}{l}{
	$\begin{aligned}
	&d\leftrightarrow f\\\midrule	&\dfx\\	&\dfy\\	&\dfz\\
	\end{aligned}$
	}
	\end{tabular}
	\caption[Cartesian components of the position, momentum, and orbital angular momentum operators in the tesseral harmonic basis]{
Cartesian components of the position, momentum, and orbital angular momentum operators in terms of the tesseral harmonic basis with $l=0,1,2,3$ ($s$, $p$, $d$, $f$). 
The operator $\hat{D}^{ll-1}$ is either the position $\hat{\bf r}$ or the momentum operator $\hat{\bf p}$, while the numbers $\delta_{ll-1}$ represent the radial integrals $\rho_{nn'll-1}$ and $\pi_{nn'll-1}$ defined in \cref{eq:RadialIntegrals}. 
	}
	\label{tab:MomentumDipole}
\end{table}
\end{landscape}
}

\end{appendices}

\backmatter
\chapter{Curriculum vit\ae}\markboth{Curriculum vit\ae}{}
\begin{itemize}
\item[]{\Large\textbf{Pasquale Marra}}
\item[]{\href{mailto:pasquale.marra@spin.cnr.it}{pasquale.marra@spin.cnr.it}}
\\[-2mm]
\item[]{\large \bf Education}
\item[]{{\bf Ph.\,D (Dr.\,Rer.\,Nat.)} at the Technische Universit\"at Dresden (Germany).
Advisor (Betreuer): Prof.\@ Jeroen van den Brink; 
Reviewer (Gutachter): Prof.\@ Berndt B\"uchner.
}
\item[]{\textbf{Master's Degree in Physics} at the {University of Salerno},
110/110 cum laude,
with the thesis \emph{``Electronic and magnetic properties of the compound LaFeAsO''}.
Advisors: Prof.\@ Canio Noce and Dr.\@ Adolfo Avella; Reviewer: Prof.\@ Roberta Citro.
}
\item[]{\textbf{Scientific High School Degree} at the ``Liceo Scientifico Giovanni da Procida'', Salerno, 60/60}
\item[]{\textbf{Harmony and Solfege Degree} at the Music School ``Conservatorio Giovanni Martucci'', Salerno}
\\[-2mm]

\item[]{\large \bf Participation at conferences and workshops}
\\[-2mm]
\item[2016]{TO-BE Spring Meeting, Towards Oxide-Based Electronics, University of Warwick (UK), {\bf poster presentation} with the title 
\emph{``Topological non-trivial phases in low-dimensional quantum pumps and nanoscopic Josephson junctions''}
}
\item[]{Current Problems in Theoretical Physics (XXII edition), Vietri~sul~Mare (Italy)}
\item[]{10th Conference of the Physics Department, University of Salerno (Italy), {\bf contribution} to the oral presentation
\emph{``Quantum many-body physics and transport in low-dimensions''}
by Roberta Citro
}
\item[2015]{FISMAT, Italian National Conference on Condensed Matter Physics, Palermo (Italy), {\bf oral presentation} with the title 
\emph{``Fractional quantization of the topological charge pumping in a one-dimensional superlattice''}
and
{\bf poster presentation} with the title 
\emph{``Detection of topological phase transitions in the Josephson current-phase relation''}
}
\item[]{EUCAS, 12th European Conference on Applied Superconductivity, Lyon (France), {\bf contribution} to the oral presentation with the title
\emph{``Stability mechanisms of high current transport in iron-chalcogenides superconducting films''}
by Antonio Leo
}
\item[]{TOP-SPIN Workshop, Spin and Topological phenomena in nanostructures, Salerno (Italy), {\bf oral presentation} with the title 
\emph{``Josephson current and Majorana bound states in 1D nanowires with spin-orbit interaction''}
}
\item[]{CNR Meeting, Naples (Italy), {\bf poster presentation} with the title \emph{``Quantum transport and topological phenomena in hybrid nanostructures''}
}
\item[]{
DPG Spring Meeting, Berlin (Germany),
{\bf oral presentation} with the title 
\emph{``Fractional quantization of the topological charge pumping in a one-dimensional superlattice''}
and
{\bf oral presentation} with the title 
\emph{``Josephson current and Majorana bound states in 1D nanowires with spin-orbit interaction''}
}
\item[2014]{
New frontiers for Majorana fermions from condensed to dark matter,
INFN Frascati National Laboratories, Frascati (Italy)
}
\item[2013]{
Photoemission and Electronic Structure of 4f and 5f Systems, International Focus Workshop, FARPES, MPI-PKS, Dresden (Germany), {\bf supporting staff} and {\bf poster presentation} with the title \emph{``Resonant inelastic x-ray scattering as a probe of the phase and excitations of the order parameter of superconductors''}
}
\item[]{
Trends, challenges and emergent new phenomena in multi-functional materials,  MAMA-Trend, Sorrento (Italy), {\bf oral presentation} with the title  \emph{``Resonant inelastic x-ray scattering as a probe of the phase and excitations of the order parameter of superconductors''}
}
\item[]{
DPG Spring Meeting, Regensburg (Germany),
{\bf oral presentation} with the title
\emph{``Establishing theoretically the capacity of resonant inelastic x-ray scattering to probe the phase and excitations of the superconducting order parameter''}
}
\item[]{
APS March Meeting, Baltimore, Maryland (USA),
{\bf contribution} to the oral presentation with the title
\emph{``Orbital physics in resonant inelastic x-ray scattering (RIXS)''} by Krzysztof Wohlfeld
}
\item[2012]{
DPG Spring Meeting, Berlin (Germany),
{\bf oral presentation} with the title 
\emph{``Resonant inelastic x-ray scattering in unconventional superconductors''}
and
{\bf contribution} to the oral presentation with the title 
\emph{``Orbitons and bi-orbitons in GdVO$_3$ and YVO$_3$''} by Luis Maeder
}
\item[2011]{
International Workshop on Resonant and Elastic X-Ray Scattering, PSI, Villigen (Switzerland),
{\bf poster presentation} with the title 
\emph{``Fingerprints of orbital physics in magnetic resonant inelastic x-ray scattering''}
}
\item[]{
DPG Spring Meeting, Dresden (Germany),
{\bf poster presentation} with the title 
\emph{``Looking for orbiton dispersion features in GdVO$_3$ with the aim of resonant inelastic x-ray scattering''}
}
\item[2010]{
7th International Conference on Inelastic X-ray Scattering,
IXS 2010, Grenoble (France),
{\bf poster presentation} with the title 
\emph{``Looking for orbiton dispersion features in GdVO$_3$ with the aim of resonant inelastic x-ray scattering''}
}
\\[2mm]
\item[]{\large \bf Talks at research institutions}
\\[-2mm]
\item[2015]{
Nordita, Stockholm (Sweden),
{\bf seminar} with the title 
\emph{``Topological non-trivial phases in low dimensions''},
invited by Prof.\@ Alexander Balatsky and Prof.\@ David Abergel
}
\item[]{
IFW Dresden (Germany),
{\bf seminar} with the title 
\emph{``Topological non-trivial phases in one dimension''},
invited by Prof.\@ Steffen Sykora
}
\item[2013]{
ITF Group Meeting, IFW Dresden (Germany),
{\bf seminar} with the title 
\emph{``Resonant inelastic x-ray scattering as a probe of the phase and excitations of the order parameter of superconductors''}
}
\item[2012]{
ITF Group Meeting, IFW Dresden (Germany),
{\bf seminar} with the title 
\emph{``Resonant inelastic x-ray scattering in unconventional superconductors''}
}
\item[2010]{
IFW Dresden (Germany), 
{\bf seminar} with the title
\emph{``Electronic properties of LaOFeAs and LaOFeP superconductors''},
invited by Dr.\@ Maria Daghofer
}
\\[2mm]
\item[]{\large \bf Schools}
\\[-2mm]
\item[2015]{
XIX Training Course in the Physics of Strongly Correlated Systems,
focus on Pnictide Superconductors, IIASS, Vietri~sul~Mare (Italy),
{\bf supporting staff} and {\bf oral presentation} with the title
\emph{``Resonant inelastic x-ray scattering as a probe of the superconducting
order parameter in iron-based superconductors''}. 
Lecturers: B.~B\"uchner, L.~de' Medici, I.~Eremin, J.~Schmalian. 
}
\item[2014]{
XVIII Training Course in the Physics of Strongly Correlated Systems,
focus on Topological Insulators, IIASS, Vietri~sul~Mare (Italy),
{\bf supporting staff} and {\bf oral presentation} with the title 
\emph{``Fractionally quantized charge pumping in a one-dimensional superlattice''}. 
Lecturers: M.~Z.~Hasan, T.~Neupert, N.~Regnault, M.~Wimmer.
}
\item[2012]{
XVII Training Course in the Physics of Strongly Correlated Systems,
IIASS, Vietri~sul~Mare (Italy),
{\bf oral presentation} with the title 
\emph{``Resonant inelastic x-ray scattering in unconventional superconductors''}. 
Lecturers: A.~Chubukov, A.~Lichtenstein, J.~M.~Tranquada, T.~Vojta.
}
\item[2011]{
XVI Training Course in the Physics of Strongly Correlated Systems, 
IIASS, Vietri~sul~Mare (Italy),
{\bf oral presentation} and {\bf proceedings} with the title 
\emph{``Fingerprints of orbital physics in resonant inelastic x-ray scattering''}. 
Lecturers: J.~A.~Mydosh, T.~Pruschke, U.~Schollw\"ock, D.~J.~Singh.
}
\item[]{
Multiband and Multiorbital Effects in Novel Materials - ICAM Summer School, Cargese (France),
{\bf poster presentation} and {\bf selected oral presentation} with the title 
\emph{``Fingerprints of orbital physics in resonant inelastic x-ray scattering''}.
Lecturers: A.~Kaminski, C.~Batista, D.~Agterberg, D.~Basov, D.~Khomskii, J.~Hoffman, J.~Lesueur, J.~Paglione, J.~P.~Brison, J.~Schmalian, J.~van~den~Brink, M.~Vojta, P.~Littlewood, S.~C.~Zhang, T.~Devereaux.
}
\item[]{
Winter school on Interfaces,
IFW-Winter School, Oberwiesental (Germany)
}
\\[2mm]
\item[]{\large \bf Teaching}
\\[-2mm]
\item[2013]{Laboratory assistant for the course \href{http://tu-dresden.de/die_tu_dresden/fakultaeten/fakultaet_mathematik_und_naturwissenschaften/fachrichtung_physik/studium/lehrveranstaltungen/praktika/grundpraktikum_iii}{\emph{``Grundpraktikum III''}} (quantum phenomena) held by Dr.\@ Andreas Schwab at the Technische Universit\"at Dresden (Germany), during Summer Semester 2013 (7 hours per week, 84 in total)}
\\[2mm]
\item[]{\large \bf Refereeing}
\\[-2mm]
\item[]{Physical Review Letters}
\\[2mm]
\item[]{\large \bf Languages}
\\[-2mm]
\item[]{
\textbf {Italian}, mother tongue
}
\item[]{
\textbf {English}, good skills in text comprehension and text production, fluently spoken
}
\item[]{
\textbf {German}, medium/basic skills in text comprehension and conversation}
\\[2mm]
\item[]{\large \bf Informatics}
\\[-2mm]
\item[]{Programming: 
Mathematica, Fortran, Scilab/Matlab, Perl, LaTeX}
\item[]{Operating systems: 
Linux (Debian, Ubuntu, Fedora), Windows}
\item[]{Scientific applications: 
Wien2k (band structure calculations), Origin, Gnuplot}
\item[]{Other applications: 
Office (LibreOffice, OpenOffice, Microsoft Office), GIMP, Inkscape}

\end{itemize}

\chapter{List of publications}\markboth{List of publications}{}
The content of this Thesis is mainly based on the following publications of the author:

\begin{itemize}

\item[{[1]}]
\emph{``Theoretical approach to resonant inelastic x-ray scattering in iron-based superconductors at the energy scale of the superconducting gap''},
Pasquale Marra, Steffen Sykora, and Jeroen van den Brink,
\href{http://dx.doi.org/10.1038/srep25386}{Sci. Rep. {\bf 6}, 25386 (2016)};

\item[{[2]}]
\emph{``Resonant inelastic x-ray scattering as a probe of the phase and excitations of the order parameter of superconductors''},
Pasquale Marra, Steffen Sykora, Krzysztof Wohlfeld, and Jeroen van den Brink,
\href{http://dx.doi.org/10.1103/PhysRevLett.110.117005}{Phys. Rev. Lett. {\bf 110}, 117005 (2013)},
\href{http://arxiv.org/abs/1212.0112}{arXiv:1212.0112};

\item[{[3]}]
\emph{``Unraveling orbital correlations with magnetic resonant inelastic x-ray scattering''}, 
Pasquale Marra, Krzysztof Wohlfeld, and Jeroen van den Brink, 
\href{http://dx.doi.org/10.1103/PhysRevLett.109.117401}{Phys. Rev. Lett. {\bf 109}, 117401 (2012)},
\href{http://arxiv.org/abs/1205.4940}{arXiv:1205.4940};

\item[{[4]}]
\emph{``Fingerprints of orbital physics in magnetic resonant inelastic x-ray scattering''},
Pasquale Marra, 
\href{http://dx.doi.org/10.1063/1.4755829}{AIP Conf. Proc. {\bf 1485}, 297 (2012)},
\href{http://arxiv.org/abs/1302.5028}{arXiv:1302.5028};

\end{itemize}

\quad\\

Other publications of the author:

\begin{itemize}

\item[{[5]}]
``\emph{Signatures of topological phase transitions in Josephson current-phase discontinuities}'', 
{Pasquale Marra}, Roberta Citro, and Alessandro Braggio, 
\href{http://arxiv.org/abs/1508.01799}{arXiv:1508.01799 (2015)};

\item[{[6]}]
``\emph{Fractional quantization of the topological charge pumping in a one-dimensional superlattice}'',
{Pasquale Marra}, Roberta Citro, and Carmine Ortix,
\href{http://dx.doi.org/10.1103/PhysRevB.91.125411}{Phys. Rev. B {\bf 91}, 125411 (2015)},
\href{http://arxiv.org/abs/1408.4457}{arXiv:1408.4457};

\item[{[7]}]
``\emph{Competition between intrinsic and extrinsic effects in the quenching of the superconducting state in Fe(Se,Te) thin films}'', 
Antonio Leo, {Pasquale Marra}, Gaia Grimaldi, Roberta Citro, Shrikant Kawale, Emilio Bellingeri, Carlo Ferdeghini, Sandro Pace, and Angela Nigro, 
\href{http://dx.doi.org/10.1103/PhysRevB.93.054503}{Phys. Rev. B {\bf 93}, 054503 (2016)},
\href{http://arxiv.org/abs/1510.05662}{arXiv:1510.05662};

\item[{[8]}]
``\emph{Stability mechanisms of high current transport in iron-chalcogenides superconducting films}'', 
Antonio Leo, Gaia Grimaldi, {Pasquale Marra}, Roberta Citro, Francesco Avitabile, Anita Guarino, Emilio Bellingeri, Shrikant Kawale, Carlo Ferdeghini, Angela Nigro, and Sandro Pace, \href{http://dx.doi.org/10.1109/TASC.2016.2542247}{IEEE Trans. Appl. Supercond. {\bf 26}, 8001104 (2016)},
\href{http://arxiv.org/abs/1603.05980}{arXiv:1603.05980};

\item[{[9]}]
``\emph{Paraconductivity of the K-doped SrFe$_2$As$_2$ superconductor}'',
{Pasquale Marra}, Angela Nigro, Zheng Li, Gen-Fu Chen, Nan-Lin Wang, Jian-Lin Luo, and Canio Noce,
\href{http://iopscience.iop.org/1367-2630/14/4/043001/}{New J. Phys. {\bf 14}, 043001 (2012)},
\href{http://arxiv.org/abs/1203.6592}{arXiv:1203.6592};

\end{itemize}

\chapter{Acknowledgments}\markboth{Acknowledgments}{}
I would like to thank Jeroen van den Brink, Steffen Sykora, and Krzysztof Wohlfeld for the precious scientific support in the realization of this thesis. 
I would also like to thank Valentina Bisogni, Claudio Mazzoli, Luuk Ament, Markus Gr\"uninger, Luis M\"ader, Marco Moretti-Sala, Hsiao-Yu Huang, Nikolay Bogdanov, Mario Cuoco, Angela Nigro, Berndt B\"uchner, and Canio Noce for fruitful discussions, ideas, and suggestions. 

Winters spent here would have been much colder, and darker, if I had not been so lucky to meet Simone and Maria. 

Ich danke meinen Spielkumpeln Edi und Josh daf\"ur, dass ich an ihrem Spielplatz sein kann.
Vielen dank auch an Frau Simone die Erste, f\"ur ihre freundliche Unterst\"utzung, Geduld und Gastfreundschaft.

E ringrazio la mia famiglia, per il loro continuo supporto. 
E per i pacchi.


\cleardoublepage
\phantomsection
\addcontentsline{toc}{chapter}{Bibliography}




\end{document}